\documentclass[12pt,twoside,openright]{book}
\pdfoutput=1
\usepackage{graphicx} 
\usepackage{amsfonts, amsmath, amssymb, bm, setspace}
\usepackage{url}
\usepackage{becthesis}
\usepackage{hyperref}
\newcommand{\twobytwo}[4]{\ensuremath{ \left(\begin{array}{cc} #1 & #2 \\ #3 & #4 \end{array}\right)       }}
\newcommand{\reff}[1]{(\ref{#1})}
\newcommand{\x}{\mathbf{x}}
\begin{document}
%%%%%%%%%%%%%%%%%%%%%%%%%%%%%%%%%%%%%%%%%%%%%%%%%%%%%%%%%%%%%%%%%
\frontmatter

%%%%%%%%%%%%%%%%%%%%%%%%%%%%%%%%%%%%%%%%%%%%%%%%%%%%%%%%%%%%%%%%%
\begin{titlepage}
\thispagestyle{empty}
\begin{center}
    \hrule
    \vspace{2pt}
    \hrule
    \vspace{1cm}
    {\huge \sc Dynamics of Quantum Vortices } \\
    \vspace{0.5cm}
    {\huge \sc at Finite Temperature}
    \vspace{1cm}
    \hrule
    \vspace{2pt}
    \hrule

    \vspace{2cm}

    \textbf{\LARGE Tod Martin Wright} \\

    \vspace{5cm}
    {\large A thesis submitted for the degree of} \\
    \vspace{0.3cm}
    {\large Doctor of Philosophy} \\
    \vspace{0.3cm}
    {\large at the University of Otago,} \\
    \vspace{0.3cm}
    {\large Dunedin, New Zealand.} 

    \vspace{2cm}

    {\centering
        \includegraphics{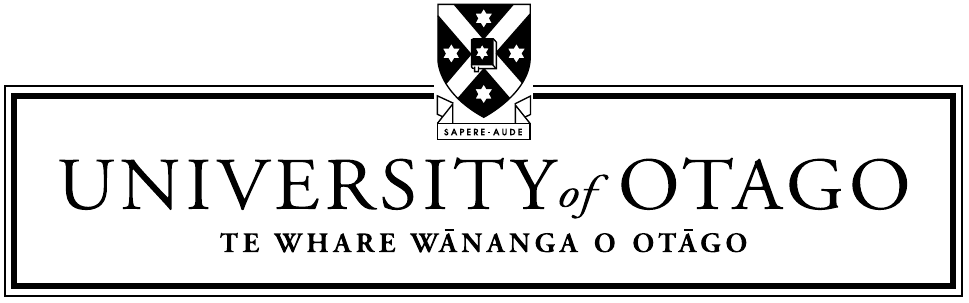}}

    \vspace{1cm}

    {\large March 2010}
\end{center}
\end{titlepage}

%%%%%%%%%%%%%%%%%%%%%%%%%%%%%%%%%%%%%%%%%%%%%%%%%%%%%%%%%%%%%%%%%
\newpage
\thispagestyle{plain}
\vspace*{10.25cm}
{\noindent}Copyright \copyright~2010 Tod Martin Wright \\
All rights reserved.\\
\\
\\
This thesis is based on work published in references~\cite{Wright08}, \cite{Wright09a}, \cite{Wright10a} and \cite{Wright10b}, 
copyright \copyright\ 2008, 2009, 2010, and 2010 the American Physical Society, respectively. \\
\\
Figure~\ref{fig:quad_quad} has been adapted with permission from reference~\cite{Blakie08}, copyright \copyright\ 2008 {\mbox Taylor~\&~Francis.}  \\
\vspace*{2.5cm}
\subsubsection*{Version history}
Original version submitted for examination: March 26, 2010.\\
Revised version post-examination: October 26, 2010.\\
Version submitted to arXiv: November 26, 2010.

%%%%%%%%%%%%%%%%%%%%%%%%%%%%%%%%%%%%%%%%%%%%%%%%%%%%%%%%%%%%%%%%%
\newpage
\thispagestyle{plain}
\section*{Abstract}
We perform investigations into the behaviour of finite-temperature degenerate Bose gases using a classical-field formalism.  Applying the projected Gross-Pitaevskii equation approach to the description of equilibrium and nonequilibrium regimes of the degenerate Bose gas, we consider the interpretation of the classical-field trajectories, and their relation to mean-field descriptions of finite-temperature Bose condensates.  We show that the coherence of the field can be characterised by its temporal correlations, and discuss how the symmetry-broken averages familiar from mean-field theories emerge from the classical-field trajectories.

\vspace{0.5\baselineskip}

The major focus of this thesis is the dynamics of quantum vortices in finite-temperature Bose fields.  We show that a finite-temperature condensate containing a precessing vortex in a cylindrically symmetric trap can be realised as an ergodic equilibrium of the classical-field theory.  We demonstrate the identification of the rotationally symmetry-broken condensate orbital from the short-time correlations of the field, and show how the thermal core-filling of the vortex emerges from the classical-field fluctuations.  We then consider the nonequilibrium dynamics that result when such a precessing-vortex configuration is subjected to a static trap anisotropy which arrests its rotation.  We decompose the nonequilibrium field into condensed and noncondensed components, and observe novel coupled relaxation dynamics of the two components.

\vspace{0.5\baselineskip}

Finally, we consider the nucleation of vortices in an initially zero-temperature quasi-two-dimensional condensate stirred by a rotating trap anisotropy.  We show that vacuum fluctuations in the initial state provide an irreducible mechanism for breaking the initial symmetries of the condensate and seeding the subsequent dynamical instability.   A rotating thermal component of the field quickly develops and drives the growth of oscillations of the condensate surface leading to the nucleation of vortices.  We study the relaxation and rotational equilibration of the initially turbulent collection of vortices, and monitor the effective thermodynamic parameters of the field during the relaxation process.  We show that the equilibrium temperature of the field is well predicted by simple arguments based on the conservation of energy, and that thermal fluctuations of the field prevent the vortices from settling into a rigid crystalline lattice in this reduced dimensionality.   We find that true condensation in the field is completely destroyed by the disordered motion of vortices, but show that the temporal correlations of the field distinguish the quasi-coherent \emph{vortex liquid} phase in the trap centre from the truly thermal material in its periphery.

%%%%%%%%%%%%%%%%%%%%%%%%%%%%%%%%%%%%%%%%%%%%%%%%%%%%%%%%%%%%%%%%%
\cleardoublepage
\thispagestyle{plain}
\section*{Acknowledgements}\enlargethispage{-\baselineskip}
I would like to express my gratitude to my supervisor, Professor Rob Ballagh.  Rob is an enthusiastic physicist, a dedicated mentor, the consummate professional, and above all a genuinely nice guy.  I have benefited greatly from his wisdom, guidance, and support over the last few years.  Despite his ever-growing workload, Rob has always found time to discuss my research, offer professional advice, and even just chat, for which I am very grateful.  I thank him for everything he has taught me, and everything he allowed me to learn for myself.  

\vspace{0.5\baselineskip}

My co-supervisor Assoc.\ Prof.\ P. Blair Blakie has always made himself available to give advice and help where he can, for which I am also very grateful.

\vspace{0.5\baselineskip}

I would particularly like to thank Dr.~Ashton Bradley, not only for giving me the code employed in much of the research of this thesis, but also for his input into and support of my research, initially from a distance, but with increased patience and dedication after his move to Otago.  I would also like to thank Prof.\ Crispin Gardiner for his input into my work, his encouragement, and his patronage.   

\vspace{0.5\baselineskip}

I have learned a lot in my brief acquaintance with Assoc.\ Prof.~Colin Fox.  Colin has a staggering knowledge and understanding of a wide range of matters physical, mathematical, and statistical.  I thank him for his assistance with mathematics, his miscellaneous wisdoms and insights, and for broadening my perspective.  I would also like to thank Terry Cole and (more recently) Peter Simpson for their help with computing, and for everything I have learned from them.   

\vspace{0.5\baselineskip}

\begin{sloppypar}Special thanks are due to Assoc.\ Prof.~Lars Bojer Madsen, Prof.~Klaus M\o lmer, Dr.~Nicolai Nygaard, Dr.~Katherine Challis, and others at Aarhus University, for their hospitality during my brief visit to Denmark.   I'm also grateful to (among others) \mbox{Dr.~Andrew} Sykes, Assoc.\ Prof.~Matthew Davis, Dr.~Thomas Hanna, Dr.~Simon \mbox{Gardiner}, and Dr.~Robin Scott for being familar faces at conferences here and abroad, and for taking the time to talk physics with me.\end{sloppypar}

\vspace{0.5\baselineskip}

I would like to thank all the other students that have shared the postgraduate experience with me, and in particular Bryan Wild, Em\'ese T\'oth, and Dr.~Alice Bezett, who have been supportive officemates throughout the majority of my time in room 524b.  Sam Lowrey deserves special mention for frequently pulling me away from my work to get some much-needed fresh air and perspective.  I am grateful also to my friends outside physics, particularly to Luke, Morgan, and Anna, for regularly taking me away from it all, and for understanding when this hasn't been feasible. 

\vspace{0.5\baselineskip}

I am grateful to my sisters Milvia and Layla, for their support and encouragement.  Finally, I would like to thank my parents, for their belief in me, their support, and their encouragement, and in particular my father, for sparking my interest in science at an early age.

\begin{center}
    \emph{ This work was supported by a University of Otago Postgraduate Scholarship; the Marsden Fund of New Zealand under Contract No. UOO509; and the New Zealand Foundation for Research, Science and Technology under Contract~No.~NERF-UOOX0703.}
\end{center}
\thispagestyle{plain}

%%%%%%%%%%%%%%%%%%%%%%%%%%%%%%%%%%%%%%%%%%%%%%%%%%%%%%%%%%%%%%%%%
\fancyhead[RO]{\bfseries\nouppercase{\rightmark}}

\tableofcontents

%%%%%%%%%%%%%%%%%%%%%%%%%%%%%%%%%%%%%%%%%%%%%%%%%%%%%%%%%%%%%%%%%
\mainmatter

\fancyhead[RO]{\nouppercase{\rightmark}}

\chapter{Introduction}
\label{chap:intro}
%%%%%%%%%%%%%%%%%%%%%%%%%%%%%%%%%%%%%%%%%%%%%%%%%%%%%%%%%%%%%%%%%%%%%%%%%%%%%%%%%%%%%%%%%%%%%%%%%%%%%%%%%%%%%%%%%%%%%%%%%%%%%%%%%%%%
%%%%%%%%%%%%%%%%%%%%%%%%%%%%%%%%%%%%%%%%%%%%%%%%%%%%%%%%%%%%%%%%%%%%%%%%%%%%%%%%%%%%%%%%%%%%%%%%%%%%%%%%%%%%%%%%%%%%%%%%%%%%%%%%%%%%
\section{Ultracold atoms and Bose-Einstein condensation}
Since the first experimental achievement of \emph{Bose-Einstein condensation} in dilute alkali gases in 1995 \cite{Anderson95,Davis95,Bradley95}, research into Bose condensates and other quantum degenerate phases of matter in cold gaseous samples of atoms has grown explosively.  The field of \emph{ultracold atomic physics} has become a unique nexus between researchers from a diverse array of subfields of modern physics, ranging from atomic physics, laser physics and quantum optics, to nuclear and condensed-matter physics, quantum information, fluid dynamics and even astrophysics and cosmology.  In recent years the confinement and manipulation of ultracold atoms has expanded to encompass the study of many different atomic species (both bosonic and \emph{fermionic} \cite{DeMarco99}), in a range of geometries \cite{Esteve06,Morsch06,Hadzibabic10}, with a variety of different interatomic interactions \cite{Chin10,Lahaye09}.  All of these studies share the common feature of investigating interacting many-body systems and their behaviour in a clean and precisely controlled environment \cite{Bloch08}.

Bose-Einstein condensates were historically the first ultracold atomic systems realised experimentally, and these systems are still intensively studied in laboratories throughout the world.  A Bose condensate forms when a macroscopic number of identical \emph{bosonic} atoms congregate into a single spatial mode.  The remarkable aspect of the phase transition to the Bose-condensed state is that it is driven by the quantum statistics of the particles, rather than the interactions between them, and may thus occur even in a noninteracting sample of bosonic atoms.  The phenomenon of Bose-Einstein condensation of massive particles was predicted in 1925 by Einstein, and London conjectured in 1938 that Bose condensation was responsible for the remarkable properties of the then recently discovered superfluid helium.  However, the advent of \emph{laser cooling} allowed researchers for the first time to observe the Bose condensation phenomenon in isolation, with the unmatched precision of control and observation developed over many decades in atomic and laser physics.  

As they are essentially composed of many atoms residing in a single spatial mode, Bose condensates provide an amplification of the underlying quantum mechanical physics of microscopic systems to an observable scale \cite{Leggett01}.  Condensates are characterised by their long-range quantum coherence, and form a close analogue of coherent laser light, despite being composed of massive, interacting particles.  The quantum coherence properties of atomic Bose condensates have been demonstrated dramatically in experiments of matter-wave interference \cite{Andrews97}, mixing \cite{Deng99}, and amplification \cite{Inouye99}.

An important property of Bose condensates is their \emph{superfluidity}: like superfluid helium, they can support persistent currents and vortices.  However, in contrast to the superfluid \emph{liquid} helium, in which only $\sim 10\%$ of the atoms form a Bose condensate due to the strong depleting effect of interatomic interactions \cite{Penrose56}, the weakly interacting nature of the dilute gases allows the formation of essentially pure condensates.  Moreover, the finite-sized and \emph{inhomogeneous} nature of these gaseous superfluids leads to new features, and behaviour quite different to the more traditional, essentially uniform superfluids \cite{Giorgini97}.  There has thus been intensive study of the thermodynamics of the trapped Bose gases, and of the critical regime associated with the Bose-condensation phase transition \cite{Ensher96,Gerbier04}.  Furthermore, additional phases of degenerate Bose gases have been studied in novel trapping geometries such as \emph{optical lattices} \cite{Morsch06}, and in reduced dimensionalities, where new quasi-coherent phases of the gas appear \cite{Stock05,Esteve06,Clade09}. 

The precision of control in atomic physics experiments also allows for the thorough investigation of the \emph{dynamic} properties of dilute condensates.  Experiments have considered not only the collective response of condensates to perturbing potentials \cite{Jin97,Marago00}, but also the formation and dynamics of topological defects such as solitons \cite{Burger99,Khaykovich02} and vortices \cite{Matthews99,Madison00} in these quantum fluids.  Experiments have even extended to consider the strongly \emph{nonequilibrium} dynamics of these systems, including turbulent regimes of the atomic field \cite{Henn09} and nonequilibrium phase transitions \cite{Weiler08}.

In addition to the precision with which condensates can be observed and manipulated experimentally, their \emph{weakly} interacting nature makes them an attractive system for theoretical investigation.  In comparison to the enormous complexity of dense liquids such as superfluid helium, these dilute gases are amenable to \emph{tractable} theoretical approaches, including essentially \emph{ab initio} numerical calculations and, in many cases, analytical calculations.  In particular, some of the theoretical tools \cite{Bogoliubov47} which could only give a qualitative level of insight into the strongly interacting superfluid helium found new life as practically \emph{exact} descriptions of the dilute-gas condensates in many scenarios.  The dilute condensates thus offer us an opportunity to understand superfluidity in terms of \emph{microscopic} models of the relevant physics.  In addition to the traditional tools of superfluidity, condensed matter and nuclear physics, the close analogy between Bose condensates and coherent light fields has lead to the broad application of methods from \emph{quantum optics} and laser physics to the description of condensates \cite{Blakie08}.  In general, however, many questions remain as to how to properly describe the dynamics of Bose-Einstein condensates in general situations, in which the atomic cloud possesses a significant thermal component, and is possibly far from equilibrium \cite{Proukakis08}.  Moreover, heavy numerical calculations are usually required to understand the properties and dynamics of Bose condensates in detail.  These numerical studies typically involve the solution of \emph{nonlinear} equations, and so small inaccuracies and errors in their implementation can easily lead to completely unphysical results.  It is thus of great importance not only to develop accurate theoretical models of Bose condensation, but also to construct accurate \emph{numerical} models and methodologies for the simulation of condensates and their dynamics.  
%%%%%%%%%%%%%%%%%%%%%%%%%%%%%%%%%%%%%%%%%%%%%%%%%%%%%%%%%%%%%%%%%%%%%%%%%%%%%%%%%%%%%%%%%%%%%%%%%%%%%%%%%%%%%%%%%%%%%%%%%%%%%%%%%%%%
\section{Quantum vortices}\label{sec:intro_vortices}
The `smoking gun' of the superfluid transition in quantum fluids is the appearance of quantum \emph{vortices}: tiny whirlpools in which the superfluid circulates about a `core' where the superfluid density vanishes \cite{Donnelly91}.  These vortex structures are topological defects, which reflect the inherent quantum-mechanical nature of a superfluid: due to the single-valued nature of the wave-function phase which determines the superfluid flow, the phase around a vortex must change by an exact multiple of $2\pi$, and thus the rotation of the superfluid is fundamentally \emph{quantised}.
Singly charged vortices (with phase circulation $\pm2\pi$) are topologically stable, in the sense that they must encounter another such defect (of opposite rotation) or a boundary of the fluid in order to decay.  Vortices with higher charges are energetically unstable to decay by fissioning into two or more vortices of lesser topological `charge' \cite{Donnelly91,Pu99}.   Under rotation, superfluid helium mimics the velocity profile of a rotating rigid body by admitting an ensemble of many such singly charged vortices, which, due to their mutual repulsion \cite{Sonin87}, form a regular crystalline \emph{lattice} at equilibrium.  

The physics of superfluid vortices are of great general interest, as analogous objects occur in a variety of physical systems including type-II superconductors \cite{Blatter94}.  Above a critical magnetic imposed magnetic field strength, such superconductors admit thin filaments of magnetic flux, about which the charged `condensate' of Cooper-paired electrons forms vortices.  Vortex structures are also thought to occur in the dense nucleon-superfluid phases of rotating neutron stars \cite{Baym69}.  In both of these systems the behaviour, and in particular the \emph{dynamics}, of vortices are crucial to understanding the physics of the quantum fluids involved.  The nonequilibrium dynamics of \emph{turbulence} in superfluids are of great interest due to the role of quantised vortices in such regimes.  Indeed, the turbulence of \emph{classical} fluids is significantly complicated by the continuous nature of vorticity in the system.  Studies of so-called \emph{quantum} or \emph{superfluid} turbulence \cite{Barenghi01} are thus actively pursued in the hope that the study of turbulence in systems with quantised vorticity will yield insights into the classical turbulence problem.

The realisation of Bose-Einstein condensates and other degenerate superfluid phases of matter in atomic-physics experiments has provided new opportunities to study the intrinsic phenomenology of quantum vortices.  The first vortices in Bose condensates were formed by coherent interconversion between two hyperfine components of a condensed atomic cloud, creating a structure in which one component circulates around a non-circulating core composed of the other component \cite{Matthews99}.  By selectively removing the core-filling component with resonant light pressure, an isolated vortex was obtained \cite{Anderson00}.  In further experiments, inspired by the \emph{rotating-bucket} experiments of superfluid helium, atomic gases were evaporatively cooled to condensation in a rotating anisotropic trapping potential \cite{Madison00}, producing both single vortices and small vortex arrays.  This was followed by investigations of the dynamical stirring of a preformed, vortex-free condensate with a rotating potential anisotropy \cite{Madison00b,Abo-Shaeer01,Hodby02}.  In this approach an initially nonrotating condensate could be transformed by a strongly nonequilibrium process into a rotating vortex lattice.  Other experiments considered the formation of a vortex lattice by cooling a rotating thermal cloud through the Bose-condensation transition in a \emph{cylindrically symmetric} trap, in order to observe the \emph{intrinsic} dynamics of vortex nucleation driven by the rotation of the normal fluid \cite{Haljan01a}.  The dynamics of vortex-lattice formation in this case are very rich, as the growth of the condensate and the formation of the vortex lattice are essentially intertwined nonequilibrium phase transitions \cite{Bradley08}.  Vortices have also been imprinted in condensates using topological (Berry) phases \cite{Leanhardt02}, and by `sweeping' a laser light potential through the condensed atomic cloud \cite{Inouye01}. Another scheme for forming a vortex involves the transfer of orbital angular momentum from a laser beam to the Bose condensate \cite{Andersen06}.  Vortices can form spontaneously in the merging of multiple Bose condensates, due to the essentially random relative phase difference that develops between the condensates \cite{Scherer07}, and can also be `trapped' in the nascent condensate during its rapid growth from a (nonrotating) quenched thermal cloud \cite{Weiler08}.  These formation processes are closely related to conjectured mechanisms of phase-defect trapping in early-universe cosmology \cite{Kibble76,Zurek85}.  

Experiments have also investigated the role of vortices in the two-dimensional superfluid transition \cite{Stock05}, and the effect of artificial pinning potentials on vortices \cite{Tung06}.  The realisation of a \emph{strongly interacting} superfluid in a degenerate \emph{Fermi} gas was conclusively demonstrated by the observation of vortices in this system \cite{Zwierlein05}.  Rotating Bose condensates containing vortices have also been proposed as a platform for investigating fundamental many-body physics, including aspects of symmetry-breaking in many-body systems \cite{Dagnino09}, and the potential realisation of \emph{strongly correlated} states of vortex matter, such as bosonic analogues of the fractional-quantum-Hall states in rapidly rotating condensates \cite{Viefers08}.  
The difficulty in attaining the extreme experimental conditions necessary to observe such states has spurred researchers to consider alternative scenarios involving (e.g.) \emph{artificial gauge fields} \cite{Lin09} in order to obtain high vortex densities in the hope of observing the \emph{quantum melting} of dense vortex lattices.

The weakly interacting nature of dilute atomic gases offers us hope of obtaining \emph{quantitative} theoretical descriptions of vortex dynamics in atomic Bose condensates.  However, the results of experiments involving strongly nonequilibrium dynamics of vortex formation and motion have often proved difficult to describe theoretically.  Conspicuously, the standard mean-field approximations for Bose condensates are inapplicable in the turbulent regimes of Bose-field dynamics arising in vortex-formation scenarios, in which a clear distinction between condensed and noncondensed material cannot be made \cite{Lobo04}.  It is therefore important to develop dynamical theories of the degenerate Bose gas which can accurately describe the nonequilibrium dynamics of vortices, their interactions with the thermal component of the field, and turbulent behaviour in the degenerate Bose gas.
%%%%%%%%%%%%%%%%%%%%%%%%%%%%%%%%%%%%%%%%%%%%%%%%%%%%%%%%%%%%%%%%%%%%%%%%%%%%%%%%%%%%%%%%%%%%%%%%%%%%%%%%%%%%%%%%%%%%%%%%%%%%%%%%%%%%
\section{This work}
%%%%%%%%%%%%%%%%%%%%%%%%%%%%%%%%%%%%%%%%%%%%%%%%%%%%%%%%%%%%%%%%%%%%%%%%%%%%%%%%%%%%%%%%
\subsection{Thesis overview}
In chapter~\ref{chap:theory} we give the general theoretical background for this thesis.  We begin by briefly reviewing the description of quantum many-body systems in first and second quantisation.  We introduce the concept of Bose-Einstein condensation, first for noninteracting systems and then for interacting systems, and derive the Gross-Pitaevskii equation for the condensate wave function and the Bogoliubov theory for its excitations.  We discuss the notion of superfluidity, and its relation to Bose-Einstein condensation.  We then consider the quantum vortices which are a characteristic feature of superfluidity, and discuss their response to imposed forces, including the frictional effects of a normal (non-superfluid) component of the fluid.  

In chapter~\ref{chap:cfield} we introduce the classical-field methodology we employ throughout this thesis.  We begin with a heuristic explanation of the fundamental classical-field approximation, before deriving the \emph{projected Gross-Pitaevskii equation} for the evolution of classical-field trajectories.  We describe how the \emph{ergodicity} of this nonlinear equation of motion for the classical atomic field gives rise to a practical means for evaluating correlations in the field, and discuss the emergence of thermodynamic behaviour from the fundamentally dynamical field equation.  We conclude this chapter with a brief review of the concept of quasiprobability distributions and phase-space methods.  Focussing in particular on the \emph{truncated-Wigner} formalism, we relate the equation of motion obtained in the truncated-Wigner approach to the projected Gross-Pitaevskii equation, and discuss the implications of the truncated-Wigner formulation for our classical-field investigations in this thesis.

In chapter~\ref{chap:numerics} we describe the techniques we use for the numerical integration of the projected Gross-Pitaevskii equation, and quantify the performance of the Gauss-Laguerre-quadrature algorithm we use for the simulation of rotating systems and vortices in this thesis.  We then contrast the harmonic-basis projected Gross-Pitaevskii approach we adopt in this thesis to other common discretisations of the continuum Gross-Pitaevskii equation.  

In chapter~\ref{chap:anomalous} we consider the temporal correlations that emerge in the classical-field description of a harmonically trapped Bose gas, and show how the condensate can be identified by its quasi-uniform phase rotation.  We show that this provides an analogue of the symmetry-breaking assumption commonly employed in traditional (mean-field) theories of Bose condensation, and we identify the condensate as the anomalous first moment of the classical field.  We then demonstrate the generality of this prescription by calculating the pair matrix which describes anomalous correlations in the thermal field, and calculate the anomalous thermal density of the field, which we find to have temperature-dependent behaviour consistent with that predicted by mean-field theories.

In chapter~\ref{chap:precess} we consider the equilibrium precession of a single vortex in a finite-temperature Bose-Einstein condensate.  We show that a precessing-vortex configuration can be obtained as an ergodic equilibrium of the classical field with finite conserved angular momentum.  The symmetry-breaking nature of this state requires us to go beyond formal ergodic averaging in order to characterise condensation in the field, and we show that an analysis of the short-time fluctuation statistics of the field reveals the symmetry-broken condensate orbital, the thermal filling of the vortex core, and the Goldstone mode associated with the broken rotational symmetry.

In chapter~\ref{chap:arrest} we consider a nonequilibrium scenario of vortex arrest.  We form an equilibrium finite-temperature precessing-vortex configuration in a cylindrically symmetric trap, and then introduce a static trap anisotropy.  The anisotropy leads to the loss of angular momentum from the thermal cloud and, consequently, to the decay of the vortex.  Our approach provides an intrinsic description of the coupled relaxation dynamics of the nonequilibrium condensate and thermal cloud.  By considering the short-time correlations of the nonequilibrium field, we separate it into condensed and noncondensed components, and identify novel dynamics of angular-momentum exchange between the two components during the arrest.  We also discuss qualitatively distinct regimes of relaxation that we obtain by varying the strength of the arresting trap anisotropy.

The remainder of this thesis is concerned with the nucleation of quantum vortices by the stirring of a condensate with a rotating trap anisotropy.  We begin in chapter~\ref{chap:stir_background} by briefly reviewing the prior literature on this topic.  We discuss the experimental approaches to condensate stirring taken by research groups at \'Ecole Normale Sup\'eriure, MIT and Oxford.  We then discuss the various theoretical and numerical approaches previously taken to describe the condensate-stirring process, and the more surprising features of the experimental results.  We conclude this chapter by outlining the remaining open questions and controversy as to the mechanisms responsible for vortex nucleation and damping in these experiments.

In chapter~\ref{chap:stirring} we present our classical-field approach to the stirring problem.   We consider the stirring of a condensate initially at zero temperature in a highly oblate (quasi-two-dimensional) trapping potential.  We add a representation of vacuum noise to the initial field configuration according to the truncated-Wigner prescription.  This noise seeds the spontaneous processes responsible for the ejection of material from the unstable condensate, and we show that the ejected material quickly forms a rotating thermal cloud which drives the nucleation of vortices at the condensate surface.  We quantify the development of the thermodynamic parameters of the field during its evolution, and show that the final temperature attained is well predicted by simple energy-conservation arguments.  We find that the equilibrium state of the atomic field in this quasi-2D scenario is a disordered vortex liquid, in which vortices are prohibited from crystallising into a regular lattice by thermal fluctuations.  We find that condensation in the field is destroyed by the vortex motion, but show that the temporal correlations in the field allow us to distinguish the quasi-coherent structure at the centre of the trap from the truly thermal material in its periphery. 

Finally, a summary and conclusion is given in chapter~\ref{chap:conclusion}, in which we outline possible directions for future research.
%%%%%%%%%%%%%%%%%%%%%%%%%%%%%%%%%%%%%%%%%%%%%%%%%%%%%%%%%%%%%%%%%%%%%%%%%%%%%%%%%%%%%%%%
\subsection{Peer-reviewed publications}
Much of the work contained in this thesis has been published in refereed journals.  The work on temporal coherence and anomalous averages in classical-field simulations presented in chapter~\ref{chap:anomalous} has appeared in \emph{Physical Review A} \cite{Wright10b}.  The work on the equilibrium precession of a vortex and on the arrest of vortex rotation presented in chapters~\ref{chap:precess}~and~\ref{chap:arrest} has also been published in \emph{Physical Review A} (references \cite{Wright09a} and \cite{Wright10a} respectively).  The discussion and simulations of the condensate-stirring dynamics presented in chapters~\ref{chap:stir_background}~and~\ref{chap:stirring} constitute an expanded and updated account of work also published in \emph{Physical Review A} \cite{Wright08}.   
%%%%%%%%%%%%%%%%%%%%%%%%%%%%%%%%%%%%%%%%%%%%%%%%%%%%%%%%%%%%%%%%%%%%%%%%%%%%%%%%%%%%%%%%%%%%%%%%%%%%%%%%%%%%%%%%%%%%%%%%%%%%%%%%%%%%
%%%%%%%%%%%%%%%%%%%%%%%%%%%%%%%%%%%%%%%%%%%%%%%%%%%%%%%%%%%%%%%%%%%%%%%%%%%%%%%%%%%%%%%%%%%%%%%%%%%%%%%%%%%%%%%%%%%%%%%%%%%%%%%%%%%%

\chapter{Background theory}
\label{chap:theory}
%%%%%%%%%%%%%%%%%%%%%%%%%%%%%%%%%%%%%%%%%%%%%%%%%%%%%%%%%%%%%%%%%%%%%%%%%%%%%%%%%%%%%%%%%%%%%%%%%%%%%%%%%%%%%%%%%%%%%%%%%%%%%%%%%%%%
%%%%%%%%%%%%%%%%%%%%%%%%%%%%%%%%%%%%%%%%%%%%%%%%%%%%%%%%%%%%%%%%%%%%%%%%%%%%%%%%%%%%%%%%%%%%%%%%%%%%%%%%%%%%%%%%%%%%%%%%%%%%%%%%%%%%
In this chapter we will review the general theoretical background for this thesis.  The key results of many-body theory, including the formalism of second quantisation, are reviewed in many standard texts on the subject, for example references \cite{Fetter71a,Abrikosov63,Blaizot86}.  Here, we give a brief review of these topics, to illustrate the fundamental complexity of many-body systems such as the degenerate gases we seek to describe in this thesis, and to give context to the \emph{classical-field} approximation to the many-body theory we introduce in chapter~\ref{chap:cfield}.  We then proceed to review some standard definitions and theoretical tools for the study of Bose-Einstein condensation.  Many excellent reviews of these topics exist in the literature (see references~\cite{Leggett01,Dalfovo99,Ketterle99,Fetter99,Pitaevskii03,Pethick02,Castin01}), and hence we will review only a small selection of topics relevant for our discussions later in this thesis.  Finally, we will discuss aspects of superfluidity and vortices relevant to our investigations in this thesis, primarily following material presented in references~\cite{Fetter99,Donnelly91,Fetter01}.
%%%%%%%%%%%%%%%%%%%%%%%%%%%%%%%%%%%%%%%%%%%%%%%%%%%%%%%%%%%%%%%%%%%%%%%%%%%%%%%%%%%%%%%%%%%%%%%%%%%%%%%%%%%%%%%%%%%%%%%%%%%%%%%%%%%%
\section{Identical-particle ensembles}
In the study of \emph{dilute} atomic gases, the atoms can usually be considered as an ensemble of $N$ identical structureless particles which interact via a pair-wise (binary) potential $V(\mathbf{x}_1 - \mathbf{x}_2)$.  In the fundamental or \emph{first-quantised} description of such an assembly of particles in quantum mechanics, the dynamics of the ensemble are governed by the many-body Hamiltonian
\begin{equation}\label{eq:back_1stQH}
	H_\mathrm{MB} = \sum_i H_\mathrm{sp}(\mathbf{x}_i) + \sum_{i<j} V(\mathbf{x}_i - \mathbf{x}_j),
\end{equation}
where the single-particle Hamiltonian
\begin{equation}\label{eq:back_Hsp}
	H_\mathrm{sp}(\mathbf{x}) = \frac{-\hbar^2\nabla^2}{2m} + V_\mathrm{ext}(\mathbf{x}),
\end{equation}
includes the kinetic energy of each atom and the effect of an \emph{external} potential $V_\mathrm{ext}(\mathbf{x})$.  The state of the atoms is described by a many-body state in the appropriate many-particle Hilbert space, which is formed as the tensor product of the $N$ individual single-particle Hilbert spaces \cite{Blaizot86}
\begin{equation}\label{eq:back_H_N}
	\mathcal{H}_N = \mathcal{H}_1^{(1)} \otimes \mathcal{H}_1^{(2)} \otimes \cdots \otimes \mathcal{H}_1^{(N)}.
\end{equation}
More specifically, the particular many-body state is required to exhibit a specific symmetry under permutations of the particle labels, depending on the spin of the particles involved.  Formally, this requirement can be deduced for elementary particles from the considerations of relativistic quantum mechanics \cite{Pauli40}.  The atoms we consider here are of course composite particles, but they may be treated simply as structureless particles with spin given by the total spin of their composite nucleons and electrons, provided that (as is always the case in this thesis) the energy scales of their interaction do not probe their true internal structure \cite{Ketterle99}.   For bosons (with integer spin in units of the reduced Planck's constant $\hbar$) the state must be symmetric with respect to the permutation of particle labels, and thus resides in the subspace $\mathcal{H}_N^S$ of $\mathcal{H}_N$ spanned by completely symmetric states, whereas in the case of Fermions (with half-integer spin) the state must be antisymmetric with respect to the permutation of labels (i.e., $\langle\cdots \x_j,\x_i,\cdots|\psi\rangle = - \langle \cdots \x_i,\x_j,\cdots|\psi\rangle$), and is thus a vector in the subspace $\mathcal{H}_N^A$ spanned by completely antisymmetric states \cite{Blaizot86}.
%%%%%%%%%%%%%%%%%%%%%%%%%%%%%%%%%%%%%%%%%%%%%%%%%%%%%%%%%%%%%%%%%%%%%%%%%%%%%%%%%%%%%%%%
\subsection{Second quantisation}\label{subsec:back_2nd_Q}
We now discuss an alternative representation of the interacting particle ensemble \cite{Fetter71a} which is known as the \emph{second-quantised} representation of the many-body system.  This formulation promotes the discussion to that of a \emph{quantum field theory}.  The concept of a quantum field theory originates in relativistic quantum mechanics, in which it is found that the quantum theory is \emph{necessarily} many-body in nature, as a consequence of mass-energy equivalence \cite{Weinberg95}.  In the ultracold scenarios we consider in this thesis, the rest-mass of atoms dwarfs all other energy scales in the system, and atom-number changing processes do not occur.  
However, in studies of non-relativistic many-body systems, the second-quantised formalism has great advantages over the first-quantised formalism.  First, it provides a convenient encapsulation of the wave-function symmetry constraints, which rapidly become unwieldy in describing ensembles of increasing particle number.  
Second, it is a convenient formulation in which to identify approximations to the (generally intractable) complete many-body description.  Although the most widespread approximations schemes are based on the techniques of diagrammatic perturbation theory \cite{Fetter71a,Abrikosov63}, in this thesis we utilise a different approximation scheme:
the second-quantised formulation of the many-body system allows us to identify a \emph{classical limit}.  Indeed in the case of bosons, the appropriate quantum field theory can be obtained by building quantum fluctuations into the dynamics of a classical field which evolves in a \emph{single-particle} space.
Significant insight into the dynamics of the quantum many-body system can thus, in certain regimes, be obtained by studying this dramatically simpler classical field, as we discuss in chapter~\ref{chap:cfield}.

To formulate the second-quantised theory, we introduce the \emph{field operator} $\hat{\Psi}(\mathbf{x})$, which obeys the relations
\begin{equation}\label{eq:back_bose_commutator}
		\left[\hat{\Psi}(\x),\hat{\Psi}(\x')\right]_\pm=\delta(\x-\x') ; \;\;\; \left[\hat{\Psi}(\x),\hat{\Psi}(\x')\right]_\pm=\left[\hat{\Psi}^\dagger(\x),\hat{\Psi}^\dagger(\x')\right]_\pm = 0,
\end{equation}
where $[\cdot,\cdot]_-$ denotes a commutator, and applies in the case of bosons, and $[\cdot,\cdot]_+$ denotes an anti-commutator, which applies in the case of fermions.
In this description, the operator $\hat{\Psi}(\mathbf{x})$ and its (Hermitian) conjugate are the fundamental operator-valued quantities, which act on the \emph{Fock space} formed as the union of many-body spaces with all (physical) particle numbers \cite{Blaizot86}
\begin{equation}
	\mathcal{F} = \mathcal{H}_0\oplus\mathcal{H}_1\oplus \mathcal{H}_2\oplus \cdots.
\end{equation}
It is important to note that the coordinates $\mathbf{x}$ in this formalism are not operators, and simply label the field operators.  In this formulation the (first-quantised) many-body Hamiltonian \reff{eq:back_1stQH} can be rewritten \emph{equivalently} as the second-quantised Hamiltonian
\begin{equation}\label{eq:back_2ndQH}
	\hat{H} = \int\!d\mathbf{x}\,\hat{\Psi}^\dagger(\mathbf{x})H_\mathrm{sp}\hat{\Psi}(\mathbf{x}) + \frac{1}{2} \int\!d\mathbf{x}\int\!d\mathbf{x}'\,\hat{\Psi}^\dagger(\mathbf{x})\hat{\Psi}^\dagger(\mathbf{x}') V(\mathbf{x}-\mathbf{x}') \hat{\Psi}(\mathbf{x}')\hat{\Psi}(\mathbf{x}).
\end{equation}
A general single-particle operator on the many-body space in the first-quantised treatment acts on \emph{all} the indistinguishable particles, and is thus of form
\begin{equation}\label{eq:back_one-body_operator}
		J = \sum_{i=1}^N J(\mathbf{x}_i),
\end{equation}
where $J(\mathbf{x}_i)$ acts on the $i^\mathrm{th}$ particle.  Such an operator is promoted to an operator on the Fock space by the correspondence \cite{Fetter71} 
\begin{equation}\label{eq:back_one-body_Fock_operator}
		J \rightarrow \hat{J} = \int\!d\mathbf{x}\,\hat{\Psi}^\dagger (\mathbf{x}) J(\mathbf{x}) \hat{\Psi}(\mathbf{x}),
\end{equation}
and two- (and higher) body operators in the first-quantised formalism can be promoted to operators on the Fock space by similar correspondences.
We stress that if the system is in a state $|\chi\rangle$ (on the Fock space) with \emph{definite} particle number $N$, then this formulation is \emph{equivalent} to that of the first-quantised many-body theory, with the many-body wave function given by \cite{Schweber61}
\begin{equation}
	\psi(\mathbf{x}_1,\mathbf{x}_2,\cdots,\mathbf{x}_N) = \frac{1}{\sqrt{N!}} \bigl\langle \mathrm{vac} \bigl| \hat{\Psi}(\mathbf{x}_N)\cdots\hat{\Psi}(\mathbf{x}_2)\hat{\Psi}(\mathbf{x}_1) \bigr|\chi\bigr\rangle,
\end{equation}
where $\langle \mathrm{vac} |$ is the bra corresponding to the state with no particles (the vacuum).  However, the second-quantised formalism also allows us to describe states with \emph{indefinite} particle numbers.  In the remainder of this thesis, we will restrict our attention to bosons.
%%%%%%%%%%%%%%%%%%%%%%%%%%%%%%%%%%%%%%%%%%%%%%%%%%%%%%%%%%%%%%%%%%%%%%%%%%%%%%%%%%%%%%%%%%%%%%%%%%%%%%%%%%%%%%%%%%%%%%%%%%%%%%%%%%%%
\section{Degenerate bosons}
%%%%%%%%%%%%%%%%%%%%%%%%%%%%%%%%%%%%%%%%%%%%%%%%%%%%%%%%%%%%%%%%%%%%%%%%%%%%%%%%%%%%%%%%
\subsection{Bose condensation}\label{subsec:back_bose_condensation}
%%%%%%%%%%%%%%%%%%%%%%%%%%%%%%%%%%%%%%%%%%%%%%%%%%%%%%%%%
\subsubsection{Noninteracting case}
In this thesis, we are concerned only with degenerate bosons.  In most regimes, the behaviour of identical-particle ensembles is very similar to that of distinguishable particles, and noticeable differences occur only as the system nears \emph{degeneracy}, which is to say, when (some) single-particle modes have occupation of order $\gtrsim 1$.  In such conditions, the statistics of the particles become very important.  In the case of identical bosons, a statistical \emph{clustering} effect is exhibited by the particles.  The origin of this effect can be seen already in a simple toy model \cite{Leggett01}: the distribution of two particles over two equivalent (e.g., equal energy) states.  In the case of distinguishable particles, there are four ways to distinguish the particles (two choices of state for each of the two particles), and so picking permissible system states at random we expect to find both particles in the same state $50\%$ of the time.  By contrast, in the case of identical bosons, there are only three possible two-body states (both particles in state A, both in state B, and one particle in each of A and B).  The probability of both particles being found in the same state is thus approximately $67\%$ in this case.  This is illustrative of a general tendency of identical bosons to exhibit an `attraction' on one another, which is \emph{purely statistical} in origin. 

This behaviour is of course reflected in the equilibrium distribution of identical bosons, which is most conveniently expressed in the grand canonical ensemble, where the occupation of a single-particle state $|j\rangle$ with energy $\epsilon_j$ is
\begin{equation}\label{eq:back_bose_dist}
	n(\epsilon_j) = \frac{1}{e^{\beta (\epsilon_j-\mu)} - 1},
\end{equation}
with $\mu$ the \emph{chemical potential} and $\beta = 1/k_\mathrm{B}T$ the inverse temperature.  
A crucial property of the Bose distribution (\ref{eq:back_bose_dist}) is that the excited state population $N_\mathrm{ex} = \sum_{j>0} n(\epsilon_j)$ exhibits a \emph{saturation} behaviour \cite{Castin01}.  This can be understood by the following argument: measuring energies relative to that of the ground state (i.e. setting $\epsilon_0=0$), for any temperature $T>0$ the ground state occupation becomes negative for chemical potentials $\mu>0$.  Hence we require $\mu\leq 0$.  The excited state population is therefore bounded $N_\mathrm{ex} \leq \sum_{k>0} [\mathrm{exp}(\beta\epsilon_k)-1]^{-1}\equiv N_\mathrm{ex}^{(\max)}$, which can be shown to be finite. 
Thus as atoms are added to the ensemble at fixed temperature $T$, eventually the excited states will admit no more atoms, and so the additional atoms accumulate in the ground state of the external potential in which they are confined.  In this limit the chemical potential approaches zero from below ($\mu\to0^-$), and the condensate occupation $n(\epsilon_0)$ diverges towards infinity.  More formally, the occupation of the ground state becomes an \emph{extensive} parameter of the system, while the occupations of all other modes remain intensive, and the ground state population is said to form a \emph{Bose-Einstein condensate} \cite{Pethick02,Castin01}. 

For a given number of particles $N$, we define the \emph{critical temperature} $T_\mathrm{c}$ as that temperature at which $N_\mathrm{ex}^{(\max)}=N$; below this temperature, atoms accumulate in the ground state, i.e., condensation occurs.  A standard approach to evaluating $N_\mathrm{ex}^{(\max)}$ is to make a \emph{semiclassical approximation}, replacing the sum over states with an integral over energies with a density of states function\footnote{An alternative approach is to employ a series expansion for the Bose function \cite{Castin01}, which allows one to obtain next-order corrections beyond the semiclassical limit.}.  This approach is valid for temperatures $k_\mathrm{B}T \gg \hbar\omega$.  One obtains \cite{Pethick02}
\begin{equation}
	N = N_\mathrm{ex}(T_\mathrm{c},\mu=0) = \int_0^\infty\!d\epsilon\,\frac{g(\epsilon)}{e^{\epsilon/k_\mathrm{B}T_\mathrm{c}}-1},
\end{equation}
where for the experimentally relevant case of 3D harmonic trapping, with external potential
\begin{equation}
    V_\mathrm{ext}^{\mathrm{(harm)}}(\mathbf{x})=\frac{m}{2}\left(\omega_x^2x^2+\omega_y^2y^2+\omega_z^2z^2\right),
\end{equation}
the density of states $g(\epsilon)=\epsilon^2/2(\hbar\bar{\omega})^3$, with the geometric-mean trapping frequency $\bar{\omega} = (\omega_x\omega_y\omega_z)^{1/3}$.  In this geometry we thus obtain the result
\begin{equation}
	k_\mathrm{B}T_\mathrm{c} = \frac{\hbar\bar{\omega}N^{1/3}}{[\zeta(3)]^{1/3}} \approx 0.94\hbar\bar{\omega}N^{1/3},
\end{equation}
where we have substituted an approximate numerical value for the \emph{Riemann zeta} function $\zeta(z)$ evaluated at $z=3$.  We can in fact go further and calculate the number of excited particles at temperatures below $T_\mathrm{c}$, where the chemical potential $\mu=0$.  We find immediately
\begin{equation}
	N_\mathrm{ex} = N\left(\frac{T}{T_\mathrm{c}}\right)^3,
\end{equation}
and we thus have for the condensate population
\begin{equation}
	N_0 = N\left[1 - \left(\frac{T}{T_\mathrm{c}}\right)^3\right]\!.
\end{equation}
%%%%%%%%%%%%%%%%%%%%%%%%%%%%%%%%%%%%%%%%%%%%%%%%%%%%%%%%%
\subsubsection{Interacting case}\label{subsubsec:theory_interacting}
As the `clustering' behaviour of bosons is an intrinsic quantum-statistical effect, independent of details of the atomic interaction and any external trapping potential, we expect similar condensation behaviour to occur in an interacting Bose gas.  While in the noninteracting case, energy eigenstates of the many-body system are obtained as (properly symmetrised) products of single-particle eigenstates, interactions between the atoms lead to correlations in the many-body wave function, so that the concept of single-particle energy levels is not meaningful for the interacting system.  In order to generalise the notion of an extensive occupation of the lowest-energy single-particle state to the interacting case, Penrose and Onsager \cite{Penrose56} therefore considered the \emph{one-body reduced density operator}   
\begin{equation}\label{eq:back_obd_opr}
	\hat{\rho}_1 = N \mathrm{Tr}_{2\cdots N}\left\{\hat{\rho}_N\right\},
\end{equation}
which we have written here as a partial trace of the $N$-body density matrix $\hat{\rho}_N$, which is in general a statistical mixture $\hat{\rho}_N = \sum_s p_s |\psi_N^s\rangle\langle\psi_N^s|$ of distinct $N$-particle states $|\psi_N^s\rangle$.  It is convenient to consider the one-body density operator in a coordinate-space representation
\begin{equation}\label{eq:back_obd_mtx}
	\rho_1(\mathbf{x},\mathbf{x}') = \left\langle \mathbf{x}' |\hat{\rho}_1| \mathbf{x} \right\rangle \equiv \left\langle \hat{\Psi}^\dagger(\mathbf{x})\hat{\Psi}(\mathbf{x}') \right\rangle.
\end{equation}
The one-body density matrix encapsulates all single-particle correlations in the system.  For example, we can see from the definition of the extended general one-body operator (\ref{eq:back_one-body_Fock_operator}) and using the cyclic property of the trace, that the expectation of any extended one-body operator $\hat{J}$ is simply given by the trace of the original single-body operator with respect to the one-body density matrix, i.e. $\langle\hat{J}\rangle = \mathrm{Tr}\{\rho_1 J\}$.  As the one-body density matrix is Hermitian, it can be diagonalised with \emph{real} eigenvalues $n_i$, and corresponding eigenvectors $\chi_i(\mathbf{x})$
\begin{equation}
	\rho_1(\mathbf{x},\mathbf{x}') = \sum_i n_i \chi_i^*(\mathbf{x})\chi_i(\mathbf{x}').
\end{equation}
It is immediately apparent that the eigenvalue $n_i$ quantifies the (mean) occupation of the mode $\chi_i(\mathbf{x})$ in the many-body state. 
An appropriate generalisation of the notion of Bose-Einstein condensation is thus obtained by regarding the most highly occupied mode $\chi_0(\mathbf{x})$ as the appropriate analogue of the lowest-energy single-particle state of the noninteracting system: when the occupation $n_0$ is an extensive parameter, with all others remaining intensive, a condensate is formed, with \emph{condensate orbital} $\chi_0(\mathbf{x})$.   In this case we can separate out the condensate and write 
\begin{equation}
	\rho_1(\mathbf{x},\mathbf{x}') = n_0\chi_0^*(\mathbf{x})\chi_0(\mathbf{x}') + \sum_{i>0} n_i\chi^*_i(\x)\chi_i(\x').
\end{equation}
In the thermodynamic limit we can replace the sum by an integral, which tends to zero for large separations $|\x-\x'|$ \cite{Pitaevskii03}.  By contrast, the contribution of the condensate remains finite, provided $\x$ and $\x'$ remain within the extent of the condensate orbital, and in the limiting case of an infinite, homogeneous system we obtain the result that $\lim_{|\x-\x'|\to\infty}\rho_1(\x,\x') \to n_0$, a property known as \emph{off-diagonal long-range order} \cite{Penrose56}.  In quantum optical terms, this property reflects the first-order coherence of the Bose field, as discussed by Naraschewski and Glauber \cite{Naraschewski99}.

It should be noted, however, that although this Penrose-Onsager definition of condensation is unambiguous in many simple scenarios, it is not difficult to construct systems for which the one-body density matrix can give misleading results.  Examples include strongly nonequilibrium systems \cite{Leggett01} and situations involving attractive interactions \cite{Wilkin98}, and indeed even the (common) case of harmonic trapping allows undamped centre-of-mass motion of the atomic cloud \cite{Dobson94}\footnote{A very general result for harmonically confined interacting-particle systems guarantees the presence of dipole oscillations of the centre-of-mass of the particle assembly at the appropriate trapping frequencies \cite{Dobson94}.  This is in close analogy with \emph{Kohn theorem} for the centre-of-mass cyclotron resonance of interacting electrons in a static magnetic field \cite{Kohn61}, and it is thus conventional to speak of the `Kohn theorem' and `Kohn modes' of trapped degenerate gases.}, which can lead to questionable results for the condensate fraction from the Penrose-Onsager approach \cite{Drummond07}.  The appropriate definition of condensation in more general situations remains an uncertain and controversial topic, 
as discussed in references~\cite{Leggett01,Wilkin98,Drummond07,Pethick00,Gajda06,Drummond07,Yamada09}.  
The characterisation of coherence and condensation in the atomic field beyond the simple Penrose-Onsager definition forms a major theme of this thesis, and we discuss alternative measures of coherence in chapters~\ref{chap:anomalous}, \ref{chap:precess}, \ref{chap:arrest}, and \ref{chap:stirring}.
%%%%%%%%%%%%%%%%%%%%%%%%%%%%%%%%%%%%%%%%%%%%%%%%%%%%%%%%%%%%%%%%%%%%%%%%%%%%%%%%%%%%%%%%
\subsection{The Gross-Pitaevskii equation}\label{subsec:back_GPE}
Having determined the appropriate definition of the condensate in terms of the condensate orbital $\chi_0(\mathbf{x})$, we wish to \emph{characterise} the orbital $\chi_0(\mathbf{x})$ and its evolution with time.  At zero temperature, the form and evolution of $\chi_0(\mathbf{x})$ are governed by the \emph{Gross-Pitaevskii} equation (GPE), which we derive here.  In the conditions of condensation in this zero-temperature limit, only low-energy, binary interactions between atoms are relevant \cite{Dalfovo99}, and these interactions are well characterised by a single parameter, the $s$-wave scattering length, independent of the details of the interatomic potential.  This motivates the replacement of the interatomic potential by a purely repulsive \emph{effective potential} \cite{Pitaevskii03} or a (regularised) delta-function \emph{pseudopotential} \cite{Huang87} which produces the same $s$-wave scattering length.  Quite aside from the fact that the precise details of the interatomic potential are unknown, this replacement is essential as the true interatomic potential supports bound states, precluding the perturbative treatment of interactions which is implicit in mean-field treatments \cite{Castin01}.  For the purposes of our discussion here it will be sufficient to assume the common case in which the inter-particle interactions are described by the simple contact interaction $V(\mathbf{x})=U_0\delta(\mathbf{x})$, where the interaction coefficient $U_0\equiv 4\pi\hbar^2a_s/m$, with $a_s$ the $s$-wave scattering length.  Substituting this effective potential into the second-quantised Hamiltonian~\reff{eq:back_2ndQH}, we obtain the cold-collision Hamiltonian
\begin{equation}\label{eq:back_cold_collision}
	\hat{H} = \int\!d\mathbf{x}\,\hat{\Psi}^\dagger(\mathbf{x})H_\mathrm{sp}\hat{\Psi}(\mathbf{x}) + \frac{U_0}{2} \int\!d\mathbf{x}\,\hat{\Psi}^\dagger(\mathbf{x})\hat{\Psi}^\dagger(\mathbf{x}) \hat{\Psi}(\mathbf{x})\hat{\Psi}(\mathbf{x}).
\end{equation}
%%%%%%%%%%%%%%%%%%%%%%%%%%%%%%%%%%%%%%%%%%%%%%%%%%%%%%%%%
\subsubsection{Hartree-Fock ansatz}\label{subsubsec:theory_hartree}
We first consider the time-independent GPE, which describes the form of the stationary condensed mode $\chi_0(\mathbf{x})$ as a nonlinear eigenvalue problem.  There are several ways to derive this equation; to best give insight into the physical content of the equation, we derive it here by assuming a Hartree-Fock ansatz for the many-body wave function in terms of the condensate orbital $\chi_0(\mathbf{x})$. Specifically, we assume the ansatz \cite{Leggett01}
\begin{equation}\label{eq:back_hartree_ansatz}
	\psi(\mathbf{x}_1,\mathbf{x}_2,\cdots,\mathbf{x}_N) = \prod_{i=1}^N \chi_0(\mathbf{x}_i).
\end{equation} 
This ansatz is clearly in accord with our expectations for a zero-temperature condensate: every atom resides in the same single-particle orbital $\chi_0(\mathbf{x})$.  The energy of the system (i.e., the expectation value of \reff{eq:back_cold_collision} in the corresponding Fock state) is
\begin{equation}
	E_\mathrm{HF} = N \int\!d\mathbf{x}\left[ \frac{\hbar^2}{2m} |\nabla \chi_0(\mathbf{x})|^2 + V_\mathrm{ext}(\mathbf{x})|\chi_0(\mathbf{x})|^2 + \frac{(N-1)U_0}{2}|\chi_0(\mathbf{x})|^4 \right]\!.
\end{equation}
By functionally minimising this energy (see, e.g., \cite{Blakie01}), subject to the constraint of normalisation of the condensate orbital ($\int\!d\mathbf{x}\,|\chi_0(\mathbf{x})|^2=1$), we obtain the time-independent GPE
\vspace{-0.80\baselineskip}
\begin{equation}\label{eq:back_TIGPE}
	-\frac{\hbar^2}{2m}\nabla \chi_0(\mathbf{x}) + V_\mathrm{ext}(\mathbf{x})\chi_0(\mathbf{x}) + (N-1)U_0|\chi_0(\mathbf{x})|^2\chi_0(\mathbf{x}) = \lambda\chi_0(\mathbf{x}),
\vspace{-0.40\baselineskip}
\end{equation}
where the parameter $\lambda$ here enters as the Lagrange multiplier associated with the normalisation constraint \cite{Leggett01,Blakie01}, and it can be shown that $\lambda=\partial E_\mathrm{HF}/\partial N$, i.e., $\lambda$ is the \emph{chemical~potential} of the condensate.  The term in $|\chi_0(\mathbf{x})|^2$ describes the \emph{mean-field potential} experienced by each atom due to the presence of all the other atoms, and we see that it vanishes (properly) in the case of a single atom.  In practice we typically approximate $N-1\approx N$ on the basis that $N$ is large, and the GPE is traditionally written in terms of the function $\phi(\mathbf{x}) \equiv \sqrt{N}\chi_0(\mathbf{x})$ with norm square equal to the total particle number \cite{Leggett01}.  The function $\phi(\mathbf{x})$ is also commonly referred to as the condensate `wave function', though one should be careful not to confuse it with the true many-body wave function~\reff{eq:back_hartree_ansatz} of the condensate in this approach.  Although this Hartree-Fock approach yields the clearest physical insight into the Gross-Pitaevskii (GP) equation, the corresponding time-dependent GP equation cannot be derived rigorously in this manner \cite{Leggett01}, and so we must consider less transparent approaches to defining the condensate wave function.
%%%%%%%%%%%%%%%%%%%%%%%%%%%%%%%%%%%%%%%%%%%%%%%%%%%%%%%%%
\subsubsection{Symmetry-breaking approach}\label{subsubsec:symmetry_breaking}
A widespread approach to deriving the Gross-Pitaevskii equation is based on the idea of \emph{spontaneous symmetry breaking} by a phase transition, which although less transparent than the Hartree-Fock approach, is more direct.  Formally, such a phase transition is described in terms of the emergence of a nonzero \emph{order parameter} below the phase transition.  The archetypal example of such an order parameter is the magnetisation $\langle \bm{M} \rangle$ which emerges spontaneously in a ferromagnetic system cooled below the Curie temperature \cite{Landau69}.  Conventionally, Bose-Einstein condensation is formulated as a similar symmetry-breaking phase transition, in which the emergent order parameter is the expectation value of the field operator $\langle \hat{\Psi}(\mathbf{x}) \rangle$, which is associated with condensation in the system.  As the first moment $\langle \hat{\Psi}(\mathbf{x}) \rangle$ of the field acquires a finite value, its phase becomes well-defined, breaking the $\mathrm{U(1)}$ (gauge) symmetry of the Hamiltonian (\ref{eq:back_cold_collision}) under global rotations of the field operator phase $\hat{\Psi}(\mathbf{x})\rightarrow\hat{\Psi}(\mathbf{x})e^{i\theta}$ (see, e.g., \cite{Leggett95}).  It is important to note that rotations of this phase are generated by the \emph{total particle number} $\hat{N} = \int\!d\mathbf{x}\,\hat{\psi}^\dagger(\mathbf{x})\hat{\psi}(\mathbf{x})$ \cite{Blaizot86}, and the breaking of this symmetry is thus tantamount to allowing the nonconservation of particle number.  This symmetry breaking is therefore \emph{not physical}, and is merely a computational convenience.  Formally this fiction is introduced by calculating expectation values in a modified ensemble \cite{Hohenberg65} of states with only a small dispersion in the phase (and amplitude) of the order parameter.  We can achieve a similar result at zero temperature by assuming that the atomic field is in a \emph{coherent state} $|\phi(\mathbf{x})\rangle$ \cite{Gardiner00}, which is a superposition of states with \emph{all} (physical) particle numbers, such that $\hat{\Psi}(\mathbf{x})|\phi(\mathbf{x})\rangle = \phi(\mathbf{x})|\phi(\mathbf{x})\rangle$. 

The time-dependent Gross-Pitaevskii equation is derived using this notion of spontaneous symmetry breaking as follows:  using the commutation relation (\ref{eq:back_bose_commutator}) we derive from the second-quantised Hamiltonian (\ref{eq:back_cold_collision}) the Heisenberg equation of motion
\begin{equation}\label{eq:back_heisenberg}
	i\hbar\frac{\partial \hat{\Psi}(\mathbf{x})}{\partial t} = \left[\hat{\Psi}(\mathbf{x}),\hat{H}\right] = \left\{ H_\mathrm{sp} + U_0 \hat{\Psi}^\dagger(\mathbf{x})\hat{\Psi}(\mathbf{x}) \right\} \hat{\Psi}(\mathbf{x}). 
\end{equation}
Taking the expectation value of \reff{eq:back_heisenberg} in the coherent state $|\phi(\mathbf{x})\rangle$ amounts to making the replacement $\hat{\Psi}(\mathbf{x})\to\phi(\mathbf{x})$, yielding the time-dependent Gross-Pitaevskii equation
\begin{equation}\label{eq:back_TDGPE}
	i\hbar\frac{\partial \phi(\mathbf{x},t)}{\partial t} = \left\{ H_\mathrm{sp} + U_0 |\phi(\mathbf{x},t)|^2 \right\} \phi(\mathbf{x},t).
\end{equation}
This equation describes the \emph{collective} motion of the condensate at zero temperature, as for example in response to a time-dependent external trapping potential $V_\mathrm{ext}(\mathbf{x},t)$.  It is important to note that this equation only describes the motion so long as the Bose field remains completely condensed.  However, \emph{dynamical instabilities} of the condensate can arise, and are marked in the GP solution by the instability of the condensate mode to exponential growth of collective excitations.  This in fact represents the runaway amplification of beyond-GP fluctuations of the Bose field, and so the appearance of a dynamical instability marks the breakdown of the Gross-Pitaevskii description, as we discuss further in section~\ref{subsec:theory_Bogoliubov}.

The derivation of the time-independent GP equation in the symmetry-breaking approach parallels the Hartree-Fock derivation of section~\ref{subsubsec:theory_hartree}, as we similarly obtain the GP equation by minimising the energy of the many-body system subject to an ansatz for the many-body state.  As we have broken the conservation of particle number, we work in the grand-canonical ensemble where the \emph{mean} atom number is conserved.  We thus consider the grand-canonical Hamiltonian 
\begin{equation}\label{eq:back_kamiltonian}
	\hat{K}=\hat{H}-\mu\hat{N},
\end{equation}
where $\hat{H}$ is the Hamiltonian (\ref{eq:back_cold_collision}), and the chemical potential $\mu$ enforces the conservation of particle number on average.  Taking the expectation of equation~\reff{eq:back_kamiltonian} in the state $|\phi(\mathbf{x})\rangle$ yields \cite{Fetter99} 
\begin{equation}\label{eq:back_kay_naught}
	K_0 = H_0 - \mu N_0 = \int\!d\mathbf{x} \left[ \phi^*(\x) (H_\mathrm{sp}-\mu) \phi(\x) + \frac{U_0}{2} |\phi(\x)|^4 \right]\!.
\end{equation}
We will refer to the functional $H_0[\phi(\x)]$ as the Gross-Pitaevskii \emph{energy functional}.  In the zero-temperature limit appropriate to our assumed coherent state $|\phi(\x)\rangle$, the free energy of the system is simply $\langle\hat{K}\rangle$, and so the system equilibrium is found by minimising equation~(\ref{eq:back_kay_naught}), i.e., by requiring that $K_0$ is stationary under variations of $\phi(\mathbf{x})$ and its complex conjugate.  The Euler-Lagrange equation resulting from variations of the latter is the time-independent GP equation 
\begin{equation}\label{eq:back_tigpe_simple}
	\left\{H_\mathrm{sp} + U_0|\phi(\mathbf{x})|^2\right\}\phi(\mathbf{x}) = \mu\phi(\mathbf{x}).	
\end{equation}
We note that this form of the time-independent GPE can also be obtained from the time-dependent GPE \reff{eq:back_TDGPE} by assuming a uniform phase rotation $\phi(\mathbf{x},t)=e^{-i\mu t}\phi(\mathbf{x})$.  We notice that whereas the time development of a true quantum-mechanical wave function is governed by its energy, of the phase of the order parameter rotates with frequency determined by its \emph{chemical potential} \cite{Pitaevskii03}.
%%%%%%%%%%%%%%%%%%%%%%%%%%%%%%%%%%%%%%%%%%%%%%%%%%%%%%%%%
\subsubsection{Healing length}
An important length scale which characterises solutions of the GP equation is the \emph{healing length}, which is the distance over which the condensate wave function, constrained to zero density by some applied potential, returns to its bulk density.  This length is thus set by the balance of the kinetic energy associated with the curvature of the wave function and the interaction energy associated with its density \cite{Pethick02}.  Equating the two we find the healing length
\begin{equation}
	\eta = \frac{\hbar}{\sqrt{2mn(\x)U_0}},
\end{equation}
where $n(\x)=|\phi(\x)|^2$ is the local density of the condensate.  In inhomogeneous condensates the healing length is thus a spatially varying quantity.  However, in characterising inhomogeneous condensates it is common to simply use the healing length corresponding to the central (i.e., peak) density of the condensate as an order-of-magnitude estimate \cite{Dalfovo99}.  This healing length characterises (for example) the core size of a \emph{quantum vortex} in the condensate (section~\ref{subsec:back_vortices}). 
%%%%%%%%%%%%%%%%%%%%%%%%%%%%%%%%%%%%%%%%%%%%%%%%%%%%%%%%%
\subsubsection{Thomas-Fermi limit}
In the limit of large condensate population, the nonlinearity in the time-independent Schr\"odinger equation dominates, and the kinetic energy becomes negligible in comparison.  In this limit we may therefore simply drop the curvature term from equation~\reff{eq:back_tigpe_simple}, leaving us with an algebraic equation for the condensate density, which we can rearrange to find the \emph{Thomas-Fermi condensate wave function} 
\begin{equation}
	\Psi_\mathrm{TF}(\x) = \sqrt{\frac{\mu_\mathrm{TF}-V_\mathrm{ext}(\x)}{U_0}}\Theta\bigl(\mu_\mathrm{TF} - V_\mathrm{ext}(\x)\bigr),
\end{equation}
where $\Theta(x)$ denotes a Heaviside function.  
The quantity $\mu_\mathrm{TF}$ is the Thomas-Fermi (TF) limit chemical potential, which for harmonic trapping becomes \cite{Norrie05a}
\begin{equation}
	\mu_\mathrm{TF} = \frac{1}{2}\left(15a\hbar^2\sqrt{m}\omega_x\omega_y\omega_zN_0\right)^{2/5},
\end{equation}
where $N_0 = \int\!d\x\,|\Psi_\mathrm{TF}|^2$ is the Thomas-Fermi condensate population.
In this geometry, the half-width (Thomas-Fermi radius) of the TF condensate orbital in the $i^\mathrm{th}$ radial direction is
\begin{equation}
	R_i^\mathrm{TF} = \frac{1}{\omega_i}\sqrt{\frac{2\mu_\mathrm{TF}}{m}}.
\end{equation} 
For large condensates the Thomas-Fermi wave function provides an excellent approximation to the exact Gross-Pitaevskii condensate mode, differing only at the condensate boundary, where the density $|\Psi_\mathrm{TF}(\x)|^2$ of the TF condensate vanishes abruptly at $x_i=R_i^\mathrm{TF}$, whereas the boundary of the true Gross-Pitaevskii condensate orbital is smoothed so as to reduce its kinetic energy.
%%%%%%%%%%%%%%%%%%%%%%%%%%%%%%%%%%%%%%%%%%%%%%%%%%%%%%%%%
\subsubsection{Hydrodynamic formulation and collective modes}
We now show that the time-dependent Gross-Pitaevskii equation~\reff{eq:back_TDGPE} can be reformulated as a hydrodynamic equation for the condensate motion \cite{Stringari96b,Dalfovo99}.  We perform the \emph{Madelung transformation}, making the substitution $\phi(\x) = \sqrt{n(\x)}e^{iS(\x)}$ for the condensate wave function.  In terms of the hydrodynamic velocity $\mathbf{v}=(\hbar/m)\nabla S$ we thus obtain the \emph{continuity equation}
\begin{equation}\label{eq:back_continuity}
	\frac{\partial}{\partial t} n + \nabla \cdot (n\mathbf{v}) = 0,
\end{equation}
expressing the conservation of the condensate population, and the \emph{Euler equation}
\begin{equation}
	m\frac{\partial}{\partial t}\mathbf{v} + \nabla\left(V_\mathrm{ext} + gn -\frac{\hbar^2}{2m\sqrt{n}}\nabla^2\sqrt{n} + \frac{mv^2}{2}\right) = 0.
\end{equation}
The term in $\nabla^2\sqrt{n}$ here is the \emph{quantum pressure} resulting from the kinetic energy term in the time-dependent Gross-Pitaevskii equation.  In the limit of large condensates we can drop the quantum pressure term, which corresponds to the Thomas-Fermi approximation discussed above, to obtain the simpler Euler equation
\begin{equation}\label{eq:back_euler}
	m\frac{\partial}{\partial t}\mathbf{v} + \nabla\left(V_\mathrm{ext} + gn + \frac{mv^2}{2}\right) = 0.
\end{equation}
One can easily show that the time-independent solution of equation~\reff{eq:back_euler} corresponds to the Thomas-Fermi solution discussed previously \cite{Dalfovo99}.  Introducing a small deviation $\delta n(\x,t) = n(\x,t) - n_\mathrm{TF}(\x)$, and linearising equations~\reff{eq:back_continuity}~and~\reff{eq:back_euler} about the TF solution, we can estimate the frequencies of collective modes of oscillation of the condensate \cite{Stringari96b}.   In the case of harmonic trapping axially symmetric along the $z$ axis (most relevant to this thesis), we obtain an equation 
\begin{equation}\label{eq:back_hydrodynamic_PDE}
	m\frac{\partial^2}{\partial t^2} \delta n = \nabla \cdot \biggl\{\left[\mu-\frac{m}{2}\left(\omega_\perp^2r_\perp^2 + \omega_z^2z^2\right)\right]\nabla \delta n\biggr\},
\end{equation}
for the linear deviations.  Substituting a time-harmonic ansatz $\delta n(\x,t) = \delta n(\x)e^{-i\omega t}$ into equation~\reff{eq:back_hydrodynamic_PDE}, we can obtain a differential equation for the spatial modes of oscillation, and deduce the corresponding oscillation frequencies \cite{Stringari96b}.
Of particular interest are the so-called \emph{surface modes}, which have no radial nodes.  In the axially symmetric geometry, each such mode is characterised by the quantum numbers $l$ and $m$ which specify the mode angular momentum and its projection along the $z$ axis, respectively.  
In particular, we find frequencies $\omega_{l,m}$ for the \emph{dipole oscillations}
\begin{equation}
	\omega_{1,1} = \omega_\perp ;\quad \omega_{1,0} = \omega_z,
\end{equation}
which are of course the \emph{exact} eigenfrequencies of the dipole modes of the harmonically trapped interacting gas by the Kohn theorem (section~\ref{subsec:back_bose_condensation}).  We also find in general
\begin{equation}
	\omega_{l,\pm l} = \sqrt{l}\omega_\perp,
\end{equation}
which is the frequency of a surface oscillation of multipolarity $l$ about the $z$ axis.  Such a surface oscillation can be excited by a rotating trap anisotropy of the same multipolarity.  In general, the efficiency of this excitation frequency will depend on the frequency at which the anisotropy is rotated.  Taking into account the multipolarity of the trap we expect \emph{resonant} excitation to occur at a frequency $\omega_\mathrm{res} = \omega_l / l$.  In the particular case of quadrupole mode excitation relevant to chapters~\ref{chap:stir_background}~and~\ref{chap:stirring} we find the resonant driving frequency
\begin{equation}
	\omega_Q = \omega_\perp/\sqrt{2}.
\end{equation}
%%%%%%%%%%%%%%%%%%%%%%%%%%%%%%%%%%%%%%%%%%%%%%%%%%%%%%%%%%%%%%%%%%%%%%%%%%%%%%%%%%%%%%%%
\subsection{Bogoliubov theory}\label{subsec:theory_Bogoliubov}
In this section we consider the character of excitations of the condensed field.  We therefore assume that a single mode ($\chi_0(\x)$) of the atomic field is condensed, and replace the corresponding annihilation operator $\hat{a}_0 = \int\!d\x\,\chi_0^*(\x) \hat{\Psi}(\x)$ with the \emph{number} $\sqrt{N_0}$ (where $N_0$ is the condensate occupation), and retain a second-quantised description for the remainder of the field.  Although this is the standard symmetry-breaking approach in the literature, we note that Castin and Dum \cite{Castin98} have provided a more rigorous and less conceptually ambiguous \emph{number-conserving} formulation of the Bogoliubov expansion (see also \cite{Gardiner97a}).  However, the additional complexity involved in the rigorous asymptotic expansion of \cite{Castin98} obscures somewhat the simple physical picture of the Bogoliubov expansion we wish to present here.  We thus make the \emph{Bogoliubov shift}
\begin{equation}\label{eq:back_bog_shift}
	\hat{\Psi}(\mathbf{x}) \rightarrow \phi(\mathbf{x}) + \hat{\delta}(\mathbf{x}),
\end{equation}
separating the field operator $\hat{\Psi}(\mathbf{x})$ into a condensed part $\phi(\mathbf{x})=\sqrt{N_0}\chi_0(\x)$ and a \emph{fluctuation operator} $\hat{\delta}(\mathbf{x})$, which is assumed to obey the standard bosonic commutation relations (equation~\reff{eq:back_bose_commutator}).  We then proceed to derive a Hamiltonian which describes the fluctuation operator $\hat{\delta}(\mathbf{x})$, and thus characterises the excitations of the condensate.  
Substituting the separation \reff{eq:back_bog_shift} into the second-quantised Hamiltonian~\reff{eq:back_cold_collision} and assuming that the condensate wave function $\phi(\mathbf{x})$ is a solution of \reff{eq:back_TIGPE}, we obtain
\begin{eqnarray}\label{eq:back_kamiltonian2}
	\hat{K} &=& \int\!d\mathbf{x}\,\hat{\delta}^\dagger \Bigl[H_\mathrm{sp} + 2U_0|\phi|^2 - \mu \Bigr] \hat{\delta} \nonumber \\
&&+\frac{U_0}{2} \int d\mathbf{x} \left[ (\phi^*)^2 \hat{\delta}\hat{\delta} + \phi^2 \hat{\delta}^\dagger\hat{\delta}^\dagger + \phi^* \hat{\delta}^\dagger \hat{\delta} \hat{\delta} + \phi \hat{\delta}^\dagger \hat{\delta}^\dagger \hat{\delta} + \hat{\delta}^\dagger\hat{\delta}^\dagger\hat{\delta}\hat{\delta} \right]\!,
\end{eqnarray}
where we have dropped a constant term $K_0$ which is simply an energy shift and does not otherwise affect the motion of the fluctuation operator\footnote{Linear terms in $\hat{\delta}$ vanish due to the stationarity of \reff{eq:back_kay_naught} under variations of $\phi(\mathbf{x})$ \cite{Fetter99}.}, and we suppress the spatial arguments of $\phi$ and $\hat{\delta}$ for notational convenience.  In general, this Hamiltonian remains, like the original second-quantised Hamiltonian \reff{eq:back_cold_collision}, intractable.  However, a Hamiltonian which is quadratic in the field operators can always be diagonalised (in a generalised sense, as we discuss below) \cite{Blaizot86}
and so to proceed from here we approximate this Hamiltonian by one in which higher powers of $\hat{\delta}$ do not occur.  The approach we take is to simply neglect terms which are cubic and quartic in $\hat{\delta}$.  This approximation is clearly appropriate in the limit that the noncondensed fraction of the field is small compared to the condensate (i.e. $\langle \hat{\delta}^\dagger\hat{\delta}\rangle \ll |\phi|^2 $), and we obtain in this case the quadratic Hamiltonian 
\begin{equation}\label{eq:back_Kquad}
	\hat{K}_\mathrm{quad} = \int\!d\mathbf{x}\,\hat{\delta}^\dagger \Bigl[H_\mathrm{sp} + 2U_0|\phi|^2 - \mu \Bigr] \hat{\delta} +\frac{U_0}{2} \int\!d\mathbf{x}\left[ (\phi^*)^2 \hat{\delta}\hat{\delta} + \phi^2 \hat{\delta}^\dagger\hat{\delta}^\dagger\right]\!.
\end{equation}
As this Hamiltonian contains terms like $\hat{\delta}\hat{\delta}$, its eigenbasis is \emph{not} a single-particle basis, but can be expressed in terms of two-component (\emph{spinor}) basis functions.  Physically, this is because (at this order) the excitations of the condensate involve \emph{pairs} of atoms:  the interactions `mix' the roles of creation and annihilation operators of single-particle modes, and the elementary excitations of the system are created by so-called \emph{quasiparticle operators}, which are superpositions of these creation and annihilation operators \cite{Blaizot86}.  Such quasiparticle operators are introduced by defining
\begin{equation}
	\hat{b_k} = \int\!d\mathbf{x}\left[u_k^*(\mathbf{x}) \hat{\delta}(\mathbf{x}) - v_k^*(\mathbf{x})\hat{\delta}^\dagger(\mathbf{x}) \right]\!.
\end{equation}
Requiring that the $\hat{b}_k$ satisfy the standard Bose commutation relations ($[\hat{b}_j,\hat{b}_k^\dagger]=\delta_{jk}$; $[\hat{b}_j,\hat{b}_k]=0$) we find that the quasiparticle spinors $(u_k(\mathbf{x}),v_k(\mathbf{x}))^T$ are orthogonal with respect to the metric  
\begin{equation}
	\sigma_z = \twobytwo{1}{0}{0}{-1}\!,
\end{equation}
i.e., we have
\begin{equation}\label{eq:back_bog_ip}
	\int\!d\mathbf{x}\,\left(u_j^*(\mathbf{x}),v_j^*(\mathbf{x})\right) \cdot \sigma_z \cdot \binom{u_k(\mathbf{x})}{v_k(\mathbf{x})} = \int\!d\mathbf{x}\, u_j^*u_k - v_j^*v_k = \delta_{jk}.
\end{equation}
Our aim is to find a \emph{diagonal Hamiltonian}, written in terms of the $\hat{b}_k$ and their Hermitian conjugates, i.e., one which takes the form (up to a constant energy)
\begin{equation}\label{eq:back_Kquad_diag}
	\hat{K}_\mathrm{quad} = \sum_k \epsilon_k\hat{b}_k^\dagger \hat{b}_k.
\end{equation}
This form for the Hamiltonian implies the commutation relation $[\hat{b}_k,\hat{K}_\mathrm{quad}]=\epsilon_k\hat{b}_k$.  Evaluating this commutator explicitly for the Hamiltonian~\reff{eq:back_Kquad}, we find that the mode functions 
$u_k(\mathbf{x}),v_k(\mathbf{x})$ satisfy the \emph{Bogoliubov-de Gennes} equations
\begin{equation}\label{eq:BdG_eqns}
	\mathcal{L} \binom{u_k(\mathbf{x})}{v_k(\mathbf{x})} = \epsilon_k \binom{u_k(\mathbf{x})}{v_k(\mathbf{x})},
\end{equation}
which we have written here in terms of the matrix
\begin{equation}\label{eq:back_bog_mtx}
	\mathcal{L} \equiv \twobytwo{H_\mathrm{sp} -\mu + 2U_0|\phi|^2}{U_0\phi^2}{-U_0(\phi^*)^2}{-\bigl[H_\mathrm{sp} - \mu + 2U_0|\phi|^2\bigr]^*}\!.
\end{equation}
%%%%%%%%%%%%%%%%%%%%%%%%%%%%%%%%%%%%%%%%%%%%%%%%%%%%%%%%%
\subsubsection{Dual modes and Goldstone mode}
\begin{sloppypar}
A symmetry property of the matrix \reff{eq:back_bog_mtx} implies that for every spinor $(u_k(\mathbf{x}),v_k(\mathbf{x}))^T$ which is a solution of equation~\reff{eq:BdG_eqns} (with energy $\epsilon_k$), there is a dual solution $(v_k^*(\mathbf{x}),u_k^*(\mathbf{x}))^T$ with energy $-\epsilon_k^*$, and which clearly has normalisation (with respect to the inner product \reff{eq:back_bog_ip}) $-1$ \cite{Castin01}.  These negative-normed modes are simply a dual representation of the positive-normed modes, and do not appear in the Bogoliubov form of the quadratic Hamiltonian.  However, these two families of modes do not in fact exhaust the eigenbasis of $\mathcal{L}$; the spinor
\begin{equation}
	\Theta_\mathrm{G} \equiv \binom{\phi(\mathbf{x})}{-\phi^*(\mathbf{x})},
\end{equation}
is an eigenvector of $\mathcal{L}$, with eigenvalue $\epsilon_G=0$, and normalisation $\int\!d\x\,\Theta_\mathrm{G}^\dagger \sigma_z \Theta_{G}=0$.  This mode does not appear in the diagonalised Hamiltonian~\reff{eq:back_Kquad_diag} due to its zero eigenvalue, and is the \emph{spurious} or \emph{Goldstone} mode associated with the breaking of phase symmetry \cite{Blaizot86}.  A more careful treatment of the problem \cite{Lewenstein96} reveals that the full quadratic Hamiltonian is of the form
\begin{equation}
	H_\mathrm{quad} = \sum_k \epsilon_k\hat{b}_k^\dagger \hat{b}_k + \frac{\hat{P}^2}{2m_G},
\end{equation}
where the `momentum' operator $\hat{P}=\int\!d\mathbf{x}\,[\chi_0^*(\x) \hat{\delta}(\x) + \chi_0(\x) \hat{\delta}^\dagger(\x)]$ is related to the phase of the condensate mode, and the `mass' $m_\mathrm{G}$ is inversely proportional to the peak density of the condensate \cite{Lewenstein96}.  The appearance here of a momentum variable without a corresponding potential signifies that the phase of the condensate diffuses over time, restoring the symmetry spuriously broken by our coherent-state approximation \cite{Lewenstein96}.  In general, such a term appears for each symmetry of the Hamiltonian which is broken by the assumed vacuum state of the approximate quadratic Hamiltonian \cite{Blaizot86}.  It is important to note, however, that this particular spurious mode arises due to the breaking of phase symmetry, which, as we have noted, is not physical.  Indeed, no such phase-diffusion term arises in the more careful number conserving approach of \cite{Castin98}.  In fact, for finite systems such as we discuss in this thesis, broken symmetries arise only as a result of approximations \cite{Blaizot86}.  The form of the Goldstone Hamiltonian term $\hat{P}^2/2m_\mathrm{G}$ appearing here as a result of phase-symmetry breaking will allow us to better understand the consequences of phase and rotational symmetry breaking in our classical-field simulations in chapters~\ref{chap:anomalous}~and~\ref{chap:precess} respectively.
\end{sloppypar}
%%%%%%%%%%%%%%%%%%%%%%%%%%%%%%%%%%%%%%%%%%%%%%%%%%%%%%%%%
\subsubsection{Relation to small oscillations in Gross-Pitaevskii theory}
We now illustrate a deep connection between the Bogoliubov theory for the elementary excitations of the condensed Bose gas in second quantisation, and the small oscillations (`normal modes') of the condensate in the time-dependent Gross-Pitaevskii formalism.
We consider the small oscillations of the condensate in the Gross-Pitaevskii approach, generalising the hydrodynamic approach to the oscillations previously discussed in section~\ref{subsec:back_GPE}.  We assume some `reference' configuration $\phi_0(\x,t)$ of the time-dependent Gross-Pitaevskii equation~\reff{eq:back_TDGPE}, and calculate to linear order the evolution of a small deviation $\delta \phi(\x,t) = \phi(\x,t) - \phi_0(\x,t)$ about this reference solution.  We thus obtain the linear evolution equation \cite{Castin01}
\begin{eqnarray}\label{eq:back_deviant_GPE}
	i\hbar\frac{\partial}{\partial t} \delta \phi(\x,t) &=& -\frac{\hbar^2}{2m}{\nabla^2}\delta\phi(\x,t) + V_\mathrm{ext}(\x)\delta\phi(\x,t) \nonumber \\
	&&+ 2U_0|\phi_0(\x,t)|^2\delta\phi(\x,t) + U_0\phi_0^2(\x,t) \delta\phi^*(\x,t).
\end{eqnarray}
We consider here the particular case in which $\phi_0(\x,t)$ corresponds to an \emph{eigenfunction} of the time-independent Gross-Pitaevskii equation, i.e., $\phi_0(\x,t) = \phi_0(\x)e^{-i\mu t/\hbar}$, and assume the ansatz
\begin{equation}\label{eq:back_bog_ansatz}
	\delta\phi(\x,t) = e^{-i\mu t/\hbar} \sum_k \left[ u_k(\x) e^{-i\omega_k t} + v_k^*(\x) e^{i\omega_k t}\right]\!.
\end{equation} 
Substituting the ansatz~\reff{eq:back_bog_ansatz} into equation~\reff{eq:back_deviant_GPE}, and equating powers of $e^{\pm i\omega_k t}$, we find that the mode functions $u_k(\x), v_k(\x)$ are solutions of 
\begin{equation}
	\mathcal{L} \binom{u_k(\mathbf{x})}{v_k(\mathbf{x})} = \hbar\omega_k \binom{u_k(\mathbf{x})}{v_k(\mathbf{x})},
\end{equation}
where $\mathcal{L}$ is given by equation~\reff{eq:back_bog_mtx}.  We thus have the interesting result that the small \emph{collective} oscillations of the condensate, as described by Gross-Pitaevskii equation, are in one-to-one correspondence with the \emph{elementary excitations} of the system in the quadratic (Bogoliubov) approximation to the second-quantised field theory, with the frequencies $\omega_k$ of the small oscillations equal to the (Heisenberg-picture) phase rotation frequencies $\epsilon_k/\hbar$ of the Bogoliubov ladder operators.  This is in some sense not surprising, as the above linearisation of the Gross-Pitaevskii equation amounts to a quadratisation of the Gross-Pitaevskii \emph{energy functional}~\reff{eq:back_kay_naught} about the Gross-Pitaevskii eigenfunction solution, in direct analogy to the quadratisation of the second-quantised Hamiltonian we performed to obtain the elementary excitations. 
%%%%%%%%%%%%%%%%%%%%%%%%%%%%%%%%%%%%%%%%%%%%%%%%%%%%%%%%%
\subsubsection{Stability of excitations}
In equilibrium, we think of the condensate as being the `ground' (or \emph{vacuum}) state of the system, with the weakly populated Bogoliubov modes constituting some excitation of the system above this ground state, due to thermal (and quantum) fluctuations.  In such a situation, the Bogoliubov excitation energies $\epsilon_k$ are therefore real and positive quantities.  However, in more general situations, the excitation energies may fail to be positive or even real, and this signifies an instability of the condensate, as we now discuss.\newline
\newline\emph{Thermodynamic instability}\newline
It is possible that the Gross-Pitaevskii wave function $\phi(\x)$, although an eigenfunction of the Gross-Pitaevskii equation, is not the lowest-energy eigenfunction (for its occupation $N$), i.e., it is a collectively excited state of the condensate.  It is then possible that one or more Bogoliubov energies $\epsilon_k$ may be negative.  In the presence of dissipation, the condensate can then lose energy as the population of the negative-energy Bogoliubov mode increases, and this can lead to a complete restructuring of the degenerate gas.  In this case the condensate is considered to be \emph{thermodynamically unstable}.  It should be noted however that a GP mode can be an excited eigenfunction but nevertheless possess no negative-energy Bogoliubov excitations, i.e., if it resides at a \emph{local} minimum of the GP energy functional~\reff{eq:back_kay_naught}.  In this case the state is referred to as \emph{thermodynamically metastable}.\newline
\newline\emph{Dynamic instability}\newline
A second case is that in which an eigenvalue $\epsilon_k$ has a nonzero imaginary part.  Clearly an energy eigenvalue with a positive imaginary part yields exponential growth of the mode population with time, and it can be shown \cite{Castin01} that if $\epsilon_k$ is an eigenvector of $\mathcal{L}$ then $\epsilon_k^*$ is also, and we thus require that there are no eigenvalues with negative imaginary parts either.  In the case that such a complex eigenvalue appears in the Bogoliubov spectrum, the condensate mode is said to be \emph{dynamically unstable}.  It is important to note that the growth of population in a dynamically unstable mode does \emph{not} require dissipation, but is rather a dynamical amplification of population in the unstable mode, fed by the condensate.  In the second-quantised picture, dynamically unstable modes will always undergo growth due to their irreducible \emph{vacuum} population.  By the correspondence between the Bogoliubov modes and the small oscillations of the condensate, the corresponding oscillations of the Gross-Pitaevskii equation will also undergo unstable growth with time, provided that they are populated by (e.g.) numerical noise.  The evolution in this case can quickly lead to field dynamics beyond the validity not only of the GP equation but also of the Bogoliubov description.  It is intuitive that a dynamical instability can only occur if the condensate mode is not the ground Gross-Pitaevskii eigenstate, and indeed it can easily can easily be shown that thermodynamic stability implies dynamic stability~\cite{Castin01}.  In fact, it can be shown that dynamical instabilities arise from a kind of generalised avoided level crossing occurring between excitations with positive and negative energies \cite{Skryabin00,Anglin03}. 
%%%%%%%%%%%%%%%%%%%%%%%%%%%%%%%%%%%%%%%%%%%%%%%%%%%%%%%%%
\subsubsection{Self-consistent mean-field theories}\label{subsubsec:back_self_consistent}
In order to describe condensates at higher temperatures in such a mean-field framework, we must (by some suitable approximation) retain the effects of terms in the grand canonical Hamiltonian~\reff{eq:back_kamiltonian2} of cubic and quadratic order in $\hat{\delta}(\x)$.  One approach is to assume essentially \emph{ad hoc} factorisations of the field operators, such as $\hat{\delta}^\dagger\hat{\delta}\hat{\delta} \rightarrow 2\langle \hat{\delta}^\dagger\hat{\delta} \rangle \hat{\delta} + \langle \hat{\delta} \hat{\delta} \rangle \hat{\delta}^\dagger$ \cite{Griffin96}.  The second expectation value in this approximation is the \emph{anomalous density} $m(\mathbf{x})\equiv\langle \hat{\delta}(\mathbf{x})\hat{\delta}(\mathbf{x})\rangle$ which represents the pairing correlations in the noncondensed component induced by the interacting condensate.   Using such Hartree-Fock-Bogoliubov (HFB) approximations, one obtains \emph{generalised} Gross-Pitaevskii and Bogoliubov-de Gennes equations, which include additional mean-field potential terms depending on the noncondensed population and anomalous density, and which must therefore be solved \emph{self-consistently} (i.e., by an iterative procedure \cite{Blaizot86,Griffin96}).  Such approaches have been applied to finite-temperature equilibrium scenarios \cite{Giorgini96,Hutchinson97,Gies04}.  There are, however, nontrivial complications introduced by the factorisation approximations \cite{Griffin96,Morgan00,Proukakis08}.  Moreover, these theories do not allow for any particle exchange between the condensate and thermal cloud \cite{Proukakis08}, which severely limits their applicability in nonequilibrium scenarios.  The most conspicuous failing of the mean-field theories, however, is the requirement that the Bose field be explicitly separated into condensed and noncondensed parts.  In general nonequilibrium scenarios (such as, e.g., turbulent dynamics of the Bose field \cite{Lobo04}) no such clear distinction can be made, and so mean-field theories are inapplicable in such regimes.
%%%%%%%%%%%%%%%%%%%%%%%%%%%%%%%%%%%%%%%%%%%%%%%%%%%%%%%%%
\subsubsection{Number-conserving approaches}
As we have noted, it is possible to reformulate the Bogoliubov theory in terms which explicitly preserve the $\mathrm{U(1)}$ symmetry of the fundamental second-quantised Hamiltonian \reff{eq:back_cold_collision}, and thus conserve particle number \cite{Girardeau59,Castin98,Gardiner97a,Morgan00,Gardiner07}, at the expense of a significant increase in complexity.  In these approaches the condensate is regarded as the dominant eigenvector of the one-body density matrix $\chi_0(\mathbf{x})$, with corresponding annihilation operator $\hat{a}_0$, and the noncondensed field is described in terms of operators which commute with the total particle number; e.g., Castin and Dum \cite{Castin98} work in terms of the noncondensate operator $\hat{\Lambda} = \hat{N}^{-1/2} \hat{a}_0^\dagger \hat{\delta}$, where $\hat{a}_0^\dagger$ creates a quanta in the condensate orbital $\chi_0(\mathbf{x})$, and similar number-conserving operators are employed in references \cite{Gardiner97a,Morgan00,Gardiner07}.  In particular, the approaches of references~\cite{Gardiner97a,Castin98} demonstrate that the time-dependent Gross-Pitaevskii equation traditionally derived under symmetry-breaking assumptions (section~\ref{subsec:back_GPE}) is indeed valid at zero temperature.  An essential aspect of these treatments is that the rigorous separation of the Bose field operator into condensate and noncondensate parts requires that the two components are \emph{orthogonal} (i.e., $\int\!d\x\,\chi_0^*(\x) \hat{\delta}(\x)=0$)\footnote{We note that in the standard symmetry-breaking approach one is free to \emph{choose} the excitations (and thus $\hat{\delta}(\x)$) to be orthogonal to the condensate \cite{Morgan98}, but this is not required by the derivation.}. Time-dependent formalisms in this number-conserving approach have been formulated both for low temperature \cite{Castin98,Gardiner97a} and moderate temperature regimes \cite{Gardiner07}.  The method of \cite{Gardiner07} requires only that the condensed atoms constitute the \emph{majority} of the field population, and allows in principle for (e.g.) the growth of the thermal component during the evolution.  However, the method is complicated significantly by the appearance of nonlocal terms which serve to maintain the orthogonality of the time-dependent condensed and noncondensed components of the field throughout their evolution, and has therefore only been applied in the linear-response regime so far \cite{Proukakis08}.  Furthermore, such an approach suffers again from the explicit separation of the field into condensed and noncondensed parts, precluding its use in strongly nonequilibrium and turbulent regimes of Bose-field dynamics. 
%%%%%%%%%%%%%%%%%%%%%%%%%%%%%%%%%%%%%%%%%%%%%%%%%%%%%%%%%%%%%%%%%%%%%%%%%%%%%%%%%%%%%%%%%%%%%%%%%%%%%%%%%%%%%%%%%%%%%%%%%%%%%%%%%%%%
\section{Superfluidity and vortices}
%%%%%%%%%%%%%%%%%%%%%%%%%%%%%%%%%%%%%%%%%%%%%%%%%%%%%%%%%%%%%%%%%%%%%%%%%%%%%%%%%%%%%%%%
\subsection{Superfluidity}\label{subsec:back_superfluidity}
The term `superfluid' is used to describe a complex of phenomena that can arise in \emph{quantum} fluids, that distinguish them from ordinary (or \emph{normal}) fluids \cite{Leggett99}.  The properties of a superfluid are fundamentally related to the presence of a single macroscopic quantum wave function in the fluid, and the nature of the excitations of the fluid relative to this quantum degenerate `ground state'.  Some of the characteristic features of superfluidity are a direct result of the flow being described by a quantum wave function, and are foreshadowed by results in single-particle quantum mechanics; for example, the motion of electrons in a rotating molecule `slips' with respect to the rigid rotation of the nuclei \cite{Wick48}. Superfluid many-body systems (such as superfluid helium) exhibit similar behaviour \cite{Hess67}.  Superfluidity is characterised by the appearance of a \emph{superfluid velocity} $\mathbf{v}_\mathrm{s}$, for which, from the transformation properties of the order parameter associated with condensation in the system, we deduce \cite{Pitaevskii03}
\begin{equation}\label{eq:back_superfluid_velocity}
	\mathbf{v}_\mathrm{s} = \frac{\hbar}{m}\nabla S,
\end{equation}
where $S(\x,t)$ is the \emph{quantum phase} of the order parameter (i.e., of the condensate). It is important to note however that while the phase of the condensate determines the velocity of the superflow, the mass current of the superflow is not, in general, directly related to the density of the condensate \cite{Pitaevskii03}; i.e., Bose-Einstein condensation and superfluidity are intimately related but distinct concepts.  
A conspicuous example of this is provided by superfluid $^4\mathrm{He}$, in which the condensate fraction is a mere $8\%$ at zero temperature, while the superfluid fraction is unity (see, e.g., \cite{Huang87}).  However in the limit of complete Bose condensation in weakly interacting gases the superfluid and condensate densities are approximately equal \cite{Leggett00}.
%%%%%%%%%%%%%%%%%%%%%%%%%%%%%%%%%%%%%%%%%%%%%%%%%%%%%%%%%
\subsubsection{Stability of superflow}
A crucial aspect of bulk superfluids is the stability of their flow, arising from the difficulty in creating excitations in the fluid, which is a direct consequence of the spectrum of excitations about the ground state.  This can be understood already from the Bogoliubov spectrum of a homogeneous Bose gas.  In such a scenario the condensate forms in the $k=0$ (i.e., constant) plane-wave mode, and the excitations are mixtures of plane waves, with the spectrum \cite{Castin01,Pitaevskii03}
\begin{eqnarray}
	\epsilon_k &=& \left[\frac{\hbar^2k^2}{2m}\left(\frac{\hbar^2k^2}{2m} + 2U_0n\right)\right]^\frac{1}{2}  \nonumber \\
			&\approx& \hbar k \sqrt{\frac{U_0n}{m}} \quad\mathrm{for}\quad \frac{\hbar^2k^2}{2m} \ll 2U_0n.
\end{eqnarray}
At long wavelengths, the spectrum, which is linear in $k$ is that of \emph{sound waves}, i.e., the wave velocity $v_k = d\omega_k / dk = \sqrt{U_0n/m} \equiv c_\mathrm{s}$ is independent of the wave number $k$.  Simple energy and momentum conservation arguments \cite{Pitaevskii80,Castin01,Pethick00} then show that an impurity moving through the fluid can not excite quasiparticle excitations of the condensate if its velocity $v < c_\mathrm{s}$, which is known as \emph{Landau's criterion} for superflow stability.  The motion of such an object through the fluid is thus not resisted by the superfluid component.  By a Galilean transformation one obtains the archetypal property of superfluidity: dissipationless flow of the fluid through a capillary \cite{Pitaevskii03}.  In general finite-temperature scenarios there is of course also a \emph{normal} component of the fluid, which can exert a drag force on the impurity or capillary wall.  In bulk scenarios, the two fluids also exert no friction on each other \cite{Pitaevskii03}, and can be described by a \emph{two-fluid model} \cite{Landau87}. This however breaks down in inhomogeneous superfluids such as the trapped condensates we consider in this thesis, where the superfluid can be excited at its `surface' (where the sound velocity vanishes with the superfluid density), and even in the homogeneous case in the presence of defects in the superfluid, such as quantum \emph{vortices}, which we discuss in section~\ref{subsec:back_vortices}.
%%%%%%%%%%%%%%%%%%%%%%%%%%%%%%%%%%%%%%%%%%%%%%%%%%%%%%%%%
\subsubsection{Landau criterion for surface modes of a trapped condensate}
A useful result for our investigations of rotating systems in this thesis is obtained by transposing the Landau criterion for the stability of superflow to the surface modes (section~\ref{subsec:back_GPE}) of a Bose condensate.  Making this transposition we obtain the critical angular frequency \cite{Dalfovo97}
\begin{equation}
	\Omega_\mathrm{L} = \min \left(\frac{\omega_m}{m}\right)\!,	
\end{equation}
where $\omega_m$ is the frequency of a surface mode with angular-momentum projection $m$.  It is important to note the physical meaning of this result: $\Omega_\mathrm{L}$ is the minimum rotation frequency at which a surface mode is shifted (by the inertial term $-\Omega L_z$ in the rotating-frame Hamiltonian) to a negative value.  That is, in a frame rotating at an angular frequency $\Omega > \Omega_\mathrm{L}$, the surface of the condensate is thermodynamically unstable with respect to the growth of surface modes.  As we will discuss in chapter~\ref{chap:stir_background}, this result is confirmed by calculations which explicitly include the dissipative effects of the thermal cloud \cite{Williams2002a,Penckwitt02}.
%%%%%%%%%%%%%%%%%%%%%%%%%%%%%%%%%%%%%%%%%%%%%%%%%%%%%%%%%%%%%%%%%%%%%%%%%%%%%%%%%%%%%%%%
\subsection{Vortices}\label{subsec:back_vortices}\enlargethispage{-\baselineskip}
Given a condensate wave function $\phi(\mathbf{x}) = |\phi(\mathbf{x})|e^{iS(\mathbf{x},t)}$, the velocity of the superflow is given by equation~\reff{eq:back_superfluid_velocity}.
An immediate consequence of this is that the superflow is irrotational, i.e., we must have $\nabla \times \mathbf{v}_\mathrm{s} = \mathbf{0}$.  Stokes' theorem then implies that the integral of the velocity $\oint_C \mathbf{v}_\mathrm{s} \cdot d \bm{\ell} =0$ around any closed path $C$ which encloses a simply-connected domain.  If the enclosed domain is not simply connected, then this result does not hold, and the only relevant constraint is that imposed by the single-valued nature of the phase, which is however only defined up to the periodicity ($2\pi$) of the complex phase.  We therefore find the Feynman-Onsager quantisation condition
\begin{equation}
	\oint_C  \mathbf{v}_\mathrm{s} \cdot d\bm{\ell} = n\kappa,
\end{equation}
where $\kappa\equiv h/m$ is the \emph{quantum of circulation}, and $n$ is some integer.
A condensate which is otherwise simply connected can therefore support a circulation by admitting a node about which the phase of the wave function wraps continuously from $0$ to $2n\pi$.  In two dimensions such a node is called a point vortex, in three dimensions the node has finite extent and forms a \emph{vortex line} (or vortex filament).  In this thesis, we will consider vortices only in quasi-two-dimensional systems, and we will therefore deal only with point vortices.  The fluid flow remains irrotational everywhere, as the phase and thus velocity are meaningless at the node, where the wave function vanishes, however, the vortex can be thought of as a point `charge' of rotation.  
%%%%%%%%%%%%%%%%%%%%%%%%%%%%%%%%%%%%%%%%%%%%%%%%%%%%%%%%%
\subsubsection{Dynamics of vortices}
In practice, vortices with circulation $|n|>1$ are energetically unstable to decay into $n$ singly charged vortices \cite{Donnelly91}.  Similarly, considering the energy of a collection of vortex lines, one can show that this energy takes the same form as the electrostatic energy of a collection of charged filaments \cite{Sonin87}, i.e., vortices of opposite sign exert an attractive force on one another, and vortices of like sign exert a repelling force on each other.  However, by Galilean invariance, one finds \cite{Fetter99} that vortices are carried along by the local flow of the superfluid, and thus by the flow of other vortices.  This somewhat counterintuitive relation between the force between the vortices and their motion is an example of the peculiar dynamics obeyed by vortices.  Hydrodynamic considerations yield the result that vortex moving (with velocity $\mathbf{v}_\mathrm{l}$) against a background superflow (velocity $\mathbf{v}_\mathrm{s}$) experiences the \emph{Magnus force} \cite{Donnelly91} $\mathbf{F}_\mathrm{m} = \rho_\mathrm{s} \bm{\kappa} \times (\mathbf{v}_\mathrm{l} - \mathbf{v}_\mathrm{s})$, where $\rho_\mathrm{s}$ is the density of the superfluid, and $\bm{\kappa}$ is a vector with magnitude $\kappa$ directed along the axis of circulation of the vortex.  As a result, the motion of the vortex subject to an applied external force $\mathbf{F}_\mathrm{ext}$ is governed by the law \cite{Sonin97}
\begin{equation}\label{eq:back_magnus}
	\rho_\mathrm{s} \left[\left(\mathbf{v}_\mathrm{l} - \mathbf{v}_\mathrm{s}\right) \times \bm{\kappa} \right] = \mathbf{F}_\mathrm{ext}.
\end{equation}
%%%%%%%%%%%%%%%%%%%%%%%%%%%%%%%%%%%%%%%
\begin{figure}
	\begin{center}
	\includegraphics[width=0.9\textwidth]{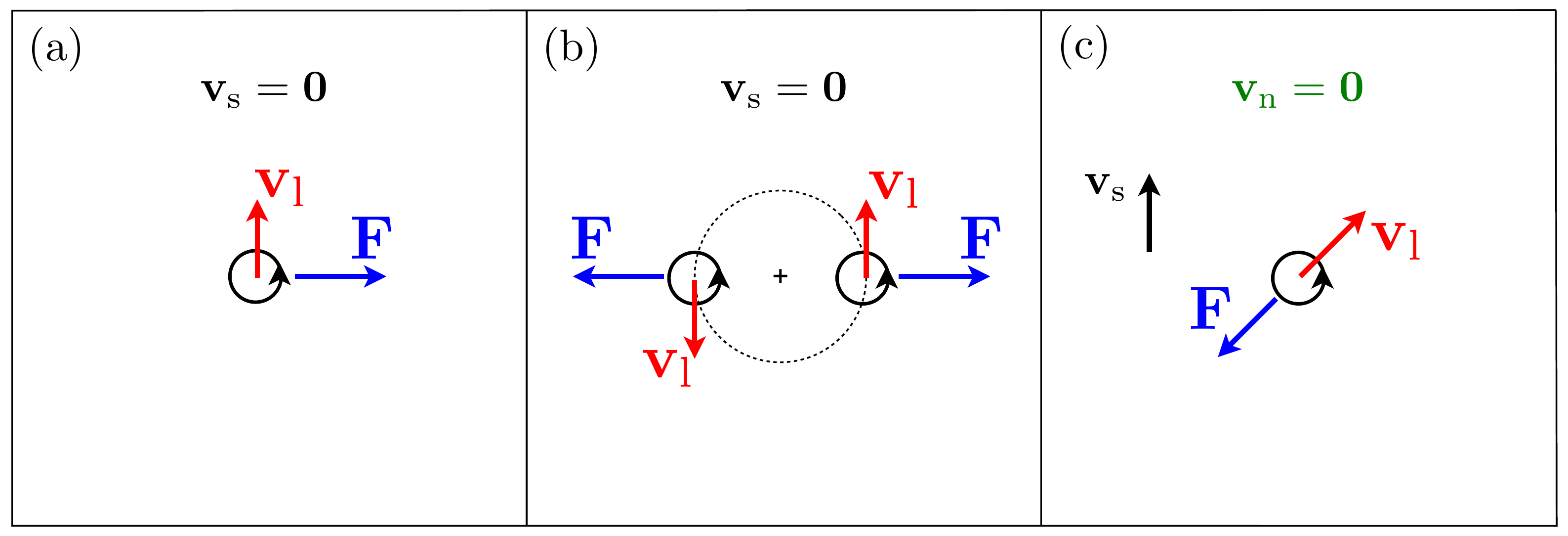}
	\caption{\label{fig:magnus_force} (a) Motion of a vortex with respect to the superfluid background: the vortex velocity is perpendicular to the imposed external force (see equation~\reff{eq:back_magnus}). (b) Motion of two like vortices.  Each vortex is advected by the flow of the other, and they therefore orbit about their geometric centre. (c) Effect of the longitudinal friction force on a vortex:  in the frame of the normal fluid, the vortex drifts at an angle to the background superflow.}
	\end{center}
\end{figure}
%%%%%%%%%%%%%%%%%%%%%%%%%%%%%%%%%%%%%%%
Equation~\reff{eq:back_magnus} shows that, in contrast to Newtonian dynamics in which the acceleration of an object is proportional to the force applied to it, the \emph{velocity} of a vortex relative to the superflow is proportional to the force acting on it \cite{Sonin87}\footnote{This non-Newtonian dynamical law has some amusing consequences.  For example, a system of three vortices in an unbounded flow in 2D is integrable, the motion only becoming chaotic upon the addition of a \emph{fourth} vortex \cite{Aref83}.}, as illustrated in figure~\ref{fig:magnus_force}(a).  Consequently two like vortices which energetically `repel' each other tend to orbit about their geometric centre (in the absence of dissipation), as illustrated in figure~\ref{fig:magnus_force}(b).  As a result of this energetic repulsion and tendency to orbit about one another, a superfluid to which a large rotation is imparted acquires a large array of singly charged vortices, and the lowest-energy configuration of such an array is a crystalline vortex \emph{lattice} (see section~\ref{subsec:back_vortices_in_conds}).

It is clear from this discussion that there is an analogy between vortices in a superfluid and charged particles (or filaments) moving in a background magnetic field.  Indeed there is a strong analogy between vortices in a (uniform) two-dimensional superfluid and a ($2+1$)-dimensional electrodynamics \cite{Arovas97}, with phonons playing the role of electromagnetic radiation.  Electrostatic and electrodynamic analogies are therefore a very powerful tool in understanding the behaviour of vortices in superfluid systems. 
%%%%%%%%%%%%%%%%%%%%%%%%%%%%%%%%%%%%%%%%%%%%%%%%%%%%%%%%%
\subsubsection{Vortices at finite temperature}
At a finite temperature, we must include the effects of thermal interactions on vortex motion.  The fundamental process of interest is the scattering of excitations by the vortex core \cite{Sonin97}, which produces a conventional (longitudinal) drag force on the vortex $F_\parallel = - D(\mathbf{v}_\mathrm{l} - \mathbf{v}_\mathrm{n})$, where $\mathbf{v}_\mathrm{n}$ is the velocity of the normal component, and $D$ is a (temperature dependent) coefficient characterising the strength of the friction.  As one might expect, this frictional force opposes the motion of the vortex, but the corresponding motion is of course determined by the Magnus law \reff{eq:back_magnus}.  For example, a vortex moving against a stationary normal fluid $\mathbf{v}_\mathrm{n}=\mathbf{0}$ moves at some angle to the background superflow, as illustrated in figure~\ref{fig:magnus_force}(c).  The existence and magnitude of a second \emph{transverse} force component due to the scattering of excitations remains a controversial issue \cite{Sonin97,Ao93,Flaig06}.  Such a force component would be nondissipative, and numerical calculations \cite{Berloff07,Jackson09} suggest that if this force is indeed nonzero, it is certainly small.

The interaction of vortices with the normal flow further complicates the dynamics of a realistic superfluid system, and the additional effects are included at a phenomenological level in the extended Hall-Vinen-Bekharevich-Khalatnikov theory \cite{Hall56}.  This model applies in the limit that the spacing between vortex lines is small, and the vorticity of the superfluid (and its interaction with the normal fluid) is treated as a continuum property of the superfluid.
%%%%%%%%%%%%%%%%%%%%%%%%%%%%%%%%%%%%%%%%%%%%%%%%%%%%%%%%%%%%%%%%%%%%%%%%%%%%%%%%%%%%%%%%
\subsection{Vortices in Bose condensates}\label{subsec:back_vortices_in_conds}
%%%%%%%%%%%%%%%%%%%%%%%%%%%%%%%%%%%%%%%%%%%%%%%%%%%%%%%%%
\subsubsection{Central-vortex states}
\begin{figure}
	\begin{center}
	\includegraphics[width=0.9\textwidth]{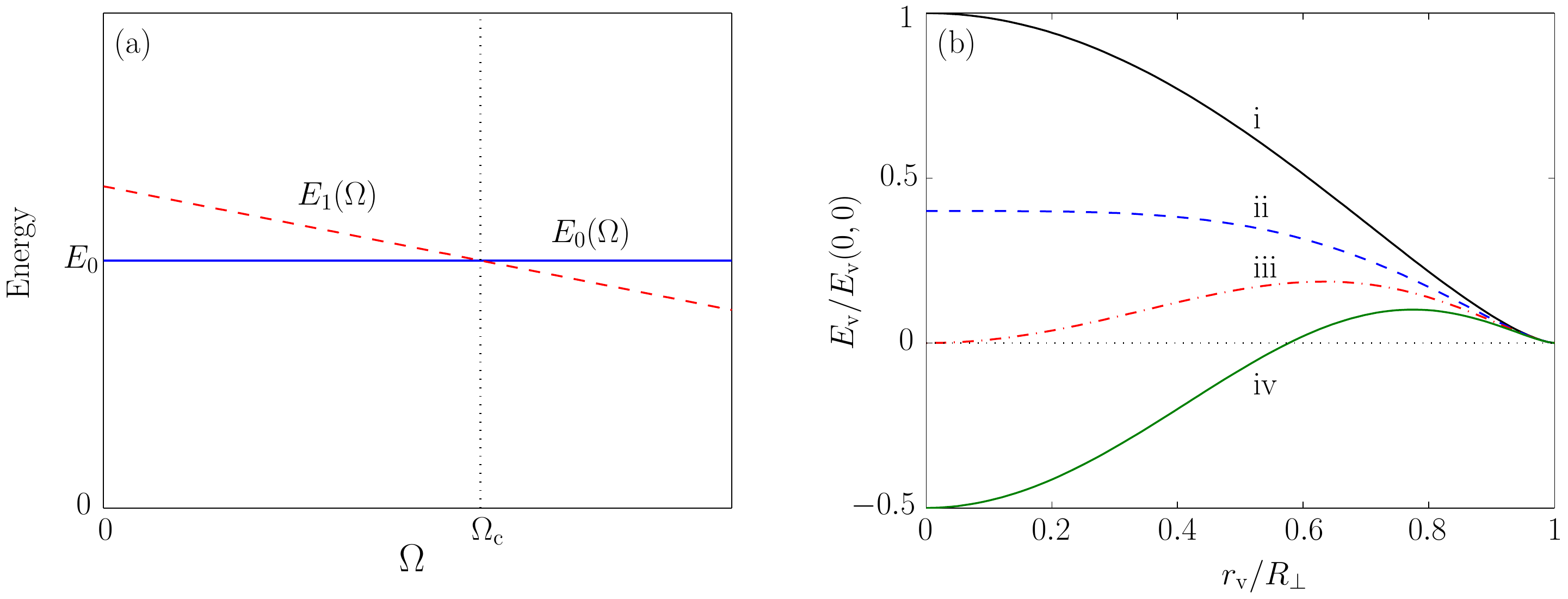}
	\caption{\label{fig:vortex_energy} (a) Rotating-frame energies of a generic vortex-free ground state and the corresponding central-vortex state.  The energy of the vortex state decreases with increasing frame angular velocity, and at a critical frequency $\Omega_\mathrm{c}$ its energy drops below that of the vortex-free state. (b) Energy of a vortex state as a function of the radial displacement of the vortex, evaluated in various rotating frames.  Cases are: (i) $\Omega=0$, thermodynamically unstable state; (ii) $\Omega=\Omega_\mathrm{m}$, onset of vortex metastability; (iii) $\Omega=\Omega_\mathrm{c}$ onset of thermodynamic stability of the vortex; (iv) $\Omega=(3/2)\Omega_\mathrm{c}$, stable state with energy barrier.}
	\end{center}
\end{figure}

A condensate in a cylindrically symmetric harmonic trap with a single, \emph{central}, singly charged vortex parallel to the $z$ axis has a Gross-Pitaevskii wave function of form
\begin{equation}
	\psi(\x) = e^{i\theta}|\psi(r_\perp,z)|,
\end{equation}
where $r_\perp=\sqrt{x^2+y^2}$ is the radial distance from the $z$ axis, and $\theta$ is the azimuthal angle around this axis.  In the noninteracting limit, this wave function is of form
\begin{equation}
	\psi_1(\x) = \frac{N}{\pi^{3/4} \sqrt{d_\perp^2 d_z}} e^{i\theta} r e^{-\frac{1}{2}(r_\perp^2/d_\perp^2 + z^2/d_z^2)},
\end{equation}
where $d_\perp=\sqrt{\hbar/m\omega_\perp}$ and $d_z=\sqrt{\hbar/m\omega_z}$ are the oscillator lengths corresponding to the confinement in the $r_\perp$ and $z$ directions, respectively.  In an inertial (laboratory) frame, this wave function has higher energy than that of the condensate consisting of $N$ atoms in the ground state of the trap, $\psi_0(\x)=[N/(\pi^{3/2} d_\perp^2 d_z)] \exp\{-\frac{1}{2}(r_\perp^2/d_\perp^2+z^2/d_z^2)\}$; i.e., it is an excited state.  However, in a rotating frame the energy of the vortex will be lowered due to its circulation.  We thus consider the relative energies of the two states in a frame rotating with angular velocity $\Omega$ about the $z$ axis.  In such a frame the GP energy is given by
\begin{equation}
	E_\Omega[\psi] = E - \Omega \langle L_\mathrm{z}\rangle = \int\!d\x\,\psi^* \left[-\frac{\hbar^2}{2m}\nabla^2 + V_\mathrm{ext} - \Omega L_z + \frac{U_0}{2}|\psi|^2\right]\psi.	 
\end{equation}
As the vortex-free ground state $\psi_0(\x)$ carries no circulation, its energy $E_0$ is the same in all rotating frames, while the energy $E_1$ of the vortex state $\psi_1(\x)$ is reduced from its lab-frame value by $N\hbar\Omega$ in the frame rotating at angular velocity $\Omega$.  As such there is a \emph{critical} angular velocity at which the energy of the vortex state becomes equal to that of the vortex-free state, and for higher angular velocities, the vortex state is a lower energy state than the vortex-free state (see figure~\ref{fig:vortex_energy}(a)).  However, in the noninteracting limit one can easily see that, as the lab-frame energy of the vortex state is $E_1=E_0+N\hbar\omega_\perp$, this crossover does not occur in the noninteracting case until $\Omega=\omega_\perp$, in which limit the trapping potential is completely cancelled by the centrifugal force resulting from the rotation \cite{Fetter01}.

In the opposite Thomas-Fermi limit of \emph{dominant} interactions, the vortical condensate has density approximately of the form \cite{Fetter01}
\begin{equation}
	n(r_\perp,z) \approx \frac{\mu}{U_0} \left(1-\frac{\eta^2}{r_\perp^2} - \frac{r_\perp^2}{R_\perp^2} - \frac{z^2}{R_z^2}\right)\Theta\!\left(1 - \frac{\eta^2}{r_\perp^2} - \frac{r_\perp^2}{R_\perp^2} - \frac{z^2}{R_z^2}\right)\!, 
\end{equation}
where $R_\perp,R_z$ are the appropriate Thomas-Fermi radii, and $\eta$ is the healing length.  As the state is a singly charged on-axis vortex, it carries angular momentum $L_\mathrm{v}=N\hbar$, and it can be shown that the lab-frame energy of this state is \cite{Lundh97}
\begin{equation}\label{eq:back_TF_vortex_energy}
	E_\mathrm{v} = \frac{4\pi}{3}\frac{n_0\hbar^2R_z}{m} \ln\left(\frac{0.67R_\perp}{\eta}\right)\!,
\end{equation}
where $n_0$ is the peak density of the corresponding vortex-free TF state, with healing length $\eta$.
Comparing the vortex-state energy equation~\reff{eq:back_TF_vortex_energy} with that of the corresponding vortex-free state, we find that the critical frequency at which the vortex state becomes the ground state is \cite{Lundh97}
\begin{equation}\label{eq:back_TF_critical_freq}
	\Omega_\mathrm{c} = \frac{5}{2} \frac{\hbar^2}{mR_\perp^2} \ln\left(\frac{0.67R_\perp}{\eta}\right)\!.
\end{equation}
We note that the vortex-state energy equation~\reff{eq:back_TF_vortex_energy} is obtained by assuming the healing length $\eta$ as short-range cutoff in the calculation of the kinetic energy due to the vortex flow pattern (cf. \cite{Ginzburg58}).  Numerical solution of the full Gross-Pitaevskii equation reveals that the true energy contribution of the vortex is somewhat higher \cite{Fetter01}.  Nevertheless equation~\reff{eq:back_TF_critical_freq} agrees well with numerically exact results \cite{Dalfovo96} for the critical frequency in the limit of large condensates \cite{Lundh97}.
%%%%%%%%%%%%%%%%%%%%%%%%%%%%%%%%%%%%%%%
%%%%%%%%%%%%%%%%%%%%%%%%%%%%%%%%%%%%%%%
%%%%%%%%%%%%%%%%%%%%%%%%%%%%%%%%%%%%%%%%%%%%%%%%%%%%%%%%%
\subsubsection{Off-axis vortices} 
So far we have shown the energy of the vortex at the centre of the trap is lower than that of the vortex-free state in a frame rotating above some critical angular velocity $\Omega_\mathrm{c}$.  Here we consider the approximate energy of the vortex state as a function of the displacement of the vortex from the trap centre, thus continuously connecting the central-vortex and vortex-free limits.   
A Thomas-Fermi limit calculation for the energy of state with vortex displaced off-axis by $r_\mathrm{v}$, in a frame rotating at angular velocity $\Omega$, leads to the result \cite{Fetter01,Kramer02}
\begin{equation}
	E_\mathrm{v}(r_\mathrm{v};\Omega) = \biggl[1 - \Bigl(\frac{r_\mathrm{v}}{R_\perp}\Bigr)^2\biggr]^\frac{3}{2} \biggl\{E_\mathrm{v}(0;0) - N\hbar\Omega \Bigl[1 - \Bigl(\frac{r_\mathrm{v}}{R_\perp}\Bigr)^2\,\Bigr]\biggr\},
\end{equation}
where $E_\mathrm{v}(0;0)$ is the lab-frame energy of the central-vortex state given by equation~\reff{eq:back_TF_vortex_energy}.
In figure~\ref{fig:vortex_energy}(b) we plot the results of this energy as a function of the vortex displacement, for several values of the rotating-frame angular velocity $\Omega$.  In the laboratory frame (case (i)), the energy of the state decreases monotonically with increasing vortex displacement, showing that the central vortex is truly thermodynamically unstable in this regime; in the presence of weak dissipation in this frame (i.e., in the presence of stationary thermal cloud), the dissipation serves to expel the vortex from the condensate.  In case (ii) we illustrate the form of the energy in a frame rotating at frequency 
\begin{equation}
	\Omega_\mathrm{m} = \frac{3}{2}\frac{\hbar}{mR_\perp^2}\ln\left(\frac{0.67R_\perp}{\eta}\right),
\end{equation}
at which the curvature of $E_\mathrm{v}(r_\mathrm{v})$ at $r_\mathrm{v}=0$ changes sign.  For greater angular velocities $\Omega > \Omega_\mathrm{m}$ the function $E_\mathrm{v}(r_\mathrm{v})$ is convex at $r_\mathrm{v}=0$, and thus in the presence of weak dissipation in this frame the vortex is pushed back towards the centre of the trap.  As the vortex energy at the trap centre is still higher than the energy of the vortex-free state, this frequency marks the onset of \emph{metastability} of the central-vortex state.  

In case (iii) we indicate the curve corresponding to the frequency $\Omega_\mathrm{c}$ given by equation~\reff{eq:back_TF_critical_freq} (note that  $\Omega_\mathrm{c} = (5/3)\Omega_\mathrm{m}$).  As we have discussed, beyond this frequency the energy of the central-vortex state is lower than that of the vortex-free state, i.e., the vortex state is the global energy minimum.  However, even for significantly higher frequencies (case (iv), $\Omega=(3/2)\Omega_\mathrm{c}$), we note the presence of an \emph{energy barrier} that must be surmounted by the vortex in order for it to migrate from the outside of the condensate to the centre.  Thus, even in the presence of weak dissipation, above the frequency $\Omega_\mathrm{c}$ additional mechanisms are necessary in order for the vortex to migrate to the trap centre, as we discuss in chapters~\ref{chap:stir_background} and \ref{chap:stirring}.  
%%%%%%%%%%%%%%%%%%%%%%%%%%%%%%%%%%%%%%%%%%%%%%%%%%%%%%%%%
\subsubsection{The anomalous mode}
An \emph{anomalous mode} is a Bogoliubov excitation of the GP state with \emph{negative energy}, the appearance of which signals the thermodynamical instability of the condensate (section~\ref{subsec:theory_Bogoliubov}).   As a central single-vortex state is thermodynamically unstable in the lab frame (figure~\ref{fig:vortex_energy}(b)), such an anomalous mode arises in the Bogoliubov expansion about such a state \cite{Dodd97}.  
In the noninteracting limit, this mode simply reduces to the ground state of the trap \cite{Rokhsar97}.  For the interacting case, in the Bogoliubov description, the $u$ and $v$ components of this anomalous mode have circulation $0$ and $4\pi$ respectively, indicating that the mode has angular momentum \cite{Mizushima04} $q=-1$ with respect to the central-vortex state.  The energy of this state (relative to that of the condensate) therefore increases with increasing angular velocity of the rotating frame, and reaches zero at the metastability frequency $\Omega_\mathrm{m}$, as we expect.  The energy of the anomalous mode of the central-vortex state therefore determines the frequency of precession of a vortex in the limit of \emph{infinitesimal} displacement of the vortex from the trap axis~\cite{Fetter01}.  
More generally, an anomalous mode appears for a \emph{finitely} off-axis vortex, and has an excitation energy related to the precession frequency of the vortex \cite{Linn00}.  However, this relationship has been shown not to hold in the extension to finite-temperature HFB theories \cite{Isoshima04,Wild09}.
%%%%%%%%%%%%%%%%%%%%%%%%%%%%%%%%%%%%%%%%%%%%%%%%%%%%%%%%%
\subsubsection{Vortex lattices}
As multiply charged vortices are unstable, the equilibrium configuration of a macroscopically rotating superfluid is expected to contain a proliferation of singly charged vortices.  In the limit of large angular momentum,
 we expect such a rotating superfluid to exhibit a \emph{coarse-grained} velocity profile which mimics that of solid-body rotation.  The latter is given by a velocity $\mathbf{v}_\mathrm{sb} = \bm{\Omega} \times \mathbf{x}$ at position $\mathbf{x}$, and yields a constant vorticity $\nabla \times \mathbf{v}_\mathrm{sb} = 2\bm{\Omega}$.  This prompted Feynman \cite{Feynman55} to estimate the density of vortices in a macroscopically rotating fluid as the ratio of this vorticity to the elemental vorticity $\kappa$, yielding
\begin{equation}
	n_\mathrm{v} = \frac{2\Omega}{\kappa} = \frac{m\Omega}{\pi\hbar},
\end{equation}
which is referred to as \emph{Feynman's rule} \cite{Donnelly91}.  Calculations performed by Tkachenko show that the minimum energy configuration is one in which the vortex lattice is \emph{triangular} \cite{Tkachenko66}, and in the literature such lattices are known as Abrikosov lattices, named for Abrikosov who predicted the analogous structure for magnetic flux lines in type-II superconductors \cite{Abrikosov57}.  We note, however, that calculations performed by Campbell and Ziff \cite{Campbell79} indicated that the rotational velocity field in fact leads to significant deviations from this ideal triangular structure in uniform, cylindrically confined superfluids.  It is at first somewhat surprising then that the lattices formed in dilute Bose condensates (e.g., \cite{Abo-Shaeer01,Haljan01a}) are so strikingly regular, especially considering that one might expect the vortex density to be lower at the centre of the trap, where the superfluid density and thus the kinetic energy associated with vortices is highest.   However, Sheehy and Radzihovsky \cite{Sheehy04,Sheehy04b} have shown in a careful analysis that this energetic effect is almost exactly balanced by the spatial dependence of the vortex `chemical potential' ($\propto \Omega\rho_\mathrm{s}$).  They obtain a general result for the vortex density, which in the TF limit (for isotropic or oblate trapping) reduces to the radius-dependent result
\begin{equation}
	n_\mathrm{v}(r) \approx \frac{m \Omega}{\pi \hbar} - \frac{1}{2\pi} \frac{R^2}{(R^2-r^2)^2}\ln \left(\frac{e^{-1}\hbar}{\eta^2\Omega m}\right),
\end{equation} 
where $R$ is the Thomas-Fermi radius, and $e$ is the base of the natural logarithm. 
%%%%%%%%%%%%%%%%%%%%%%%%%%%%%%%%%%%%%%%%%%%%%%%%%%%%%%%%%%%%%%%%%%%%%%%%%%%%%%%%%%%%%%%%%%%%%%%%%%%%%%%%%%%%%%%%%%%%%%%%%%%%%%%%%%%%
%%%%%%%%%%%%%%%%%%%%%%%%%%%%%%%%%%%%%%%%%%%%%%%%%%%%%%%%%%%%%%%%%%%%%%%%%%%%%%%%%%%%%%%%%%%%%%%%%%%%%%%%%%%%%%%%%%%%%%%%%%%%%%%%%%%%

\chapter{Classical-field theory}
\label{chap:cfield} 
%%%%%%%%%%%%%%%%%%%%%%%%%%%%%%%%%%%%%%%%%%%%%%%%%%%%%%%%%%%%%%%%%%%%%%%%%%%%%%%%%%%%%%%%%%%%%%%%%%%%%%%%%%%%%%%%%%%%%%%%%%%%%%%%%%%%
%%%%%%%%%%%%%%%%%%%%%%%%%%%%%%%%%%%%%%%%%%%%%%%%%%%%%%%%%%%%%%%%%%%%%%%%%%%%%%%%%%%%%%%%%%%%%%%%%%%%%%%%%%%%%%%%%%%%%%%%%%%%%%%%%%%%
In this chapter, we introduce the classical-field formalism we use throughout this thesis.  The term `classical-field theory' in fact refers to a \emph{family} of theoretical and computational formalisms for the degenerate Bose gas, with generally different regimes of applicability and utility, which have been recently reviewed in \cite{Blakie08}\footnote{We note that the authors of \cite{Blakie08} coined the term `c-field' to describe the collection of classical-field methods discussed in that work, so as to avoid the potential confusion associated with the phrase `classical field', which can give the misleading impression that these methods are not quantum mechanical.}.  However, they all share the feature of allowing for the description of many-body physics beyond the reach of a simple Gross-Pitaevskii treatment, while retaining the computational tractability of a description in terms of a single-particle state space.  In this thesis, we focus on the Hamiltonian (i.e., deterministic) dynamics of classical fields described by the \emph{projected Gross-Pitaevskii equation}, and in particular on the description of \emph{thermal} Bose fields using this formalism, in both equilibrium and nonequilibrium scenarios.  This chapter serves to explain the physical content and formulation of this particular methodology, and to place it in the broader context of classical-field methods for the degenerate Bose gas.

We begin this chapter in section~\ref{sec:cfield_basic} with a heuristic introduction to the basic notions of the classical-field treatment, and explain its validity and utility for the finite-temperature scenarios we will consider in this thesis.  In section~\ref{sec:cfield_projection} we motivate the low-energy \emph{projection} of the field that is a central aspect of the method, and present the derivation of the projected Gross-Pitaevskii equation from the second-quantised Hamiltonian for the dilute Bose gas previously introduced in chapter~\ref{chap:theory}.  In section~\ref{sec:cfield_ergodicity} we describe the ergodic character of solutions of the projected GP equation, and discuss the \emph{thermodynamics} of the classical field that emerge dynamically from the field trajectories.  Finally in section~\ref{sec:cfield_twa} we introduce the closely related \emph{truncated-Wigner} approximation for the degenerate Bose gas, and discuss its relation and relevance to the classical-field formalism we employ in this thesis. 
%%%%%%%%%%%%%%%%%%%%%%%%%%%%%%%%%%%%%%%%%%%%%%%%%%%%%%%%%%%%%%%%%%%%%%%%%%%%%%%%%%%%%%%%%%%%%%%%%%%%%%%%%%%%%%%%%%%%%%%%%%%%%%%%%%%%
\section{Nature of the classical-field approximation}\label{sec:cfield_basic}
At its simplest level, classical-field theory results from the replacement of a second-quantised (Bose) field Hamiltonian with an analogous classical field Hamiltonian.  This is motivated by the observation that in regimes of thermal behavior, the dominant characteristic of the atomic Bose-field system is its multimode, self-interacting nature, rather than the quantised nature of the atomic field.  Indeed,  under conditions of high mode occupation, field commutators become relatively unimportant, and a satisfactory description of the bosonic field can be obtained from a classical-field model.  It is important to note that while this replacement is superficially the same as the replacement of the field operator $\hat{\Psi}(\mathbf{x})$ by a classical (c-number) field in the Gross-Pitaevskii approximation (section~\ref{subsec:back_GPE}) the interpretation is very different. For example, the GP wave function $\phi(\mathbf{x})$ describes a condensate, which carries no entropy, while by contrast, the classical field in our models is typically a strongly fluctuating field, and through its fluctuations, describes \emph{thermal} behaviour of the field which is \emph{by definition} outside the regime of validity of the GPE.  

In general, we will think of the classical-field descriptions of the systems we study as approximations to the full second-quantised field theory.  However, we can also view the classical-field system as the `original' classical-dynamical system, from which the quantum field theory is obtained by quantising \emph{once} \cite{Peskin95}.  Taking this viewpoint, it makes sense to characterise the behaviour of this classical-field system first, before layering on the complexity of field quantisation.

\begin{figure}
	\begin{center}
	\includegraphics[width=0.9\textwidth]{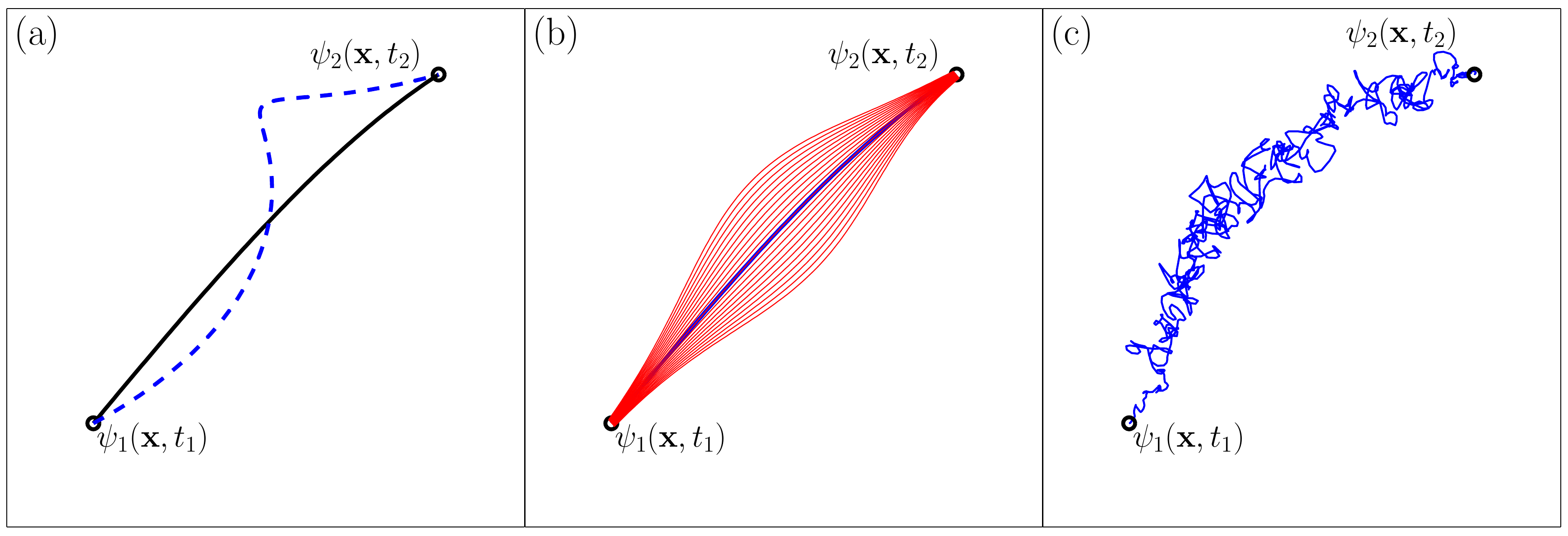}
	\caption{\label{fig:path_integral} (a) Evolution of a simple classical field in the Lagrangian viewpoint: the physical trajectory (solid line) is that for which the action is minimised. (b) Path integral interpretation of quantum field theory: all possible paths contribute to the probability of the transition from $\psi_1$ to $\psi_2$; strongly nonclassical trajectories interfere destructively. (c) Self-interacting classical field: the path of least action is itself strongly fluctuating, and samples many different field configurations during the classical evolution.}
	\end{center}
\end{figure}

This viewpoint also lends us more insight into the nature of the classical-field approximation, when we consider an alternative formulation of the quantum field theory: the \emph{path integral} formulation \cite{Feynman48,Zee03}.  In this formulation, the transition probability for the field to evolve from a configuration $\psi_1(\mathbf{x},t_1)$ at time $t_1$ to another configuration $\psi_2(\mathbf{x},t_2)$ at a later time $t_2$ is given by a sum (path integral) over phase factors $e^{iS[\psi(\mathbf{x}, t)]/\hbar}$, where $S[\psi(\mathbf{x}, t)]$ is the \emph{action} evaluated for each conceivable path between the end configurations (figures~\ref{fig:path_integral}(a-b)).  In this way, quantum fluctuations (the physical content of the commutation relations \reff{eq:back_bose_commutator}) are built into the theory.  The classical field trajectory is regained in the limit that the contribution of these quantum fluctuations to the path integral is small\footnote{In single-particle quantum mechanics this limit is of course $\hbar\to0$.  However, our classical field equation contains an explicit $\hbar$, so the limit here is more accurately regarded as that of infinite particle number \cite{Polkovnikov10}.}, in which case the path of least action (i.e., the classical trajectory) dominates the integral (figure~\ref{fig:path_integral}(a)).  In the situations we consider in this thesis, we evolve a \emph{single} classical trajectory of the field, which exhibits strong fluctuations (figure~\ref{fig:path_integral}(c)) driven by the kinetics of its own self-interaction.  This strongly fluctuating character is more significant than the `spread' of contributing paths included in the quantum theory, and so we do not consider the contributions of these other paths.  We thus obtain a \emph{nonperturbative} description of the field dynamics, at the cost of sacrificing the Bose commutation relations \reff{eq:back_bose_commutator} and the results which follow from them (such as the exact Bose-Einstein thermodynamic distribution in equilibrium).  Another important feature of the fluctuating classical trajectory of the field illustrated in figure~\ref{fig:path_integral}(c) is that we can extract meaningful statistics by averaging over the fluctuations \emph{along} the path of the field (rather than over multiple neighbouring paths).  In equilibrium this reflects the fact that the classical field is \emph{ergodic}, and samples from the appropriate thermal distribution over time, as we discuss in section~\ref{sec:cfield_ergodicity}.  However, this also provides a representation of thermal fluctuations of the field in scenarios away from equilibrium.  A strength of the classical-field method is that it implicitly includes the effect of these thermal fluctuations on the \emph{dynamics} of the field, thus providing a formalism for modelling \emph{finite-temperature nonequilibrium} behaviour of the atomic Bose field.
%%%%%%%%%%%%%%%%%%%%%%%%%%%%%%%%%%%%%%%
%%%%%%%%%%%%%%%%%%%%%%%%%%%%%%%%%%%%%%%
%%%%%%%%%%%%%%%%%%%%%%%%%%%%%%%%%%%%%%%%%%%%%%%%%%%%%%%%%%%%%%%%%%%%%%%%%%%%%%%%%%%%%%%%%%%%%%%%%%%%%%%%%%%%%%%%%%%%%%%%%%%%%%%%%%%%
\section{Projection}\label{sec:cfield_projection}
%%%%%%%%%%%%%%%%%%%%%%%%%%%%%%%%%%%%%%%%%%%%%%%%%%%%%%%%%%%%%%%%%%%%%%%%%%%%%%%%%%%%%%%%
\subsection{Motivation}\label{sec:cfield_motivation}
A crucial aspect of the classical-field formulation we use in this work is the \emph{projection} of the classical field onto a low-energy region.  In practice we simulate a field which differs from the second-quantised Bose field $\hat{\Psi}(\mathbf{x})$ not only in that its quantum-operator character is neglected, but also in that it comprises an expansion over a \emph{finite} number of single-particle basis modes, which are typically chosen as the lowest-energy eigenmodes of some single-particle Hamiltonian.  In this section we briefly outline some of the motivations for the restriction of the field in this manner, before introducing the projection formally in section~\ref{subsec:cfield_proj_cft}.
\newline\newline\emph{Numerical implementation of field theory}\newline
The first motivation is a practical one: a true `field theory' is a theory of the dynamics of a system with an infinite number of degrees of freedom, i.e., the field is obtained as the continuum limit of a (classical or quantum) dynamical system with a discrete set of dynamical variables \cite{Sudarshan74}.  Obviously, it is not in general possible to describe such a system with finite computational resources.  In practice, every computational realisation of field theory must be defined in terms of only some subset of the true degrees of freedom of the field theory, and so we are compelled to consider carefully which degrees of freedom we include in our numerical model and which we discard.  This is doubly important as naive discretisations of the theory can introduce serious pathologies into the dynamics, as we discuss further in section~\ref{sec:numerics_cfschemes}, when we compare the projected formalism to other common discretised formulations of the Gross-Pitaevskii theory.
\newline\newline\emph{Long-wavelength nature of the theory}\newline
The second motivation is that the quantum field theory corresponding to the Hamiltonian \reff{eq:back_cold_collision} can only be regarded as an \emph{effective field theory}, describing the behaviour of the atomic field on wavelengths which are long compared to some \emph{cutoff} length scale.  This is evidenced by the use of a contact ($s$-wave) interaction potential, which implicitly assumes we are only describing the dynamics of wavelengths of the field large compared to the range $r_0$ of the interatomic potential \cite{Gardiner03}, such that we can neglect the true structure of this potential.  Moreover, the contact potential is independent of the relative momentum, and thus describes the same interaction strength for all collisions, whereas physically the efficiency of scattering by the true inter-atomic potential is curtailed dramatically for relative momenta $k>1/a_s$.  Note that this issue is already present in the second-quantised Hamiltonian \reff{eq:back_2ndQH} from which we derive the Gross-Pitaevskii equation, but does not lead to problems in that case as we consider only the condensate mode, which is essentially the `longest-wavelength' mode\footnote{In particular, (repulsive) interactions act to `smooth' the GP mode so that it is slowly-varying in space, and contains only long-wavelength components.}, and for which the $s$-wave assumption holds true.   It does, however, lead to ultraviolet divergences in the self-consistent mean-field theories which generalise the GP description to finite temperatures.  Morgan \cite{Morgan00} has shown that by formally restricting the description of the atomic field to its long-wavelength components, the correct description of the interparticle interaction experienced by this field appears as precisely the `contact' potential used in the continuum theory, but this interaction potential is now formally regular due to the long-wavelength nature of the field, and this is the scenario we consider in this thesis.  Formally the scattering coefficient is then a cutoff-dependent parameter \cite{Morgan00,Sinatra02,Blakie08}, however in practice the correction due to the cutoff is small \cite{Morgan00,Norrie06a}, and we can simply use the experimentally measured value for $a_s$ in our models.  
\newline\newline\emph{Validity of the classical-field approximation}\newline
The third motivation is that by choosing the cutoff at an appropriate height in energy, it is possible (in some scenarios) to ensure that all single-particle modes in the classical-field description are highly occupied, i.e., that their classical occupations are $\gtrsim 1$, dominating the neglected `vacuum occupation' of the modes.  In such a regime the vacuum fluctuations have little effect on the dynamics and the classical-field approximation is formally justified.  At thermal equilibrium of the classical field in this limit, the thermodynamics of the classical field can be identified with the high-occupation limit of the true Bose-field system, and highly accurate \cite{Davis06} estimates for the correlations and thermodynamics of the quantum many-body system can be obtained by including a description of \emph{above-cutoff} atoms into the theory.  However, in general scenarios (away from equilibrium, for example), this restriction of the domain of the classical field so that all modes are highly occupied may not be practical, or even possible.  In this thesis, we focus on the \emph{intrinsic} phenomenology of the classical field system, and so we will frequently relax this requirement.  As a result, our results are not expected to be quantitatively accurate predictions for the true bosonic field theory.  Nevertheless, we find that the behaviour of the classical field yields great insight into the behaviour of the atomic Bose field, in scenarios which are essentially inaccessible to other contemporary finite-temperature formalisms, built on alternative approximations to the underlying many-body theory.
%%%%%%%%%%%%%%%%%%%%%%%%%%%%%%%%%%%%%%%%%%%%%%%%%%%%%%%%%
\subsubsection{Expression in single-particle energy levels}\enlargethispage{-\baselineskip}
As we have already noted, the way in which the atomic interaction is represented in our model requires us to restrict the field so as to exclude high-momentum components.  However, in the systems we consider in this thesis, the atomic field is \emph{confined} by an external trapping potential, which breaks the translational symmetry of the system.  Momentum is thus no longer a conserved quantity of the system, and so it is not appropriate to impose the classical-field cutoff in such terms.

In typical conditions, we expect all single-particle modes of the classical field below the cutoff to be populated, including those which `border' the cutoff.  Due to the complexity of the interacting field theory, requisite conditions for the cutoff are best formulated by considering thermal equilibrium (or quasi-equilibrium) scenarios.  In such cases, we expect the transfer of population to obey the principle of detailed balance \cite{Reif65}: the transfer of population from a mode $j$ to a mode $k$ is exactly balanced by the transfer of population from $k$ to $j$.  By carefully applying a cutoff we enforce this behaviour for transfer between modes on opposite sides of the cutoff: detailed balance is ensured in the trivial sense that the transfer in both directions is identically zero.  Applied correctly, this cutoff should then have no effect on the dynamics, at least close to equilibrium.  However, it is important that the cutoff is defined in a way which respects the symmetries of the system as otherwise the resulting equilibrium of the below-cutoff field will not be, in general, consistent with an equilibrium of the full system.  In such cases the effective detailed-balance condition resulting from the cutoff will be spurious.  For this reason, the cutoff should be defined in terms of the \emph{energy} of the system, so that the cut can be made with minimal disruption to the dynamics and thermodynamics of the below-cutoff field.  Equivalently, we require that the constant-energy surface of the cutoff system most closely matches a constant-energy surface of the full interacting system. 

Of course, in general the surface of constant energy for the interacting system will be a very complicated function of the field variables, which is impossible to calculate.  However, we do know that for an equilibrium system containing a condensate, the energy of the system can be approximately diagonalised by the Bogoliubov transformation (section~\ref{subsec:theory_Bogoliubov}).  Moreover, we know that at high energies the Bogoliubov modes return to single particle states, i.e. $v_k \rightarrow 0$ as $\epsilon_k\to\infty$ \cite{Dalfovo99}, and in this limit the $u_k$ approach the energy eigenstates of the appropriate single-particle Hamiltonian (with shifted energies).  In fact, in harmonic trapping, for energies $\gtrsim 3\mu$, where $\mu$ is the chemical potential of the condensate, the energies of the Bogoliubov modes and those of the single-particle energy eigenstates agree to within about $3\%$ \cite{Gardiner97b}.  At such energies the presence of the condensate simply results in a small upwards shift of the single-particle mode energies.  Therefore, effecting the cutoff in terms of the single-particle energy in the trapping potential, we can enforce a cutoff which is \emph{approximately} diagonal in the energy of the interacting system.  In practice, this cutoff is implemented by an explicit \emph{projection} onto single-particle eigenstates with energies below the cutoff.
%%%%%%%%%%%%%%%%%%%%%%%%%%%%%%%%%%%%%%%%%%%%%%%%%%%%%%%%%%%%%%%%%%%%%%%%%%%%%%%%%%%%%%%%
\subsection{Projected classical-field theory}\label{subsec:cfield_proj_cft}
%%%%%%%%%%%%%%%%%%%%%%%%%%%%%%%%%%%%%%%%%%%%%%%%%%%%%%%%%
\subsubsection{The classical-field energy functional}
%%%%%%%%%%%%%%%%%%%%%%%%%%%%%%%%%%%%%%%
\begin{figure}
	\begin{center}
	\includegraphics[width=0.5\textwidth]{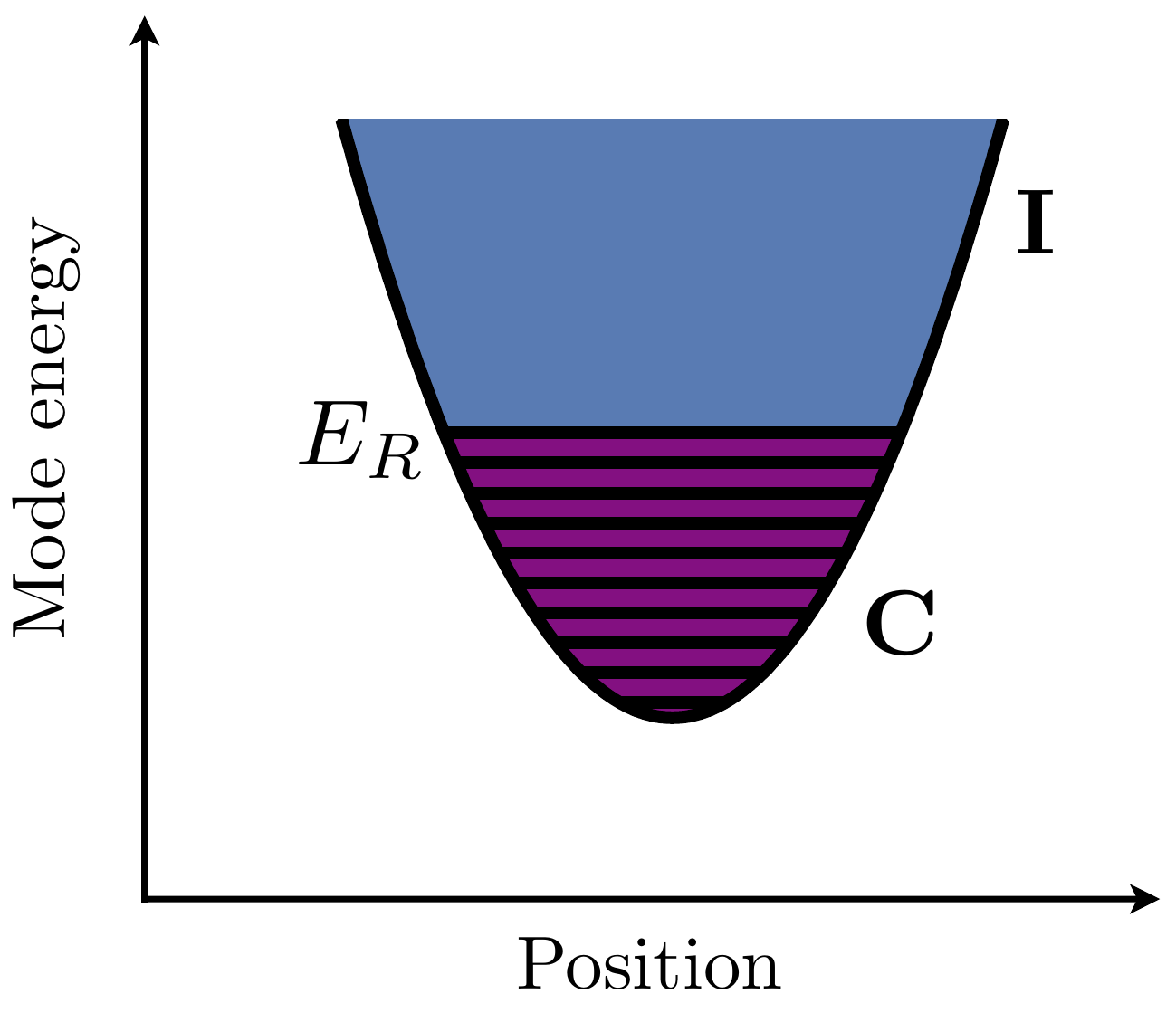}
	\caption{\label{fig:harmonic} Schematic view of the coherent region (condensate band) $\mathbf{C}$ and incoherent region (noncondensate band) $\mathbf{I}$ for a harmonic trap.}
	\end{center}
\end{figure}
%%%%%%%%%%%%%%%%%%%%%%%%%%%%%%%%%%%%%%%
In our projected classical-field theory, we introduce a cutoff energy $E_R$ to separate the system into a low-energy \emph{coherent region} ($\mathbf{C}$) consisting of single-particle modes energies satisfying $\epsilon_n<E_R$, and a high-energy \emph{incoherent region} ($\mathbf{I}$) containing the remaining modes, as indicated in figure~\ref{fig:harmonic}.  These regions are also sometimes referred to as the condensate and noncondensate \emph{bands}, respectively \cite{Gardiner02,Gardiner03,Bradley08}.  Where possible, it is most convenient to carry out the separation in the basis which diagonalises the single-particle Hamiltonian of the system, since at sufficiently high energies ($\epsilon_n\sim E_R$ \cite{Gardiner97b,Gardiner03}) the many-body Hamiltonian \reff{eq:back_cold_collision} is approximately diagonal in this basis, as discussed in section~\ref{sec:cfield_motivation}.  For certain potentials, an appropriate numerical quadrature method exists which allows for the efficient integration of the resulting equation of motion.  More generally, the single-particle Hamiltonian is divided into a \emph{base} Hamiltonian and a remaining potential part which is possibly time dependent
\begin{equation}\label{eq:cfield_split}
	H_\mathrm{sp} = H_0 + \delta V(\mathbf{x},t).
\end{equation}
The total potential in general comprises a contribution $V_0(\mathbf{x})$ from the base Hamiltonian and the perturbation potential, i.e. $V(\mathbf{x},t) = V_0(\mathbf{x}) + \delta V(\mathbf{x},t)$, while the base Hamiltonian may contain inertial terms associated with the transformation to a rotating frame. 

The validity of implementing the cutoff in terms of the single-particle energy corresponding to $H_0$ requires that $\delta V(\mathbf{x},t)$ be a small perturbation to $H_0$ (so that, e.g. the discrepancy between the constant-energy surfaces of $H_0$ and $H_\mathrm{sp}$ at the cutoff is of the same order as that between the single-particle and Bogoliubov Hamiltonians).
We then introduce the projected field operator
\begin{equation}
	\hat{\psi}(\mathbf{x}) \equiv \sum_{n \in \mathbf{C}}\hat{a}_n\phi_n(\mathbf{x}) ={\cal P}\hat{\Psi}(\mathbf{x}), 
\end{equation}
where the projector is defined by
\begin{equation}\label{eq:cfield_Pdef}
	{\cal P}f(\mathbf{x})\equiv\sum_{n \in \mathbf{C}}\phi_n(\mathbf{x})\int\!d\mathbf{x}'\, \phi_n^*(\mathbf{x}')f(\mathbf{x}').
\end{equation}
The summation is over all modes satisfying $\epsilon_n\leq E_R$, where $H_0\phi_n(\mathbf{x})=\epsilon_n\phi_n(\mathbf{x})$ defines the modes in terms of the base Hamiltonian.  The number of single-particle modes included in the coherent region is the \emph{condensate-band multiplicity}, denoted by $\mathcal{M}$.  The projected field operator thus obeys the commutation relation
\begin{equation}
	\left[\hat{\psi}(\mathbf{x}),\hat{\psi}^\dagger(\mathbf{x}')\right] = \delta_\mathcal{P}(\mathbf{x},\mathbf{x}'),
\end{equation}  
where the \emph{projected delta function}
\begin{equation}
	\delta_\mathcal{P}(\mathbf{x},\mathbf{x}') = \sum_{n\in\mathbf{C}}\phi_n(\mathbf{x})\phi_n^*(\mathbf{x}').
\end{equation}

In the $s$-wave limit, the effective low energy Hamiltonian for the system is then given by
\begin{equation}\label{eq:cfield_Heff}
	H_{\mathrm{eff}}=\int\!d\mathbf{x}\,\hat{\psi}^\dag(\mathbf{x}) H_{\mathrm{sp}} \hat{\psi}(\mathbf{x}) +\frac{U_0}{2} \hat{\psi}^\dag(\mathbf{x})  \hat{\psi}^\dag(\mathbf{x}) \hat{\psi}(\mathbf{x})\hat{\psi}(\mathbf{x}).
\end{equation}
We now obtain a classical theory by making the replacement 
\begin{equation}
	\hat{\psi}(\mathbf{x})\to \psi(\mathbf{x})\equiv \sum_{n\in \mathbf{C}} \alpha_n \phi_n(\mathbf{x}),
\end{equation}
in $H_{\mathrm{eff}}$ which yields the classical-field \emph{energy functional}
\begin{equation}\label{eq:cfield_HCF}
	H_{\mathrm{CF}}=\int\!d\mathbf{x}\,\psi^*(\mathbf{x}) H_{\mathrm{sp}} \psi(\mathbf{x}) +\frac{U_0}{2} |\psi(\mathbf{x})|^4 \equiv E[\psi].
\end{equation}
%%%%%%%%%%%%%%%%%%%%%%%%%%%%%%%%%%%%%%%%%%%%%%%%%%%%%%%%%
\subsubsection{Classical mechanics interpretation}
The energy functional~\reff{eq:cfield_HCF} is in fact a classical Hamiltonian, from which we can derive the evolution of the projected field $\psi(\mathbf{x},t)$.  To see this, we write $H_\mathrm{CF}$ explicitly in terms of the classical-field amplitudes $\{\alpha_j,\alpha_j^*\}$, and obtain
\begin{eqnarray}
	H_\mathrm{CF} &=& \sum_{n\in\mathbf{C}} \epsilon_n \alpha_n^*\alpha_n + \sum_{mn\in\mathbf{C}} \int\!d\mathbf{x}\,\phi_m^*(\mathbf{x}) \delta V(\mathbf{x},t) \phi_m(\mathbf{x})\alpha_m^*\alpha_n \nonumber \\
&&+ U_0\!\sum_{mnpq\in\mathbf{C}}\int\!d\mathbf{x}\,\phi_m^*(\mathbf{x})\phi_n^*(\mathbf{x})\phi_p(\mathbf{x})\phi_q(\mathbf{x}) \alpha_m^*\alpha_n^*\alpha_p\alpha_q.
\end{eqnarray}
By analogy to the relation between the quantum ladder operators $\{\hat{a}_j,\hat{a}_j^\dagger\}$ and the (abstract) position and momentum operators $\{\hat{x}_j,\hat{p}_j\}$ of the individual field modes, we introduce canonical position and momentum variables \cite{Davis03,Blakie08}
\begin{eqnarray}\label{eq:cfield_canon_varQ}
	Q_n &=& \frac{1}{\sqrt{2\epsilon_n}}(\alpha_n+\alpha_n^*) \\ 
	P_n &=& i\sqrt{\frac{\epsilon_n}{2}}(\alpha_n^*-\alpha_n). \label{eq:cfield_canon_varP}
\end{eqnarray}
The usual Hamilton's equations \cite{Goldstein50,Sudarshan74} for $Q_n$ and $P_n$ then yield $i d \alpha_n / dt = \partial H/\partial \alpha_n^*$, and the complex-conjugate expression.  Introducing the \emph{projected functional derivative} operator \cite{Gardiner03,Blakie08}
\begin{eqnarray}\label{eq:cfield_func_deriv}
	\frac{\bar{\delta}}{\bar{\delta}\psi(\mathbf{x})} &=& \sum_{n\in\mathbf{C}} \phi_n^*(\mathbf{x})\frac{\partial}{\partial\alpha_n}, 
\end{eqnarray}
and its conjugate, 
the evolution equations for the classical-field amplitudes can be summarised by the functional equation 
\begin{equation}
	i\hbar\frac{\partial\psi(\mathbf{x})}{\partial t} = \frac{\bar{\delta}H_{\mathrm{CF}}}{\bar{\delta}\psi^*(\mathbf{x})},
\end{equation}
for the projected classical field.
The resulting equation of motion
\begin{equation}\label{eq:cfield_pgpe} 
	i\hbar\frac{\partial\psi(\mathbf{x})}{\partial t}={\cal P}\Bigl\{\bigl(H_{\mathrm{sp}}+U_0|\psi(\mathbf{x})|^2\bigr)\psi(\mathbf{x})\Bigr\},
\end{equation}
is called the \emph{projected} Gross-Pitaevskii equation (PGPE) (see, e.g., \cite{Blakie08}).  
%%%%%%%%%%%%%%%%%%%%%%%%%%%%%%%%%%%%%%%%%%%%%%%%%%%%%%%%%
\subsubsection{Conservation laws and Ehrenfest relations}
We now remark on some conservation laws satisfied by the evolution of the PGPE.  
In general, the there will be multiple \emph{first integrals} of the PGPE evolution.  The first integrals are functionals of the field which are conserved by the Hamiltonian evolution.  In general applications of the PGPE, there are always at least two such functionals: the field energy $E[\psi]$ and normalisation $N[\psi]$.  More generally, there may be other conserved quantities which are conjugate to symmetries of the classical-field Hamiltonian~\reff{eq:cfield_HCF} (e.g., conserved angular momentum in the case of a rotationally invariant single-particle Hamiltonian).  

Here we discuss the appearance of the normalisation and energy first integrals in the PGPE formalism.  First, we note that the Hamiltonian~\reff{eq:cfield_HCF} is invariant under (global) phase rotations $\psi(\mathbf{x})\to\psi(\mathbf{x})e^{i\theta}$.  The group of such transformations is generated by the classical-field normalisation $N=\int\!d\mathbf{x}\,|\psi(\mathbf{x})|^2$, which is therefore conserved \cite{Goldstein50}.  Introducing the projected functional Poisson bracket \cite{Blakie08}
\begin{equation}
	\{F,G\} = \int\!d\mathbf{x}\left[\frac{\bar{\delta}F}{\bar{\delta}\psi(\mathbf{x})}\frac{\bar{\delta}G}{\bar{\delta}\psi^*(\mathbf{x})} - \frac{\bar{\delta}F}{\bar{\delta}\psi^*(\mathbf{x})}\frac{\bar{\delta}G}{\bar{\delta}\psi(\mathbf{x})}\right]\!, 
\end{equation}
it can be shown explicitly that $\{N,H_\mathrm{CF}\}=0$.  This conservation of the field normalisation as a result of the invariance of the Hamiltonian $H_\mathrm{CF}$ under such phase transformations is the classical precursor to the `gauge' symmetry of the quantised field discussed in section~\ref{subsubsec:symmetry_breaking}.
Furthermore, if the potential $\delta V(\mathbf{x})$ is time-independent it follows immediately \cite{Goldstein50,Sudarshan74} that the energy~\reff{eq:cfield_HCF} of the classical field is conserved by the PGPE evolution.  

We can also derive \emph{Ehrenfest} relations for the evolution of `expectation values' of single-particle operators, obtained by averaging over the (instantaneous) classical-field configuration
\begin{equation}
	\overline{A} \equiv \int\!d\mathbf{x}\,\psi^*(\mathbf{x}) A \psi(\mathbf{x}).
\end{equation}
The standard (non-projected) Gross-Pitaevskii equation satisfies the conventional Ehrenfest relations for the quantities of immediate interest: the position $\mathbf{x}$, momentum $\mathbf{p}$ and angular momentum $\mathbf{L}$ (see, e.g., \cite{Caradoc00}).  The corresponding results for the PGPE also contain corrections due to the projection operation, which are most straightforwardly derived by rewriting the PGPE as \cite{Bradley05,Bradley_PhD}
\begin{equation}
	i\hbar\frac{\partial \psi}{\partial t} = (1 - \mathcal{Q}) \Bigl\{\bigl(H_{\mathrm{sp}}+U_0|\psi(\mathbf{x})|^2\bigr)\psi(\mathbf{x})\Bigr\},
\end{equation}
where we have introduced the projector $\mathcal{Q}$ which is the orthogonal complement of $\mathcal{P}$, i.e.
\begin{equation}
	\mathcal{Q}f(\mathbf{x}) \equiv (1-\mathcal{P})f(\mathbf{x}) = \sum_{n\in \mathbf{I}} \phi_n(\mathbf{x}) \int\!d\mathbf{x}'\,\phi_n^*(\mathbf{x}')f(\mathbf{x}').
\end{equation}
We then find the generalised Ehrenfest relations 
\begin{eqnarray}\label{eq:cfield_ehrenfest_x} 
	\frac{d \overline{\mathbf{x}}}{dt} &=& \frac{\overline{\mathbf{p}}}{m} + Q_\mathbf{x} \\ \label{eq:cfield_ehrenfest_p}
	\frac{d \overline{\mathbf{p}}}{dt} &=& - \overline{\nabla V(\mathbf{x},t)} + Q_\mathbf{p} \\ \label{eq:cfield_ehrenfest_L} 
	\frac{d \overline{\mathbf{L}}}{dt} &=& -\frac{i}{\hbar} \overline{\mathbf{L} V(\mathbf{x},t)} + Q_\mathbf{L} \\ \label{eq:cfield_ehrenfest_H} 
	\frac{d H_\mathrm{CF}}{dt} &=& \overline{\frac{\partial V(\mathbf{x},t)}{\partial t}},
\end{eqnarray}
where the \emph{boundary terms} are
\begin{equation}\label{eq:cfield_QA}
	Q_A = \frac{2}{\hbar} \mathrm{Im}\Bigl[\big(\delta V(\mathbf{x},t) + U_0|\psi(\mathbf{x})|^2\bigr)\mathcal{Q}\left\{A\psi(\mathbf{x})\right\}\Bigr].
\end{equation}
It is clear from~\reff{eq:cfield_QA} that the boundary terms vanish when $[\mathcal{Q},A]\equiv-[\mathcal{P},A]=0$, in which case we regain the Ehrenfest relations of the continuum field theory.  This result will be of particular importance when we apply the PGPE to rotating systems in later chapters.
%%%%%%%%%%%%%%%%%%%%%%%%%%%%%%%%%%%%%%%%%%%%%%%%%%%%%%%%%%%%%%%%%%%%%%%%%%%%%%%%%%%%%%%%%%%%%%%%%%%%%%%%%%%%%%%%%%%%%%%%%%%%%%%%%%%%
\section{Ergodicity}\label{sec:cfield_ergodicity}
The PGPE \reff{eq:cfield_pgpe} is, fundamentally, an equation of motion which describes the \emph{dynamics} of the classical field $\psi(\mathbf{x},t)$.  However, an important feature of the dynamical system described by the Hamiltonian~\reff{eq:cfield_HCF} is that it is \emph{ergodic} \cite{Lebowitz73}, which is to say 
that trajectories of the PGPE sample the microcanonical density of the field densely.  For given values $\{I_i\}$ of the first integrals $\{F_i\}$ of the PGPE system, the microcanonical density is the probability distribution over field configurations which has a constant nonzero value on the sub-manifold defined by $F_i = I_i$, and is zero elsewhere.  
In the simple case in which $E[\psi]$ and $N[\psi]$ are the only first integrals, with values $E_0$ and $N_0$ respectively, the microcanonical density is
\begin{equation}\label{eq:cfield_mu_density}
	P_\mu[\psi;E_0;N_0] = \left\{
	\begin{array}{ll}
		\!\mathrm{const} &\qquad E[\psi] = E_0,\; N[\psi] = N_0 \\
		\!0 &\qquad \mathrm{else},  
	\end{array} \right.  % \}
\end{equation}
where the constant is chosen such that the integral of the density over all field configurations is unity.  
In principle, all equilibrium properties of the microcanonical classical-field system are given by averages in the density \reff{eq:cfield_mu_density}.  As the evolution is ergodic, almost any initial condition with $E[\psi]=E_0$ and $N[\psi]=N_0$ will evolve to a configuration (microstate) which is \emph{typical} of the appropriate thermal equilibrium of the field\footnote{We use the phrase `almost any' to signify that there are some field configurations (a set of measure zero \cite{Sethna06}) which will not evolve in an ergodic manner.  An important case is initial conditions which exhibit a symmetry which is preserved by the Hamiltonian evolution.}.  After this \emph{equilibration} of the field, the field essentially samples from the density~\reff{eq:cfield_mu_density} uniformly over time.  This provides a powerful methodology for calculating the equilibrium properties of the classical field: we simply allow the system to attain its own equilibrium, and then \emph{observe} its equilibrium behaviour directly.

It is worth elaborating somewhat on the nature of this equilibration, as the way this apparently irreversible behaviour arises from the purely deterministic equation~\reff{eq:cfield_pgpe} is a potential source of confusion.  An initial field configuration that we choose may appear far from the typical configurations of the field we observe in equilibrium.  However, in equilibrium the system will eventually fluctuate so that all such unlikely configurations occur, as their probability in the microcanonical density is of course the same as any other microstate consistent with $P_\mu[\psi]$ \footnote{They are \emph{unlikely} configurations, however, in the sense that their number of such configurations (or more accurately the volume of microcanonical density that they occupy) is comparatively small.}.  The subsequent `relaxation' we observe is simply the system continuing with its evolution after such a fluctuation, and is formally reversible.  In particular, the information describing the initial state is preserved in the precise details of the fluctuations of the field once it has reached equilibrium \cite{Connaughton05}, and the initial state can be regained by reversing the evolution of the system.  However, in practical applications this reversibility is inevitably lost after some time due to the buildup of numerical error \cite{Lebowitz73}.
%%%%%%%%%%%%%%%%%%%%%%%%%%%%%%%%%%%%%%%%%%%%%%%%%%%%%%%%%%%%%%%%%%%%%%%%%%%%%%%%%%%%%%%%
\subsection{Correlation functions}\label{subsec:cfield_correl}
A particularly useful application of the ergodicity of the PGPE system is the calculation of \emph{correlation functions} of the classical field, which take the form of averages in the microcanonical density $P_\mu[\psi]$ (which we denote by $\langle\cdots\rangle_\mu$) of a functional $F[\psi(\mathbf{x})]$ of the classical field.  Such averages provide various characterisations of the properties of the classical field in equilibrium, and are analogous to the quantum-statistical-mechanical expectations of field variables obtained by tracing over the quantum density matrix $\langle F[\hat{\psi}(\mathbf{x})]\rangle = \mathrm{Tr}\{\hat{\rho}F[\hat{\psi}(\mathbf{x})\}$.  The (hypothesised) ergodicity is formalised by the identification of the microcanonical average $P_\mu[\psi]$ with the infinite-time limit of an average along any given trajectory of the field \cite{Lebowitz73}:
\begin{equation}\label{eq:cfield_ergodic_avg}
	\langle F[\psi(\mathbf{x})] \rangle_\mu = \lim_{\theta\to\infty} \left\{\frac{1}{\theta} \int_0^\theta\!dt\,F[\psi(\mathbf{x},t)]\right\}.
\end{equation}
This is a very useful identity, as identifying the full set of classical-field configurations which satisfy the first-integral constraints is almost certainly impossible, due to the nonlinearity of the classical-field energy functional~\reff{eq:cfield_HCF} \cite{Blakie08}.  In practice, we average the functional $F[\psi(\mathbf{x},t)]$ over a finite period of field evolution.  The trajectories of the PGPE thus provide for a kind of \emph{Monte Carlo} sampling of the microcanonical density of the field, with the distinction that the `randomness' of the sampling arises from the ergodicity of the deterministic trajectories, rather than any explicit stochastic element.  Intuitively, we expect the results of the averaging to be accurate estimates of the corresponding microcanonical averages, provided that we average over a period which is long compared to the slowest time scale in the problem (at equilibrium), which is of the order of the slowest harmonic trapping period.
%%%%%%%%%%%%%%%%%%%%%%%%%%%%%%%%%%%%%%%%%%%%%%%%%%%%%%%%%
\subsubsection{Field covariance}
\begin{sloppypar}
A correlation function of the classical field of particular interest is the \emph{covariance matrix}
\begin{equation}\label{eq:cfield_density_mtx} 
	G(\mathbf{x},\mathbf{x}') = \langle \psi^*(\mathbf{x})\psi(\mathbf{x}')\rangle_\mu.
\end{equation}
This quantity is the obvious classical-field analogue of the one-body density matrix $\rho_1(\mathbf{x},\mathbf{x}')$ introduced in section~\ref{subsec:back_bose_condensation}\footnote{In light of this analogy, in the classical-field context we will sometimes simply refer to the covariance matrix $G(\mathbf{x},\mathbf{x}')$ as the one-body density matrix.}.  In particular, $G(\mathbf{x},\mathbf{x}')$ is, like $\rho_1(\mathbf{x},\mathbf{x}')$, Hermitian, and can be written in the diagonal form
\begin{equation}
	G(\mathbf{x},\mathbf{x}') = \sum_i n_i \chi_i^*(\mathbf{x})\chi_i(\mathbf{x}'),
\end{equation}
where the real constants $\{n_j\}$, indexed in order of decreasing magnitude, are the occupations of the corresponding modes $\{\chi_j(\mathbf{x})\}$.  By analogy to the Penrose-Onsager \cite{Penrose56} definition of condensation in the quantum-many-body system (section~\ref{subsec:back_bose_condensation}), condensation in the classical field is signaled by the most highly occupied mode $\chi_0(\mathbf{x})$ having an occupation $n_0$ which is significantly larger than all other occupations $n_i$.  This definition in terms of correlations in the microcanonical density is an unambiguous measure of condensation in the simple equilibrium regimes in which it is applicable.   
\end{sloppypar}
%%%%%%%%%%%%%%%%%%%%%%%%%%%%%%%%%%%%%%%%%%%%%%%%%%%%%%%%%
\subsubsection{Beyond the microcanonical density}
A major theme of this thesis is the generalisation of the notion of correlation functions of the field to situations in which the correlations of interest are in some way \emph{concealed} in the formal microcanonical density itself.  For example, the microcanonical density of the field inherits the $\mathrm{U(1)}$ phase symmetry of the classical-field Hamiltonian (section~\ref{subsec:cfield_proj_cft}).  However, we show in chapter~\ref{chap:anomalous} that over comparatively short time periods, the classical field effectively samples a modified ensemble which breaks the $\mathrm{U(1)}$ phase symmetry of the Hamiltonian~\reff{eq:cfield_HCF}, and so by considering the fluctuation statistics of the field over such short periods we can obtain nonzero values for \emph{symmetry-broken} correlation functions, in analogy to the symmetry-broken approach to the second-quantised field discussed in section~\ref{subsubsec:symmetry_breaking}.  Similarly in chapter~\ref{chap:precess}, we consider a scenario in which the classical-field condensate breaks the \emph{rotational} symmetry of the classical-field Hamiltonian.  By considering the fluctuations of the field on short time scales we can extract the symmetry-broken condensate orbital, which is concealed in the formal rotationally-invariant microcanonical density.  Another scenario which necessitates considering field correlations beyond the microcanonical density is the case of \emph{nonequilibrium} fields, which we consider in chapters~\ref{chap:arrest}~and~\ref{chap:stirring}.  In those scenarios, we are interested in the \emph{relaxation} of the classical field to a (new) thermal equilibrium, and we thus require means of characterising the field during this relaxation period.  We will find in such cases that the separation of time scales between the slow `macroscopic' relaxation of the field and the rapid motion of thermal fluctuations of the field is such that we can obtain meaningful characterisations of the field from averages taken over intermediate time scales. 
%%%%%%%%%%%%%%%%%%%%%%%%%%%%%%%%%%%%%%%%%%%%%%%%%%%%%%%%%%%%%%%%%%%%%%%%%%%%%%%%%%%%%%%%
\subsection{Thermodynamics}\label{subsec:cfield_thermo}
%%%%%%%%%%%%%%%%%%%%%%%%%%%%%%%%%%%%%%%%%%%%%%%%%%%%%%%%%
\subsubsection{Microcanonical thermodynamics}
The second important application of the ergodicity of the field is to the calculation of the thermodynamic parameters of the system: the temperature and chemical potential.
As the PGPE system is Hamiltonian and ergodic, its thermodynamics are in principle described completely by the microcanonical \emph{entropy function}
\begin{equation}\label{eq:cfield_entropy}
	S = k_\mathrm{B} \ln \int\!d\bm{\Gamma}\,\delta(H(\bm{\Gamma})-E)\prod_i \delta(F_i(\bm{\Gamma}) - I_i),
\end{equation}
where $H(\bm{\Gamma})$ is the classical-field Hamiltonian \reff{eq:cfield_HCF} written as a function of the phase-space variables $\bm{\Gamma}$, and the $F_i$ are the other conserved first integrals of the Hamiltonian evolution (e.g., the field normalisation, see section~\ref{subsec:cfield_proj_cft}).
The equilibrium temperature of the microcanonical system is formally given by
\begin{equation}
	\frac{1}{k_\mathrm{B}T} = \left(\frac{\partial S}{\partial E}\right)_{\!\!F_i},
\end{equation}
where the subscript $F_i$ indicates that the derivative is taken with all other constants of motion held fixed. 
Calculating the temperature using this definition however requires the calculation of the entropy function as a function of energy, which is again an intractable task.  However, a comparatively recent advance in statistical mechanics was provided by Rugh \cite{Rugh97}, who showed that the temperature can instead be expressed as an average in the microcanonical density
\begin{equation}
	\frac{1}{T(E)} = \left\langle \mathcal{F} [H] \right\rangle_\mu,
\end{equation}
where $\mathcal{F}[H]$ is a given functional of the system Hamiltonian \cite{Rugh97}.  This provides a great advantage as it allows us to calculate the temperature by an ergodic averaging procedure.  However, the evaluation of the functional $\mathcal{F}$ in terms of the classical-field canonical variables $\{Q_j,P_j\}$ is very complicated, and so this technology has only been implemented for the cases of plane-wave \cite{Davis03} and Hermite-basis projection \cite{Davis05}.  In this thesis we consider some scenarios with an additional conserved first integral (angular momentum) for which the Rugh approach would require significant extension, and we also consider nonequilibrium fields, for which the Rugh approach is not applicable.  We therefore need also to consider other approaches to estimating the temperature, based on the principle of equipartition of energy in the microcanonical system. 
%%%%%%%%%%%%%%%%%%%%%%%%%%%%%%%%%%%%%%%%%%%%%%%%%%%%%%%%%
\subsubsection{Equipartition and the classical-field distribution function}
Although the thermodynamics of the interacting classical field system can only be evaluated numerically, we can gain some insight into the thermodynamic behaviour of the classical-field system by considering the noninteracting limit.  In this limit, the Hamiltonian becomes quadratic in the field variables $\{\alpha_n,\alpha_n^*\}$, and so the principle of \emph{equipartition} of energy applies.  It is most convenient to consider the situation in the grand-canonical ensemble. The equipartition theorem \cite{Pathria96} then implies that a single-particle mode with energy $\epsilon_k$ has mean occupation
\begin{equation}\label{eq:cfield_equipartition}
	n_k = \frac{k_\mathrm{B}T}{\epsilon_k-\mu}.
\end{equation}
A crucial point is that this distribution is obtained from the formal Bose-Einstein distribution~\reff{eq:back_bose_dist} in the limit of high occupations $n_k \gg 1$, which occurs when the parameter $(\epsilon_k-\mu)/k_\mathrm{B}T \ll 1$.  Formally, the equipartition distribution is obtained as the first term in the expansion of the Bose distribution~\reff{eq:back_bose_dist} in this parameter \cite{Blakie07}.  As is well known, the distribution \reff{eq:cfield_equipartition}, when applied to a continuum field theory (with an infinite number of modes) leads to an ultraviolet catastrophe: each mode of the field contains equal (finite) energy, leading to an infinite total energy.  Of course, in the classical field theory with a finite field energy, we face the inverse problem \cite{Blakie08}: as the cutoff is raised, the energy in the field will be divided over an increasing number of modes, and the occupation of each individual mode tends to zero as the cutoff is raised to infinity\footnote{In practice this should not occur for the Hamiltonian field dynamics, as the perturbing effect of field interactions and thus the coupling to higher modes diminishes rapidly for increasing mode energies, so the `creep' of population to higher energies will eventually terminate.}.  In the second-quantised field theory, the occupation of each mode is quantised, and so at high energies where $n_k\ll 1$ the decay becomes exponential in the energy (approaching a Boltzmann distribution), regularising this divergent behaviour.

The distribution~\reff{eq:cfield_equipartition} provides a useful guide to the thermodynamic behaviour of the classical-field system, and allows us to understand how a classical-field model can be used to emulate the statistical mechanics of a Bose-field system.  The central point is that the distribution~\reff{eq:cfield_equipartition} shares the \emph{saturation} property of the Bose distribution.  For a given temperature, the total excited-state population is bounded from above, as for the Bose distribution, except that the bound in this case depends crucially on the fact that there are only a finite number of excited states below the cutoff.  As a result, the classical-field distribution describes condensation in the lowest-energy mode, with a critical temperature depending explicitly on the cutoff energy $E_R$.  In appendix~\ref{app:cfield_transition} we derive the general (geometry-independent) form of the transition temperature.  For the most experimentally relevant case of three-dimensional harmonic trapping (with density of states $g(\epsilon)=\epsilon^2/2(\hbar\overline{\omega})^3$, where $\overline{\omega}=(\omega_x\omega_y\omega_z)^{1/3}$ is the geometric mean trapping frequency) we obtain
\begin{equation}
	k_\mathrm{B}T = \frac{N}{(\hbar\overline{\omega})^3E_R^2}.
\end{equation}
Going beyond the noninteracting case, we expect the equipartition principle to apply to the noncondensed modes of the classical field, to the extent to which they can be considered to be weakly interacting.  Lobo \emph{et al.} \cite{Lobo04} have used this principle to estimate the temperature of a classical field, by assuming that the energy of the field above that of the corresponding ground state is equipartitioned over $\mathcal{M}-1$ noninteracting modes.  In later chapters we take a related approach, and construct a spatial fitting function based on the assumption of equipartition over the thermal modes of the field, and a Hartree-Fock model for the field interactions.  Fitting this function to the spatial distribution of the field allows us to estimate effective thermodynamic parameters of the field in nonequilibrium scenarios.
%%%%%%%%%%%%%%%%%%%%%%%%%%%%%%%%%%%%%%%%%%%%%%%%%%%%%%%%%%%%%%%%%%%%%%%%%%%%%%%%%%%%%%%%%%%%%%%%%%%%%%%%%%%%%%%%%%%%%%%%%%%%%%%%%%%%
\section{Truncated-Wigner approximation}\label{sec:cfield_twa}
In the discussion above, we have obtained a classical-field equation of motion simply by replacing the quantised field operator $\hat{\psi}$ with a classical-valued analogue $\psi$.  Here we discuss how the same equation of motion can be obtained by more formal means.  In this \emph{truncated-Wigner} approach, the classical equation of motion \reff{eq:cfield_pgpe} is obtained as an approximate equation of motion for classical trajectories which sample the quantum evolution of the system.  In this approach, formal quantum expectation values can be obtained through simple correspondences with averages over an ensemble of the classical trajectories.  
In this thesis we consider scenarios of strongly thermal dynamics, and the relaxation of atomic fields over long time scales, beyond the strict validity of the formal truncated-Wigner method.  Nevertheless, considering the truncated-Wigner approximation lends us further insight into the content of the classical-field method, and in particular shows us how to avoid some potential ambiguities and pathologies in our classical-field simulations. 
%%%%%%%%%%%%%%%%%%%%%%%%%%%%%%%%%%%%%%%%%%%%%%%%%%%%%%%%%%%%%%%%%%%%%%%%%%%%%%%%%%%%%%%%
\subsection{The Wigner distribution}
%%%%%%%%%%%%%%%%%%%%%%%%%%%%%%%%%%%%%%%%%%%%%%%%%%%%%%%%%
\subsubsection{Phase-space methods}
An important tool for the study of quantum-mechanical systems developed and frequently employed in quantum optics research is the concept of a \emph{quantum-classical correspondence}.  The basic idea \cite{Gardiner00} is to establish a correspondence between the quantum density matrix $\hat{\rho}$ and a \emph{classical function} $P(\mathbf{z})$ of \emph{phase-space} variables (e.g., $x$ and $p$), which we will denote generically by a vector $\mathbf{z}$.  The function $P(\mathbf{z})$ is a kind of probabilistic distribution over these phase-space variables, however, in order to completely describe the quantum state of the system, the function $P(\mathbf{z})$ generally fails to exhibit all the familiar properties of a classical probability distribution; e.g., $P(\mathbf{z})$ may be non-positive-definite, or even singular.  An important feature of this approach is that the \emph{evolution equation} for $\hat{\rho}$ can be transformed into a classical partial differential equation for $P(\mathbf{z})$.  This allows us to apply many techniques of classical statistical mechanics to the analysis and solution of the evolution equation for $P(\mathbf{z})$ \cite{Carmichael99}.

\begin{sloppypar}
A particular advantage of the corresponding classical formulation of a quantum-dynamical system arises in scenarios in which the classical equation of motion for $P(\mathbf{z})$ is (or can be approximated by) a \emph{Fokker-Planck equation}
%\begin{equation}\label{eq:cfield_fokker_planck}
%	\frac{dP(\mathbf{z},t)}{dt} = -\sum_j \frac{\partial}{\partial z_j} [A_j(\mathbf{z},t) P(\mathbf{z},t)] + \frac{1}{2}\sum_{jk}\frac{\partial^2}{\partial z_j \partial z_k} \left\{ \left[\bar{\mathbf{B}}(\mathbf{z},t)\bar{\mathbf{B}}^T(\mathbf{z},t)\right]_{jk} P(\mathbf{z},t)\right\},
%\end{equation}
\begin{eqnarray}\label{eq:cfield_fokker_planck}
	\frac{dP(\mathbf{z},t)}{dt} &=& -\sum_j \frac{\partial}{\partial z_j} [A_j(\mathbf{z},t) P(\mathbf{z},t)] \nonumber \\
&&\quad+ \frac{1}{2}\sum_{jk}\frac{\partial^2}{\partial z_j \partial z_k} \left\{ \left[\bar{\mathbf{B}}(\mathbf{z},t)\bar{\mathbf{B}}^T(\mathbf{z},t)\right]_{jk} P(\mathbf{z},t)\right\},
\end{eqnarray}
which we have expressed in terms of the \emph{drift vector} $\mathbf{A}(\mathbf{z},t)$ and \emph{diffusion matrix} $\bar{\mathbf{B}}(\mathbf{z},t)$.  In this case we can exploit the correspondence between such a Fokker-Planck equation and the (Ito) \emph{stochastic differential equation} (SDE)
\begin{equation}\label{eq:cfield_generic_SDE}
	d\mathbf{z} = \mathbf{A}(\mathbf{z},t) dt + \bar{\mathbf{B}}(\mathbf{z},t) d\mathbf{W}(t),
\end{equation}
where $d\mathbf{W}(t)$ is a vector of independent Wiener processes \cite{Gardiner04}.  The SDE \reff{eq:cfield_generic_SDE} includes the effects of fluctuations such that an ensemble of independent stochastic solutions $\{\mathbf{z}_i(t)\}$ exhibits the statistics described by the solution of the Fokker-Planck equation \reff{eq:cfield_fokker_planck}\footnote{The relation between the Fokker-Planck and SDE formulations is thus closely analogous to that between the Schr\"odinger and Heisenberg pictures of classical and quantum mechanics \cite{Sudarshan74,Gardiner04}.}. The primary advantage of this formulation is the comparative numerical simplicity of integrating such a stochastic equation, as opposed to integrating the Fokker-Planck equation for $P(\mathbf{z},t)$ itself.   Moreover, as individual stochastic trajectories (corresponding to distinct realisations of the noise process $d\mathbf{W}(t)$) are such that their \emph{ensemble} statistics correspond to the statistics of the quantum-mechanical ensemble, individual trajectories can be interpreted (with caution) as representative of particular observed trajectories of the true quantum system \cite{Blakie08}.
\end{sloppypar}
%%%%%%%%%%%%%%%%%%%%%%%%%%%%%%%%%%%%%%%%%%%%%%%%%%%%%%%%%
\subsubsection{The Wigner distribution}
\begin{sloppypar}We consider here a particular quasiprobability distribution which is relevant to the classical-field approximation used in this thesis.  It is in fact historically the first such distribution, introduced by Wigner in 1932 \cite{Wigner32}.  Its meaning is best understood by first introducing it for a single-particle system, in terms of the fundamental dynamical variables $x$ and $p$.  We define
\begin{equation}\label{eq:cfield_wignerxp}
	W(x,p) = \frac{1}{\pi\hbar}\int\!dy\,\left\langle x-y | \hat{\rho} | x+y \right\rangle e^{2ipy/\hbar},
\end{equation}
which is a kind of phase-space density in $x$ and $p$, analogous to that in Hamiltonian mechanics and statistical mechanics \cite{Sudarshan74}. In particular, we find for the marginal probabilities $\int\!dp\,W(x,p) = \langle x|\hat{\rho}|x\rangle$ and $\int\!dx\,W(x,p) = \langle p|\hat{\rho}|p\rangle$ \cite{Gardiner00}.  However, unlike a classical probability density, $W(x,p)$ can legitimately acquire a negative value at a given point in phase space\footnote{The interpretation of quantum mechanics as a theory which admits such negative probabilities is discussed, for example, by Feynman \cite{Feynman91}.}.  Non-positivity of the Wigner function is associated with quantum interference and other nonclassical effects.  Regarded as a probability distribution, it can be shown that the moments of the Wigner distribution correspond to the quantum expectation values of \emph{symmetrised} products of the variables $x$ and $p$, i.e.
\begin{equation}
	\left\langle\{x^m p^n\}_\mathrm{sym}\right\rangle = \int\!dx\int\!dp\;W(x,p) x^m p^n.	
\end{equation}
\end{sloppypar}

In the second-quantised description of electromagnetic and atomic fields, each mode of the field is dynamically equivalent to a harmonic oscillator, and so in quantum and atom optics applications, it is most convenient to define the Wigner function in terms of the ladder operators $\hat{a}$ and $\hat{a}^\dagger$ of these oscillators.  This representation is straightforwardly related to the $(x,p)$ representation \cite{Hillery84}, due to the simple canonical relation between the two representations (cf. equations~\reff{eq:cfield_canon_varQ}~and~\reff{eq:cfield_canon_varP}).  We find for the Wigner function of a single mode of the field
\begin{equation}\label{eq:cfield_wigner_single_mode}
	W(\alpha,\alpha^*) = \frac{1}{\pi^2} \int\!d^2\lambda\,e^{-\lambda\alpha^* + \lambda^*\alpha}\,\mathrm{Tr}\bigl\{\hat{\rho}e^{\lambda \hat{a}^\dagger - \lambda^* \hat{a}}\bigr\},
\end{equation}
and the relation of moments of this function to quantum expectation values is 
\begin{equation}\label{eq:cfield_moments_relation}
	\left\langle \{(a^\dagger)^ma^n\}_\mathrm{sym} \right\rangle  = \int\!d^2\alpha\,(\alpha^*)^m\alpha^n W(\alpha,\alpha^*) \equiv \left\langle (\alpha^*)^m\alpha^n \right\rangle_\mathrm{W}.
\end{equation}	
Furthermore we can deduce correspondences between the $pre-$ or $post-$ multiplication of the density matrix by a ladder operator and the action of a \emph{classical} operator on the Wigner function.  We find~\cite{Gardiner00}
\begin{eqnarray}\label{eq:cfield_opr_corresp}
	\hat{a}\hat{\rho} &\leftrightarrow& \left(\alpha+\frac{1}{2}\frac{\partial}{\partial \alpha^*}\right)W(\alpha,\alpha^*) \\
	\hat{a}^\dagger\hat{\rho} &\leftrightarrow& \left(\alpha^*-\frac{1}{2}\frac{\partial}{\partial \alpha}\right)W(\alpha,\alpha^*) \\
	\hat{\rho}\hat{a} &\leftrightarrow& \left(\alpha-\frac{1}{2}\frac{\partial}{\partial \alpha^*}\right)W(\alpha,\alpha^*) \\
	\hat{\rho}\hat{a}^\dagger &\leftrightarrow& \left(\alpha^*+\frac{1}{2}\frac{\partial}{\partial \alpha}\right)W(\alpha,\alpha^*).
\end{eqnarray}
These correspondences allow us to convert a quantum (von Neumann or master) equation for the density matrix into a classical partial-differential equation for the Wigner function.
%%%%%%%%%%%%%%%%%%%%%%%%%%%%%%%%%%%%%%%%%%%%%%%%%%%%%%%%%%%%%%%%%%%%%%%%%%%%%%%%%%%%%%%%
\subsection[Truncated-Wigner approximation for the interacting Bose gas]{Truncated-Wigner approximation for the interacting \\ Bose gas} 
%%%%%%%%%%%%%%%%%%%%%%%%%%%%%%%%%%%%%%%%%%%%%%%%%%%%%%%%%
\subsubsection{Equations of motion}
We now discuss the equations of motion that result from the application of the Wigner formalism to the low-energy effective Hamiltonian~\reff{eq:cfield_Heff} of the dilute Bose gas.  
We begin with the Schr\"odinger-picture representation of the quantum dynamics, where the density-matrix $\hat{\rho}$ evolves according to the von Neumann equation
\begin{eqnarray}
	i\hbar\frac{\partial\hat{\rho}}{\partial t} &=& \left[\hat{H},\hat{\rho}\right] \nonumber \\
	&=& \int\!d\mathbf{x}\left\{ \hat{\psi}^\dagger H_\mathrm{sp} \hat{\psi}\hat{\rho} - \hat{\rho} \hat{\psi}^\dagger H_\mathrm{sp} \hat{\psi} + \frac{U_0}{2}\left( \hat{\psi}^\dagger\hat{\psi}^\dagger\hat{\psi}\hat{\psi}\hat{\rho} - \hat{\rho}\hat{\psi}^\dagger\hat{\psi}^\dagger\hat{\psi}\hat{\psi} \right)\right\}\!.
\end{eqnarray}
As the quantum system contains a finite number of modes (those spanning the low-energy region $\mathbf{C}$), the Wigner function \reff{eq:cfield_wigner_single_mode} must be generalised to the multimode case. Fortunately this generalisation is straightforward, and is discussed (for e.g.) in \cite{Blakie08}.  For notational compactness we express the multimode Wigner function as a \emph{functional} distribution $W(\{\psi,\psi^*\})$; using the projected functional differentiation operators~\reff{eq:cfield_func_deriv}, 
the (projected) functional operator correspondences are obtained from \reff{eq:cfield_opr_corresp} by replacing $\hat{a}\to\hat{\psi}$, $\alpha\to\psi$ and $\partial/\partial\alpha\to\bar{\delta}/\bar{\delta}\psi$, \emph{etc}.  We thus find the Wigner-function evolution equation \cite{Steel98}
\begin{eqnarray}\label{eq:cfield_wigner_full}
	\frac{\partial W\bigl(\{\psi,\psi^*\}\bigr)}{\partial t} &=& \frac{i}{\hbar} \int\!d\mathbf{x} \Bigg\{ \frac{\bar{\delta}}{\bar{\delta}\psi} \Bigl[H_\mathrm{sp} + U_0\bigl(|\psi|^2 - \delta_\mathcal{P}(\mathbf{x},\mathbf{x})\bigr)\Bigr]\psi \nonumber \\
&&\quad-\frac{\bar{\delta}}{\bar{\delta}\psi^*}\Bigl[H_\mathrm{sp} + U_0\bigl(|\psi|^2 - \delta_\mathcal{P}(\mathbf{x},\mathbf{x})\bigr)\Bigr]\psi^* \nonumber \\
&&\quad-\frac{U_0}{4} \left(\frac{\bar{\delta}^3}{\bar{\delta}\psi^2\bar{\delta}\psi^*}\psi - \frac{\bar{\delta}^3}{\bar{\delta}\psi \bar{\delta}\psi^{*2}}\psi^*\right) \Bigg\} W\bigl(\{\psi,\psi^*\}\bigr).
\end{eqnarray}
We note that there are no second-derivative terms in this evolution equation.  However, \emph{third}-order derivative terms appear, which correspond to the effects of dynamical quantum fluctuations.  These quantum effects are expected to be negligible for modes with high occupations, as discussed in section~\ref{sec:cfield_basic}.  
Neglecting these terms we thus obtain an equation for $W(\{\psi,\psi^*\},t)$ which describes \emph{nondiffusive} (classical Liouvillian) evolution of the Wigner function.  Consequently, the corresponding `stochastic' differential equation which samples the evolution of this distribution (see equation~\reff{eq:cfield_generic_SDE}) is in fact the \emph{deterministic} differential equation 
\begin{equation}\label{eq:cfield_wigner_pgpe}
	i\hbar\frac{\partial\psi(\mathbf{x},t)}{\partial t} = H_\mathrm{sp} \psi(\mathbf{x}) + \mathcal{P}\Bigl\{U_0\bigl(|\psi(\mathbf{x})|^2 - \delta_\mathcal{P}(\mathbf{x},\mathbf{x}) \bigr)\psi(\mathbf{x})\Bigr\}.
\end{equation}
The truncated-Wigner approximation is valid provided that the dynamical quantum corrections encoded in the neglected third-derivative terms play little role in the evolution of the underlying physical system.  An unambiguous validity criterion for the truncated-Wigner method is thus that all modes in the region $\mathbf{C}$ are highly occupied.  Indeed in the `classical' limit $U_0\to 0$, $N[\psi]\to\infty$, $U_0N[\psi]=\mathrm{const}$, the third-derivative terms in equation~\reff{eq:cfield_wigner_full} vanish and the equation of motion~\reff{eq:cfield_wigner_pgpe} is thus the asymptotically exact equation of motion for the system in this limit \cite{Blakie08}.  In this limit, the projected delta function which appears in equation~\reff{eq:cfield_wigner_pgpe} as an additional (external) potential term also vanishes, and so this term is typically neglected \cite{Norrie06a,Bradley_PhD}, and we regain the PGPE (equation~\reff{eq:cfield_pgpe}).  In fact, Norrie \emph{et al.} \cite{Norrie06a} have derived a weaker condition for the validity of the evolution in terms of the local density of the system in position space.  However, the \emph{formal} interpretation of the stochastic trajectories of equation~\reff{eq:cfield_wigner_pgpe} as samples of the Wigner distribution evolution is not valid on long time scales, due to the ergodic thermalisation of the field trajectories, as we discuss 
in section~\ref{subsec:cfield_relevance_to_PGPE}.
%%%%%%%%%%%%%%%%%%%%%%%%%%%%%%%%%%%%%%%%%%%%%%%%%%%%%%%%%
\subsubsection{Initial conditions}
\begin{sloppypar}
Although the evolution equation~\reff{eq:cfield_wigner_pgpe} is itself deterministic, a stochastic element (corresponding to quantum fluctuations) does enter into the truncated Wigner description through the \emph{initial conditions} of the classical-field trajectories.  Generalising the relation~\reff{eq:cfield_moments_relation} to the multimode case \cite{Blakie08} we find for example 
\begin{eqnarray}
	\langle \psi^*(\mathbf{x})\psi(\mathbf{x}')\rangle_\mathrm{W} &\equiv& \int\! \mkern 4mu\overline{\mkern-4mu \mathcal{D}\mkern-2mu}\mkern2mu^2\psi\,\psi^*(\x)\psi(\x') W\bigl(\{\psi,\psi^*\}\bigr) \nonumber  \\
	&=&  \left\langle \hat{\psi}^\dagger(\mathbf{x}) \hat{\psi}(\mathbf{x}') \right\rangle + \frac{1}{2}\delta_\mathcal{P}(\mathbf{x},\mathbf{x}'), 
\end{eqnarray}
where we have introduced the projected functional integration operator $\int\!\mkern 4mu\overline{\mkern-4mu \mathcal{D}\mkern-2mu}\mkern2mu^2\psi = \prod_{n\in\mathbf{C}} \int\!d^2\alpha_n$.  Thus even in the absence of any population ($\langle\hat{\psi}^\dagger\hat{\psi}\rangle=0$), the density of the field in the Wigner distribution is finite ($\langle|\psi(\mathbf{x})|^2\rangle_\mathrm{W} = (1/2)\delta_\mathcal{P}(\mathbf{x},\mathbf{x})$).  This density corresponds to the \emph{vacuum fluctuations} of the field, i.e., the zero-point occupation of each mode of the field\footnote{In the limit that the cutoff height is raised to infinity, the projected delta function becomes a true delta function, and we regain the well-known result that there is infinite population (and thus energy) contained in the zero-point occupation of a (continuum) bosonic field.}.  For noninteracting modes, this statistical spread in the Wigner function is included simply by populating the mode amplitudes $\alpha_k$ according to $\langle \alpha_k^*\alpha_l \rangle = (n_k + 1/2) \delta_{kl}$, with $\langle \alpha_k\rangle = \langle \alpha_k \alpha_l \rangle = 0$, where $n_k$ is the (real) thermal population.  This can easily be achieved by, for example, sampling the $\{\alpha_j\}$ from an appropriate Gaussian distribution.  In the case of the interacting fields such as we consider in this thesis, a single mode basis is not an appropriate basis for adding noise and, in general,  we can only attempt to populate the field with noise in a way that best approximates the true many-body state we wish to represent.  There are many different approaches to constructing the noise, depending on the system and state being modelled \cite{Steel98,Gardiner02,Sinatra02,Polkovnikov04,Bradley_PhD,Blakie08}, as we discuss in chapter~\ref{chap:stirring}. 
\end{sloppypar}
%%%%%%%%%%%%%%%%%%%%%%%%%%%%%%%%%%%%%%%%%%%%%%%%%%%%%%%%%%%%%%%%%%%%%%%%%%%%%%%%%%%%%%%%
\subsection{Relevance to the PGPE formalism}\label{subsec:cfield_relevance_to_PGPE}
Perhaps the greatest importance of the truncated-Wigner derivation of the projected Gross-Pitaevskii equation is that it shows that the classical-field equation of motion can be obtained from the full many-body theory in a \emph{controlled} approximation.  In fact, Polkovnikov \cite{Polkovnikov03,Polkovnikov10} has shown how the full quantum dynamics can be formally expanded about a Gross-Pitaevskii equation-like description, and that the truncated-Wigner method appears as the inclusion of quantum corrections to leading order in this expansion.

Another important result obtained from the truncated-Wigner formalism is the appearance of the initial noise associated with quantum fluctuations.  In general, this noise allows for the inclusion of spontaneous processes in the GP-type description, which otherwise describes only coherent processes.  This also allows us to avoid some pathologies of a pure-GP description of the field.  In fact, the symmetries of the pure-GP condensate mode can be such that the deterministic evolution of the PGPE is constrained from thermal equilibration.  This can be equivalently viewed as a state which is unstable to decay by spontaneous processes being unable to do so in the absence of added fluctuations, or as the initial GP state lying on the `set of measure zero' of microcanonical trajectories which do not exhibit ergodic behaviour.  The truncated-Wigner removes this issue unambiguously, by showing that for a  classical field to be sensibly interpreted as a beyond-mean-field description of the quantum Bose field, it must include an irreducible level of fluctuations, which \emph{break} all symmetries of the naive GP state, and seed all spontaneous transitions of the field.  This aspect will be important in chapter~\ref{chap:stirring} when we consider the evolution of an initially zero-temperature condensate following a \emph{dynamical instability}, in which unstable excitations of the condensate mode grow exponentially and spontaneously.

It is important to note that the truncated-Wigner method is only an approximation, and in particular, the error associated with the neglect of the cubic-noise terms grows with time during the system evolution.  Indeed we have already seen this from our consideration of the PGPE trajectories:  as the time $t\to\infty$, the Hamiltonian trajectories come to an ergodic equilibrium.  Thus the moments of the Wigner function at long times correspond to moments of the equilibrium microcanonical density of the classical field, and even in the high-occupation regime, the formal operator expectations $\langle \hat{a}_k^\dagger \hat{a}_k\rangle$ we obtain from the Wigner ensemble will be distorted by the subtraction of the half quanta of `noise' from each mode of the classical-field distribution.  We thus face again the fundamental issue that the neglect of the Bose commutation relations in the evolution of the field leads to the loss of the exact Bose statistics of the field.  A potentially problematic consequence of this behaviour is that the energy and normalisation added to the field as a representation of the vacuum noise leads in general to a spurious raising of the temperature of the field.  This consequently affects damping rates in the field, which are strongly dependent on the temperature \cite{Fedichev98a}.  Sinatra \emph{et al.}~\cite{Sinatra02} thus suggested the additional condition $T_\mathrm{class} - T \ll T$ on the temperature of the field, in relation to the temperature $T_\mathrm{class}$ of the Hamiltonian field at microcanonical equilibrium.  However, this condition does not necessarily apply in strongly nonequilibrium scenarios in which the temperature changes dramatically,  and in chapter~\ref{chap:stirring} we consider the rethermalisation of a classical field following a dynamical instability at \emph{zero temperature}, but nevertheless find that the energy (and angular momentum) added to the system in the form of vacuum noise is small compared to that `liberated' from the mean field in the transition to the new equilibrium.
%%%%%%%%%%%%%%%%%%%%%%%%%%%%%%%%%%%%%%%%%%%%%%%%%%%%%%%%%
\subsubsection{The stochastic Gross-Pitaevskii equation}
Here we briefly discuss the \emph{stochastic} Gross-Pitaevskii equation (SGPE) developed by Gardiner and coworkers \cite{Gardiner02,Gardiner03}.  This formalism is closely related to the finite-temperature projected Gross-Pitaevskii equation, and the formal truncated-Wigner method.  The essence of the treatment is that an \emph{open systems} approach is taken to deriving an equation of motion for the condensate-band density operator, including the dissipative effect of its coupling to the above-cutoff region.  The above-cutoff atoms are assumed to be at thermal equilibrium, and form a \emph{thermal bath} for the below cutoff atoms.  The evolution equation for the below-cutoff atoms is then a \emph{master equation}, derived using a high-temperature expansion.  Mapping this master equation onto an evolution equation for the Wigner function we find proper diffusive terms (i.e., second order derivatives).  The appropriate stochastic equations for the classical field then include explicit noise and damping terms, resulting from the coupling to the thermal bath.  The coupling terms include the exchange of atoms between the condensate band and the bath, and the stochastic Gross-Pitaevskii equation is thus a \emph{grand-canonical} method for the Bose field.   To ensure the validity of the approximations made in the derivation, it is limited to the study of comparatively high-temperature Bose fields ($T\gtrsim0.7T_\mathrm{c}$), but allows in particular for the study of the response of the Bose field to \emph{quenches} of the thermal parameters of the bath~\cite{Bradley08,Weiler08}.
%%%%%%%%%%%%%%%%%%%%%%%%%%%%%%%%%%%%%%%%%%%%%%%%%%%%%%%%%%%%%%%%%%%%%%%%%%%%%%%%%%%%%%%%%%%%%%%%%%%%%%%%%%%%%%%%%%%%%%%%%%%%%%%%%%%%
%%%%%%%%%%%%%%%%%%%%%%%%%%%%%%%%%%%%%%%%%%%%%%%%%%%%%%%%%%%%%%%%%%%%%%%%%%%%%%%%%%%%%%%%%%%%%%%%%%%%%%%%%%%%%%%%%%%%%%%%%%%%%%%%%%%%

\chapter{Numerical methods}
\label{chap:numerics}
%%%%%%%%%%%%%%%%%%%%%%%%%%%%%%%%%%%%%%%%%%%%%%%%%%%%%%%%%%%%%%%%%%%%%%%%%%%%%%%%%%%%%%%%%%%%%%%%%%%%%%%%%%%%%%%%%%%%%%%%%%%%%%%%%%%%
%%%%%%%%%%%%%%%%%%%%%%%%%%%%%%%%%%%%%%%%%%%%%%%%%%%%%%%%%%%%%%%%%%%%%%%%%%%%%%%%%%%%%%%%%%%%%%%%%%%%%%%%%%%%%%%%%%%%%%%%%%%%%%%%%%%%
In this chapter, we outline the numerical integration of the projected Gross-Pitaevskii equation (PGPE).  We discuss the reformulation of the PGPE in terms of dimensionless quantities, interaction-picture methods for the temporal integration of the PGPE, the spectral-Galerkin approach to its spatial integration, and Gaussian quadrature rules for the efficient evaluation of the integrals that arise in this formulation.  We then contrast the PGPE approach with other numerical discretisations of the Gross-Pitaevskii theory, and explain why the projection operation is necessary for the accurate simulation of the classical-field dynamics.
%%%%%%%%%%%%%%%%%%%%%%%%%%%%%%%%%%%%%%%%%%%%%%%%%%%%%%%%%%%%%%%%%%%%%%%%%%%%%%%%%%%%%%%%%%%%%%%%%%%%%%%%%%%%%%%%%%%%%%%%%%%%%%%%%%%%
\section[Numerical integration of the projected Gross-Pitaevskii equation]{Numerical integration of the projected \\ Gross-Pitaevskii equation}\label{sec:numerics_integration}
%%%%%%%%%%%%%%%%%%%%%%%%%%%%%%%%%%%%%%%%%%%%%%%%%%%%%%%%%%%%%%%%%%%%%%%%%%%%%%%%%%%%%%%%
\subsection{Dimensionless formalism}\label{subsec:dimless}
In implementing the numerical solution of the projected Gross-Pitaevskii equation, we reformulate it in terms of \emph{dimensionless quantities}.  This serves two purposes: first, it ensures that the numerical quantities manipulated during the integration are of reasonable magnitude, i.e., we avoid carrying around factors of very large or (as is typically the case in describing quantum-mechanical systems in SI units) very small numbers that can lead to numerical round-off error.  The second reason is that by identifying appropriate `natural' units of the system we often find that the behaviour of the system depends on an irreducible set of parameters which is smaller than the set of physical (dimensioned) parameters.  The systems we treat in this thesis are harmonically trapped, and this trapping provides for obvious definitions for natural units.  In general, we consider systems which are oblate, and so we base our units on the in-plane or \emph{radial} trapping frequency~$\omega_r=\omega_x=\omega_y$ \footnote{In chapters \ref{chap:arrest} and \ref{chap:stirring} we include an elliptical deformation of the trap in the $xy$ plane.  In this case $\omega_r$ refers to the `average' radial trapping frequency, i.e. $\omega_r = \sqrt{(\omega_x^2+\omega_y^2)/2}$ \cite{Hodby02}.} .
For definiteness, we begin with the PGPE expressed in terms in dimensioned quantities (measured in, e.g., SI units).  Denoting such quantities with a bar (e.g. $\overline{x}$), the PGPE is written
\begin{equation}
	i\hbar \frac{\partial \overline{\psi}(\overline{\x})}{\partial \overline{t}} = \overline{\mathcal{P}}\Bigl\{\Bigl(\mkern2mu\overline{\mkern-2muH}_\mathrm{sp} + \overline{U}_0 \left|\overline{\psi}(\overline{\x})\right|^2\Bigr)\overline{\psi}(\overline{\x})\Bigr\},
\end{equation}
where $\mkern2mu\overline{\mkern-2muH}_\mathrm{sp}$ is the single-particle Hamiltonian
\begin{equation}\label{eq:numeric_Hsp}
	\mkern2mu\overline{\mkern-2muH}_\mathrm{sp} = \frac{-\hbar^2\overline{\nabla}^2}{2m} + \frac{m}{2}\bigl(\omega_x^2\overline{x}^2 + \omega_y^2\overline{y}^2 + \omega_z^2\overline{z}^2\bigr) + i\hbar\overline{\Omega}\Bigl(\overline{x}\frac{\partial}{\partial\overline{y}} - \overline{y}\frac{\partial}{\partial\overline{x}}\Bigr),
\end{equation}
and the projector
\begin{equation}
	\overline{\mathcal{P}}\,\overline{f}(\overline{\x}) = \sum_{n\in\mathbf{C}} \mkern1mu\overline{\mkern-1mu\phi}_n(\overline{\x}) \int\!d\overline{\x}'\,\mkern1mu\overline{\mkern-1mu\phi}_n\mkern-5mu^*(\overline{\x}') \overline{f}(\overline{\x}'),
\end{equation}
where the $\mkern1mu\overline{\mkern-1mu\phi}_n(\overline{\x})$ are eigenstates of the dimensioned base Hamiltonian, i.e. $\mkern2mu\overline{\mkern-2muH}_0\mkern1mu\overline{\mkern-1mu\phi}_n(\overline{\x}) = \overline{\epsilon}_n\mkern1mu\overline{\mkern-1mu\phi}_n(\overline{\x})$.  We now introduce units
\begin{eqnarray}
	r_0 &=& \sqrt{\frac{\hbar}{m\omega_r}} \\
	t_0 &=& \frac{1}{\omega_r} \\
	\varepsilon_0 &=& \hbar\omega_r,
\end{eqnarray}
such that the dimensioned quantities are related to the dimensionless ones by $\overline{t}=t_0t$, $\overline{\x}=r_0\x$, etc.  In particular, the general correspondence for wave functions is $\overline{\chi}(\overline{\x}) = r_0^{-3/2} \chi(\overline{\x}/r_0) \equiv r_0^{-3/2} \chi(\x)$.  We thus obtain the dimensionless PGPE
\begin{equation}
	i\frac{\partial\psi}{\partial t} = \mathcal{P}\Bigl\{\left(H_\mathrm{sp} + U_0|\psi(\x)|^2\right)\psi(\x)\Bigr\},
\end{equation}
where the dimensionless single-particle Hamiltonian has the form
\begin{equation}
	H_\mathrm{sp} = \frac{-\nabla^2}{2} + \frac{1}{2}\bigl(x^2+y^2+\lambda_z^2z^2\bigr) + i\Omega\Bigl(x\frac{\partial}{\partial y} - y\frac{\partial}{\partial x}\Bigr),
\end{equation}
with $\Omega \equiv \overline{\Omega}/\omega_r$ and $\lambda_z\equiv\omega_z/\omega_r$; the dimensionless interaction strength is
\begin{equation}
	U_0 = \frac{\overline{U}_0}{\hbar\omega_rr_0^3};
\end{equation}
and the dimensionless projector $\mathcal{P}$ is such that
\begin{equation}
	\mathcal{P}f(\x) = \sum_{n\in\mathbf{C}} \phi_n(\x) \int\!d\x'\,\phi_n^*(\x') f(\x'),
\end{equation}
where the dimensionless basis modes $\phi_n(\x)$ are eigenstates of the dimensionless base Hamiltonian ($H_0\phi_n(\x) = \epsilon_n\phi_n(\x)$) with dimensionless energy eigenvalues $\epsilon_n = \overline{\epsilon}_n/\varepsilon_0$.
%%%%%%%%%%%%%%%%%%%%%%%%%%%%%%%%%%%%%%%%%%%%%%%%%%%%%%%%%
\subsubsection{Unit-normalised field}
In most of this thesis we will work with a classical field $\psi(\x)$ normalised to the total coherent-region population, $\int\!d\x\,|\psi(\x)|^2 =\int\!d\overline{\x}\,|\overline{\psi}(\overline{\x})|^2 \equiv N_\mathrm{c}$.  However, in chapter~\ref{chap:anomalous} we consider the generic behaviour of the PGPE without explicit reference to physical units.  In this chapter (only) we will therefore work in terms of a classical field which is normalised to unity, i.e., we define
\begin{equation}
	\psi(\x) = \frac{\overline{\psi}(\overline{\x}) r_0^{3/2}}{\sqrt{N_\mathrm{c}}}.
\end{equation}
In that case the factor $N_\mathrm{c}$ is absorbed into the nonlinear coefficient, i.e., we have the PGPE
\begin{equation}\label{eq:num_unit_pgpe} 
	i\frac{\partial\psi(\x)}{\partial t} = \mathcal{P}\Bigl\{\left(H_\mathrm{sp} + C_\mathrm{nl}|\psi(\x)|^2\right)\psi(\x)\Bigr\},
\end{equation}
where $C_\mathrm{nl} = N_\mathrm{c}\overline{U}_0/\hbar\omega_rr_0^3$.
%%%%%%%%%%%%%%%%%%%%%%%%%%%%%%%%%%%%%%%%%%%%%%%%%%%%%%%%%
\subsubsection{Quasi-two-dimensional system}
In chapters~\ref{chap:precess}-\ref{chap:stirring} we will consider quasi-two-dimensional systems, which are physically three-dimensional systems tightly confined in the axial direction (see, e.g. \cite{Stock05,Clade09}).  These are realised formally in our classical-field model by choosing a cutoff energy less than the oscillator spacing in the tightly confined ($z$) direction, i.e., $E_R < \hbar\omega_z$.  We thus have
\begin{equation}
	\mathcal{P}f(\x) = \sum_{n^\in\mathbf{C}^\perp} \phi_n^\perp(x,y)\zeta_0(z) \int\!d^3\x'\left[\phi_n^\perp(x',y')\right]^*\zeta_0^*(z') f(\x'),
\end{equation}
where the ground oscillator state $\zeta_0(z)=(\lambda_z/\pi)^{1/4}\exp(-\lambda_z^2 z^2/2)$ obeys
\begin{equation}
	H_z\zeta_0(z) = \biggl[-\frac{1}{2}\frac{\partial^2}{\partial z^2} + \frac{1}{2}\lambda_z^2z^2\biggr]\zeta_0(z) = \frac{\lambda_z}{2}\zeta_0(z),
\end{equation}
the set of functions $\{\phi_n^\perp(x,y)\}$ are the eigenstates of the transverse single-particle Hamiltonian $H_\perp = H_\mathrm{sp} - H_z$ with eigenvalues $\{\epsilon_n^\perp\}$, and the region $\mathbf{C}^\perp \equiv \{ n : \epsilon_n^\perp + \lambda_z/2 \leq E_R\}$.
Defining $\psi(\x) = \psi_\perp(x,y)\zeta_0(z)$, we therefore have 
\begin{equation}
	i\frac{\partial\psi_\perp(x,y)}{\partial t} = \mathcal{P}_\perp \Bigl\{\Bigl(H_\mathrm{\perp} + \frac{\lambda_z}{2} + U_\mathrm{2D} |\psi_\perp(x,y)|^2\Bigr)\psi_\perp(x,y) \Bigr\},
\end{equation}
where the effective two-dimensional interaction strength is
\begin{equation}
	U_\mathrm{2D} = \frac{\overline{U}_0}{r_0^2\hbar\omega_r \sqrt{2\pi}z_0} = \frac{\overline{U}_\mathrm{2D}}{r_0^2\hbar\omega_r},
\end{equation}
where $z_0=\sqrt{\hbar/m\omega_z}$ is the oscillator length in the tightly-confined dimension, and $\overline{U}_\mathrm{2D} = \overline{U}_0/z_0\sqrt{2\pi}$ is the SI interaction strength appropriate to bosons in a quasi-2D geometry \cite{Petrov00}, and the two-dimensional projector is defined 
\begin{equation}
	\mathcal{P}_\perp f(x,y) \equiv \sum_{n\in\mathbf{C}^\perp} \phi_n^\perp(x,y) \int\!dx' dy'\left[\phi_n^\perp(x',y')\right]^* f(x',y').
\end{equation}
%%%%%%%%%%%%%%%%%%%%%%%%%%%%%%%%%%%%%%%%%%%%%%%%%%%%%%%%%%%%%%%%%%%%%%%%%%%%%%%%%%%%%%%%
\subsection{Spectral-Galerkin formulation}\label{subsec:galerkin}
Now, as we have noted, the single-particle Hamiltonian can be separated as $H_\mathrm{sp}=H_0+\delta V(\mathbf{x},t)$.  As the projector $\mathcal{P}$ is defined in terms of eigenfunctions of $H_0$, we have immediately
\begin{equation}
	[\mathcal{P},H_0] = 0,
\end{equation}
and so the equation of motion can be written as
\begin{equation}\label{eq:numerics_pgpe_separated}
	i\frac{\partial \psi(\x)}{\partial t} = H_0\psi(\x) + \mathcal{P}\Bigl\{\bigl(\delta V(\x) + \lambda |\psi(\x)|^2\bigr)\psi(\x)\Bigr\}.
\end{equation}
Moreover, as the Hamiltonian $H_0$ is \emph{diagonal} in the basis of eigenmodes $\{\phi_n(\mathbf{x})\}$, it is natural to take a \emph{spectral Galerkin} approach, i.e., to derive from the PGPE~\reff{eq:numerics_pgpe_separated} a set of coupled equations for the coefficients $\{\alpha_n\}$ in the expansion of $\psi$.  This is achieved by calculating the overlap of each side with the general basis mode $\phi_m(\mathbf{x})$, i.e., by applying the operation $\int\!dx\,\phi_m^*(\mathbf{x})$ to each side.  As the projector $\mathcal{P}$ is an identity operator for all basis modes $\phi_n$, we have
\begin{equation}
	\int\!d\mathbf{x}\,\phi_m^*(\mathbf{x}) \mathcal{P}\{f(\mathbf{x})\} = \int\!d\mathbf{x}\,\phi_m^*(\mathbf{x})f(\mathbf{x}),
\end{equation}
and thus
\begin{equation}\label{eq:numerics_pgpe_galerkin}
	i\frac{\partial \alpha_m}{\partial t} = \epsilon_m \alpha_m + F_m[\psi(\mathbf{x})] + H_m[\psi(\mathbf{x})],
\end{equation}
where
\begin{eqnarray}
	F_m[\chi(\mathbf{x})] &=& U_0\int\!d\mathbf{x}\,\phi_m^*(\mathbf{x}) |\chi(\x)|^2 \chi(\mathbf{x}), \\
	H_m[\chi(\mathbf{x})] &=& \int\!d\mathbf{x}\,\phi_m^*(\mathbf{x}) \delta V(\x,t) \chi(\mathbf{x}).
\end{eqnarray}
We have used the functional notation to emphasise that the nonlinear term $F_m[\psi(\x)]$ depends on the basis coefficients $\{\alpha_m\}$ through the whole projected field $\psi(\x)$.  By contrast, the perturbing potential term can be expressed as a matrix multiplication $H_m[\psi(\mathbf{x})]= \sum_{n} H_m[\alpha_n\phi_n(\mathbf{x})] = \sum_{n} H_{mn} \alpha_n$.  However, in practice we will evaluate both terms using Gaussian quadratures, so we will not make this distinction, and we will in fact often group the two projected terms together as $G_m[\psi(\x)] = F_m[\psi(\x)] + H_m[\psi(\x)]$.
%%%%%%%%%%%%%%%%%%%%%%%%%%%%%%%%%%%%%%%%%%%%%%%%%%%%%%%%%%%%%%%%%%%%%%%%%%%%%%%%%%%%%%%%
\subsection{Interaction-picture methods}
The PGPE as written in equation~\reff{eq:numerics_pgpe_galerkin} immediately lends itself to an `interaction picture' approach \cite{Ballagh00b}.  The basic idea is to exploit the diagonal representation of the base Hamiltonian $H_0$ to transform the field (or more precisely its amplitude representation $\{\alpha_m\}$) to an interaction picture with respect to this operator.  The effect of the remaining operators on the field is then evaluated efficiently in this interaction picture by traditional techniques for numerical integration, such as the \emph{Runge-Kutta} family of integrators \cite{Press92}.  By eliminating the time-dependence of the field due to the base Hamiltonian $H_0$ from the problem, we are able to take much longer integration steps for the same level of numerical accuracy.

In this approach, the field is transformed to the interaction picture with respect to $H_0$ and relative to some time origin $t_0$, so that
\begin{equation}
	\psi^\mathrm{I}(\x,t) = e^{i(t-t_0)H_0} \psi(\x,t) \equiv U_\mathrm{IP}\psi(\x, t),
\end{equation}
which yields the equation of motion 
\begin{equation}
	i\frac{\partial \psi^\mathrm{I}(\x, t)}{\partial t} = \mathcal{P}\bigg\{U_\mathrm{IP}\Big[\delta V + U_0\big|U_\mathrm{IP}^\dagger\psi^\mathrm{I}(\x, t)\big|^2\Big] U_\mathrm{IP}^\dagger \psi^\mathrm{I}(\x, t) \bigg\}.
\end{equation}
The spectral representation is again obtained by applying the operation $\int\!d\x\,\phi_m^*(\x)$ to both sides of the equation, from which we obtain
\begin{equation}
	i\frac{\partial \alpha^\mathrm{I}_m}{\partial t} = G_m^\mathrm{I}\left[\psi^\mathrm{I}(\x, t)\right],
\end{equation}
where 
\begin{eqnarray}
	G_m^\mathrm{I}[\chi(\mathbf{x}, t)] &=& \int\!d\x\,\phi_m^*(\x) U_\mathrm{IP}\Big[\delta V(\x,t) + U_0\big|U_\mathrm{IP}^\dagger\chi(\x,t)\big|^2\Big]U_\mathrm{IP}^\dagger\chi(\x,t) \\
	&=& e^{i\epsilon_m(t-t_0)}G_m\Big[U_\mathrm{IP}^\dagger \chi(\mathbf{x}, t)\Big].
\end{eqnarray}

This forms the basis for integration by a Runge-Kutta approach in the interaction picture, where the origin $t_0$ is judiciously chosen to minimise the number of interaction-picture transformations.   In practice we use an \emph{adaptive} Runge-Kutta variant \cite{Davis_DPhil}, to control the accuracy of our solutions.  This approach is briefly reviewed in appendix~\ref{app:IP_algorithms}.
%%%%%%%%%%%%%%%%%%%%%%%%%%%%%%%%%%%%%%%%%%%%%%%%%%%%%%%%%%%%%%%%%%%%%%%%%%%%%%%%%%%%%%%%
\subsection{Gaussian-quadrature evaluation of projected operators}
In general, the basis modes we deal with are of the form
\begin{equation}
	\phi_n(\x) = \mathcal{W}(\x)P_n(\x),
\end{equation}
where $P_n(\x)$ is a polynomial of some degree $\mathcal{D}_{P_n}$, and $\mathcal{W}(\x)$ is some general \emph{weight} function, which is common to all the basis modes $\phi_n(\x)$.  For example, in the case of the Hermite basis modes of the one-dimensional harmonic oscillator, $\mathcal{W}(x) = e^{-x^2/2}$.  Consequently, the classical field can be expressed
\begin{equation}
	\psi(\x) = \sum_{n\in\mathbf{C}} \alpha_n \phi_n(\x) = \mathcal{W}(\x) Q(\x),
\end{equation}
where $Q(\x)$ is a polynomial of \emph{at most} some degree $\mathcal{D}_Q$, which is set by the cutoff energy and thus the allowed basis indices $n\in\mathbf{C}$.  We now restrict our attention to the case $\delta V(\x,t)=0$, and consider the evaluation of the nonlinear matrix element $F_m[\psi(\x)]$
\begin{eqnarray}
	F_m[\psi(\x)] &=& U_0\int\!d\x\,\phi_m^*(\x) \mathcal{W}^3(\x) |Q(\x)|^2Q(\x) \\
				  &=& U_0\int\!d\x\,\mathcal{W}^4(\x) R_m(\x),
\end{eqnarray}
where $R_m(\x)$ is a polynomial of degree at most $\mathcal{D}_{R_m}$.  The relevant case is that in which $\mathcal{W}^4(\x)\equiv W(\x)$, where $W(\x)$ is the \emph{weight factor} of a Gaussian quadrature rule~\cite{Press92}
\begin{equation}\label{eq:num_quad_rule}
	\int\!d\x\,W(\x) f(\x) \approx \sum_{j=1}^{N_Q} w_j f(\x_j),
\end{equation}
for approximating the integral.  The quadrature \emph{abscissas} $\x_j$ form a spatial grid which is in general not uniformly spaced, and the quadrature \emph{weights} $w_j$ are the corresponding numeric weighting values appropriate to the quadrature rule.  The crucial point is that given the bound $\mathcal{D}_{R_m}$ on the degree of polynomial $R_m(\x)$, the order $N_Q$ of the quadrature rule~\reff{eq:num_quad_rule} can be chosen such that the integral $F_m[\psi(\x)]$ can be evaluated \emph{exactly}:
\begin{equation}
	F_m\left[\psi(\x)\right] = U_0\sum_{j=1}^{N_Q} w_j R_m(\x_j).
\end{equation}

To use this approach in practice, the roots $\x_j$ and weights $w_j$ appropriate to the quadrature rule of the requisite order are precomputed before the evolution begins, as is the \emph{transformation matrix} $U_{kn} \equiv \phi_n(\mathbf{x}_k)$ between the spectral and quadrature-grid representations of the classical field.
The general procedure for evaluating $F_m[\psi(\x)]$ is then (figure~\ref{fig:quad_quad}):
%%%%%%%%%%%%%%%%%%%%%%%%%%%%%%%%%%%%%%%
\begin{figure}
	\begin{center}
	\includegraphics[width=0.9\textwidth]{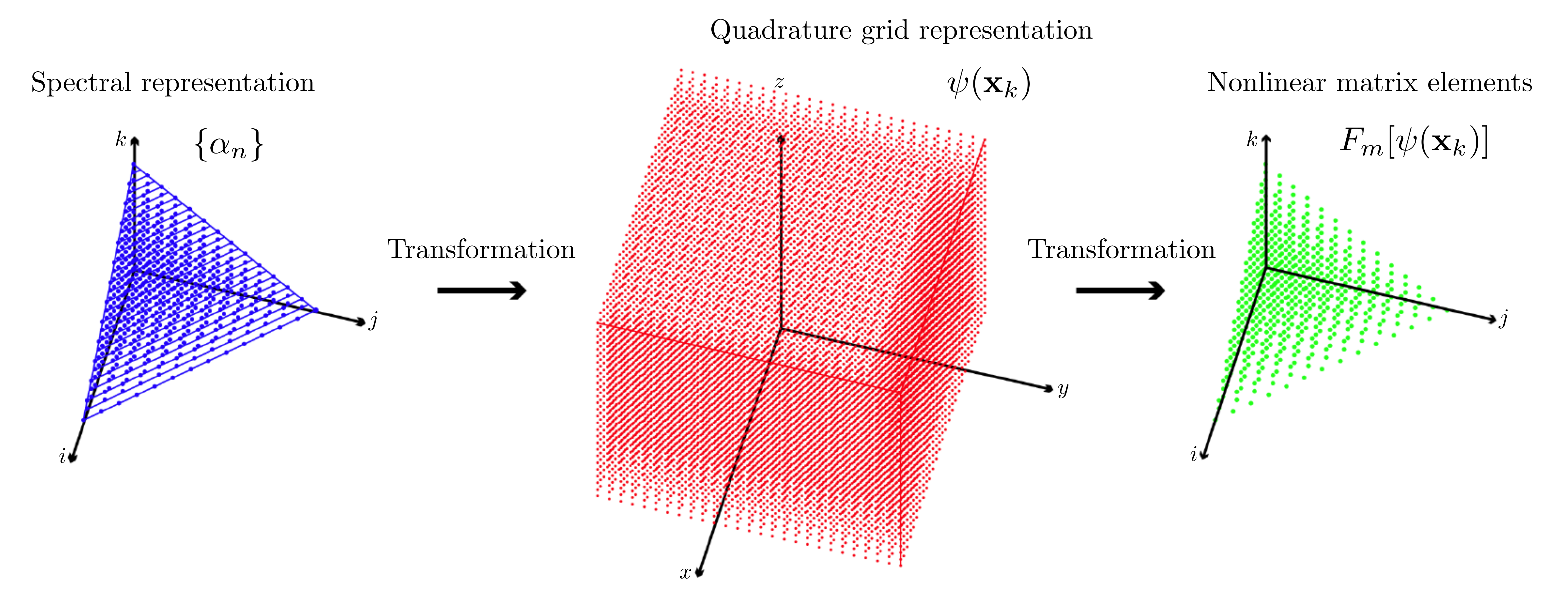}
	\caption{\label{fig:quad_quad} Visual representation of the Gaussian-quadrature method for evaluating the projected nonlinear term.  (Figure adapted from reference~\cite{Blakie08}.)}
	\end{center}
\end{figure}
%%%%%%%%%%%%%%%%%%%%%%%%%%%%%%%%%%%%%%%

\begin{enumerate}
	\item{Transform from the coefficient representation to a spatial representation on the quadrature grid
		\begin{equation}
			\psi(\x_k) = \sum_{n\in\mathbf{C}} U_{kn} \alpha_n.
		\end{equation}
	}
	\item{Construct the integrand on the quadrature grid by local multiplication
		\begin{equation}
			R(\x_k) = w_k \left|\psi(\x_k)\right|^2\psi(\x_k).
		\end{equation}
	}
	\item{Obtain the desired matrix elements by performing the inverse transformation
		\begin{equation}
			F_m\left[\psi(\x)\right] = U_0\sum_{k=1}^{N_\mathrm{Q}} U^*_{km} R(\x_k).
		\end{equation}
	}
\end{enumerate}

%%%%%%%%%%%%%%%%%%%%%%%%%%%%%%%%%%%%%%%%%%%%%%%%%%%%%%%%%%%%%%%%%%%%%%%%%%%%%%%%%%%%%%%%%%%%%%%%%%%%%%%%%%%%%%%%%%%%%%%%%%%%%%%%%%%%
\section{Performance of the Gauss-Laguerre algorithm}\label{sec:numerics_performance}
In this section we consider the numerical performance of the Gauss-Laguerre-quadrature algorithm for the integration of the rotating-frame PGPE.  The efficiency of the algorithm is a crucial concern, and indeed limits the size of system and duration of evolution we can simulate in a given period of real (wall) time.  The execution time of numerical algorithms depends in general both on the speed of the processor, and the bandwidth of the memory in the computer on which it is executed.  All the results we present in this section are for the C-optimised MATLAB implementation of the algorithm\footnote{In order to increase the efficiency of the algorithm, we have written its core processor-intensive operations (i.e., matrix multiplications and calls to the fast Fourier transform library \cite{fftw_org}) in the C language, using MATLAB's MEX interface \cite{matlab}.}, timed on workstation with a 2.66GHz Intel Xeon processor and 667MHz DDR2 RAM, which is of comparable performance to the compute nodes of the cluster on which the majority of production simulations were run\footnote{The actual performance on the multi-core compute nodes will depend on the system load conditions during execution, and so we consider the performance on an isolated workstation for clarity.}.
%%%%%%%%%%%%%%%%%%%%%%%%%%%%%%%%%%%%%%%%%%%%%%%%%%%%%%%%%%%%%%%%%%%%%%%%%%%%%%%%%%%%%%%%
\subsection{Base rotating-frame algorithm}
We begin by considering the performance of the base algorithm developed by Bradley \cite{Bradley_PhD,Bradley08}.  In this algorithm (section~\ref{subsec:quad_app_lag_nl}), the quadrature grid (cf. figure~\ref{fig:quad_quad}(b)) is a set of polar coordinates $\{r_k,\theta_j\}$, with $N_x$ points in the radial direction and $N_\theta$ points in the azimuthal direction.  In general, both the multiplicity $\mathcal{M}$ of the basis modes (cf. figure~\ref{fig:quad_quad}(a)) and the number of points in each direction on the quadrature grid will depend both on the cutoff energy and the angular velocity of the rotating frame in which the projection is effected.  As noted by Bradley, by considering the operations involved in the evaluation of the nonlinear term (cf.~\cite{Dion03}), one finds that the computational load of the evaluation of the nonlinear term by Gauss-Laguerre quadrature should scale as $O(NN_x\ln N_\theta)$, where $N=\max(N_x,\ln N_\theta)$ \cite{Bradley08}\footnote{In fact for nontrivial cutoff parameters we always have $N_x>\ln N_\theta$, and so the algorithmic scaling becomes simply $O(N_x^2N_\theta)$.}.  As the evaluation of this term forms the most numerically intensive part of the Runge-Kutta algorithms, we expect similar scaling for the total integrator. Here, we consider the `real world' performance of the algorithm, including the dependence of the integrator step time on the energy cutoff and frame-rotation frequency, and the difference between the fixed-step-size RK4IP implementation and the adaptive ARK45-IP variant (appendix~\ref{app:IP_algorithms}).  
%%%%%%%%%%%%%%%%%%%%%%%%%%%%%%%%%%%%%%%%%%%%%%%%%%%%%%%%%
\subsubsection{Dependence on energy cutoff}
\begin{figure}
	\begin{center}
	\includegraphics[width=0.9\textwidth]{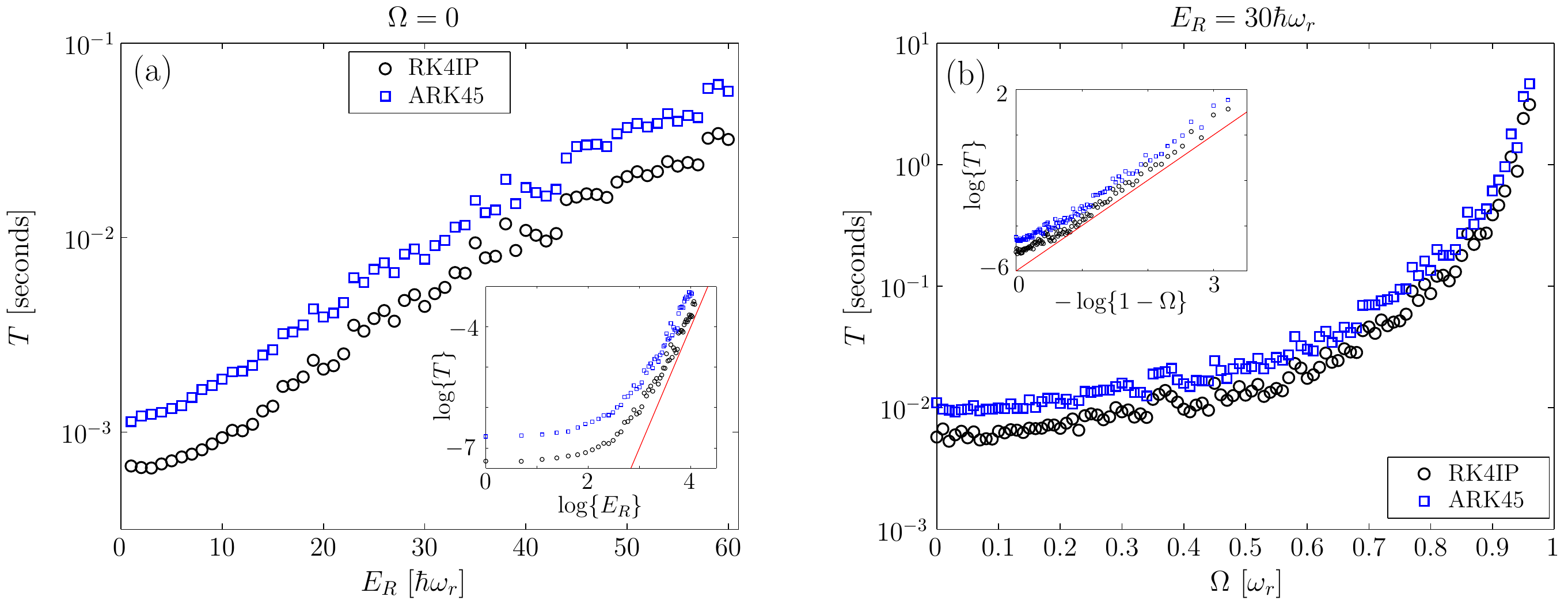}
	\caption{\label{fig:rk4vsark} Measured step times for RK4IP and ARK45-IP implementations of the rotating-frame PGPE.  a) Dependence of the step time $T$ on the cutoff energy $E_R$ at fixed rotation rate $\Omega=0$.  The inset log-log plot shows that the algorithm approaches the expected $O(E_R^3)$ scaling as $E_R$ increases.  b) Dependence of the step time on the angular velocity $\Omega$ of the rotating frame at fixed cutoff energy $E_R=30\hbar\omega_r$.  The inset log-log plot shows that the algorithm approaches a scaling $O((1-\Omega)^{-2})$ as $\Omega\to1$.}
	\end{center}
\end{figure}

We first consider the average step time of the two Runge-Kutta algorithms as a function of the cutoff energy $E_R$, for the nonrotating case $\Omega=0$.  For each value of the cutoff energy we plot in figure~\ref{fig:rk4vsark}(a) the mean of 100 measured step times for both the RK4 (circles) and ARK (squares) algorithms.  
%%%%%%%%%%%%%%%%%%%%%%%%%%%%%%%%%%%%%%%
%%%%%%%%%%%%%%%%%%%%%%%%%%%%%%%%%%%%%%%
In both cases the scaling of the step time $T$ with cutoff energy $E_R$ is comparatively flat for low cutoffs $E_R \lesssim 10\hbar\omega_r$, as the step time is dominated by fixed overhead factors in this small-basis limit.  For larger quadrature grids, the step-time scaling is reasonably close to the $O(E_R^3)$ scaling we expect\footnote{In the nonrotating limit $\Omega=0$, we have $N_x=\overline{N}+1$ and $N_\theta = 4\overline{N}+1$, thus $NN_xN_\theta = N_x^2N_\theta \approx 4\overline{N}^3$.}, as indicated in the inset to figure~\ref{fig:rk4vsark}(a), in which we plot $\log(T)$ against $\log(E_R)$. We include a line $y=mx+c$ with slope $m=3$ for comparison.  We note that the ARK steps take on average 1.8 times as long to execute as the RK4 steps, which is reasonable given that six evaluations of the nonlinear term are required for each step of the ARK algorithm, versus four evaluations for the RK4 algorithm, but also suggests that a significant amount of time is involved in the additional diagonal multiplications and copying operations in the ARK implementation.  

We also note that the increase in step time with the cutoff height is not monotonic.  Indeed it is clear from the fact that `glitches' in the upwards progression of step times appear simultaneously in both the RK4 and ARK data sets that these variations are reproducible, and not merely statistical fluctuations.  This non-monotonicity is a consequence of the fast Fourier transform algorithms used to evaluate the azimuthal quadrature\footnote{In practice we use the FFTW library to evaluate discrete Fourier transforms \cite{fftw_org, Frigo05}.}, for which the efficiency of the transform depends not only on the number $N_\theta$ of azimuthal grid points, but on the factorisation of this quantity as a product of prime numbers.  In particular, quadrature grids for which the number $N_\theta$ of azimuthal quadrature points is prime can be significantly slower than comparable grids with more favourable factorisations.
%%%%%%%%%%%%%%%%%%%%%%%%%%%%%%%%%%%%%%%%%%%%%%%%%%%%%%%%%
\subsubsection{Dependence on rotation rate}
As the rotation frequency of the projector is increased, the coherent region of the classical field becomes increasingly skewed towards modes with higher angular momentum projections.  The size of quadrature grid required to evaluate the nonlinear integral increases dramatically (in both the azimuthal and radial dimensions) as more high angular momentum modes are included in the calculation.  We consider here the dependence of the step time on the rotation frequency $\Omega$ of the frame in which the PGPE solution is propagated, by comparing the execution times of step evaluations at fixed energy cutoff $E_R=30\hbar\omega_r$ (figure~\ref{fig:rk4vsark}(b)).  In general we expect the step time to scale like\footnote{We have $N_x \approx \overline{N}/(1-\Omega)$ and $N_\theta = 2(\overline{l}_-+\overline{l}_+)+1 \approx 4\overline{N}/(1-\Omega)$, thus $N_x^2N_\theta \approx 4\overline{N}/(1-\Omega)^{-3}$.} $T \propto (1-\Omega)^{-3}$.  However in practice we find $T\propto (1-\Omega)^{-2}$, as shown in the inset, where we plot $\log(T)$ against $-\log(1-\Omega)$, and include a line with slope $m=2$ for comparison.  It is not clear why we observe this scaling (which is \emph{more favourable} than the theoretical prediction), rather than the expected $T\propto (1-\Omega)^{-3}$.  It is possibly due to some compiler optimisation arising in the multiplication of the triangular transformation matrices involved in the quadrature rule, which have many zero entries and become increasingly skewed as the rotation rate $\Omega$ increases.
%%%%%%%%%%%%%%%%%%%%%%%%%%%%%%%%%%%%%%%%%%%%%%%%%%%%%%%%%%%%%%%%%%%%%%%%%%%%%%%%%%%%%%%%
\subsection{Perturbing potential term}
\begin{figure}
	\begin{center}
	\includegraphics[width=0.9\textwidth]{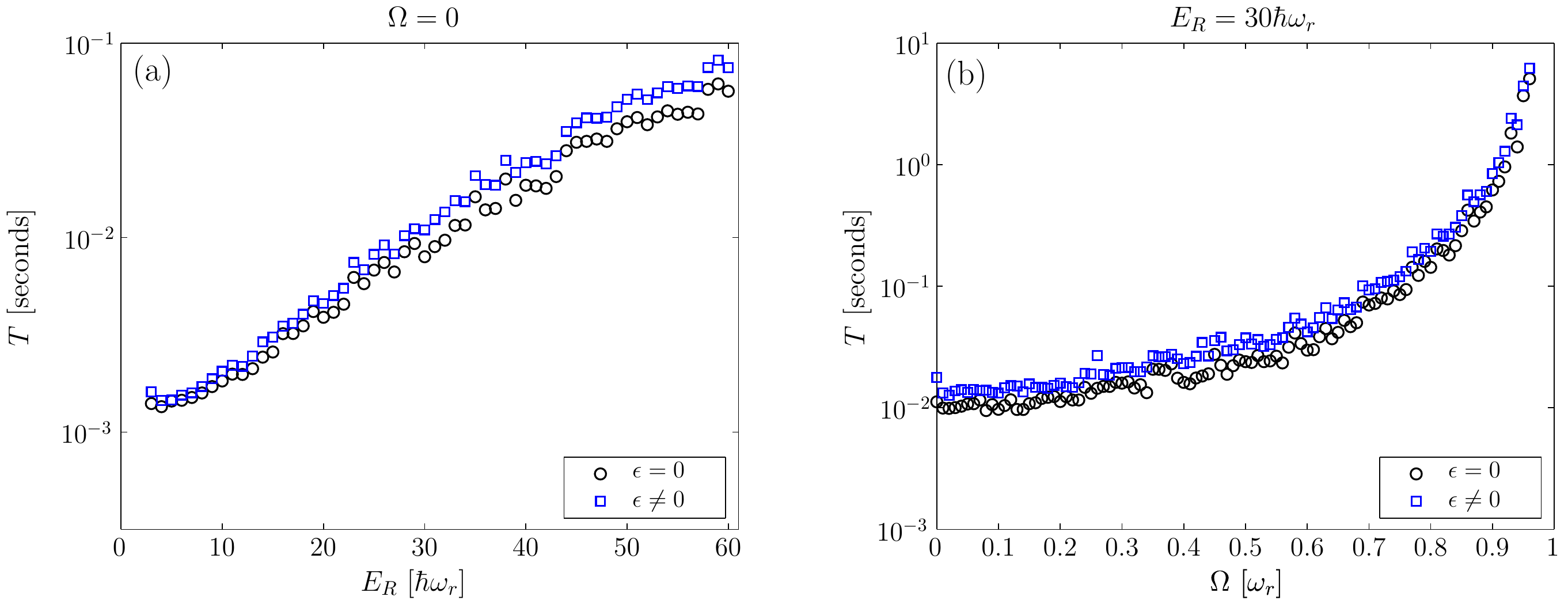}
	\caption{\label{fig:epsvsnoeps} Effect of the perturbing potential term $\delta V(\x)$. a) ARK45-IP step times with and without the additional quadrature rule for the perturbing trap anisotropy, as a function of cutoff energy $E_R$ at fixed rotation rate $\Omega=0$. b) Measured step times as a function of rotation rate $\Omega$ at fixed cutoff energy $E_R=30\hbar\omega_r$.}
	\end{center}
\end{figure}

The significant extension of the Gauss-Laguerre quadrature algorithm employed in this thesis over that used in other work \cite{Bradley08} is the addition of the elliptical potential term (appendix~\ref{app:app_quad}).  It is thus of interest to quantify how the inclusion of this term affects the step times.  In this section we compare the `base' algorithm (with ellipticity $\epsilon=0$) against the algorithm with $\epsilon\neq0$.  For simplicity, we only compare results for the adaptive (ARK) algorithm, this being the algorithm we use for all simulations in practice.
Plotting the step times in the presence of the perturbation alongside those obtained in its absence against the cutoff energy $E_R$ with $\Omega=0$, (figure~\ref{fig:epsvsnoeps}(a)), we see that the algorithm is fractionally slower, with the step time increasing from $\approx5\%$ to $\approx40\%$ of that of the base algorithm, over the range of cutoff heights considered.   
%%%%%%%%%%%%%%%%%%%%%%%%%%%%%%%%%%%%%%%
%%%%%%%%%%%%%%%%%%%%%%%%%%%%%%%%%%%%%%%
Plotting the dependence of step times on the rotation rate $\Omega$ (figure~\ref{fig:epsvsnoeps}(b)), we observe that the step time associated with the perturbing potential is $\approx 40\%$ of the base step time for all $\Omega$ we consider, in agreement with the $\Omega=0$ values obtained for this cutoff height ($E_R = 30\hbar\omega_r$).  
%%%%%%%%%%%%%%%%%%%%%%%%%%%%%%%%%%%%%%%%%%%%%%%%%%%%%%%%%%%%%%%%%%%%%%%%%%%%%%%%%%%%%%%%%%%%%%%%%%%%%%%%%%%%%%%%%%%%%%%%%%%%%%%%%%%%
\section{Comparison to other discretisation schemes}\label{sec:numerics_cfschemes}
As we have already noted, in general any numerical implementation of GP-like classical field evolution must involve some discretisation of the continuum field.  In this section we compare the projected approach to other common discretisation schemes applied to the integration of the GP equation.  We discuss the various pathologies and ambiguities that arise in the common Cartesian-grid discretisation schemes, and argue that the projected-GPE approach we adopt in this thesis is the unique approach which is free of these issues.  This is of particular importance as in chapters~\ref{chap:stir_background} and \ref{chap:stirring} we contrast the results of our PGPE calculations with those of other simulations which use Cartesian-grid methods.
%%%%%%%%%%%%%%%%%%%%%%%%%%%%%%%%%%%%%%%%%%%%%%%%%%%%%%%%%%%%%%%%%%%%%%%%%%%%%%%%%%%%%%%%
\subsection{Phase-space covering}\label{subsec:num_phase_space_covering}
%%%%%%%%%%%%%%%%%%%%%%%%%%%%%%%%%%%%%%%%%%%%%%%%%%%%%%%%%
\subsubsection{One-dimensional harmonic oscillator}
\begin{figure}
	\begin{center}
	\includegraphics[width=0.45\textwidth]{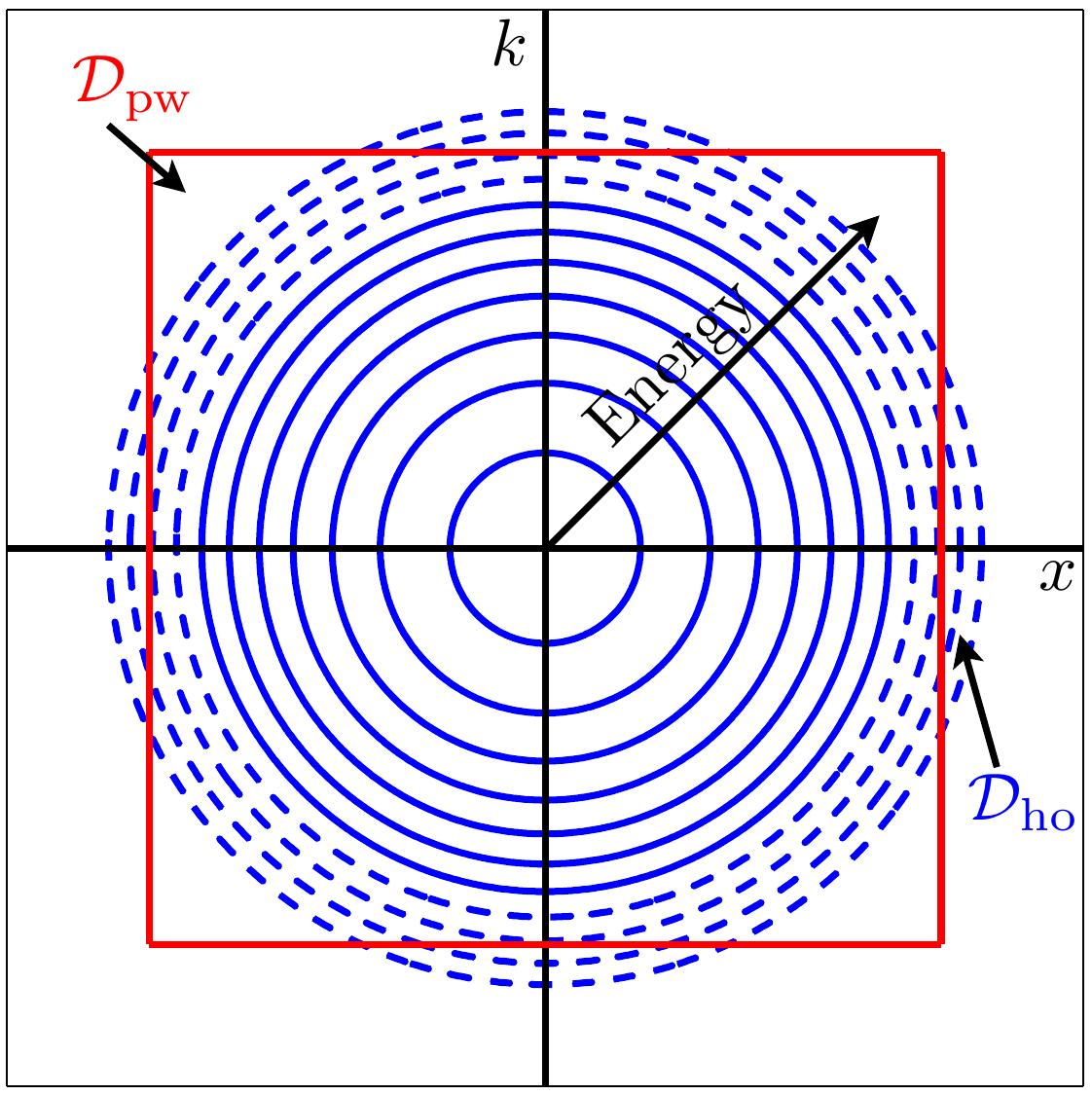}
	\caption{\label{fig:phasespace} Phase-space region $\mathcal{D}_\mathrm{ho}$ covered by the first $\bar{n}+1=11$ harmonic oscillator modes (circles).  The square indicates the region $\mathcal{D}_\mathrm{pw}$ enclosed by the corresponding optimal plane-wave basis with $\bar{n}+1$ grid points (see \cite{Bradley05}).  Dashed circles indicate harmonic oscillator modes which are not accurately resolved in the plane-wave basis.}
	\end{center}
\end{figure}

In this section we follow Bradley \emph{et al.} \cite{Bradley05} in considering the phase space of a one-dimensional harmonic oscillator (of frequency $\omega$), bounded by an energy cutoff $E_\mathrm{c} = (\bar{n}+1/2)\hbar\omega$.  In terms of appropriately scaled quantities $x$ and $k$ (see section~\ref{subsec:dimless})\footnote{The units of momentum and wave number corresponding to the system of units introduced in section~\ref{subsec:dimless} are $p_0=\sqrt{\hbar m\omega_r}$ and $k_0=1/r_0=p_0/\hbar$, respectively.  In this system of units momentum and wave vector are thus equivalent, and so we use the two concepts interchangeably in this section.}, the phase space is thus bounded by a circle, as illustrated in figure~\ref{fig:phasespace}.
%%%%%%%%%%%%%%%%%%%%%%%%%%%%%%%%%%%%%%%
%%%%%%%%%%%%%%%%%%%%%%%%%%%%%%%%%%%%%%%

In the projected formalism, the restriction of the system to this region of phase space is effected \emph{exactly} by restricting the basis of the field to the first $\bar{n}+1$ harmonic oscillator modes.  An alternative discretisation of the system is in terms of a regularly spaced grid in position space.  In this case, there are now \emph{two} cutoffs we must choose: the spatial extent $L$ of the grid, and the spacing $\Delta x$ between grid points, which determines the \emph{Nyquist wave number} $k_\mathrm{c} = \pi/\Delta x$, which is the magnitude of the largest wave vector resolvable on the grid.  Alternatively, we can determine the grid completely by the number of points $N_x$ and the grid extent $L$.  We note that the $x$-space and $k$-space limits of the grid are independent, i.e., the same wave-number cutoff $k_\mathrm{c}$ applies for all spatial positions $|x|\leq L/2$, and so the corresponding region of phase space spanned by this plane-wave basis is a \emph{rectangle} in phase space.  It is important to note that, as the wave-number cutoff $k_\mathrm{c} = \pi N_x /L$, the (2D) phase-space `volume' spanned by the rectangular region $V_\mathrm{pw} = 2\hbar k_\mathrm{c} L = h N_x$; i.e., each mode corresponds to a region of volume $h$, as is the case for the harmonic oscillator basis, and as it should be for the quantum system (Weyl's law \cite{Baltes76}).  Bradley \emph{et al.} \cite{Bradley05} showed that for a given number of points $N_x$, the grid with extent $L_\mathrm{opt}=\sqrt{2\pi\hbar N_x / m\omega}$ is \emph{optimal}, in the sense that the boundary of this phase space is \emph{square} (in the appropriate units), and is thus the closest match to the circular phase-space region spanned by the corresponding harmonic basis.  This region is indicated by its square boundary in figure~\ref{fig:phasespace}.  We note that the energy corresponding to the extremes of this region in each individual direction is $E_\mathrm{c}^\mathrm{(pw)} = (m/2)\omega^2 L^2 = \hbar^2 k_\mathrm{c}^2 / 2m = (\pi / 4) N_x \hbar\omega$. 

The authors of \cite{Bradley05} then proceeded to diagonalise the single-particle Hamiltonian of the harmonic oscillator in both the harmonic basis, and the plane-wave basis with an equal number of modes ($N_x=\bar{n}+1$).  In the harmonic basis this is trivial, as the basis is, by definition, comprised of the first $\bar{n}+1$ oscillator modes.  By contrast, the diagonalisation of the Hamiltonian in the plane-wave basis accurately reproduces the lowest modes, but with increasing energy the modes become less well resolved in this basis.  As the eigenmodes of the Hamiltonian in this basis essentially correspond to variational approximations to the true modes subject to the `ansatz' that they are comprised of linear combinations of the $N_x$ plane-wave modes, it is intuitive that the energies of these modes increasingly overestimate the true harmonic-oscillator mode energies.  In the example of $N_x=\bar{n}+1=16$ considered in \cite{Bradley05}, approximately the top one-third of modes were poorly represented in the \emph{optimal} plane-wave basis, while other choices of plane-wave basis in general did a worse job of representing the modes.

In general, it is hard to quantify exactly what effect such error in the mode energies has on the classical-field dynamics.  Some insight can be gained by using the principle of equipartition (section~\ref{subsec:cfield_thermo}).  At first sight, one might consider that a change in the energies would have little overall effect on the thermal equilibrium, as for a given total energy $E$, equipartition predicts that $E=\sum_j \epsilon_j n_j = \mathcal{M} k_\mathrm{B}T$, so that the temperature is determined solely by the number of modes $\mathcal{M}$ and the total thermal energy $E$, and does not depend on the actual values of the particular mode energies $\{\epsilon_j\}$.  However for a fixed total energy (and thus temperature), the individual mode occupations $n_j$ must vary in inverse proportion to the energies $\epsilon_j$, and so the total thermal population $N=\sum_j n_j$ will vary with the $\{\epsilon_j\}$.  We therefore conclude that while the relation between energy and temperature in general depends on the number of classical-field modes $\mathcal{M}$ \cite{Lobo04}, the relationship between energy, temperature and \emph{chemical potential} will in general depend on the precise energies of the modes, and will therefore be corrupted for non-ideal discretisations of the classical field.

The arguments here simply consider the single-particle modes, i.e., the thermal equilibrium of a noninteracting system.  In general we are of course concerned with the thermal equilibrium of an interacting system.  To the extent that a Bogoliubov description of the normal modes is valid, the relevant issue is the accuracy with which the normal modes are resolved within the propagation basis.  Again, to the extent that the Bogoliubov modes return to a single-particle description at high energies, we expect the most accurate representation to be obtained in the appropriate harmonic-oscillator basis.  However, even in this basis we must recognise that the highest-energy Bogoliubov modes will not be well represented in the harmonic basis, and in chapter~\ref{chap:stirring} we show that a signature of this breakdown of the mode basis can be observed in the results of an interacting classical-field simulation.  Nevertheless, as the field we evolve exists in a single-particle space, the basis of single-particle energy eigenstates is the optimal choice, and we must accept some error in the description of the highest normal modes of the interacting system as the minimal penalty for effecting the cutoff necessary for a classical-field treatment of the low-energy dynamics of the field.
%%%%%%%%%%%%%%%%%%%%%%%%%%%%%%%%%%%%%%%%%%%%%%%%%%%%%%%%%
\subsubsection{Higher dimensionalities}
We have shown that in one spatial dimension, where the corresponding classical phase space is two-dimensional, we can reasonably well approximate the boundary of phase space defined by a cutoff in single-particle (harmonic-oscillator) energy by a plane-wave (equidistant grid) discretisation.  In higher dimensionalities this is no longer the case, as we now discuss.  In one dimension, defining the cutoff by a single parameter (the energy) leads to the result that the momentum cutoff is \emph{local}, i.e., the cutoff wave number is a function of the position coordinate $x$.  For example, for a cutoff defined in terms of energy of a harmonic oscillator, the range of available momentum components below the cutoff is greatest at the centre of the trap, and smoothly vanishes as $|x|$ approaches the \emph{classical turning point} associated with the cutoff energy.  In higher dimensionalities, defining a cutoff in terms of the energy means that the cutoff in each of the $2\times d$ dimensions of phase space is dependent on the coordinates in \emph{all} the other dimensions.   Thus, in two spatial dimensions, in a harmonic basis, we can define a cutoff in terms of an energy $E_\mathrm{c} = (\bar{n}+1)\hbar\omega$, leading to the constraint on the excitation numbers of the $x$ and $y$ oscillators $n_x+n_y \leq \bar{n}$, yielding a total of $\mathcal{M}_\mathrm{ho}^\mathrm{2D} \approx \bar{n}^2/2$ modes (corresponding to a phase-space volume $V_\mathrm{ho}^\mathrm{2D} \approx \bar{n}^2h^2/2$).  It is not clear how to best approximate this region of phase space with a plane-wave cutoff.  One approach is to take $N_x=N_y=n$, with $L_x=L_y=L_\mathrm{opt}(N_x)$, so that the boundary of the plane-wave space most closely matches that of the harmonic-oscillator space in the $k_xx$ and $k_yy$ planes.  
This yields a total number of modes $M_\mathrm{pw}^\mathrm{2D} = \bar{n}^2$ (corresponding to phase-space volume $V_\mathrm{pw}^\mathrm{2D} = \bar{n}^2h^2$).  This plane-wave basis thus contains \emph{twice} as many modes as the harmonic basis.  A simple equipartition argument then predicts that the equilibrium temperature will be $\sim 1/2$ of the correct value, in the plane-wave case.  The reader can easily deduce that in three dimensions, a similarly constructed plane-wave discretisation now encloses approximately \emph{six} times the volume of the appropriate harmonic-basis discretisation, leading to an equilibrium temperature one-sixth of that of the ideal harmonic-basis description\footnote{Note that reversing these arguments one finds immediately that the plane-wave basis chosen to most accurately match the \emph{volume} of the harmonic-oscillator phase space will result in dramatically smaller cutoffs in each dimension of phase space, and will therefore unnaturally constrain the physics of the system.}.  Assuming (as in the example of \cite{Bradley05}) that the top one-third of harmonic oscillator modes in each direction are spuriously resolved under the plane-wave cutoff implies that the top $\approx10\%$ of the 3D-oscillator modes are poorly resolved, with overestimated energies.

In summary, in three-dimensional calculations, the plane-wave basis which `best' matches the (hyperspherical) energy-cutoff surface in phase space contains \emph{six} times the correct number of modes, a significant fraction of which are seriously ill-resolved in the available phase space, and have spuriously high energies.  Any larger `box' in phase space would include yet more spurious degrees of freedom, while the use of any smaller box would lead to severe corruption of a greater fraction of the modes contained in it.  It is therefore clear that the thermodynamics of the system would be strongly corrupted by this poor representation of the phase space, regardless of the precise dimensions of the plane-wave grid.  An important aspect of a classical-field defined in terms of single-particle \emph{energy eigenmodes} in two and three dimensional calculations is therefore that the resulting cutoff properly accounts for the local shape of the energy surface in all dimensions of phase space, as this surface can not be well approximated by the cutoff surface resulting from a plane-wave or Cartesian-grid basis. 
%%%%%%%%%%%%%%%%%%%%%%%%%%%%%%%%%%%%%%%%%%%%%%%%%%%%%%%%%%%%%%%%%%%%%%%%%%%%%%%%%%%%%%%%
\subsection{Implementing spatial derivatives}\label{subsec:numeric_spatial_derivs}\enlargethispage{-\baselineskip}
Thus far we have simply discussed how the phase space defined by a Cartesian grid compares with the phase space spanned by the harmonic oscillator basis.  Here we consider in more detail how the spatial integration of a GP-like partial differential equation is implemented, i.e., how the various operators of the GP equation are evaluated on the computational grid.  In the case of a regular Cartesian grid in position space, the potential and nonlinear terms are \emph{local}, and thus trivial to evaluate, and the difficulty arises in evaluating \emph{derivatives} of the field, i.e., the Laplacian $-\nabla^2$ appearing in the kinetic-energy term, as well as others that may result from the transformation to a translating or rotating~frame.
%%%%%%%%%%%%%%%%%%%%%%%%%%%%%%%%%%%%%%%%%%%%%%%%%%%%%%%%%
\subsubsection{Finite-difference and Fourier derivatives}
One approach to calculating derivatives of the classical field on a position-space grid is by the use of \emph{finite-difference} approximations to the relevant derivative operator.  A thorough account of such finite-difference approximations is given by Blakie \cite{Blakie01}, and such approximations are employed in (e.g.) the Crank-Nicolson method for evolving (nonlinear) Schr\"odinger equations (see, e.g., \cite{Parker_PhD}).  Usually one approximates derivatives by \emph{central} finite difference stencils.  For example, indexing the discretised field in terms of the grid points at which it is evaluated, $\psi_i \equiv \psi(x_i)$, the simplest central finite-difference approximation to the derivative $\partial_x \psi$ is given by \cite{Blakie01} 
\begin{equation}
	(\delta_x \psi)_i = \frac{1}{2\Delta x}\left[\psi_{i+1} - \psi_{i-1}\right].
\end{equation}
The key point is that $(\delta_x \psi)_i$ is a good approximation to $\partial_x \psi$ at $x_i$ for variations of $\psi$ on scales \emph{large} compared to $\Delta x$, or equivalently, for wave numbers much smaller than the \emph{cutoff} (Nyquist) wave number $k_\mathrm{c}$.  However, this leaves a large number of intermediate wave vectors for which the derivatives are \emph{not} accurately evaluated.   The most important example is provided by the approximation to the Laplacian $-\nabla^2$, which we here consider in this low-order scheme, in one dimension where it becomes \cite{Parker_PhD}
\begin{equation}
	(-\delta_x^2 \psi)_i = -\frac{1}{(\Delta x)^2}\left[ \psi_{i+1} - 2\psi_i + \psi_{i-1} \right].
\end{equation}
If we assume periodic boundary conditions (so that $\psi_{N_x+1}=\psi_1$), then in the basis of position eigenstates $\zeta_j(x_i) = \delta_{ij}$ the operator $-\delta_x^2$ is a sparse (band-diagonal) circulant matrix which is symmetric and has a complete basis of eigenvectors, which are the plane waves $\phi_i= e^{\pm ik_ix}$ (or equivalently the pairs $\sin(k_ix)$ and $\cos(k_ix)$), where the $k_i$ are the momentum modes resolvable on the Cartesian grid.  These are of course the correct eigenvectors, in the sense that they are eigenstates of the continuum Laplacian $-\nabla^2$ we wish to approximate.  However, their corresponding eigenvalues differ from those of $-\nabla^2$.  In fact, we find that $-\delta_x^2 e^{ik_ix} = \lambda_i e^{ik_ix}$ where the eigenvalues
\begin{equation}\label{eq:fd_dispersion}
	\lambda_{k_i} = \frac{2}{(\Delta x)^2}\left[1 - \cos\left(\frac{\pi k_i}{k_\mathrm{c}}\right)\right]. 
\end{equation}
Clearly for wave numbers which are \emph{small} compared to $k_\mathrm{c}$, this dispersion relation reduces to the continuum result $\lambda_k = k^2$.  However, for $k$ as small as $k_\mathrm{c}/2$ the error in the eigenvalue $\lambda_k$ is of order $20\%$.  In fact, the dispersion relation~\reff{eq:fd_dispersion} exhibits the peculiar feature of \emph{negative mass} for excitations with $k>k_\mathrm{c}/2$, as is well known for lattice systems (e.g., in solid-state physics)\footnote{Note that the issue here is that of a poor definition of the operator, not a restriction on the space on which it acts (as we considered in the previous section), so the eigenvalues need not be overestimated with respect to their `true' values.}.  

An alternative definition of the Laplacian on the Cartesian grid is obtained from the fundamental relation \cite{Davis_DPhil}
\begin{equation}
	-\nabla^2\psi(x) = \int dk k^2 e^{-ikx} \times \frac{1}{2\pi} \int dy e^{iky} \psi(y).
\end{equation}
The analogous discrete form of this relation (written in terms of discrete Fourier transforms) has \emph{exact} eigenvalues $\lambda_{k_i}=k_i^2$.  Thus while both approximations of the Laplacian approach the same continuum limit as $\Delta_x\to0$ and thus $k_\mathrm{c}\to\infty$, only the Fourier-defined Laplacian yields the `correct' eigenvalues throughout the computational domain.  Of course one expects the accuracy of the finite-difference Laplacian to be improved by using higher-order finite-differencing stencils, and we ultimately expect convergence to the Fourier-defined Laplacian in the limit that the order of the stencil approaches the total number of points on the grid.  In this limit, however, the finite-difference Laplacian is completely nonlocal, and acts as a \emph{dense} matrix on the configuration space.  Although the Fourier-transform operator is fundamentally also a dense matrix on the configuration space, \emph{fast Fourier transform} algorithms allow for the application of the transform using $O(N\log N)$ floating-point operations, as compared to $O(N^2)$ for a standard matrix multiplication, so the numerical advantage of using the Fourier-defined Laplacian is obvious.  
%%%%%%%%%%%%%%%%%%%%%%%%%%%%%%%%%%%%%%%%%%%%%%%%%%%%%%%%%
\subsubsection{Aliasing}
Another important concern with the use of grid methods is that of momentum-mode aliasing.  This is so named because momentum modes $e^{ikx_i}$ can only be properly resolved on the grid if the magnitude of the wave vector is not greater than the Nyquist wave number ($|k| \leq k_\mathrm{c}$).  Mathematically, the plane wave with $k=k_j+2nk_\mathrm{c}$ is indistinguishable from that with $k=k_j$ on the grid $\{x_i\}$, as both waves have the same values on all points on the grid.  Thus the momentum mode $k_j$ acts as an \emph{alias} for all modes $k_j+2nk_\mathrm{c}; n\in \mathbb{Z}$.  In digital signal processing, one must therefore be aware that any frequency components above the Nyquist cutoff will be indistinguishable from their aliases below the cutoff.  In the study of nonlinear Schr\"odinger equations, the issue is that the nonlinear potential $U_0|\psi|^2$ mixes momentum components, and thus (potentially) attempts to generate momentum components which are not resolvable on the grid, resulting in spurious coupling to their mode aliases.

A common misconception is that this phenomenon is introduced by the explicit use of the Fourier transform in GP propagation algorithms.  This is not the case, and the spurious momentum-nonconserving processes simply result from the action of the (local) nonlinear potential attempting to populate unresolvable modes.  To demonstrate this we show that the states are spuriously connected by the nonlinear potential.  Usually a potential $V$ causes a transition from a state $|\zeta_a\rangle$ to state $|\zeta_b\rangle$ if the ket $V|\zeta_a\rangle$ has some component along $|\zeta_b\rangle$.  The behaviour resulting from the nonlinear potential $|\psi|^2$ can similarly be deduced, except that the potential of course now depends on the ket it is applied to.  We consider the wave function consisting of two momentum modes $k_a$ and $k_b$, defined on the $\{x_i\}$ grid, $\psi(x_i)=ae^{ik_ax_i} + be^{ik_bx_i}$.  Acting on this state with the \emph{local} potential $V_\mathrm{nl}(x_i)=|\psi(x_i)|^2$, yields 
\begin{eqnarray}
	\left[V_\mathrm{nl}\psi\right](x_i) &=& a^2b^*e^{i(2k_a-k_b)x} + a^*b^2e^{i(2k_b-k_a)x} \nonumber \\
    &&\quad + \left(|a|^2 + 2|b|^2\right)ae^{ik_ax_i} + \left(2|a|^2 + |b|^2\right)be^{ik_bx_i},
\end{eqnarray}
thus the nonlinear potential transfers population to the plane wave $e^{i(2k_a-k_b)x}$.  In the particular `worst case' considered by Norrie  \cite{Norrie05a}: $k_a=-k_\mathrm{c}$, $k_b=k_\mathrm{c}-\Delta k$, we see that the nonlinear potential attempts to transfer population to the mode $-3k_\mathrm{c}+\Delta k$, resulting in spurious transfer to the alias mode $-k_\mathrm{c}+\Delta k$.

We note that grid methods which are \emph{not} based on Fourier-transform evaluation of derivatives are frequently implemented with \emph{Dirichlet} boundary conditions, i.e., the wave function is constrained to specified values (e.g., zero) at the extremes of the grid~\cite{Parker_PhD}.  One can easily show that a wave function consistent with such a constraint (i.e., expanded in terms of $\cos(k_ix)$ and $\sin(k_ix)$) suffers from the same aliasing problem under the action of the local nonlinear potential. 

We note that the Fourier-based grid methods are thus more self-consistent than finite-difference-based methods in that they explicitly take into account the periodicity in both $x$ and $k$ space, the $k$-space periodicity being an implicit feature of all grid methods, due to the mode aliasing.  Ultimately, the problem with Fourier-based methods is that the phase space spanned by the plane-wave basis modes is not only geometrically ill-formed, but is also topologically incorrect: due to the periodic nature of the position and momentum coordinates, the phase space has no real boundary, and is instead a toroidal structure.  Momentum-mode aliasing can be eliminated for Fourier-based simulations of homogeneous systems with periodic boundary conditions in space \cite{Davis_DPhil} and simulations in free space \cite{Norrie05a}, by a projected generalisation of the method.  However, there is no simple way to accurately match the correct phase space of (e.g.) a harmonically trapped system using a plane-wave basis.
%%%%%%%%%%%%%%%%%%%%%%%%%%%%%%%%%%%%%%%%%%%%%%%%%%%%%%%%%
\subsubsection{Spectral and discrete-variable-representation methods}
An alternative approach to the discretisation of the continuous nonlinear Schr\"odinger equation is through the use of a \emph{spectral representation}.  In this approach, one expands the Schr\"odinger wave function over a set of basis functions with known derivatives, so that the action of (e.g.) the Laplacian can be evaluated exactly.  The difficulty then is in evaluating the matrix elements of potential terms between these basis functions.  In a \emph{discrete variable representation} \cite{Schneider99}, the basis functions are chosen such that an associated quadrature rule exists, and the functions form a complete basis for functions evaluated at the quadrature points $\{x_k\}$.  A transformation between the basis expansion and a set of corresponding \emph{position eigenfunctions} then allows for accurate evaluation of operators which are diagonal in position space, while retaining a simple matrix representation for the Laplacian.

The authors of \cite{Dion03} presented a technique for evolving the time-dependent GPE for a harmonically confined condensate in a basis of the Hermite-Gaussian eigenmodes of the harmonic-oscillator Hamiltonian.  Such a \emph{spectral-Galerkin} approach combines the accuracy of the Gaussian quadrature rule for the evaluation of the nonlinear potential, while \emph{exactly} diagonalising the whole single-particle Hamiltonian.  The Gauss-quadrature methods presented in appendix~\ref{app:app_quad} are a refinement of this approach, with the distinction that the precise set of basis modes used is carefully chosen to \emph{define} the classical-field system, according to a prescribed single-particle energy cutoff, and to ensure the optimal representation of all modes with energies below this cutoff.
%%%%%%%%%%%%%%%%%%%%%%%%%%%%%%%%%%%%%%%%%%%%%%%%%%%%%%%%%%%%%%%%%%%%%%%%%%%%%%%%%%%%%%%%%%%%%%%%%%%%%%%%%%%%%%%%%%%%%%%%%%%%%%%%%%%%
%%%%%%%%%%%%%%%%%%%%%%%%%%%%%%%%%%%%%%%%%%%%%%%%%%%%%%%%%%%%%%%%%%%%%%%%%%%%%%%%%%%%%%%%%%%%%%%%%%%%%%%%%%%%%%%%%%%%%%%%%%%%%%%%%%%%

\chapter{Anomalous moments in classical-field theory}
\label{chap:anomalous}
%%%%%%%%%%%%%%%%%%%%%%%%%%%%%%%%%%%%%%%%%%%%%%%%%%%%%%%%%%%%%%%%%%%%%%%%%%%%%%%%%%%%%%%%%%%%%%%%%%%%%%%%%%%%%%%%%%%%%%%%%%%%%%%%%%%%
%%%%%%%%%%%%%%%%%%%%%%%%%%%%%%%%%%%%%%%%%%%%%%%%%%%%%%%%%%%%%%%%%%%%%%%%%%%%%%%%%%%%%%%%%%%%%%%%%%%%%%%%%%%%%%%%%%%%%%%%%%%%%%%%%%%%
Equilibrium classical-field methods make no assumptions about the fluctuation statistics of the atomic field beyond the neglect of \emph{quantum} fluctuations.  In high-temperature regimes, they therefore \emph{implicitly} describe all relevant physical processes that shape the correlations of the field.  This is in stark contrast to self-consistent mean-field theories (section~\ref{subsec:theory_Bogoliubov}), in which one \emph{begins} with assumptions about the fluctuation statistics of the Bose field: the separation of the field into condensed and noncondensed (thermal) components, and the factorisation of the moments of the thermal component.  Although classical-field methods provide an accurate, nonperturbative description of the field, due to the lack of such constraining assumptions we frequently face difficulties in \emph{interpreting} the results that they describe, and specifically in relating them to more traditional descriptions of Bose-Einstein condensation, such as the mean-field theories.

In this chapter, we demonstrate that meaningful information is encoded in the \emph{temporal} behaviour of classical-field trajectories.  
We show that equilibrium classical-field trajectories exhibit a kind of symmetry-breaking phenomenon, in the sense that the condensate mode exhibits a well-defined phase at any point in time.  Due to the phase symmetry of the classical-field Hamiltonian, the condensate phase diffuses over time \cite{Sinatra08}, restoring the symmetry.  
However, averages of the classical field in an appropriate phase-rotating frame taken on time scales \emph{short} compared to that of the phase diffusion yield finite values for moments of the field which are formally zero in its microcanonical density.  These moments correspond to the \emph{anomalous moments} of the Bose field in symmetry-breaking mean-field theories.  Applying this methodology to PGPE simulations of the harmonically trapped Bose gas, we show that the first moment of the field obtained in this approach agrees well with the condensate defined by the Penrose-Onsager criterion based on one-body correlations (section~\ref{subsec:cfield_correl}), over a wide range of temperatures.  We demonstrate the generality of this prescription for evaluating anomalous moments of the classical-field variables by calculating the anomalous thermal density of the field, which we find to have form and temperature dependence consistent with the results of mean-field theory calculations.  

This chapter is organised as follows: in section~\ref{sec:anom_system} we introduce the PGPE system we study in this chapter.  In section~\ref{sec:temporal_coherence} we discuss how a nonzero first moment of the classical field arises from its temporal coherence, and make a quantitative comparison to the condensate defined by the Penrose-Onsager measure of one-body (spatial) coherence.  In section~\ref{sec:energy_dependence} we consider the dependence of this first moment and its associated frequency on the energy of the classical field. In section~\ref{sec:pairing_correlations}, we consider the anomalous second moments which comprise the classical-field \emph{pair matrix}, and construct the anomalous thermal density of the field. In section~\ref{sec:conclusions} we summarise and present our conclusions.  
%%%%%%%%%%%%%%%%%%%%%%%%%%%%%%%%%%%%%%%%%%%%%%%%%%%%%%%%%%%%%%%%%%%%%%%%%%%%%%%%%%%%%%%%%%%%%%%%%%%%%%%%%%%%%%%%%%%%%%%%%%%%%%%%%%%%
\section{System}\label{sec:anom_system}
In this chapter we consider the behaviour of a nonrotating, three-dimensional harmonically trapped classical field.  Our interest here is in characterising the correlations of the field, rather than to provide a direct comparison to an experimental system, and we therefore study a \emph{unit-normalised} classical field, which evolves according to equation~\reff{eq:num_unit_pgpe}.  We take the trap anisotropy parameter $\lambda_z=\sqrt{8}$, corresponding to a typical, moderately oblate three-dimensional trap geometry.  We choose the nonlinearity parameter $C_\mathrm{nl}=\sqrt{2}\times500$.  The corresponding ground (Gross-Pitaevskii) eigenstate of this system thus has energy $E_\mathrm{g}\approx 9N_\mathrm{c}\hbar\omega_r$, and GP eigenvalue $\lambda_\mathrm{g}\approx12\hbar\omega_r$.  Using the criteria discussed in section~\ref{sec:cfield_projection} we therefore set the energy cutoff $E_R=31\hbar\omega_r$.  We form initial states with energies in the range $E\in[9.5,24.0]N_\mathrm{c}\hbar\omega_r$, by in each case mixing an admixture of a random excited state with the Thomas-Fermi ground state of the unit-normalised field \cite{Blakie05,Blakie08}.  We then generate equilibrium states with a range of temperatures by evolving these randomised initial states in real time, exploiting the ergodicity of the PGPE (see section~\ref{sec:cfield_ergodicity}).  In practice we find that the classical-field configurations are well equilibrated after $120\omega_r^{-1}$ of real-time evolution, and we perform our analysis on the subsequent evolution of the field.
%%%%%%%%%%%%%%%%%%%%%%%%%%%%%%%%%%%%%%%%%%%%%%%%%%%%%%%%%%%%%%%%%%%%%%%%%%%%%%%%%%%%%%%%%%%%%%%%%%%%%%%%%%%%%%%%%%%%%%%%%%%%%%%%%%%%
\section[Temporal coherence and the emergence of a nonzero first moment]{Temporal coherence and the emergence of a \\ nonzero first moment}\label{sec:temporal_coherence}
%%%%%%%%%%%%%%%%%%%%%%%%%%%%%%%%%%%%%%%%%%%%%%%%%%%%%%%%%%%%%%%%%%%%%%%%%%%%%%%%%%%%%%%%
\subsection{Identification of the first moment}\label{subsec:id_first_moment}
We begin by quantifying the coherence of the time-dependent field $\psi(\mathbf{x},t)$ via its temporal power spectrum\footnote{We note that the power spectrum is related to the autocorrelation function of the field by the well-known \emph{Wiener-Khinchin} theorem \cite{Press92}.}, evaluated at different spatial locations $\mathbf{x}$. We define the temporal power spectrum of the classical field $\psi$ at position $\mathbf{x}$, evaluated over a period of length $T$ as 
\begin{equation}\label{eq:power_spec}
	H(\mathbf{x};\Omega) = \left|\mathfrak{F}^T\left\{\psi(\mathbf{x},t)\right\}\right|^2,
\end{equation}
where $\mathfrak{F}^T\{f(t)\}$ denotes the Fourier coefficient taken from some arbitrary time origin
\begin{equation}\label{eq:four_coeff}
	\mathfrak{F}^T\{f(t)\} \equiv \frac{1}{T}\int_0^Tf(t)e^{i\Omega t}dt.
\end{equation}

\begin{figure}
	\begin{center}
	\includegraphics[width=0.75\textwidth]{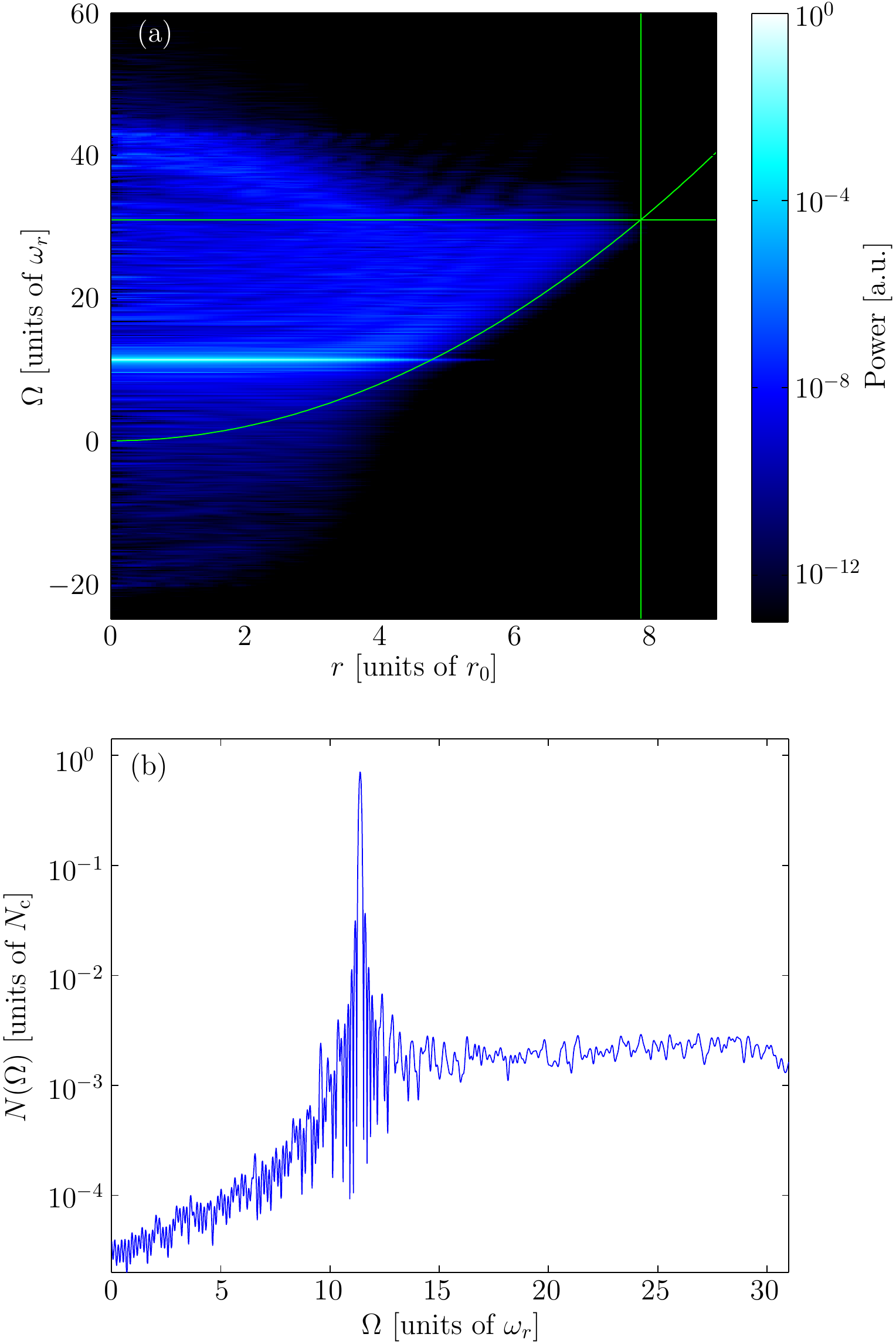}
	\caption{\label{fig:power_spectrum_in_trap}  (a) Power spectrum $H(\mathbf{x};\Omega)$ of the classical field on the plane $z=0$ (azimuthally averaged). Green (grey) lines indicate the trapping potential, cutoff energy and corresponding classical turning point of the trap. (b)  Space-integrated power spectrum $N(\Omega)$ of the field (see equation~(\ref{eq:N_omega})) as a function of the phase-rotation frequency $\Omega$.}
	\end{center}
\end{figure}

Here we calculate the power spectrum for a classical field evolving with the trapping and interaction parameters of section~\ref{sec:anom_system}, and energy $E=12.0N_\mathrm{c}\hbar\omega_r$.  Using the Penrose-Onsager (PO) approach introduced in section~\ref{subsec:cfield_correl}, we find that the condensate fraction for this classical field is $f_\mathrm{c}\equiv n_0/\sum_k n_k=0.70$.  We choose a sampling period of $40\omega_r^{-1}$, and approximate the integral in equation~(\ref{eq:four_coeff}) by a discrete sum over $1000$ equally spaced samples of the classical field.  In practice, we calculate the power spectrum at points in the $z=0$ plane, and average it over the azimuthal angle in this plane to smooth out fluctuations.  We thus obtain the averaged power spectrum as a function of the radius $r$, which we present in figure~\ref{fig:power_spectrum_in_trap}(a).  The oscillation frequencies $\Omega$ that we measure in the time-dependent field correspond, of course, to energies $\epsilon=\hbar\Omega$ in the quantum mechanical system.  
%%%%%%%%%%%%%%%%%%%%%%%%%%%%%%%%%%%%%%%
%%%%%%%%%%%%%%%%%%%%%%%%%%%%%%%%%%%%%%%

For comparison, on the same figure we also plot the profile of the harmonic trapping potential $V(r)/\hbar$ in this plane (parabolic green (grey) line) and the cutoff energy $E_R/\hbar$ (horizontal line), and the classical turning point (vertical line) of the condensate band defined by their intersection.  The most prominent feature in this plot is the strong peak in the power spectrum centred on $\Omega=11.4\omega_r$, which is a signature of the long-lived temporal phase coherence of the condensate formed in the classical field.  We identify the frequency $\lambda_0$ where this peak is located as the \emph{condensate frequency}.  The broad, lower intensity background spectrum represents the thermally occupied excitations in the classical-field system.  It is worth pointing out that, in the central region of the trap $(r\lesssim 4r_0$), the background spectrum is strongly distorted by the presence of the condensate, with positive frequency components extending to approximately $\lambda_0+E_R/\hbar$, and negative frequency components extending down to approximately $\lambda_0-E_R/\hbar$~\footnote{Note that the cutoff energy refers only to the \emph{single-particle} modes that the field is expanded on.  The energies of excitations in the interacting field are raised because of the mean-field potential they experience.}.  This behaviour represents the restructuring of the excitation spectrum of the trap by the interacting condensate, which distorts the single-particle excitations of the system into the familiar Bogoliubov particle-hole pairs (section~\ref{subsec:theory_Bogoliubov}).

The temporal coherence we observe results from the quasi\-uniform phase rotation of the condensate: the phase of the condensate exhibits a uniform rotation at frequency $\lambda_0$, superposed with a slow diffusion.  The width of the power-spectrum peak here is thus determined by the rate of this global condensate-phase diffusion.  On time scales short compared with the characteristic time scale of phase diffusion, the condensate has an approximately constant phase in a frame co-rotating at frequency $\lambda_0$, i.e., short-time averages in this frame yield a nonzero \emph{first moment} $\langle \psi \rangle$ of the classical field.  A key observation of this chapter is that time averages constructed in this way are analogous to the \emph{anomalous averages} which arise in symmetry-broken descriptions of Bose condensation \cite{Griffin96}, where the appearance of nonzero values for expectations of non-gauge-invariant quantities (i.e., the breaking of the phase symmetry) signals the presence of condensation in the field.
We thus consider the classical field frequency-shifted by $\Omega$
\begin{equation}
	\widetilde{\psi}(\mathbf{x},t;\Omega) = e^{i\Omega t} \psi(\mathbf{x},t),
\end{equation}
and consider time-averages of this quantity formed from the same set of samples used to construct the power spectrum in figure~\ref{fig:power_spectrum_in_trap}(a).  We define the time-averaged field 
\begin{eqnarray}\label{eq:time_avgd_field}
	 \phi(\mathbf{x};\Omega) &\equiv& \left\langle \widetilde{\psi}(\mathbf{x},t;\Omega) \right\rangle_t \nonumber \\
	\Big(\! &=& \mathcal{F}^T\{\psi(\mathbf{x},t)\} \;\Big), 
\end{eqnarray} 
where $\langle \cdots \rangle_t$ denotes a time average over a given period $T$ ($40\omega_r^{-1}$ in this case). 
The time-averaged field $\phi(\mathbf{x};\Omega)$ is therefore the component of the classical field which phase-rotates like $e^{-i\Omega t}$, and its norm square quantifies the total (i.e., space-integrated) power contained in the field at frequency $\Omega$, i.e.,
\begin{equation}\label{eq:N_omega}
	N(\Omega) \equiv \int\!d\mathbf{x}\,|\phi(\mathbf{x};\Omega)|^2 = \int\!d\mathbf{x}\,H(\mathbf{x};\Omega).
\end{equation}
In figure~\ref{fig:power_spectrum_in_trap}(b), we plot this power as a function of the frequency $\Omega$, and note that it exhibits a prominent peak at $\Omega=11.38\omega_r$.  We identify the frequency at which the norm square of the time-averaged field (equivalently, the space-integrated power of the field) is maximised as the condensate frequency $\lambda_0$, and the corresponding time-averaged field $\phi(\mathbf{x};\lambda_0)$ as the classical-field condensate or \emph{mean field}\footnote{The identification of the condensate and its eigenfrequency in this manner bears some resemblance to the method of Feit \emph{et al.} \cite{Feit82} for the identification of Schr\"odinger eigenstates and eigenvalues from a spectral analysis of trajectories of the time-dependent Schr\"odinger equation.  In contrast to that work, however, the condensate `eigenmode' we consider here emerges from the trajectories of a \emph{nonlinear} equation of motion, and can only be obtained from an analysis of the real-time field trajectories (or some other sampling of the PGPE microcanonical density equation~\reff{eq:cfield_mu_density}).}.  A nonzero time-averaged field occurs because the condensate has a reasonably well-defined phase on short time periods.  We identify this quasi-definite phase as an analogue of the condensate phase which emerges in symmetry-broken descriptions of Bose-Einstein condensation; in this viewpoint, the first moment $\phi(\mathbf{x};\lambda_0)$ is the analogue of the condensate wave function $\langle \hat{\Psi}(\mathbf{x}) \rangle$ in such mean-field theories of Bose condensation.  For notational convenience, we introduce the norm square $N_0\equiv N(\lambda_0)$ of the mean field,  and the normalised mean-field mode function $\phi_0(\mathbf{x})\equiv\phi(\mathbf{x};\lambda_0) / \sqrt{N_0}$.   
The norm square $N_0$ corresponds to the \emph{population} of the mean-field condensate mode, and we indeed find $N_0/N_\mathrm{c}=0.706$, in close agreement with the PO value for the condensate fraction ($f_\mathrm{c}=0.70$).  To further compare this temporal-coherence method of identifying the condensate with the PO approach, we calculate the overlap of $\phi_0(\mathbf{x})$ with the eigenvector $\chi_0(\mathbf{x})$ obtained by the PO procedure.  We find $1- |\langle \phi_0 | \chi_0 \rangle| \approx 1.4\times 10^{-4}$, i.e., the condensate orbitals obtained by the two different procedures agree to a very high accuracy.
%%%%%%%%%%%%%%%%%%%%%%%%%%%%%%%%%%%%%%%%%%%%%%%%%%%%%%%%%%%%%%%%%%%%%%%%%%%%%%%%%%%%%%%%
\subsection{Temporal coherence and sample length}\label{subsubsec:temporal_coherence}
The results obtained for the mean field have an important dependence on the averaging time.  As discussed by Sinatra and Castin, the condensate phase exhibits diffusive evolution with time in the classical microcanonical ensemble \cite{Sinatra08}.  As a consequence, the results of the spectral analysis introduced in section~\ref{subsec:id_first_moment} will depend on the length of time over which we sample the classical field.  To gain some insight into the consequences of this phase diffusion, we first consider a simple analytical model of the phase-diffusive condensate mode.  We then present and interpret the results we obtain for the classical-field condensate as a function of the time over which we sample the field.
%%%%%%%%%%%%%%%%%%%%%%%%%%%%%%%%%%%%%%%%%%%%%%%%%%%%%%%%%
\subsubsection{Single phase-diffusive mode}
We first consider a single-mode model of the condensate, in which the amplitude $a_0(t)$ of the condensate mode (with condensate frequency $\lambda_0$) exhibits phase diffusion.  We assume that the mode does not exhibit any number fluctuations, which is precisely the condition that $g_0^{(2)}=\langle |a_0|^4 \rangle_t / \langle |a_0|^2 \rangle^2_t = 1$, which is well satisfied away from the critical regime associated with the transition to the noncondensed state \cite{Bezett09b}.  We thus have $a_0=|a_0|e^{i\theta(t)}$ and, defining $\varphi(t) \equiv \theta(t) - \theta(0)$, we assume $\mathrm{var}\{\varphi(t)\} \equiv \langle \varphi(t)^2\rangle - \langle \varphi(t) \rangle^2 = 2\gamma t$ \cite{Sinatra08}, where $\gamma$ is the (phase) \emph{diffusion coefficient}, and $\langle \cdots \rangle$ denotes an average over \emph{realisations} of the amplitude $a_0(t)$ (i.e., an ensemble average).   This is precisely the behaviour of the \emph{Kubo oscillator} \cite{Gardiner04} stochastic process, which obeys the (Ito) stochastic differential equation 
\begin{equation}\label{eq:kubo_oscillator}
	da_0 = \left[(-i\lambda_0 - \gamma)dt + i\sqrt{2\gamma}dW(t)\right]a_0(t),
\end{equation}
where $dW(t)$ is a \emph{real} Wiener increment, which satisfies $\langle dW(t) dW(t') \rangle = \delta(t-t')dt$.  By studying this simple model we hope to gain insight into the behaviour of our diffusive condensate mode.

We consider the power spectrum of the mode obtained over a period $T$,
\begin{equation}
	N^{(0)}(\Omega;T) = \left|\frac{1}{T}\int_0^Tdt\;e^{i\Omega t} a_0(t) \right|^2.
\end{equation}
This power spectrum is itself a stochastic process (developing in $T$), i.e., it varies between realisations of the oscillator.  We therefore consider its mean $\langle N^{(0)}(\Omega;T)\rangle$.  Using the known result $\langle a_0(t) a_0^*(s) \rangle = |a_0|^2\exp[-i\lambda_0(t-s) -\gamma|t-s|]$ \cite{Gardiner04}, we find 
\begin{eqnarray}\label{eq:N0_mean_general}
	\langle N^{(0)}(\Omega;T) \rangle &=& \frac{1}{T^2\left(\gamma^2+\Delta^2\right)^2} \Big\{ \gamma T \left(\gamma^2 + \Delta^2\right) \nonumber \\
&&\quad+\left[e^{-\gamma T}\cos(\Delta\;T) - 1\right]\left(\gamma^2 - \Delta^2\right) - 2\gamma \Delta \sin(\Delta\;T)  \Big\}, 
\end{eqnarray}
which we have written in terms of $\Delta\equiv\Omega-\lambda_0$ for compactness.  In the limit of no diffusion ($\gamma \rightarrow 0$) we regain the result $N^{(0)}(\Omega;T) = |a_0|^2 \mathrm{sinc}^2[\frac{1}{2}(\Omega-\lambda_0)T]$ resulting from the convolution of the delta-function spectrum of the deterministic oscillator with the Fourier transform of the boxcar function representing the finite measurement time $T$ \cite{Press92}.
In the limit of a measurement made on a time scale long compared with the characteristic diffusion time (i.e. $\gamma T \gg 1$), we regain the Lorentzian spectrum of the Kubo oscillator $N^{(0)}(\Omega;T)=(2|a_0|^2\gamma/T)/[\gamma^2+(\Omega-\lambda_0)^2]$.   
%%%%%%%%%%%%%%%%%%%%%%%%%%%%%%%%%%%%%%%
\begin{figure}    
	\begin{center}
	\includegraphics[width=0.65\textwidth]{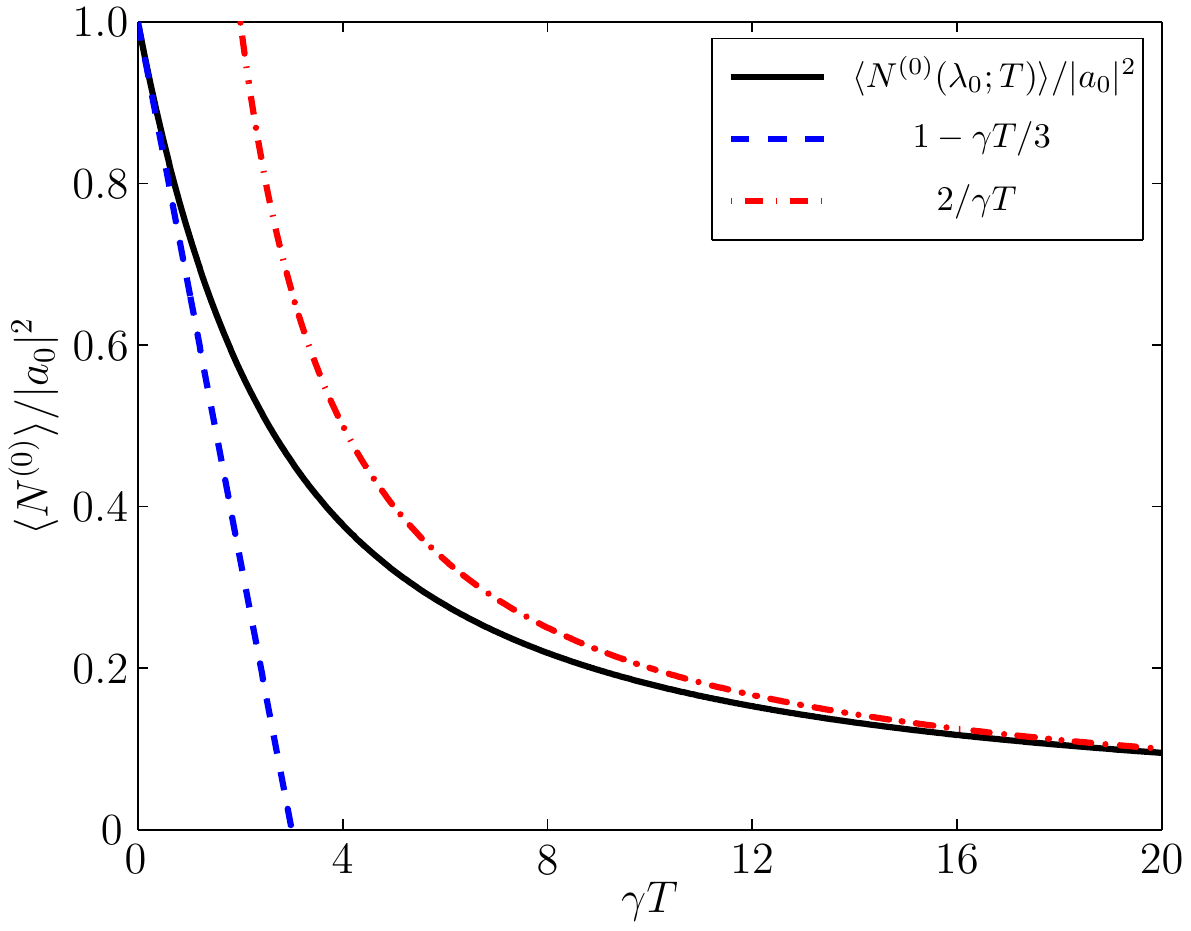}    
	\caption{\label{fig:kubo} Decay of the norm square of the mean field (i.e. the power measured at the condensate frequency) in a single-mode model, as a function of the averaging time.  The asymptotic short-time and long-time behaviours of the function are included for comparison.}
	\end{center}
\end{figure}
%%%%%%%%%%%%%%%%%%%%%%%%%%%%%%%%%%%%%%%

From equation~(\ref{eq:N0_mean_general}), the power measured at the underlying frequency $\lambda_0$ of the oscillator can be obtained by setting $\Delta=0$, giving
\begin{equation}\label{eq:mean_decay}
	\langle N^{(0)}(\lambda_0;T) \rangle = |a_0|^2\frac{2}{(\gamma T)^2} \left[\gamma T - \left(1 - e^{-\gamma T}\right)\right].
\end{equation}
For short time periods $T \ll 1/\gamma$ (such as we consider in the main text), the norm square of the mean field decays like $\sim 1 - \gamma T/3$, while at long times it decays like $\sim 2/\gamma T$ (figure~\ref{fig:kubo}).   It is important to note that this same functional form would be exhibited by (e.g.) a complex Ornstein-Uhlenbeck process \cite{Gardiner04}, which one might reasonably assume as a model for a thermally occupied mode \cite{Gardiner00,Stoof01} in a classical-field approximation.  The `bare' (i.e., infinite sampling time) power spectrum of such a mode is similarly Lorentzian, and so we expect the same behaviour for the condensate and the thermal mode, both for two-time correlations ($|\langle a^*(t) a(0) \rangle| \sim e^{-\gamma t}$) and for the measured power $N(\Omega;T)$.  The two cases (i.e. condensate and thermal mode) are thus distinguished only by the \emph{time scales} on which the power decays.  \emph{Qualitative} differences between the two types of mode only appear in second-order (and higher) correlation functions, which are sensitive to \emph{number} fluctuations.
%%%%%%%%%%%%%%%%%%%%%%%%%%%%%%%%%%%%%%%%%%%%%%%%%%%%%%%%%
\subsubsection{Multimode description}
In general, the condensate mode is only one mode in a multimode field which contains other, thermally occupied modes.  We expect the thermal field to be well described in the basis of Bogoliubov modes $\{(u_i,v_i)^T\}$ orthogonal to the condensate mode \cite{Castin01}, and thus assume
\begin{equation}
	\left\langle \frac{1}{T} \int\!dt\,e^{i\lambda_0 t} \psi(\mathbf{x},t) \right\rangle = \frac{1-e^{-\gamma T}}{\gamma T} |a_0|\left\langle e^{i\theta(0)}\right\rangle \chi_0(\mathbf{x}),
\end{equation}
where, to gain simple insight into our measurements of the field, we assume that the $\{b_j(t)\}$ are complex Ornstein-Uhlenbeck processes which are uncorrelated with one another and with the condensate process.  The space-integrated power spectrum of the field is thus 
\begin{eqnarray}
	\left\langle N(\Omega;T) \right\rangle &=& \left\langle N^{(0)}(\Omega;T)\right\rangle \int\!d\mathbf{x}\,|\chi_0(\mathbf{x})|^2  \nonumber \\
	&&\quad+ \sum_j \left\langle N^{(j)}(\Omega;T) \right\rangle \int\!d\mathbf{x}\,|u_j(\mathbf{x})|^2 + |v_j(\mathbf{x})|^2, 
\end{eqnarray}
where 
\begin{eqnarray}
	\left\langle N^{(j)}(\Omega;T) \right\rangle = \left|\frac{1}{T}\int_0^T\!dt\,e^{i\Omega t} b_j(t) \right|^2,
\end{eqnarray}
behave similarly to $\langle N^{(0)}(\Omega;T) \rangle$, except that they are centred on the frequencies $\epsilon^\mathrm{B}_j/\hbar$ of the Bogoliubov modes, and attenuate much more rapidly with $T$ ($\gamma_j \gg \gamma_0$).  There is therefore power in the field at a range of frequencies, however, for times $T\gg 1/\gamma_j$ we have $\langle N(\Omega;T) \rangle \approx N^{(0)}(\Omega;T) \int\!d\mathbf{x}\,|\chi_0(\mathbf{x})|^2$ and, moreover, 
\begin{equation}
	\left\langle \widetilde{\psi}(\mathbf{x};\lambda_0;T) \right\rangle = \frac{1-e^{-\gamma T}}{\gamma T} |a_0|\left\langle e^{i\theta(0)}\right\rangle \chi_0(\mathbf{x}),
\end{equation} 
where the appearance of the expectation $\langle e^{i\theta(0)} \rangle$ of the initial complex phase emphasises that the condensate phase varies randomly between ensemble members, `breaking' the $\mathrm{U(1)}$ symmetry in any particular realisation.
%%%%%%%%%%%%%%%%%%%%%%%%%%%%%%%%%%%%%%%%%%%%%%%%%%%%%%%%%
\subsubsection{Measured results}
\begin{figure}
	\begin{center}
	\includegraphics[width=0.65\textwidth]{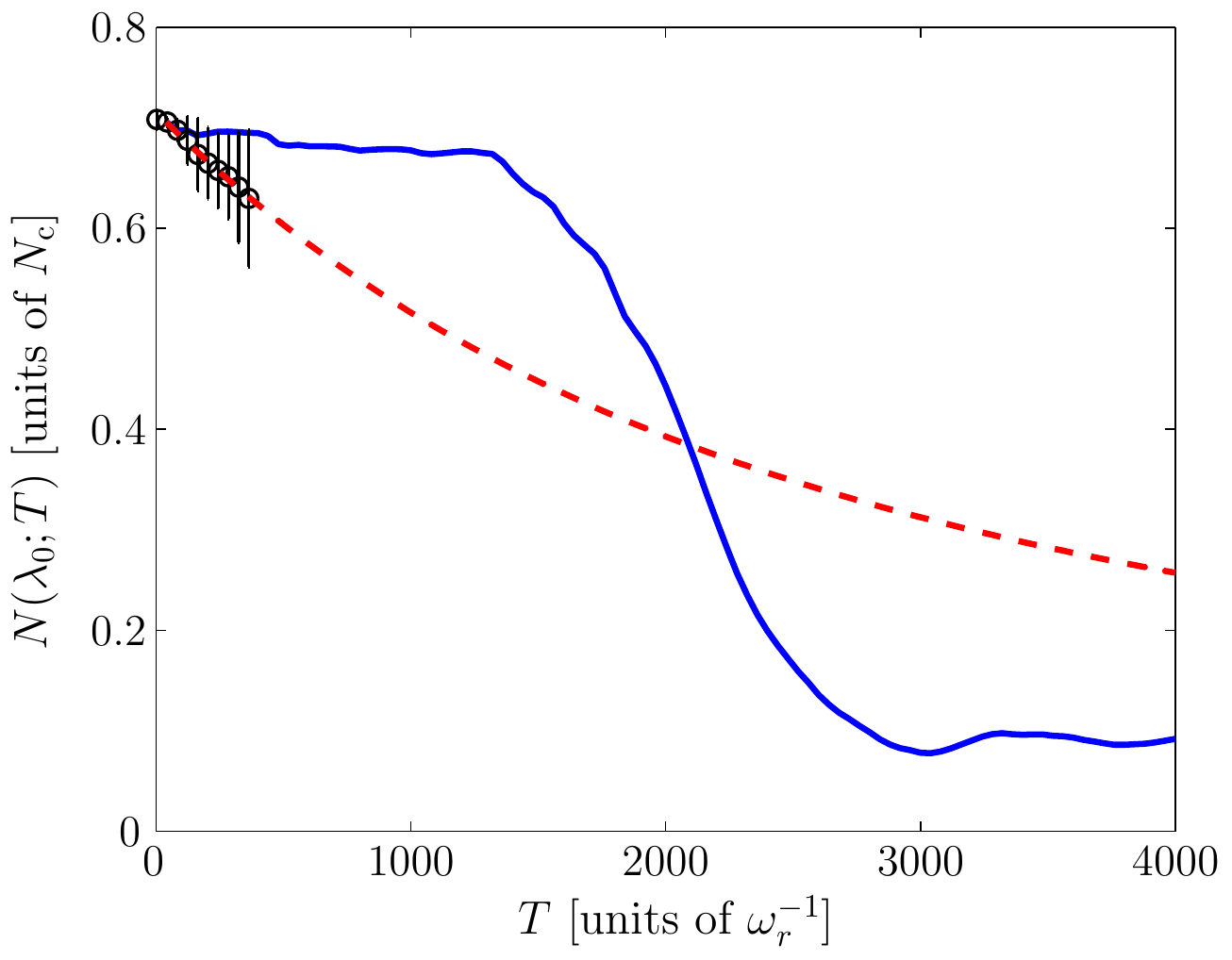}
	\caption{\label{fig:decay}  Norm square $N_0$ of the time-averaged field (i.e., space-integrated power of the classical field measured at the condensate frequency) as a function of the sampling period $T$. The solid line shows the value obtained from a single contiguous sampling of the classical field over period $T$.  Circles (with error bars) show the mean (and standard deviation) of estimates $N_0$ measured on $10$ individual $400\omega_r^{-1}$ sub-periods of the $4000\omega_r^{-1}$ time series.  The dashed line shows the expected (ensemble averaged) power, extrapolated from a least-squares fit to the means of the short-time estimates.}
	\end{center}
\end{figure}

From the foregoing discussion, we expect the power in the classical field measured at the condensate frequency to decay with time, exhibiting a power-law tail $N(\lambda_0;T) \sim 2/\gamma T$ at long times.  We illustrate this issue using the same simulation ($E=12N_\mathrm{c}\hbar\omega_r$) as in the previous section.  Increasing the sampling period to $T\gtrsim50\omega_r^{-1}$ the condensate frequency is more accurately resolved as $\lambda_0=11.39\omega_r^{-1}$.  We assume this value as a best estimate for the condensate frequency, and calculate the power at this frequency as a function of the measurement period $T$, up to a maximum measurement period of $4000\omega_r^{-1}$.  
In figure~\ref{fig:decay} we plot the power measured at frequency $\lambda_0$ (solid line), and find that it decays in a nonuniform way with increasing $T$.  However, the (normalised) mean-field orbital $\phi_0(\mathbf{x})$ we obtain at frequency $\lambda_0$ satisfies $1-|\langle\phi_0|\chi_0\rangle| \lesssim 10^{-4}$ for all averaging periods $T$ we consider; i.e., although the measured \emph{occupation} of the condensate decays with increasing averaging period due to the diffusion of the condensate phase, the \emph{mode shape} we obtain is relatively unaffected.
%%%%%%%%%%%%%%%%%%%%%%%%%%%%%%%%%%%%%%%
%%%%%%%%%%%%%%%%%%%%%%%%%%%%%%%%%%%%%%%
The nonuniform decay of the mean-field orbital's occupation we observe is to be expected for a single trajectory, whereas we expect the scaling $N \sim 2/\gamma T$ to emerge from an average over a large ensemble of similarly prepared classical-field trajectories (cf. \cite{Sinatra08}).  It is possible, however, to infer $\gamma$ from a single trajectory, as we now show.  We divide the total $4000\omega_r^{-1}$ ($10^5$-sample) period of the classical-field trajectory into $10$ consecutive sub-periods of length $400\omega_r^{-1}$ (each of $10^4$ samples), and regard these sub-periods as an ensemble of 10 distinct trajectories.  For each member of the ensemble we calculate the power $N(\lambda_0;T)$ (using the best estimate $\lambda_0=11.39\omega_r^{-1}$) as a function of $T \leq 400\omega_r^{-1}$.  We then average over these 10 ensemble members to obtain a mean power estimate for each sampling period $T$.  The means and standard deviations of these measurements are indicated by circles with error bars in figure~\ref{fig:decay}, and by performing a least-squares fit of the expected power $\langle N(\lambda_0;T) \rangle$ at the condensate frequency (equation~(\ref{eq:mean_decay})) to these mean power estimates, we estimate the phase-diffusion coefficient $\gamma\approx10^{-4}\omega_r$~\footnote{We calculated (and fitted to) estimates of the powers at $T=4,8,\cdots,400 \omega_r^{-1}$, but for clarity indicate only every tenth such estimate in figure~\ref{fig:decay}.}. The dashed line in figure~\ref{fig:decay} extrapolates the expected behaviour of $\langle N(\lambda_0;T) \rangle$ to later times.  Given this decay of the peak power with $T$, a rigorous estimate of the condensate population would in principle be obtained by forming estimates $\langle N(\lambda_0;T_i) \rangle$ for multiple sampling period lengths $T_i$, and extrapolating the resulting trend back to $T=0$ to estimate the `true' condensate population.   However, due to the weak linear decay of the power spectrum peak at short sampling periods, we can accurately estimate the condensate population as the magnitude of the dominant peak in the power spectrum obtained over a short sampling period, for all but the smallest condensate fractions, as we discuss in section~\ref{subsec:phase_freq}.
%%%%%%%%%%%%%%%%%%%%%%%%%%%%%%%%%%%%%%%%%%%%%%%%%%%%%%%%%%%%%%%%%%%%%%%%%%%%%%%%%%%%%%%%%%%%%%%%%%%%%%%%%%%%%%%%%%%%%%%%%%%%%%%%%%%%
\section[Dependence of the first moment on the field energy]{Dependence of the first moment on the field \\ energy}\label{sec:energy_dependence}
In the ergodic classical-field (PGPE) method, equilibrium field configurations of different temperatures can be formed simply by varying the (conserved) energy of the random initial field configuration \cite{Davis01}.  In each case the ergodicity of the field evolution leads to the appropriate microcanonical thermal equilibrium (section~\ref{subsec:cfield_thermo}).  In this section we investigate the behaviour of the first moment $\phi(\mathbf{x};\lambda_0)$ introduced in section~\ref{subsec:id_first_moment} as the energy (and thus temperature) of the classical-field equilibrium is varied, and compare its mode shape $\phi_0(\mathbf{x})$ and occupation $N_0$ with the Penrose-Onsager condensate orbital $\chi_0(\mathbf{x})$ and occupation $n_0$, respectively.  We further compare the condensate frequency $\lambda_0$ to the microcanonical chemical potential $\mu$ of the field obtained using the Rugh methodology introduced in section~\ref{subsec:cfield_thermo}.
%%%%%%%%%%%%%%%%%%%%%%%%%%%%%%%%%%%%%%%%%%%%%%%%%%%%%%%%%%%%%%%%%%%%%%%%%%%%%%%%%%%%%%%%
\subsection{Condensate fraction}\label{subsec:cfrac}
\begin{figure}
	\begin{center}
	\includegraphics[width=0.75\textwidth]{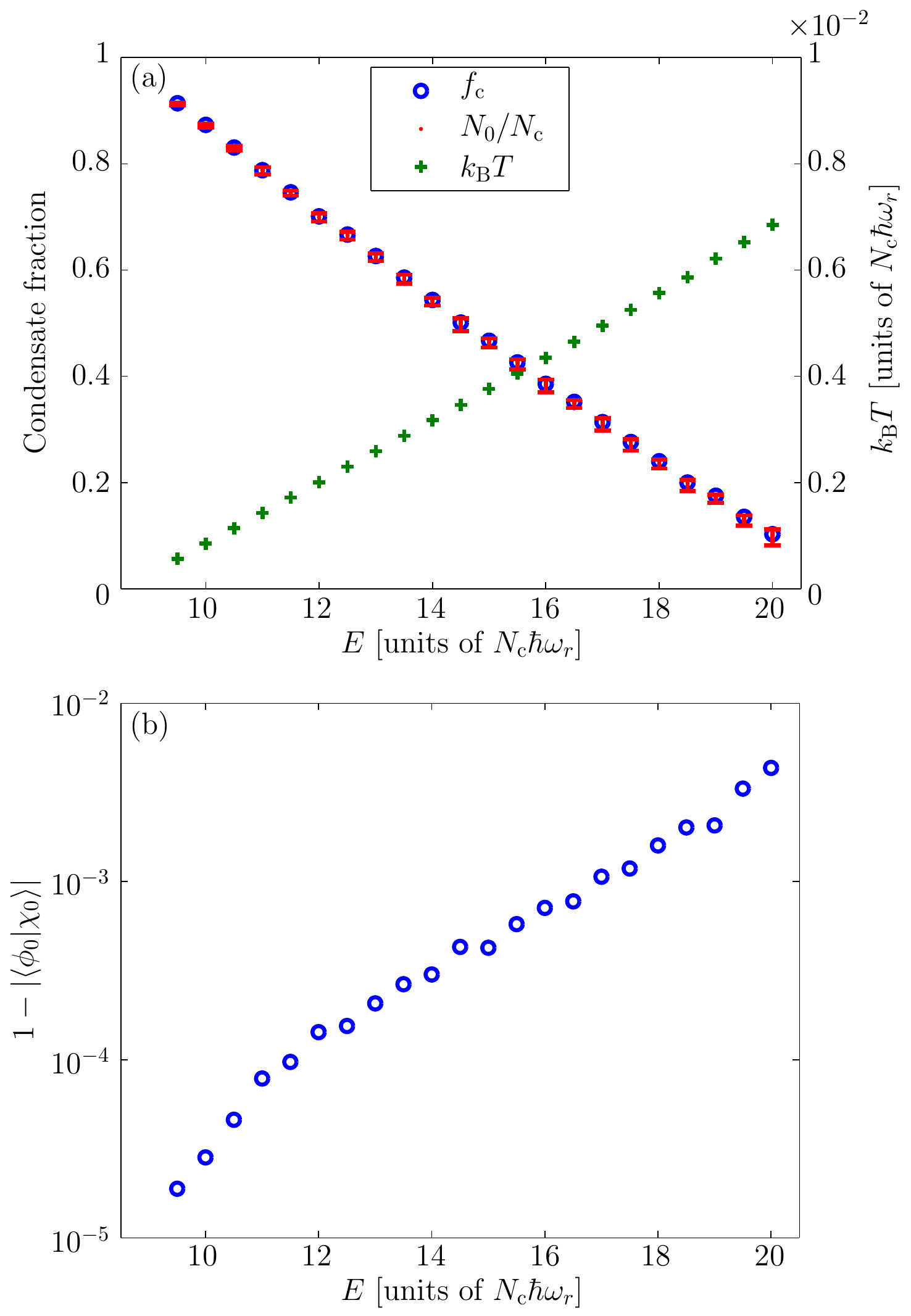}
	\caption{\label{fig:cfrac_mu}  (a) Condensed fraction of the classical field, as determined by the Penrose-Onsager procedure (circles) and by the time-averaging procedure (dots with error bars). Plusses indicate the microcanonical temperature of the field.  (b) Discrepancy $1-|\langle\phi_0|\chi_0\rangle|$ between the (unit-normalised) first moment $\phi_0(\mathbf{x})$ and the most highly occupied natural orbital $\chi_0(\mathbf{x})$ of the one-body density matrix.}
	\end{center}
\end{figure}

We consider here the norm square $N_0$ of the first moment $\phi(\mathbf{x};\lambda_0)$ defined as in section~\ref{subsec:id_first_moment}, for various values of the classical-field energy $E[\psi]$.  In figure~\ref{fig:cfrac_mu}(a) we present estimates $N_0$ for a range of classical-field energies, and compare them with the condensate occupations calculated by the PO approach.    
%%%%%%%%%%%%%%%%%%%%%%%%%%%%%%%%%%%%%%%
%%%%%%%%%%%%%%%%%%%%%%%%%%%%%%%%%%%%%%%
The corresponding classical-field temperatures, calculated using the Rugh methodology, are also included in the figure.  In practice, we calculated the PO condensate by constructing the one-body density matrix (equation~(\ref{eq:cfield_density_mtx})) from 3000 equally-spaced samples of the classical field taken from a period of $1200\omega_r^{-1}$ of the field evolution.  We then divided this period into $30$ consecutive sub-periods of length $40\omega_r^{-1}$ which we sampled at a higher resolution (1000 samples per sub-period), from which we obtained $30$ separate estimates of the classical-field first moment $\phi(\mathbf{x};\lambda_0)$.  In each sub-period we obtain the mean field as the time-averaged field with the greatest norm (as $\lambda_0$ is varied), and we obtain (generally) distinct estimates of $N_0$, $\phi_0(\mathbf{x})$ and $\lambda_0$ from each series.  In this way we exploit the ergodic character of the classical field to emulate sampling from an ensemble of similarly prepared trajectories, as in section~\ref{subsubsec:temporal_coherence}.  The red (grey) data points in figure~\ref{fig:cfrac_mu}(a) and their error bars represent in each case the mean and standard deviation, respectively, of the norm squares of the $30$ estimates of the mean-field.  We observe that these estimates agree very closely with the PO condensate fractions $f_\mathrm{c}$ (blue circles) throughout the range of energies presented.  

We also compare the mean-field orbitals obtained from the time-averaging procedure with the condensate orbitals obtained from the PO approach (see section~\ref{subsec:cfield_correl}). In figure~\ref{fig:cfrac_mu}(b) we plot the quantity $1-|\langle \phi_0|\chi_0\rangle|$ (averaged over the 30 estimates) as a measure of the discrepancy between the two orbitals.  We observe that for the energies presented ($E\leq20N_\mathrm{c}\hbar\omega_r$) the mean discrepancy is $<10^{-2}$. At higher energies (corresponding to condensate fractions $f_\mathrm{c}<0.1$), our temporal-coherence approach to identifying the condensate begins to break down: the mean-field orbital $\phi_0(\mathbf{x})$ fails to match the PO orbital $\chi_0(\mathbf{x})$ (i.e. $|\langle \phi_0|\chi_0\rangle|<0.9$) in an increasing fraction of estimates as the condensate fraction $f_\mathrm{c}\rightarrow0$, and so for clarity we have not presented estimates for these energies here.  This point is discussed further in section~\ref{subsec:phase_freq}. 
%%%%%%%%%%%%%%%%%%%%%%%%%%%%%%%%%%%%%%%%%%%%%%%%%%%%%%%%%%%%%%%%%%%%%%%%%%%%%%%%%%%%%%%%
\subsection{Condensate frequency}\label{subsec:phase_freq}
\begin{figure}
	\begin{center}
	\includegraphics[width=0.75\textwidth]{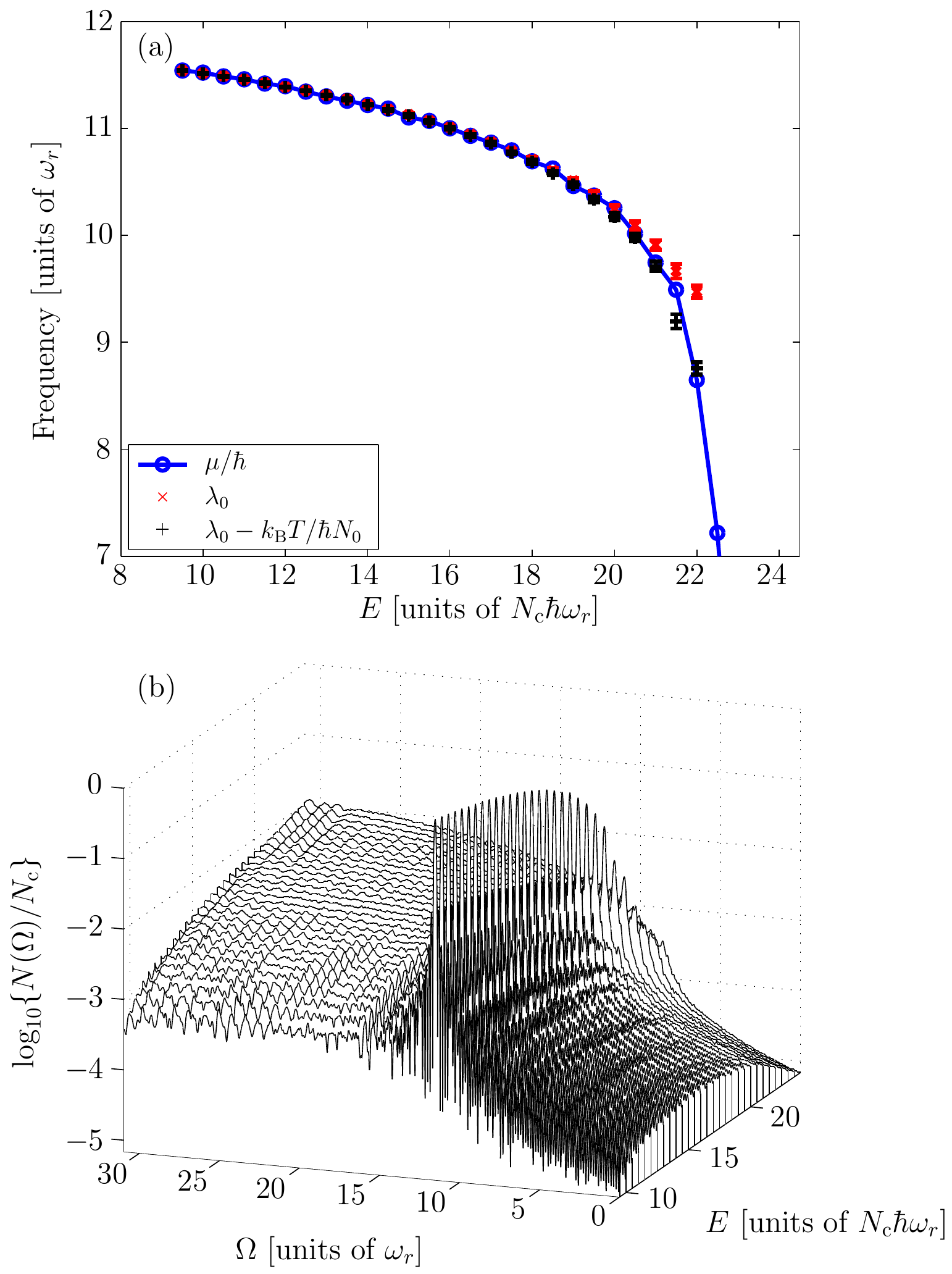}
	\caption{\label{fig:energy_frequency_carpet}  (a) Condensate frequency obtained from the time-averaging procedure (red crosses), and microcanonical chemical potential of the classical field (blue circles), for classical-field equilibria with different energies. Black plusses plot the right-hand side of equation~\reff{eq:mu_lambda_relation}. (b) Norm square of the first moment as a function of the phase-rotation frequency of the frame in which it is constructed (i.e. space-integrated power spectrum of the field), for classical-field simulations with different energies.}
	\end{center}
\end{figure}

We consider here the dependence of the condensate frequency $\lambda_0$ on the energy of the classical field.  By our analogy between the first moment of the classical field and the condensate wave function in mean-field theories (section~\ref{subsec:id_first_moment}), we associate this condensate frequency with the \emph{condensate eigenvalue} appearing in such theories, which is itself closely related to the thermodynamic chemical potential of the degenerate Bose-gas system \cite{Morgan00}.  In figure~\ref{fig:energy_frequency_carpet}(a) we plot estimates of the condensate frequency (red crosses), together with the thermodynamic chemical potential $\mu$ (blue circles with connecting line) of the classical field obtained from the Rugh analysis.  
At the very highest energies, we present results only for ensemble members for which our first-moment analysis and the PO approach agree (i.e. $\phi_0$ and $\chi_0$ overlap to within 10\%).  We observe that the condensate frequencies~$\lambda_0$ and the chemical potentials $\mu$ agree very well for energies $E \lesssim 20N_\mathrm{c}\hbar\omega_r$.  
Above this energy, the condensate frequencies $\lambda_0$ are consistently greater than the chemical potentials.  This is expected behaviour, as at a fixed \emph{total} number of system particles, the two quantities differ by a factor of order $1/N_\mathrm{cond}$, where $N_\mathrm{cond}$ is the condensate occupation \cite{Proukakis98,Morgan00}.  Davis \emph{et al.} \cite{Davis02} argued that equipartition of energy in the classical-field model predicts the relationship 
\begin{equation}\label{eq:mu_lambda_relation}
	\mu = \hbar\lambda_0 - \frac{k_\mathrm{B}T}{N_0}.
\end{equation}
In figure~\ref{fig:energy_frequency_carpet}(a), we plot the quantity $\lambda_0 - k_\mathrm{B}T/\hbar N_0$ (black plusses), where the temperature $T$ is that obtained from the method of Rugh, and find that our results are in reasonable agreement with the prediction of equation~(\ref{eq:mu_lambda_relation}). 
%%%%%%%%%%%%%%%%%%%%%%%%%%%%%%%%%%%%%%%
%%%%%%%%%%%%%%%%%%%%%%%%%%%%%%%%%%%%%%%

We now consider how the total power spectrum of the classical field varies as a function of the field energy. In figure~\ref{fig:energy_frequency_carpet}(b) we plot the power spectrum $N(\Omega)$, averaged over the 30 individual $40\omega_r^{-1}$ sampling periods, for field energies in the range $E\in[9.5,24]N_\mathrm{c}\hbar\omega_r$.  At the lowest energies the behaviour of $N(\Omega)$ is as in figure~\ref{fig:power_spectrum_in_trap}(b): the function exhibits a prominent peak which we identify with the condensate, and a broad background that we associate with thermal excitations.  As the energy (and thus temperature) of the field is increased, the condensate peak decays, and the `wing' of thermal excitations grows until it is of the same magnitude as the condensate peak, and at the highest temperatures only the thermal background remains. This explains why our approach to identifying the condensate begins to fail as the temperature approaches the phase transition: although a temporally coherent condensate may still be present in the field, it becomes increasingly likely that the peak power in any particular estimate of the power spectrum corresponds instead to thermally occupied modes, which eventually swamp the condensate completely.
%%%%%%%%%%%%%%%%%%%%%%%%%%%%%%%%%%%%%%%%%%%%%%%%%%%%%%%%%%%%%%%%%%%%%%%%%%%%%%%%%%%%%%%%%%%%%%%%%%%%%%%%%%%%%%%%%%%%%%%%%%%%%%%%%%%%
\section{Pairing correlations}\label{sec:pairing_correlations}
In the previous section, we have identified that the condensate present in the classical field can be well characterised as the time-average of the appropriately frequency-shifted field.  We now show that more general anomalous moments of the field can be obtained from time-averages in the same phase-rotating frame.  In this approach, condensation in the classical field is thus accompanied by the appearance of anomalous moments of all orders, in direct analogy to the emergence of general anomalous correlation functions in symmetry-breaking accounts of Bose-Einstein condensation \cite{Griffin96,Proukakis96}.

In terms of the Fock-space decomposition $\hat{\Psi}(\mathbf{x}) = \sum_i \hat{a}_i Y_i(\mathbf{x})$, the emergence of a mean field in the second-quantised formalism is equivalent to the appearance of nonzero first moments $\{\langle \hat{a}_i \rangle \}$.  The next-simplest anomalous averages, the quadratic moments~$\{\langle \hat{a}_i \hat{a}_j \rangle \}$ (and their conjugates), arise because of the effect of interactions which `mix' the single-particle creation and annihilation operators to form quasiparticle operators $\hat{b} \sim u \hat{a} + v^* \hat{a}^\dagger$ \cite{Blaizot86}.  Consequently, the occupation of quasiparticle modes results in the appearance of nonzero moments of single-particle operators of the form $\langle \hat{a}_i \hat{a}_j \rangle$, which represent correlations between \emph{pairs} of particles.  Like the mean field itself, these moments are formally zero in a state of fixed total particle number, although analogous quantities can be defined in particle-conserving terms \cite{Morgan00}.  Due to the appearance of these pairing correlations, in order to accurately characterise the weakly interacting Bose gas and its excitations, one must consider not only the one-body density matrix $\rho_{ij}=\langle \hat{a}_j^\dagger \hat{a}_i \rangle$, but also the \emph{pair matrix} $\kappa_{ij}=\langle \hat{a}_j\hat{a}_i \rangle$ \cite{Blaizot86}.

In the remainder of this section, we will demonstrate the application of our temporal averaging procedure to the evaluation of quadratic anomalous moments of the classical field.  By estimating the pair matrix $\kappa(\mathbf{x},\mathbf{x}')=\langle \psi(\mathbf{x})\psi(\mathbf{x}')\rangle$ of the noncondensed component of the field, we calculate the \emph{anomalous density}, which characterises pairing correlations in the thermal component of the field.  We note that a temporal signature of the anomalous density has previously been observed \cite{Brewczyk04} in homogeneous classical-field simulations, in which the anomalous density is uniform. 
%%%%%%%%%%%%%%%%%%%%%%%%%%%%%%%%%%%%%%%%%%%%%%%%%%%%%%%%%%%%%%%%%%%%%%%%%%%%%%%%%%%%%%%%
\subsection{Methodology}\label{subsec:methodology}
We seek here to characterise pairing correlations in the thermal component of the classical field, i.e., the component of the field \emph{orthogonal} to the condensate \cite{Hutchinson00}, which is obtained by projecting out the condensed component of $\psi(\mathbf{x},t)$, i.e. 
\begin{equation}\label{eq:project}
	\psi^\perp(\mathbf{x},t) = \psi(\mathbf{x},t) - \phi_0(\mathbf{x})\int\!d\mathbf{x}'\,\phi_0^*(\mathbf{x}')\psi(\mathbf{x}',t).
\end{equation}
It is important to note that we form $\psi^\perp(\mathbf{x},t)$ on a given ($40\omega_r^{-1}$) time period by projecting out the mean field obtained over the \emph{same} period, so that (anomalous) averages constructed from $\psi^\perp(\mathbf{x},t)$ over this period are formed on the same footing as the mean field itself.  We transform $\psi^\perp(\mathbf{x},t)$ to the same phase-rotating frame as the condensate, forming $\widetilde{\psi}^\perp(\mathbf{x},t)=e^{i\lambda_0t}\psi^\perp(\mathbf{x},t)$ (where $\lambda_0$ is in general different for each period), and then calculate the pair matrix 
\begin{equation}\label{eq:pair_matrix}
	\kappa^\perp(\mathbf{x},\mathbf{x}') = \left\langle \widetilde{\psi}^\perp(\mathbf{x})\widetilde{\psi}^\perp(\mathbf{x}')\right\rangle_{\!t}.
\end{equation}
  The most well-known characterisation of the anomalous correlations described by the pair matrix is given by the \emph{anomalous density} \cite{Griffin96}, which we identify as the diagonal part of the pair matrix:
\begin{equation}\label{eq:anomalous_density}
	m(\mathbf{x}) = \left\langle \widetilde{\psi}^\perp(\mathbf{x})\widetilde{\psi}^\perp(\mathbf{x})\right\rangle_{\!t} \equiv \kappa^\perp(\mathbf{x},\mathbf{x}). 
\end{equation}
We find that the general form of $m(\mathbf{x})$ is apparent from a single estimate of $\kappa^\perp(\mathbf{x},\mathbf{x}')$, formed over a temporal period $40\omega_r^{-1}$.  However, large fluctuations are present in such a single estimate, which is to be expected, as the correlations we seek to resolve here are rather subtle as compared, for example, to the coherence of the condensate.  In order to resolve the anomalous density more clearly, we therefore average over multiple estimates of $m(\mathbf{x})$;  i.e., for each of $30$ consecutive $40\omega_r^{-1}$ periods, we form both the mean field $\phi_0(\mathbf{x})$ and the corresponding anomalous density $m(\mathbf{x})$.  The phase of $m(\mathbf{x})$ is only meaningful in relation to the phase of the mean field $\phi_0(\mathbf{x})$ itself, and so for convenience, we choose the overall phase of the classical field in each sampling period such that $\phi_0(\mathbf{x})$ is maximally real.  This choice of the phase of $m(\mathbf{x})$ relative to a real and positive condensate wave function corresponds to the traditional choice in mean-field theories.  Forming multiple estimates of $m(\mathbf{x})$ in this way allows us to calculate both the mean and the variance of this quantity, as we show in section~\ref{subsubsec:anom_avg_results}.  
%%%%%%%%%%%%%%%%%%%%%%%%%%%%%%%%%%%%%%%%%%%%%%%%%%%%%%%%%%%%%%%%%%%%%%%%%%%%%%%%%%%%%%%%
\subsection{Anomalous density}\label{subsubsec:anom_avg_results}
\begin{figure}
	\begin{center}
	\includegraphics[width=0.65\textwidth]{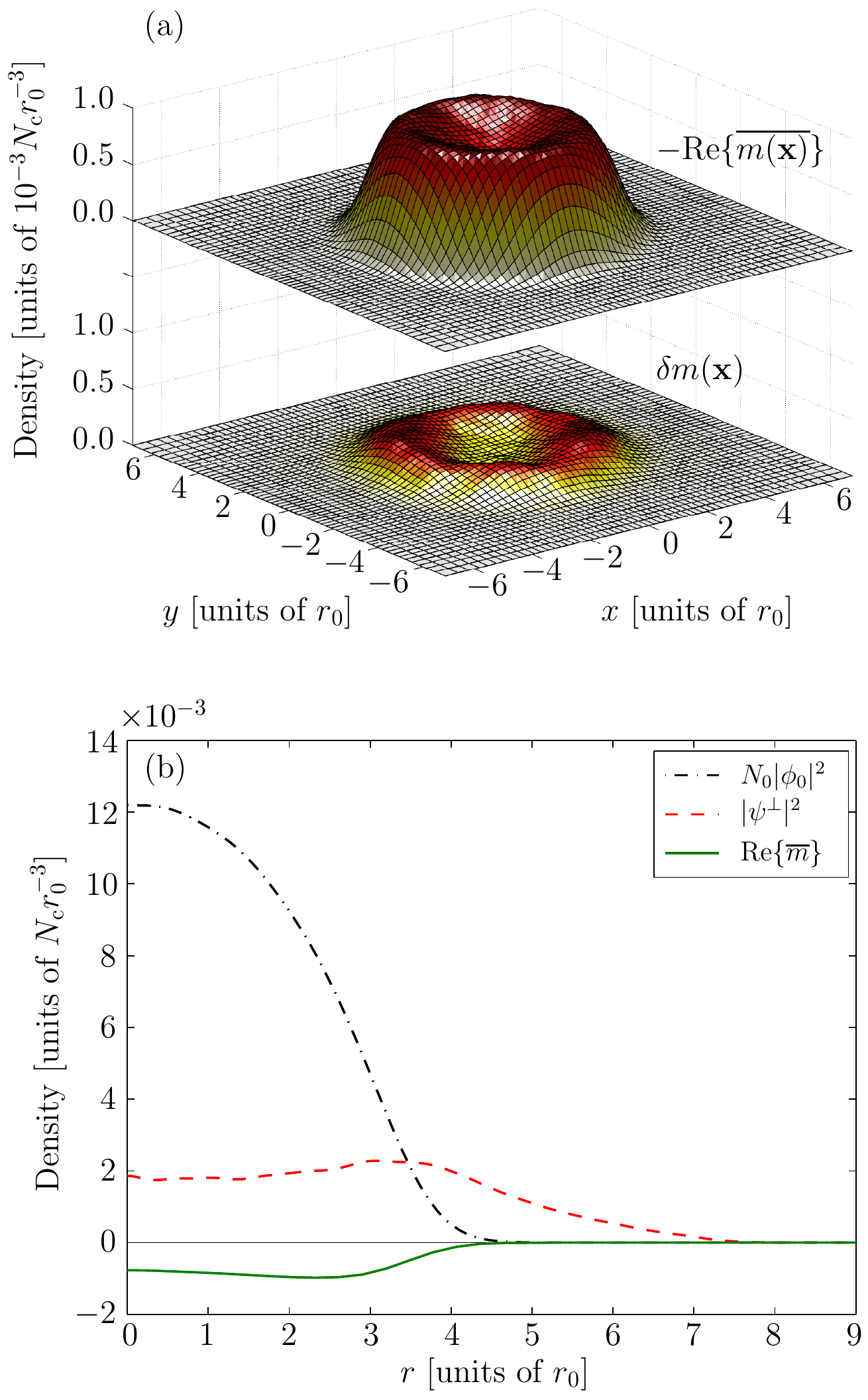}
	\caption{\label{fig:anomalous_average}  Anomalous density $m(\mathbf{x})$ of the field, for the case $E=14.5N_\mathrm{c}\hbar\omega_r$. (a) Shape of $-\mathrm{Re}\{\overline{m(\mathbf{x})}\}$ on a slice through the plane $z=0$ (upper surface), and standard deviation $\delta m(\mathbf{x})$ of anomalous-density estimates on the same plane (lower surface).  (b) Azimuthally averaged density of the condensate mode (as determined by the time averaging), complementary (orthogonal) thermal component of the field, and anomalous density, in the plane $z=0$.}
	\end{center}
\end{figure}

The mean anomalous density calculated by the procedure described in section~\ref{subsec:methodology} is mostly real and negative (i.e., has phase opposite to that of the mean field), in agreement with the results of mean-field-theory calculations \cite{Hutchinson00,Proukakis98,Bergeman00}, but exhibits some small complex-valued fluctuations due to the finite ensemble size.  Denoting averages over estimates by an overbar, we plot in figure~\ref{fig:anomalous_average}(a) the negative $-\mathrm{Re}\{\overline{m(\mathbf{x})}\}$ of the mean anomalous density on the $z=0$ plane, and the local standard deviation in estimates $\delta m(\mathbf{x}) = [\overline{|m(\mathbf{x})|^2} - |\overline{m(\mathbf{x})}|^2]^{1/2}$ of the anomalous density on this plane, calculated for a simulation with $E=14.5N_\mathrm{c}\hbar\omega_r$ (for which the condensate fraction $f_\mathrm{c}=0.50$).
%%%%%%%%%%%%%%%%%%%%%%%%%%%%%%%%%%%%%%%
%%%%%%%%%%%%%%%%%%%%%%%%%%%%%%%%%%%%%%%
The anomalous density has the spatial structure expected from mean-field calculations \cite{Hutchinson00,Proukakis98,Bergeman00}: it resides primarily in the region where the condensate exists, and its absolute value exhibits a shallow `dip' in the centre of the trap.  The standard deviation $\delta m(\mathbf{x})$ indicates that the greatest variance in density estimates occurs around the (circular) maximum of $|m(\mathbf{x})|$, while much less variation occurs in estimates of the density in the central dip. 

The anomalous density shown here exhibits a very high degree of rotational symmetry about the $z$~axis, but in general the anomalous density we obtain is distorted (the central `dip' in its absolute value becomes saddle shaped along some random axis).  
We identify this as a result of dipole-oscillation (Kohn mode) \cite{Dobson94} excitations of the field, which are immune to thermal damping (section~\ref{subsec:back_bose_condensation})\footnote{This property of the many-body system persists in the classical-field model, but note however the effect of the projector on large dipole oscillations \cite{Bradley05}.}, and are thus `frozen in' during the thermalisation of the field.
More generally, one might regard the classical field as having condensed into an excited centre-of-mass mode, and consider the correlations of the field in a frame following this motion \cite{Pethick00}.  In figure~\ref{fig:anomalous_average}(b) we plot the azithumally averaged anomalous density on the plane $z=0$, together with the similarly averaged densities of the condensate ($N_0|\phi_0(\mathbf{x})|^2$) and the orthogonal thermal component of the field ($|\psi^\perp(\mathbf{x})|^2$), for comparison.  We observe that the magnitude of the anomalous density in the centre of the trap is an appreciable fraction of that of the (normal) thermal component of the field, in agreement with references~\cite{Hutchinson00,Proukakis98,Bergeman00}.   
%%%%%%%%%%%%%%%%%%%%%%%%%%%%%%%%%%%%%%%%%%%%%%%%%%%%%%%%%%%%%%%%%%%%%%%%%%%%%%%%%%%%%%%%
\subsection{Dependence on field energy}
\begin{figure}
	\begin{center}
	\includegraphics[width=0.65\textwidth]{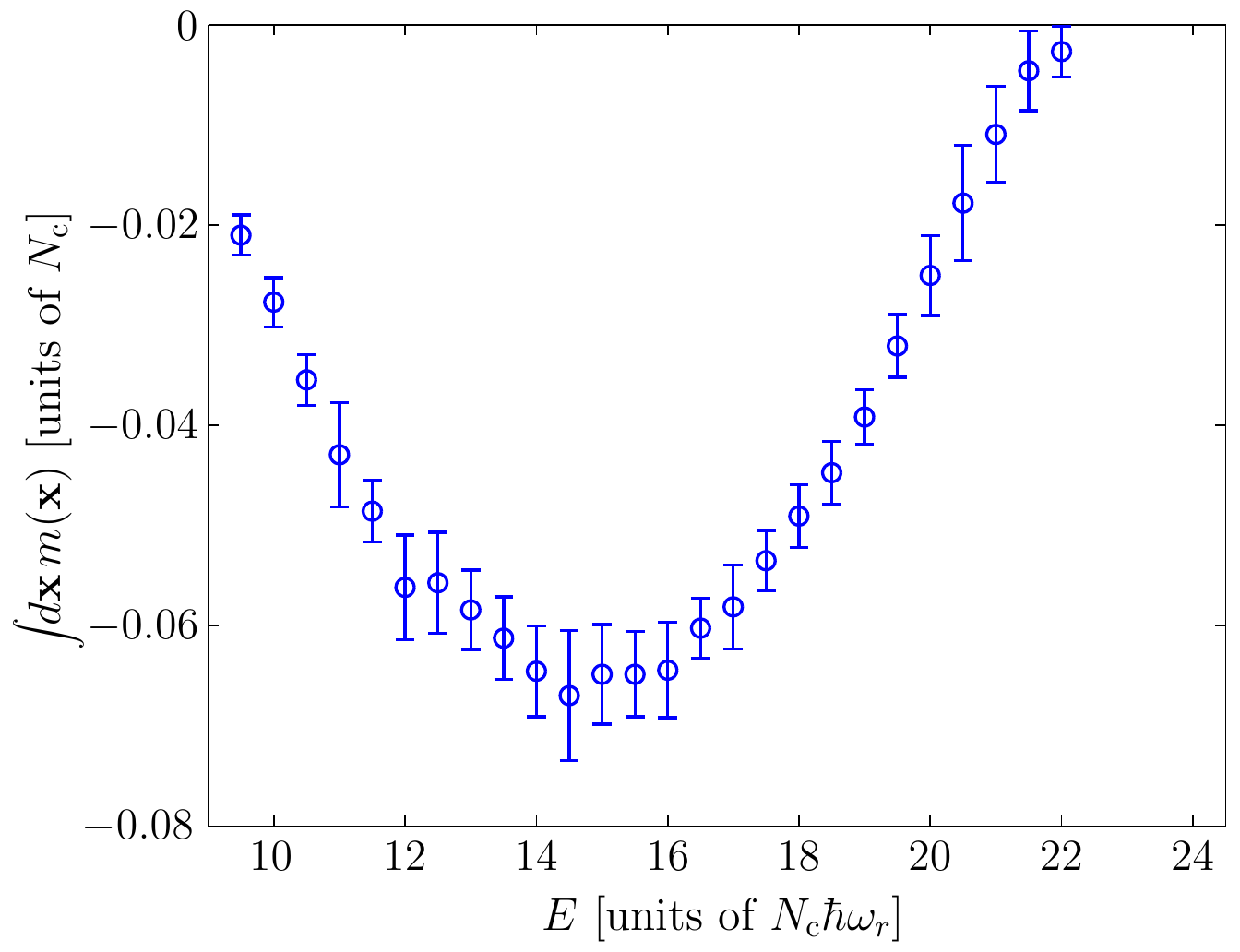}
	\caption{\label{fig:integrated_m}  Integrated value of the anomalous density $\int\!d\mathbf{x}\,m(\mathbf{x})$ as a function of field energy.}
	\end{center}
\end{figure}

Finally, we consider the dependence of the anomalous density on the energy (or equivalently, the temperature) of the projected classical field.  In mean-field theories, the anomalous density (after any renormalisation \cite{Hutchinson00,Proukakis98,Hutchinson98}) becomes small as the temperature of the system approaches zero (due to the weak occupation of quasiparticle modes in this limit), and also as it approaches the critical temperature (due to the quasiparticle modes becoming more single-particle-like as the condensate is depleted).  This behaviour is often cited as a justification for the neglect of the anomalous density in self-consistent theories (the so-called Popov approximation \cite{Griffin96}) in these two limits.  In order to characterise more fully the temperature-dependent behaviour of the anomalous density, we follow \cite{Hutchinson00} and calculate its integrated value $M\equiv \int \! d\mathbf{x}\, m(\mathbf{x})$.  In figure~\ref{fig:integrated_m} we plot the real part of $M$ (neglecting a small imaginary part that arises from incomplete convergence of the averaging; see section~\ref{subsubsec:anom_avg_results}) as a function of the classical-field energy.  
%%%%%%%%%%%%%%%%%%%%%%%%%%%%%%%%%%%%%%%
%%%%%%%%%%%%%%%%%%%%%%%%%%%%%%%%%%%%%%%
As in our consideration of the condensate eigenvalue in section~\ref{sec:energy_dependence}, we present for each energy the mean and standard deviation (error bars) of only those estimates obtained from averaging periods which produced an accurate condensate mode (section~\ref{subsec:id_first_moment}).  We observe that the behaviour of $M(E)$ is consistent with the results of mean-field theories \cite{Hutchinson00,Bergeman00}, with its absolute value $|M|$ reaching its maximum at intermediate energies (temperatures), and rapidly decreasing as we approach both the zero-temperature and critical regimes.

We note that the well-known issues of ultraviolet divergence of the anomalous density in mean-field theories arise from the \emph{zero-point} occupation of quasiparticle modes \cite{Hutchinson00,Proukakis98,Hutchinson98}, which is of course not present in the classical-field model.  Also, although the results of classical-field calculations are, in general, dependent on the cutoff energy,  the contribution to the anomalous density from successive quasiparticle modes rapidly decreases with increasing energy of the modes, as the modes return to a single-particle structure.  The requirement (for our treatment) that the anomalous density is well-contained in the low-energy region (condensate band) described by the PGPE is thus precisely the requirement that the cutoff is effected at such an energy that the interacting Hamiltonian has become approximately diagonal in the single-particle basis $\{\phi_k(\mathbf{x})\}$ (section~\ref{subsec:cfield_proj_cft}).

Finally, we note that the maximum (absolute) value of the integrated anomalous density occurs when $f_\mathrm{c}\approx0.5$, but remind the reader that this refers only to the proportion of the \emph{below-cutoff} field that is condensed.  Although the entire anomalous density should be well-described by the low-energy Hamiltonian PGPE dynamics, one would have to include the contribution of above-cutoff atoms to the \emph{normal} thermal density of the field \cite{Blakie08} in order to draw quantitative comparisons with (e.g.) the mean-field-theory calculations of \cite{Hutchinson00}.
%%%%%%%%%%%%%%%%%%%%%%%%%%%%%%%%%%%%%%%%%%%%%%%%%%%%%%%%%%%%%%%%%%%%%%%%%%%%%%%%%%%%%%%%%%%%%%%%%%%%%%%%%%%%%%%%%%%%%%%%%%%%%%%%%%%%
\section{Summary}\label{sec:conclusions}
We have demonstrated that in the Hamiltonian PGPE theory, classical-wave condensation is accompanied by long-range temporal coherence limited only by the slow diffusion of the condensate phase.  This gives rise to the appearance of a nonzero first moment of the field, as defined by short-time averages in an appropriate phase-rotating frame.  We identified the emergence of this moment with the concept of $\mathrm{U(1)}$-symmetry breaking that is central to self-consistent mean-field theories.  We showed that the mean field obtained by short-time averaging agrees well with the condensate identified by the standard Penrose-Onsager approach, except for close to the critical regime associated with the transition to the normal phase.  We found that the condensate eigenfrequency we obtain by this temporal analysis exhibits the behaviour predicted for the condensate eigenvalue in the most sophisticated mean-field approaches \cite{Morgan00}; i.e., it agrees closely with the thermodynamic chemical potential at low energies and diverges away from it in inverse proportion to the condensate occupation as the phase transition is approached.  By calculating the pair matrix and anomalous density of the noncondensed component of the field, we demonstrated explicitly that time averages in the frame rotating at the condensate frequency allow the calculation of more general anomalous moments.  We observed the anomalous density to exhibit the expected behaviour \cite{Hutchinson00}, with its magnitude reaching its maximum at intermediate temperatures and decreasing as both the $T=0$ and critical regimes are approached. 
%%%%%%%%%%%%%%%%%%%%%%%%%%%%%%%%%%%%%%%%%%%%%%%%%%%%%%%%%%%%%%%%%%%%%%%%%%%%%%%%%%%%%%%%%%%%%%%%%%%%%%%%%%%%%%%%%%%%%%%%%%%%%%%%%%%%
%%%%%%%%%%%%%%%%%%%%%%%%%%%%%%%%%%%%%%%%%%%%%%%%%%%%%%%%%%%%%%%%%%%%%%%%%%%%%%%%%%%%%%%%%%%%%%%%%%%%%%%%%%%%%%%%%%%%%%%%%%%%%%%%%%%%

\chapter{Equilibrium precession of a single vortex}
\label{chap:precess} 
%%%%%%%%%%%%%%%%%%%%%%%%%%%%%%%%%%%%%%%%%%%%%%%%%%%%%%%%%%%%%%%%%%%%%%%%%%%%%%%%%%%%%%%%%%%%%%%%%%%%%%%%%%%%%%%%%%%%%%%%%%%%%%%%%%%%
%%%%%%%%%%%%%%%%%%%%%%%%%%%%%%%%%%%%%%%%%%%%%%%%%%%%%%%%%%%%%%%%%%%%%%%%%%%%%%%%%%%%%%%%%%%%%%%%%%%%%%%%%%%%%%%%%%%%%%%%%%%%%%%%%%%%
At zero temperature, an off-axis vortex in a Bose-Einstein condensate orbits (or \emph{precesses}) about the trap axis.  Such a state represents the simplest nontrivial example of vortex dynamics in a Bose condensate.  The precession of such a vortex can be viewed as the vortex being carried by its own \emph{self-induced} superflow \cite{Donnelly91}, arising due to the effective boundary conditions imposed on the superflow by the inhomogeneity of the condensate orbital density \cite{Guilleumas01}\footnote{Corrections to this picture due to the finite extent of the vortex core have been investigated in references \cite{Nilsen06} and \cite{Jezek08}.}.   Further complexity is attained in finite-temperature scenarios, in which the vortex may be subject to additional forces resulting from its interaction with the thermal field, as discussed in section~\ref{subsec:back_vortices}.  
In this chapter we consider a finite-temperature Bose-Einstein condensate in a quasi-two-dimensional trap containing a single precessing vortex.  We find that such a configuration arises naturally as an ergodic equilibrium of the projected Gross-Pitaevskii equation, when constrained to a finite conserved angular momentum.  In an isotropic trapping potential the condensation of the classical field into an off-axis vortex state breaks the rotational symmetry of the system.  We present here a methodology to identify the condensate and the Goldstone mode associated with the broken rotational symmetry in the classical-field model, and we examine the variation in vortex trajectories and thermodynamic parameters of the field as the energy of the microcanonical field simulation is varied.  

This chapter is organised as follows:  in section~\ref{sec:prec_Simulation_procedure}, we discuss our simulation procedure, introducing in turn the parameters of our system, and the details of our numerical approach.  In section~\ref{sec:prec_Results} we present results of representative simulations, and characterise the equilibrium trajectory of the vortex.  In section~\ref{subsec:prec_Penrose_Onsager_analysis} we discuss the identification of the condensate from the fluctuation statistics of the classical field.  In section~\ref{subsec:prec_rotational_properties} we consider the rotational properties of the condensate and the complementary thermal component of the field.  In section~\ref{sec:prec_Energy_dependence} we discuss the dependence of the system behaviour on the internal energy of the field, and in section~\ref{sec:prec_projector_frame} we consider the dependence of our results on the angular velocity of the rotating frame in which the energy cutoff is effected.  In section~\ref{sec:prec_Conclusions} we summarise and present our conclusions.   
%%%%%%%%%%%%%%%%%%%%%%%%%%%%%%%%%%%%%%%%%%%%%%%%%%%%%%%%%%%%%%%%%%%%%%%%%%%%%%%%%%%%%%%%%%%%%%%%%%%%%%%%%%%%%%%%%%%%%%%%%%%%%%%%%%%%
\section{Simulation procedure}\label{sec:prec_Simulation_procedure}
%%%%%%%%%%%%%%%%%%%%%%%%%%%%%%%%%%%%%%%%%%%%%%%%%%%%%%%%%%%%%%%%%%%%%%%%%%%%%%%%%%%%%%%%
\subsection{System}\label{sec:prec_System}
We consider here a system harmonically trapped with a trapping potential isotropic in the $xy$~plane:
\begin{equation}
	V(\mathbf{x}) = \frac{m}{2}\Bigl[\omega_r^2\left(x^2+y^2\right) + \omega_z^2z^2\Bigr].
\end{equation}
In this chapter we will restrict our attention to an effective two-dimensional system. Such systems have been experimentally realised (\cite{Stock05,Clade09}) by setting the trapping frequency $\omega_{z}$ sufficiently high that no modes are excited in the $z$ direction. A 2D description of our system is thus valid provided that \cite{Petrov00} 
\begin{equation}
	 \mu + k_\mathrm{B}T  \ll  \hbar\omega_{z}, 
\end{equation}
 where $\mu$ and $T$ are the system chemical potential and temperature respectively.  Provided that the oscillator length  $l_z = \sqrt{\hbar/ m\omega_z}$ corresponding to the $z$ confinement greatly exceeds the $s$-wave scattering length $a_s$, the scattering is well described by the usual 3D contact potential  \cite{Petrov00}, and we obtain an effective two-dimensional  equation in which  the collisional interaction parameter is simply rescaled to $U_\mathrm{2D} = 2\sqrt{2\pi} \hbar^2 a_s/ml_z$  \cite{Petrov00}. 
Restricting the system to 2D permits the dynamics of a \emph{point} vortex interacting with the thermal component of the field to be studied, while removing complexities such as line bending and Kelvin-wave excitations of a vortex filament of finite extent, which provide additional mechanisms for vortex dissipation \cite{Kozik04}.  In this work we consider  $^\mathrm{23}$Na atoms  confined in a strongly oblate trap, with trapping frequencies $(\omega_r,\omega_z)=2\pi\times(10,2000)\;\mathrm{rad/s}$.  The $s$-wave scattering length is $a_s=2.75~\mathrm{nm}$ and the oscillator length  $l_z =  468~\mathrm{nm}$.  Choosing the cutoff such that $E_R < \hbar \omega_z$, we thus obtain a formally two-dimensional projected field theory, as discussed in section~\ref{subsec:dimless}.

In our simulations we will often compare our systems to what we will refer to as our principal \emph{ground state}, which we take as the ground Gross-Pitaevskii (GP) eigenmode \cite{Dalfovo99} of the trapped system in a frame rotating at angular velocity $\Omega_0=0.35\omega_r$, with eigenvalue $(\mu_\mathrm{g})_{\Omega_0}=10\hbar\omega_r$ in that frame. This state consists of $N_0=1.072\times10^4$ atoms, and contains a single on-axis vortex carrying unit angular momentum per atom (angular momentum and rotation will always refer to that about the $z$~axis in the remainder of this thesis).  In an inertial (nonrotating) frame this state has eigenvalue $\mu_\mathrm{g}=10.35\hbar\omega_r$, and energy $E_\mathrm{g}=7.646\times10^4\hbar\omega_r$.  In this chapter we will refer generically to any such inertial frame as a laboratory frame (lab frame).
%%%%%%%%%%%%%%%%%%%%%%%%%%%%%%%%%%%%%%%%%%%%%%%%%%%%%%%%%%%%%%%%%%%%%%%%%%%%%%%%%%%%%%%%
\subsection{Microcanonical evolution}\label{subsec:prec_Microcanonical_evolution}
%%%%%%%%%%%%%%%%%%%%%%%%%%%%%%%%%%%%%%%%%%%%%%%%%%%%%%%%%
\subsubsection{Microcanonical formalism}\label{subsubsec:prec_Microcanonical_formalism}
As discussed in section~\ref{sec:cfield_ergodicity}, the PGPE can be used to evolve a general configuration of the classical field to a thermal equilibrium.  In contrast to the simulations of chapter~\ref{chap:anomalous}, in which the initial field configurations were parameterised solely by their energies $E[\psi]$ (equation~\reff{eq:cfield_HCF}), in this chapter we also fix the conserved angular momentum   
\begin{equation}\label{eq:conserved_ang_mom}
	L[\psi] \equiv \int\!d\mathbf{x}\,\psi^*(\mathbf{x})\hat{L}_z\psi(\mathbf{x}),
\end{equation}
to a prescribed nonzero value\footnote{We note that the geometry and projector definition of reference \cite{Blakie05} yield a conserved angular momentum about the $x$~axis; however, any nonzero angular momenta in the fields evolved in that work were incidental, and of essentially negligible magnitude, as noted by the authors of that paper.}.  We therefore expect that the field evolves under the action of the PGPE to a thermal equilibrium consistent with the choices of conserved energy (equation~(\ref{eq:cfield_HCF})) and angular-momentum (equation~(\ref{eq:conserved_ang_mom})) first integrals.  In this chapter we choose $L[\psi]=N[\psi]\hbar=N_0\hbar\equiv L_0$, and thus explore the manifold of classical-field configurations with unit angular momentum per particle.  At zero temperature (corresponding to complete condensation in the classical-field model employed here, in which no representation of quantum depletion is included), the corresponding microstate (unique up to a choice of phase) is our principal ground state, containing a singly charged, on-axis vortex, with energy $E[\psi]=E_\mathrm{g}$.  By choosing initial configurations with $E[\psi]>E_\mathrm{g}$ and $L_z[\psi]=L_0$, we investigate the thermal equilibria that result when atoms are thermally excited out of the ground condensate mode, while conserving the total angular momentum. 
%%%%%%%%%%%%%%%%%%%%%%%%%%%%%%%%%%%%%%%%%%%%%%%%%%%%%%%%%
\subsubsection{Choice of frame}\label{subsubsec:prec_Choice_of_frame}
In applying our classical-field method, we must make a choice of the rotation rate of the frame in which the PGPE is defined (see equation~\reff{eq:numeric_Hsp}), which we will denote by $\Omega_\mathrm{p}$ in the remainder of this chapter.  The only nontrivial consequence of this choice\footnote{The appearance of the inertial correction $(\Delta H)_\mathrm{inertia} \equiv -\Omega_\mathrm{p}L_z$ to the single-particle Hamiltonian~\reff{eq:back_Hsp} serves only to ensure that the rotating-frame equations of motion faithfully describe the same dynamics as the fundamental inertial-frame equations.} is the precise definition of the projector $\mathcal{P}=\mathcal{P}(E_R;\Omega_\mathrm{p})$, and thus of the set of modes which span the low-energy space $\mathbf{C}$.  This dependence of the projector on $\Omega_\mathrm{p}$ generalises the familiar concept of an energy cutoff in classical-field theory.  As is well known \cite{Blakie08}, the thermodynamic parameters of the classical field at equilibrium depend both on the conserved first integrals of the system and the imposed cutoff.  In the present work, we expect that the equilibrium of the system minimises the free energy 
\begin{equation}\label{eq:free_energy}
	F = E - TS - \Omega L,
\end{equation}
where $\Omega=(\partial E/\partial L)_S$ is the thermodynamic angular velocity \cite{Landau69}.  The utility of the rotationally invariant Gauss-Laguerre projector (section~\ref{sec:quad_app_laguerre}) used here is that the classical-field evolution is not \emph{dynamically} biased towards a particular rotating frame, i.e., the (generalised) Ehrenfest relation~\reff{eq:cfield_ehrenfest_L} for the angular momentum becomes
\begin{equation}\label{eq:Ehrenfest_Lz}
	\frac{d\overline{L_z}}{dt} = -\frac{i}{\hbar}\overline{L_zV(\mathbf{x})}.
\end{equation}
This ensures that for the isotropic trapping we consider here, angular momentum is conserved, and further that the field is free to rotate at the angular velocity $\Omega$ which minimises equation~\reff{eq:free_energy}, which will in general \emph{not} be equal to the projector angular velocity $\Omega_\mathrm{p}$.  The relationship between the first integrals $(E,L)$ and the thermodynamic variables $(T,\Omega)$ of course depends on $\Omega_\mathrm{p}$, as different choices of cutoff yield different Hamiltonian systems.  We demonstrate this dependence by explicit calculation in section~\ref{sec:prec_projector_frame}, and note that the use of a harmonic basis allows the PGPE theory to be extended to include a mean-field description of above-cutoff atoms, whereby insensitivity to (moderate) changes in the cutoff is regained (see \cite{Blakie08}).  However, as a first application to the precessing-vortex system, we restrict our attention to a strictly Hamiltonian `PGPE system' \cite{Blakie05}, with low-energy region $\mathbf{C}$ defined by application of the projector in the frame in which we obtained the vortical ground state (i.e., we take $\Omega_\mathrm{p}=\Omega_0=0.35\omega_r$), and investigate the behaviour of this particular Hamiltonian system by varying its energy at fixed angular momentum.
%%%%%%%%%%%%%%%%%%%%%%%%%%%%%%%%%%%%%%%%%%%%%%%%%%%%%%%%%
\subsubsection{Procedure}\label{subsubsec:prec_Procedure}
As discussed in section~\ref{sec:cfield_ergodicity}, for almost any initial configuration far from equilibrium, we expect the field to approach equilibrium over time \cite{Lebowitz73}, due to the ergodicity of the system.  
We thus form our initial states as randomised field configurations under the projector defined by cutoff energy $E_R=3(\mu_\mathrm{g})_{\Omega_0}=30\hbar\omega_r$ and rotation frequency $\Omega_\mathrm{p}=0.35\omega_r$, with the same values of normalisation $N$ and angular momentum $L$ as our principal ground state, but with higher energies.  In general, forming such states is a highly nontrivial task, due to the nonlinearity of the energy functional.  In practice, we employ an algorithm which produces a field configuration with energy $E$ and angular momentum $L$ within prescribed tolerances of the desired values.  This algorithm is based on the simpler procedure presented in \cite{Blakie08} for the formation of random states \emph{not} subject to the additional angular momentum constraint.  Given the ground state $\zeta_\mathrm{g}(\x)$, and a randomised high-energy state $\zeta_\mathrm{r}(\x)$, both of normalisation $N$, we mix the two to form an intermediate configuration
\begin{equation}
	\zeta_\mathrm{m}(\x) = p_0\zeta_\mathrm{g}(\x) + p_1\zeta_\mathrm{r}(\x),
\end{equation}
with $p_1 = \sqrt{1- |p_0|^2}$, and then renormalise this field to $N$.  In general this field will not have $L[\zeta_m]=L_0$.  We therefore randomly adjust the amplitudes of pairs of `opposite' angular-momentum modes from the set $\{\alpha_\mathrm{nl}\}$ which comprise $\zeta_m(\x)$, rescaling 
\begin{eqnarray}
	\alpha_{nl} &\to& \sqrt{1-\xi}\alpha_{nl} \\
	\alpha_{n(-l)} &\to& \sqrt{1 + \xi \frac{|\alpha_{nl}|^2}{|\alpha_{n(-l)}|^2}} \alpha_{n(-l)},
\end{eqnarray}
where $n$ and $l$ are chosen randomly.
We typically choose the mixing parameter $\xi=0.1$, and the factor $|\alpha_{nl}|^2/|\alpha_{n(-l)}|^2$ ensures that the total population of the field is unaltered by this adjustment.  We repeat this adjustment for random mode pairs $((n,l),(n,-l))$ under the projector until the angular momentum of the field $L=L_0$ to within the desired tolerance.  As a result of these adjustments, the energy of the field generally increases. Thus, after obtaining convergence in $L$, we adjust the energy-mixing parameter $p_0$ appropriately, and repeat the angular-momentum adjusting procedure.  We repeat this process until both $L$ and $E$ have converged to the desired values to within the desired tolerances.  This process is obviously probabilistic, and we cannot be certain \emph{a priori} that the result will converge to an acceptable state in any given time-frame.  However, we find that the initial conditions we require for our investigations in this chapter are typically found to within satisfactory tolerances ($\Delta L/L_0 = 10^{-6}$, $\Delta E/E_\mathrm{g} = 10^{-4}$) in $\sim 5$ minutes on a standard desktop computer\footnote{As a source of `random' initial states for our ergodic simulations this procedure is certainly adequate.  However, were we to require any particular \emph{distribution} for the initial states (e.g., for explicit ensemble calculations), we would need to adopt a more sophisticated approach \cite{Fox09}.}.

We form initial states in this manner with a range of energies.  In each case the central vortex of the principal ground state remains, however the density profile of each state is severely distorted.  As all our states have $L = N_0\hbar$, the inertial-frame and projector-frame energies are in each case related by a constant shift resulting from the fixed rotational kinetic energy of the field: $(E)_0 = (E)_{\Omega_\mathrm{p}} + \hbar\Omega_\mathrm{p} N_0$.  This shift is of order $5\%$ for the ground-state wave function.  In this chapter we will discuss the energies of states in the \emph{lab frame}, and quote them as a multiple of the ground-state energy $E_\mathrm{g}$ in that frame.  

Having formed the initial states we evolve them in the projector frame for a period of $10^4$ trap cycles\footnote{The term `trap cycle' here refers to the period of oscillations in the radial trapping potential, i.e. $T_\mathrm{osc} = 2\pi/\omega_r$.} (abbreviated cyc), using the adaptive ARK45-IP algorithm (section~\ref{sec:rk4_app_ark45}) with accuracy chosen such that the relative change in field normalisation is $\leq 4\times10^{-9}$ per time step taken.  By $9\times10^3$ cyc the states have reached equilibrium, as evidenced by the settling of the distribution of particles in momentum space \cite{Davis01}\footnote{In practice we average momentum-space densities taken over consecutive 10 cycle periods.  Aside from fluctuations close to the origin $\mathbf{k}=0$ arising from the vortex precession, we observe the densities to settle to their (fluctuating) values well before $t=9000$ cyc.}, and we perform our analysis of steady-state properties on the final $10^3$ trap cycles of evolution data. 
%%%%%%%%%%%%%%%%%%%%%%%%%%%%%%%%%%%%%%%%%%%%%%%%%%%%%%%%%%%%%%%%%%%%%%%%%%%%%%%%%%%%%%%%%%%%%%%%%%%%%%%%%%%%%%%%%%%%%%%%%%%%%%%%%%%%
\section{Equilibrium behaviour of the field}\label{sec:prec_Results}
\begin{figure}
	\begin{center}
	\includegraphics[width=1.0\textwidth]{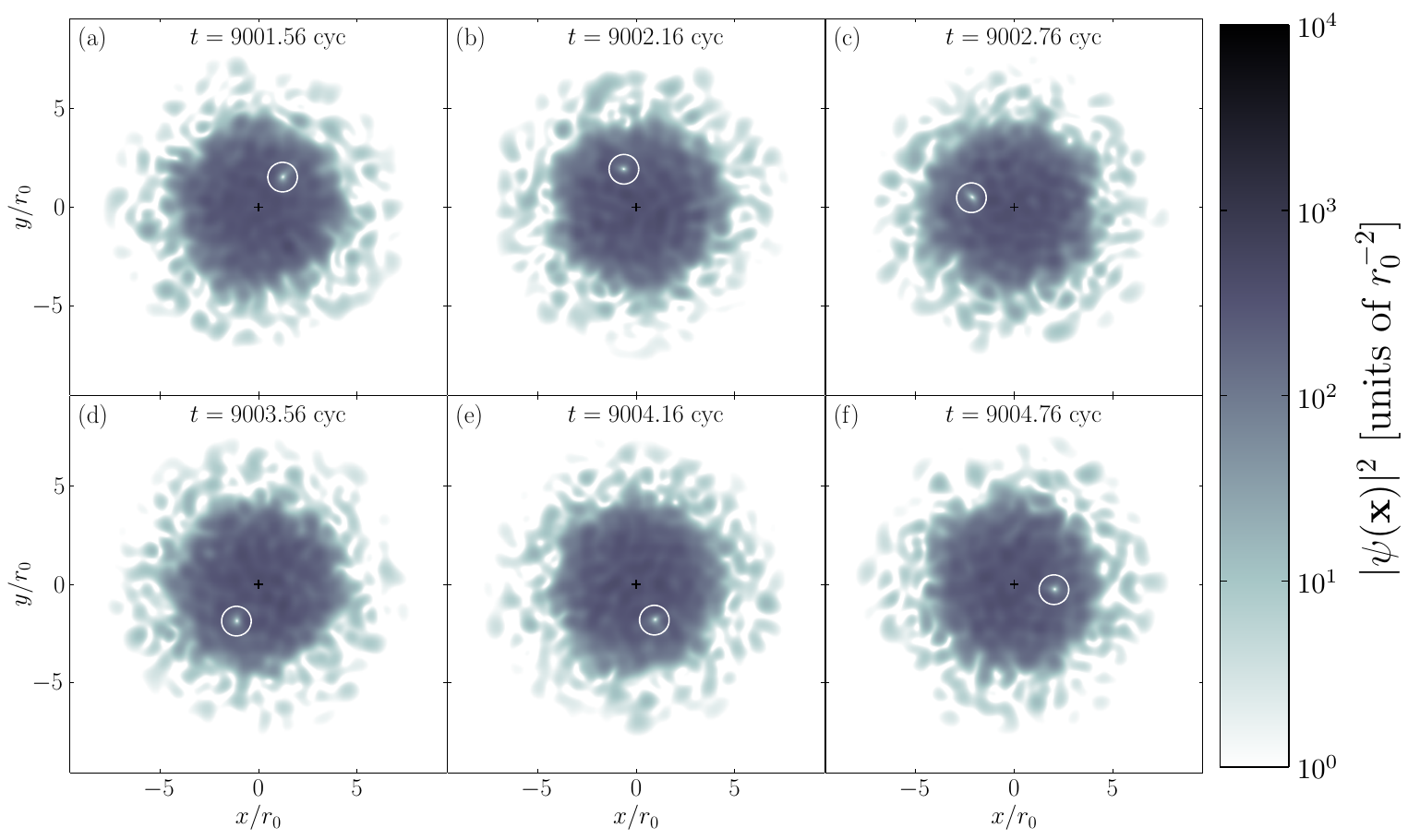}
	\caption{\label{fig:prec_density_plots1}  Classical-field density during a single orbit of the vortex, as viewed in the lab frame.  The white circle indicates the vortex position, and $+$ marks the coordinate origin (trap axis).  Parameters of the classical field are given in the text.}
	\end{center}
\end{figure}

After the equilibration period, the behaviour of the classical field simulations is qualitatively similar for energies in the range $E\in[1.04,1.15]E_\mathrm{g}$.  Each exhibits a central high-density region containing a vortex which is displaced off and precessing about the trap axis.  Viewed in the lab frame, the precession is in the same sense as the vortex (i.e. positive rotation sense, corresponding to angular-momentum numbers $m > 0$), and is stochastic in nature, with the vortex position fluctuating about a circular mean orbit, in the presence of strong density fluctuations of the background field.  As noted in section~\ref{subsubsec:prec_Choice_of_frame}, the frequency of the vortex precession is in general not equal to the angular velocity of the projector defining the microcanonical system. The condensate is located entirely in the central bulk (see section~\ref{subsec:prec_Penrose_Onsager_analysis}); outside this region the field exhibits the high-energy turbulence characteristic of purely thermal atoms in the classical-field model.  In figures~\ref{fig:prec_density_plots1}(a-f) we plot classical-field densities spanning a single orbit of the vortex in a simulation with conserved energy $E=1.10E_\mathrm{g}$.  The behaviour presented in figure~\ref{fig:prec_density_plots1} is representative of all simulations in which the equilibrium condensate contains a precessing vortex.
%%%%%%%%%%%%%%%%%%%%%%%%%%%%%%%%%%%%%%%
%%%%%%%%%%%%%%%%%%%%%%%%%%%%%%%%%%%%%%%
From a range of simulations we observe that the radius at which the vortex precesses increases as the energy of the field is increased.  At higher energies it approaches the edge of the condensate and at $E\approx1.16E_\mathrm{g}$ it is ultimately lost into the violently evolving peripheral region of surface excitations and short-lived phase defects.  As the field energy is decreased, the vortex approaches the trap centre.  However, we expect that as the energy of the field is lowered towards that of the ground state, the excitation becomes too weak to reliably generate thermal statistics on the time scales of interest, and we suspect that for low energies the background field `seen' by the vortex may not appear thermal on the time scale of its precession.  Indeed in such a low temperature regime the effects of quantum fluctuations neglected in our model will become important.  We therefore exclude simulations with energies $E<1.04E_\mathrm{g}$ from our analysis.  A detailed discussion of the dependence of system observables on the internal energy of the field is presented in section~\ref{sec:prec_Energy_dependence}.
%%%%%%%%%%%%%%%%%%%%%%%%%%%%%%%%%%%%%%%%%%%%%%%%%%%%%%%%%%%%%%%%%%%%%%%%%%%%%%%%%%%%%%%%
\subsection{Vortex precession}\label{subsec:prec_Vortex_precession}
\begin{figure}
	\begin{center}
	\includegraphics[width=1.00\textwidth]{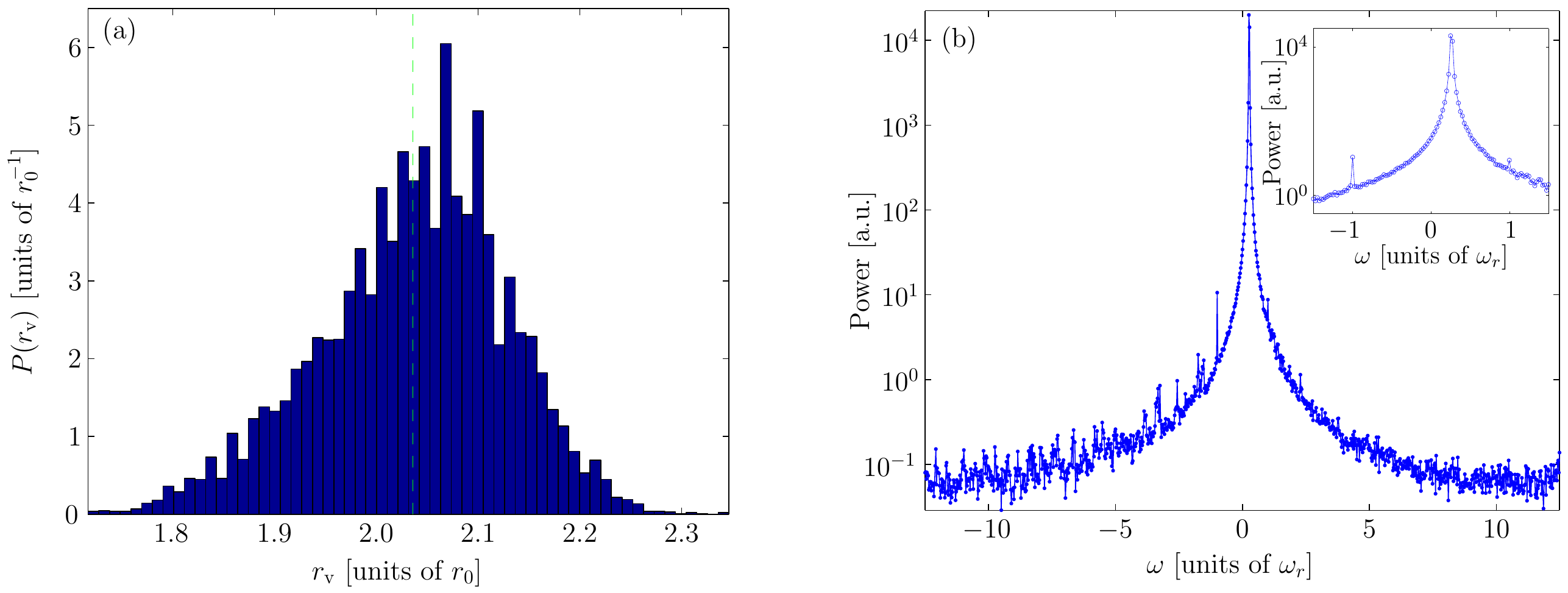}
	\caption{\label{fig:equil_vortex_motion}  Quantities characterising the motion of the vortex. (a) Normalised histogram of measured vortex-displacement magnitudes.  The vertical dashed line indicates the mean. (b) Power spectrum of the complex vortex coordinate $z_\mathrm{v} \equiv x_\mathrm{v} + iy_\mathrm{v}$, revealing the precession of the vortex as viewed in the laboratory frame.}
	\end{center}
\end{figure}

\begin{figure}
	\begin{center}
	\includegraphics[width=0.55\textwidth]{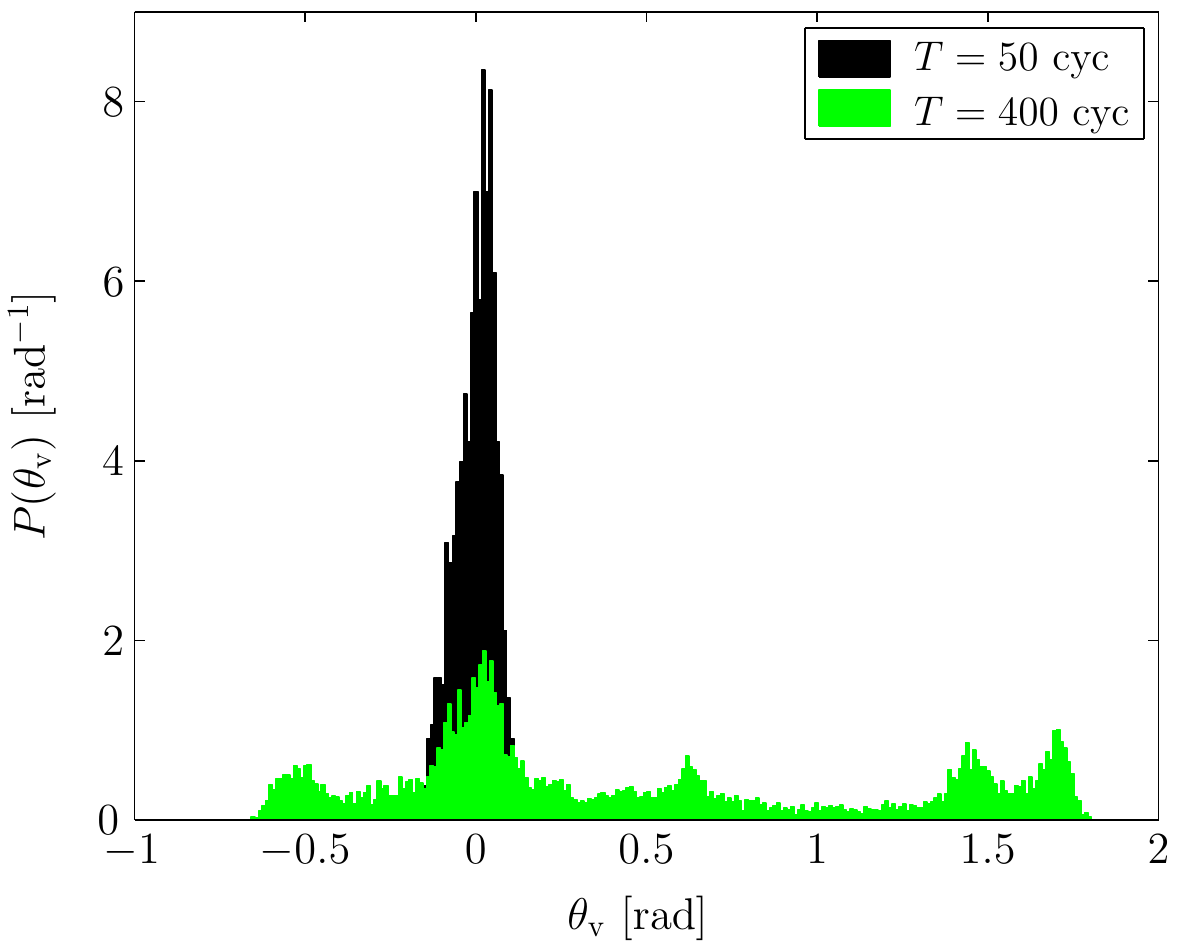}
	\caption{\label{fig:theta_hist} Normalised histograms of vortex phases $\theta_\mathrm{v}$ measured in a rotating frame (see main text) over periods of length 50 cyc and 400 cyc, starting from t=9000 cyc.}
	\end{center}
\end{figure}

In order to characterise the vortex motion, we track the vortex location as observed in the laboratory frame.  In practice we recorded vortex locations in the simulation with $E=1.10E_\mathrm{g}$, at a frequency of 25 samples per cycle for a period of 400 cycles, starting at $t=9000$~cyc.  We find that the radial distance of the vortex from the trap centre fluctuates about a mean value of $\overline{r_\mathrm{v}}=2.04r_0$, and that the series of radial distances measured (from the trap centre) has standard deviation $\sigma_{r_\mathrm{v}}=0.09r_0$, which is of the same order as the extent of the vortex core (which is approximately given by the healing length $\eta=0.20r_0$ \cite{Dalfovo99}\footnote{We use the peak density of our principal ground state to estimate the healing length $\eta$.}).  The normalised distribution of vortex-displacement magnitudes measured over the 400 cyc period is displayed in figure~\ref{fig:equil_vortex_motion}(a).
%%%%%%%%%%%%%%%%%%%%%%%%%%%%%%%%%%%%%%%
%%%%%%%%%%%%%%%%%%%%%%%%%%%%%%%%%%%%%%%
The plotted distribution shows some negative skew but this is not a consistent feature across the simulations performed.
To determine temporal characteristics of the vortex trajectory, we define a complex vortex coordinate $z_\mathrm{v}(t) \equiv x_\mathrm{v}(t) + iy_\mathrm{v}(t)$, where the coordinate pair $(x_\mathrm{v}(t_j),y_\mathrm{v}(t_j))$ specifies the vortex position at time $t_j$, and we calculate the power spectrum of this time series.  As direct periodograms are subject to large variances in the estimates they provide at each frequency \cite{Press92} we split the $10^4-$sample time series into 10 consecutive segments of equal length and average the power spectra obtained from these shorter series (cf. section~\ref{subsubsec:temporal_coherence})\footnote{In the signal processing literature this is known as \emph{Bartlett's method} \cite{Hayes96} for estimating the power spectrum.}.  In figure~\ref{fig:equil_vortex_motion}(b) we plot the power spectrum of the vortex coordinate $z_\mathrm{v}$ over the (angular) frequency range $\omega \in [-12.5,12.5] \omega_r$ permitted by the sample frequency of our data\footnote{The Cartesian-grid vortex-tracking algorithm employed here (see \cite{Caradoc00}) produces a systematic bias in the vortex-position estimate on a sub-grid-point scale (much less than $r_0$).  We have subtracted this dc component of the series $\{z_\mathrm{v}\}$ before calculating the power spectrum presented in figure~\ref{fig:equil_vortex_motion}(b).}.  The most conspicuous feature of the spectrum is the sharp peak (note the logarithmic scale of figure~\ref{fig:equil_vortex_motion}(b)) corresponding to the precession of the vortex about the trap axis.  
The next most pronounced feature of the power spectrum is a clearly resolved spike at frequency $\omega=-1\omega_r$.  A closer inspection of the power spectrum reveals an accompanying spike at $\omega=1\omega_r$ (see inset).  We identify $\omega=1\omega_r$ as the universal frequency of the persistent dipole-oscillation mode of the harmonically trapped gas (see section~\ref{subsec:back_bose_condensation}).  The frequencies $\pm1\omega_r$ thus result from a small dipole oscillation of the classical field as a whole.  Smaller features can be seen in the wings of the power-spectrum distribution, which we associate with thermal fluctuations of the classical field.
From the polar form $z_\mathrm{v}(t_j)=|z_\mathrm{v}(t_j)|e^{i\theta_\mathrm{v}(t_j)}$ of the vortex coordinate, we observe that the vortex phase $\theta_\mathrm{v}$ does not increase linearly with time, but fluctuates randomly as the vortex moves in the dynamically evolving thermal field.  Over time the vortex phase diffuses, so that the rotational symmetry of the system is restored in the ergodic density of classical-field configurations, in direct analogy to the restoration of the phase symmetry of the system by the diffusion of the condensate phase discussed in chapter~\ref{chap:anomalous}.  This is illustrated in figure~\ref{fig:theta_hist}, where we plot the distribution of vortex phases $\theta_\mathrm{v}$ as measured in a frame rotating at frequency $\Omega=0.2610\omega_r$.  
%%%%%%%%%%%%%%%%%%%%%%%%%%%%%%%%%%%%%%%
%%%%%%%%%%%%%%%%%%%%%%%%%%%%%%%%%%%%%%%
This frequency corresponds to the peak presented in figure~\ref{fig:equil_vortex_motion}(b), and has been chosen because it yields the minimum variation in vortex phase over the 400 cyc period we consider.  Histograms of the normalised vortex-phase distribution over the first 50 cycles of this period (black bars) and over the full 400 cyc period (green (grey) bars) are plotted.  We observe that at short times the fluctuations in $\theta_\mathrm{v}$ are small and reasonably Gaussian in nature.  Over longer time periods the vortex phase drifts significantly.  This diffusion of the vortex phase produces the distinctive Lorentz-like shape of the power-spectrum distribution plotted in figure~\ref{fig:equil_vortex_motion}(b).  In section~\ref{subsec:prec_PO_rotating_frames} we discuss the effects of this diffusion on the `one-body' correlations of the classical field obtained from time-averaging the classical-field trajectory, and the implications for the identification and characterisation of the condensate in these simulations.
%%%%%%%%%%%%%%%%%%%%%%%%%%%%%%%%%%%%%%%%%%%%%%%%%%%%%%%%%%%%%%%%%%%%%%%%%%%%%%%%%%%%%%%%%%%%%%%%%%%%%%%%%%%%%%%%%%%%%%%%%%%%%%%%%%%%
\section{Field-covariance analysis}\label{subsec:prec_Penrose_Onsager_analysis}
We turn now to the issue of identifying the condensate in the classical-field solutions.  In vortex-free\footnote{Vortices may of course occur in all such simulations; we use the phrase vortex-free here to refer to simulations in which any appearance of vortices is incidental, and sufficiently transient as to not affect the statistics under consideration significantly.} equilibrium scenarios the method of condensate definition based on sampling the classical-field correlations discussed in section~\ref{subsec:cfield_correl} has proven successful.  In practice, the method is performed in the spectral representation, i.e., we construct the covariance matrix 
\begin{equation}\label{eq:covar_mtx}
	G_{ij} = \langle \alpha_j^*\alpha_i \rangle_t
\end{equation}
from the propagation-basis coefficients (section~\ref{subsec:cfield_proj_cft}) of a single classical-field trajectory. The indices $i,j$ each index a full set of quantum numbers which label the basis modes (i.e., $n$ and $l$ in the present case of Gauss-Laguerre basis modes), and $\langle\cdots\rangle_t$ denotes an average over time. 
As discussed in section~\ref{sec:cfield_ergodicity}, ergodicity of the field evolution \cite{Lebowitz73} implies that averages taken over sufficiently long times approach averages over the microcanonical ensemble of field configurations, in which the coherence of a classical-field condensate \cite{Connaughton05} manifests as a single dominant eigenmode of $G_{ij}$.  However, this method of condensate identification has some significant ambiguities in more general scenarios, in particular when symmetries of the Hamiltonian are broken by the appropriate GP state \cite{Castin04}.  For example, the standard PO criterion itself may prove inappropriate in situations in which the condensate exhibits centre-of-mass motion (see section~\ref{subsec:back_bose_condensation}), prompting workers to introduce revised definitions of the density matrix appropriate to such systems, in which the centre-of-mass motion is eliminated \cite{Pethick00,Yamada09}.  

Issues of condensate definition also arise in the presence of vortices.  In scenarios in which several vortices exist, breaking the rotational symmetry of an otherwise rotationally invariant many-body Hamiltonian, an uncountably infinite degeneracy of the one-body density matrix appears, even in the idealised Gross-Pitaevskii limit of a zero-temperature system \cite{Seiringer08}.  
In this chapter we have chosen one of the simplest scenarios of rotational symmetry breaking in a finite-temperature classical field: the precession of a single vortex.  Because the presence of the vortex distinguishes different rotational orientations of the condensate, we find that in order to determine the condensate mode we must construct time averages in an appropriate rotating frame.  The fact that the (azimuthal) phase of the vortex diffuses over time in any such uniformly rotating frame means that we must abandon the notion of ergodic reconstruction of the microcanonical density by time averages.  However, we expect intuitively \cite{Blakie05} that in order to quantify the condensation in the field, the averaging time need only be long enough to distinguish the Gaussian fluctuations of thermal or chaotic modes \cite{Glauber65,Mandel95} from the non-Gaussian statistics of a quasi-uniform phase evolution characteristic of a condensed component.  We find therefore that we can successfully quantify the condensation of the field from such short-time averages, while exploiting the ergodic character of the evolution over longer time scales.   In the remainder of this chapter, we will use the term Penrose-Onsager (PO) procedure generically to refer to the construction of the covariance matrix (density matrix) equation~(\ref{eq:covar_mtx}) by a time-averaging procedure, the precise details of which will be in each case specified in the context.
%%%%%%%%%%%%%%%%%%%%%%%%%%%%%%%%%%%%%%%%%%%%%%%%%%%%%%%%%%%%%%%%%%%%%%%%%%%%%%%%%%%%%%%%
\subsection{Lab-frame analysis}\label{subsubsec:prec_lab-frame_PO}
\begin{figure}	
	\begin{center}
	\includegraphics[width=1.0\textwidth]{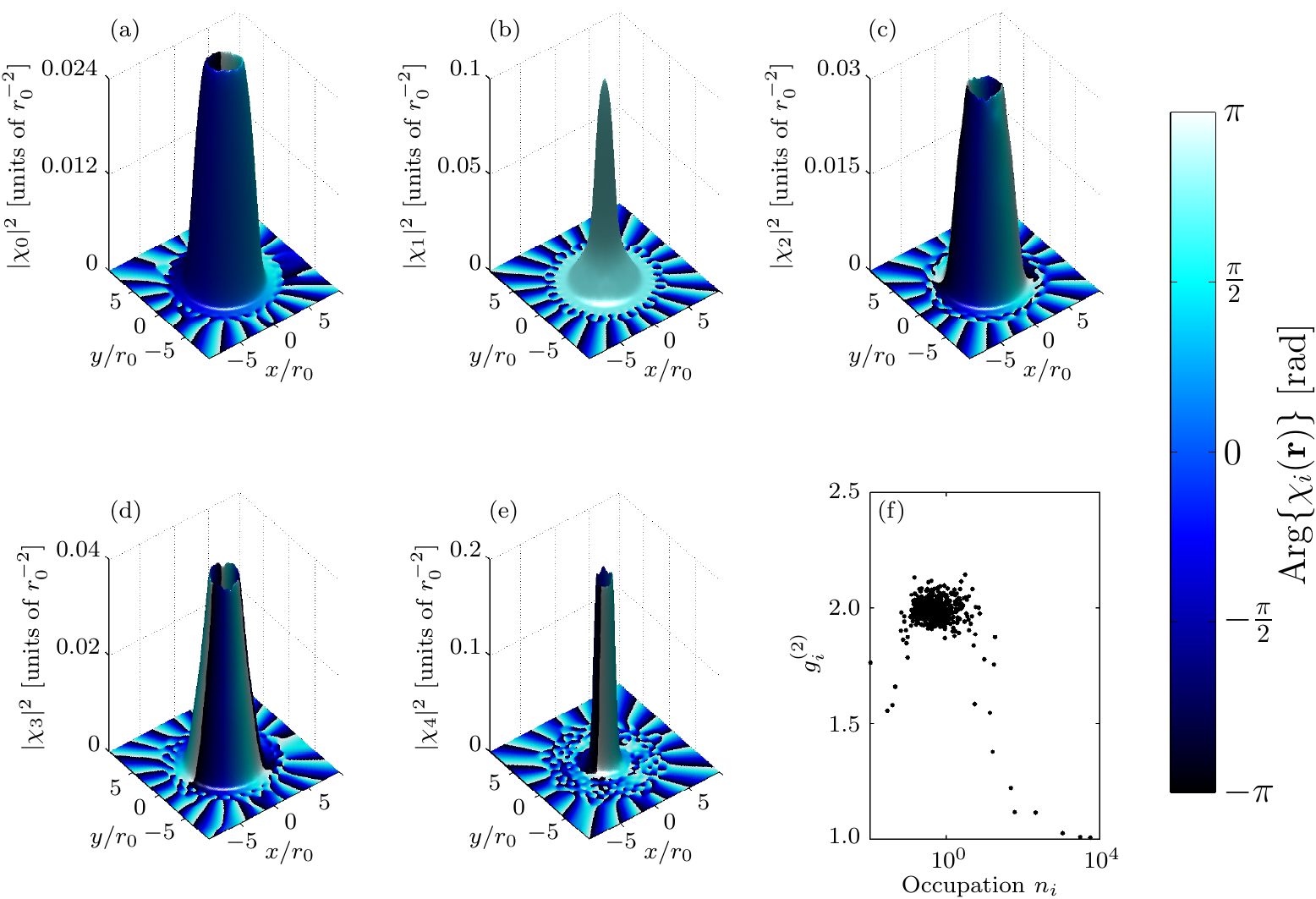}
	\caption{\label{fig:POsurfs_E105}  (a-e) Density and phase of the five most highly occupied eigenmodes of the covariance matrix, for case $E=1.05E_\mathrm{g}$. (f) Second-order coherence functions of the covariance-matrix eigenmodes versus mode occupation.}
	\end{center}
\end{figure}

We consider a simulation with energy $E=1.05E_\mathrm{g}$, which exhibits a vortex precessing at a frequency $\omega_\mathrm{v}\approx0.23\omega_r$, at a radius $\overline{r_\mathrm{v}}\approx1.3r_0$, which should be contrasted with the extent of the central density bulk $r_\mathrm{b}\approx5r_0$.  The vortex in this case is displaced from its axis by significantly more than the extent of its core (healing length $\eta=0.20r_0$), while the core remains comparatively close to the centre of the region of significant density.  We proceed by forming the covariance matrix by averaging over $2501$ equally spaced samples of the \emph{lab-frame} representation of the classical-field evolution, between $t=9000$ and $9100$~cyc.  We diagonalise the density matrix~\reff{eq:covar_mtx} to obtain its eigenvectors, the position-space representations of which we denote by $\chi_i(\mathbf{x})$, and the corresponding occupations $n_i$ (section~\ref{subsec:cfield_correl}).  In figures~\ref{fig:POsurfs_E105}(a-e) we plot the five most highly occupied eigenmodes of $G_{ij}$, which together contain some $95.4\%$ of the total field population.   

%%%%%%%%%%%%%%%%%%%%%%%%%%%%%%%%%%%%%%%
%%%%%%%%%%%%%%%%%%%%%%%%%%%%%%%%%%%%%%%
The density and phase of the most highly occupied mode ($\chi_0(\mathbf{x})$) is displayed in figure~\ref{fig:POsurfs_E105}(a) and contains a singly positively charged \emph{central} vortex (with phase circulation of $2\pi$ around its azimuth direction).  The second most highly occupied mode ($\chi_1(\mathbf{x})$, figure~\ref{fig:POsurfs_E105}(b)) is a rotationally symmetric mode with uniform phase, which forms a pronounced peak at the origin.  The third most highly occupied mode ($\chi_2(\mathbf{x})$, figure~\ref{fig:POsurfs_E105}(c)) is another vortex mode, with vanishing central density.  However, this mode is doubly (positively) charged, with the phase varying by $4\pi$ around its azimuth.  

We note that these three modes bear a strong resemblance to a central-vortex GP eigenmode $\phi_0$ and the $u$ and $v^*$ spinor components \cite{Castin98}\footnote{We adopt here the convention for Bogoliubov-mode component labelling of references \cite{Castin98, Morgan00}.} of its so-called \emph{anomalous} Bogoliubov excitation (also referred to as the lowest core-localised state by some authors \cite{Isoshima99}).  We recall from section~\ref{subsec:back_vortices_in_conds} that this is the lowest-energy excitation of the central single-vortex state in the Bogoliubov approximation, and it has \emph{negative} energy in the laboratory frame, which signals the thermodynamic instability of the vortex state in that frame.  The angular-momentum numbers $m_{u,v}=(0,2)$ of the particle ($u$) and hole ($v$) functions result naturally from the transposition of the Bogoliubov pairing ansatz to the $m=1$ vortex scenario \cite{Fetter71,Dodd97}.  The collective excitation of the anomalous mode results in the displacement of the vortex off-axis in a direction determined by the phase of the Bogoliubov excitation relative to the condensate (see \cite{Rokhsar97}) in the linear expansion\footnote{We note that the next most highly occupied eigenmodes, $\chi_3(\mathbf{x})$ and $\chi_4(\mathbf{x})$, have the form of vortex modes with charges $q=+3$ and $q=-1$ respectively.  These circulations correspond respectively to those of the $u$ and $v^*$ components of the linear excitation of the GP vortex with angular momentum $m=+2$ relative to the GP state \cite{Mizushima04}.}
\begin{equation}\label{eq:anomalous_mode_bog_expansion}
	\psi = a_0\phi_0 + bu + b^*v^*.
\end{equation}
In the (zero-temperature) Bogoliubov approximation the energy of the anomalous mode determines the frequency of precession of a displaced vortex under its own induced velocity field, in the appropriate linear-response limit (section~\ref{subsec:back_vortices_in_conds}).  

\begin{figure}	
	\begin{center}
	\includegraphics[width=0.65\textwidth]{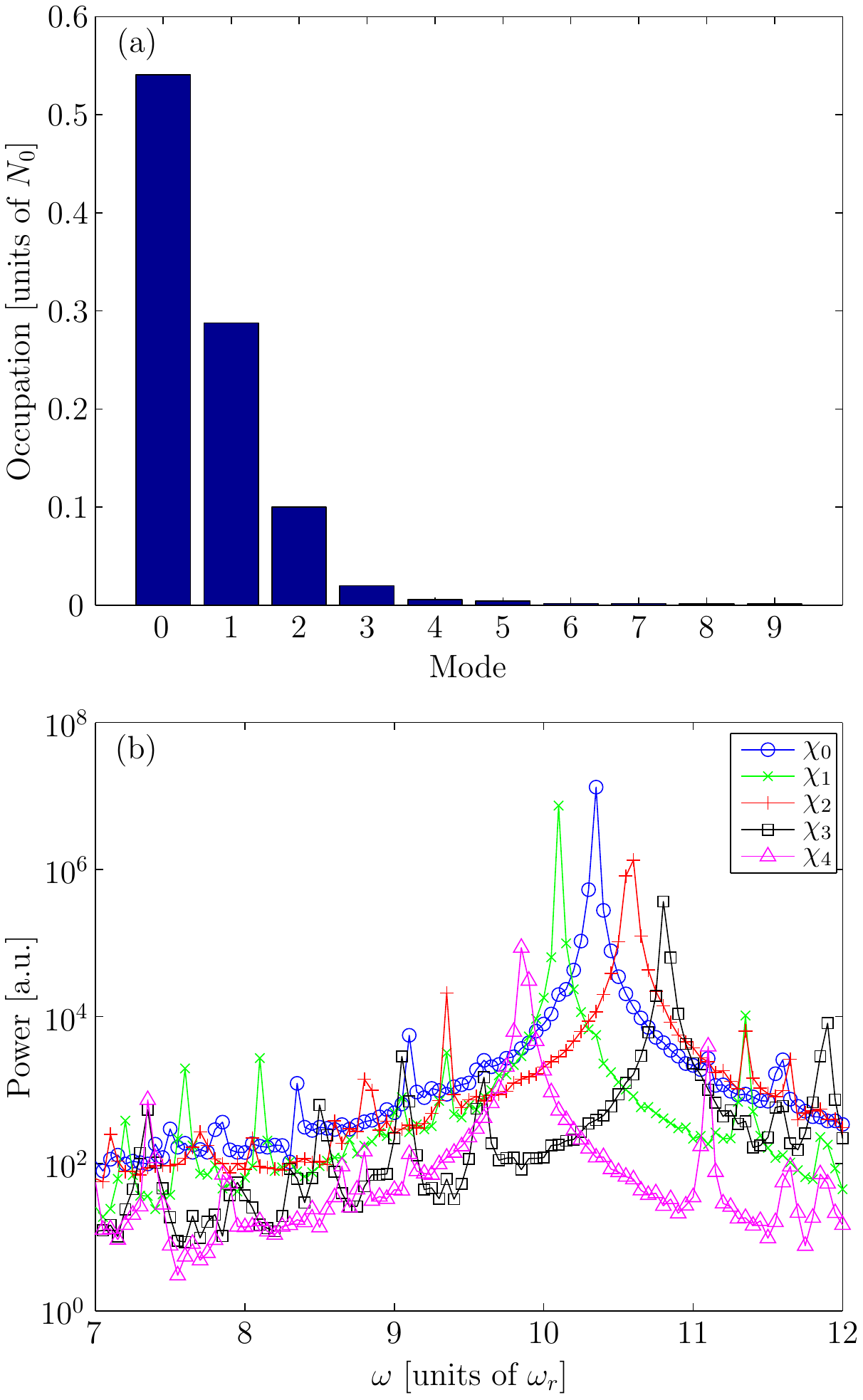}
	\caption{\label{fig:occs_spect_E105}  (a) Mean occupations of covariance matrix eigenmodes.  (b) Temporal power spectra of coefficients $\beta_i(t)$ of the five most highly occupied modes.  Quantities presented in each subplot are calculated for the case $E=1.05E_\mathrm{g}$.}
	\end{center}
\end{figure}

Returning to our classical-field simulation, we calculate the time-dependent expansion coefficients of the eigenmodes $\chi_i(\mathbf{x})$ over the 100 cyc analysis period, defined as 
\begin{equation}\label{eq:covar_mode_coefficients}
	\beta_i(t) \equiv \int\!d\mathbf{x}\,\chi_i^*(\mathbf{x})\psi(\mathbf{x},t).
\end{equation}
The average values $\langle|\beta_i|^2\rangle_t$ of course yield simply the mode occupations $n_i$, which are plotted for the $10$ most highly occupied modes in figure~\ref{fig:occs_spect_E105}(a).  
%%%%%%%%%%%%%%%%%%%%%%%%%%%%%%%%%%%%%%%
%%%%%%%%%%%%%%%%%%%%%%%%%%%%%%%%%%%%%%%
From the coefficients $\{\beta_i\}$ we form the classical correlation functions \cite{Gardiner00,Blakie05}
\begin{equation}\label{eq:define_g2}
	g_i^{(2)} = \frac{\left\langle|\beta_i|^4\right\rangle_t}{\left(\left\langle|\beta_i|^2\right\rangle_t\right)^2},
\end{equation}
which indicate the coherence of the modes over the analysis period.  The (local) correlation functions $g^{(n)}$ adopt values of $g^{(n)}=n!$ and $g^{(n)}=1$ for purely chaotic or thermal fields and purely coherent fields, respectively \cite{Glauber65,Blakie05}.  The distribution of $g^{(2)}$ values over occupation numbers is shown in figure~\ref{fig:POsurfs_E105}(f).  We find that the $g^{(2)}$ values calculated for the five most highly occupied modes $\chi_i(\mathbf{x}) :\;0\leq i\leq 4$ are all $\lesssim1.1$.  By contrast, the vast majority of the remaining, less populated modes $(i>4)$ have $g^{(2)}\approx2$.  This suggests that these five most highly occupied modes are \emph{coherent}, while the majority of modes $\chi_i$ are to a large degree incoherent \emph{with respect to the averaging procedure employed here}.  These coherent modes thus appear to comprise a \emph{fragmented} condensate \cite{Leggett01}; i.e., a field configuration in which more than one `condensed' single-particle mode is present.
In chapter~\ref{chap:anomalous}, the temporal power spectrum of the classical field was used to analyse the coherence in the field.  Here we apply a similar technique to extract information from the mode coefficients $\beta_i(t)$.  In figure~\ref{fig:occs_spect_E105}(b) we plot the spectra obtained from these samples, each of which was formed by averaging spectra estimates from five consecutive time series (see section~\ref{subsec:prec_Vortex_precession}).  
We observe that the spectrum of each mode coefficient presented in figure~\ref{fig:occs_spect_E105}(b) exhibits several peaks, the highest of which is in each case several orders of magnitude greater than the others.  We interpret the largest peak in each case as an indication of the quasi-uniform phase rotation associated with coherence in the classical field (see chapter~\ref{chap:anomalous}).  We also note that the frequency peaks of modes $\chi_1$ and $\chi_2$ are spaced evenly about the peak of $\chi_0$.  Specifically we find for the frequencies of peak power $(\omega_\mathrm{p,0},\omega_\mathrm{p,1},\omega_\mathrm{p,2})=(10.35,10.10,10.60)\omega_r$, corresponding to a quasiparticle energy of $\varepsilon_\mathrm{q}=\hbar(\omega_\mathrm{p,1}-\omega_\mathrm{p,0})\approx-0.25\hbar\omega_r$, in agreement with the vortex precession frequency $\omega_\mathrm{v}\approx0.23\omega_r$ (determined by the technique of section~\ref{subsec:prec_Vortex_precession}) to within the frequency resolution $\Delta\omega=0.05\omega_r$.   

A precessing-vortex GP state can of course be understood as a linear combination of such angular-momentum eigenmodes.  In a frame rotating at the vortex precession frequency ($\Omega_\mathrm{v}$), the phases of the various components all rotate at a single common frequency (the condensate eigenvalue $\mu/\hbar$).  In the laboratory frame these components have frequencies $\omega_m = \mu/\hbar + \Omega_\mathrm{v}m$, and their consequent dephasing over time leads to the apparent fragmentation observed in the lab frame\footnote{We note that the smaller frequency peaks in the spectrum of each coefficient correspond to the quasi-uniform phase rotation of other angular-momentum components.  This arises due to the imperfect nature of the PO decomposition resulting from (e.g.) the finite sample size used to construct $G_{ij}$ \cite{Blakie05}.}.  Simulations of classical fields with higher energies, which contain vortices precessing at larger radii, when subjected to the above analysis, again yield a decomposition into angular-momentum eigenmodes.  However, in such cases we find that the most highly occupied mode has $m=0$, as the anomalous particle component $u$ has grown, pushing the vortex further from the trap centre and replacing the $m=1$ vortex mode as most highly occupied.
%%%%%%%%%%%%%%%%%%%%%%%%%%%%%%%%%%%%%%%%%%%%%%%%%%%%%%%%%%%%%%%%%%%%%%%%%%%%%%%%%%%%%%%%
\subsection{Rotating frames}\label{subsec:prec_PO_rotating_frames}
\begin{figure}
	\begin{center}
	\includegraphics[width=1.0\textwidth]{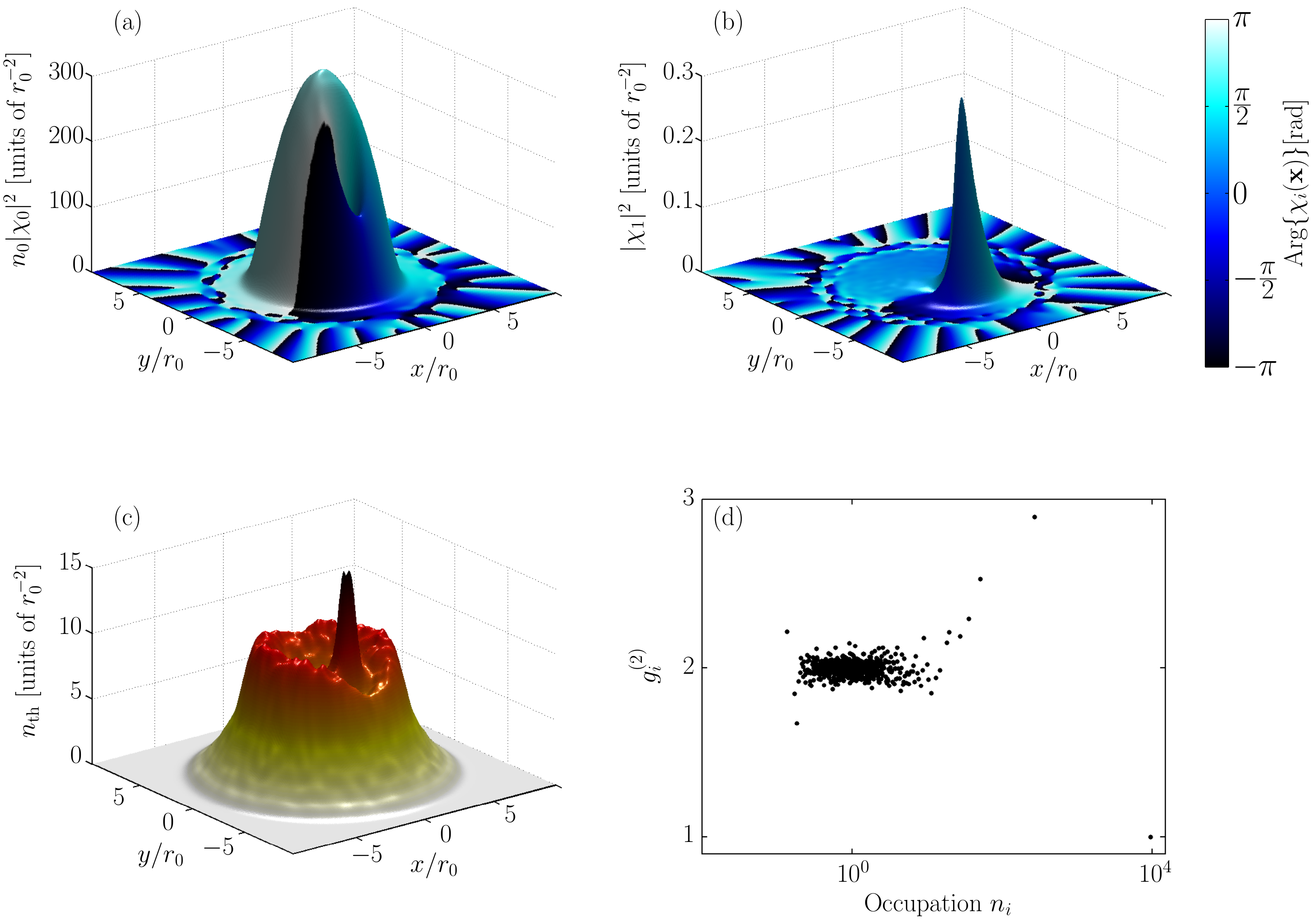}
	\caption{\label{fig:POsurfs_E110_corot}  Density and phase of (a) the condensate and (b) the Goldstone mode, as determined from the covariance matrix formed in the co-rotating frame.  The condensate density is the density of mode $\chi_0(\mathbf{x})$, shown normalised to its occupation $n_0$. (c) Thermal density of the field. (d) Second-order coherence functions of the covariance-matrix eigenmodes versus mode occupation.} 
	\end{center}
\end{figure}

The above results of a naive application of the PO procedure in the laboratory frame suggest that the classical field condenses in a rotating frame.  The mean phase rotation of the various angular-momentum components is such that upon transformation to an appropriate rotating frame these modes collectively exhibit quasi-uniform phase rotation at a common frequency.  We therefore expect that transforming the field to its representation in such a rotating frame, before applying the PO procedure, will yield a single condensate mode.  We consider now a simulation with $E=1.10E_\mathrm{g}$, where in equilibrium a single vortex precesses with mean displacement of $\overline{r_\mathrm{v}}\approx2.0r_0$ from the trap centre.  After transforming the field coefficients $\{\alpha_{nl}\}$ to a frame (the \emph{measurement} frame) with fixed rotation frequency $\Omega_\mathrm{m}$, we construct the covariance matrix in the same manner as described in section~\ref{subsubsec:prec_lab-frame_PO}.  In figures~\ref{fig:POsurfs_E110_corot}(a-b) we plot the two most highly occupied modes obtained from this covariance matrix, having performed the PO procedure in a frame with $\Omega_\mathrm{m}=0.2610\omega_r$.  We discuss below the reasons for this choice of rotation frequency.  
%%%%%%%%%%%%%%%%%%%%%%%%%%%%%%%%%%%%%%%
%%%%%%%%%%%%%%%%%%%%%%%%%%%%%%%%%%%%%%%
The most highly occupied mode is now a vortical state in which the vortex is \emph{displaced} off-axis, similarly to the case for the full classical field.  The second most highly occupied mode is now a peaked function which partially `fills' the density dip associated with the vortex in $\chi_0$.  The prominent peak of this mode exhibits a distinctive amplitude and phase pattern, which we identify as that of the \emph{Goldstone mode} associated with the breaking of rotational symmetry by the off-axis vortex state.  Such a mode arises in scenarios in which the condensate mode $\phi_0$ breaks a symmetry of the system Hamiltonian (see section~\ref{subsec:theory_Bogoliubov}).  In the case of broken rotational symmetry \cite{Lobo05}, the Goldstone mode has the form $(u_\mathrm{G},v_\mathrm{G})\sim(L_z \phi_0,-L_z^*\phi_0^*)$, and corresponds to small rotations of the symmetry-broken condensate mode.  As such a Goldstone mode is of a fundamentally different nature to the `proper' normal modes (quasiparticles) of the system \cite{Blaizot86}, we regard this mode as separate from the thermal or noncondensate fraction of the field\footnote{Note that the Goldstone mode only happens to be the second most highly occupied mode in the present case as its spurious `population' is larger than that of all other noncondensate modes.}.  We therefore form the thermal density as the total density of all modes other than the condensate mode $(\chi_0)$ and Goldstone mode $(\chi_1)$,
\begin{equation}\label{eq:thermal_density}
	n_\mathrm{th}(\mathbf{x}) = \sum_{i\geq2} n_i |\chi_i(\mathbf{x})|^2.
\end{equation}\enlargethispage{-\baselineskip}
We plot this density in figure~\ref{fig:POsurfs_E110_corot}(c).  We note that the thermal density exhibits a sharp peak at the vortex location, resulting from the erratic motion of the vortex about its mean trajectory in the classical-field evolution.  Thus a careful application of the PO procedure yields a core-filling thermal component, as is predicted by the Hartree-Fock-Bogoliubov family of self-consistent mean-field theories \cite{Virtanen01}.  This is in contrast to the instantaneous classical field, which, due to its essential nature as a single-particle wave function (i.e., vector in the projected single-particle Hilbert space) contains only `bare' vortices, with no possibility of core filling.  In the microcanonical-ergodic approach of the PGPE (section~\ref{sec:cfield_ergodicity}) the two (and higher) particle correlations emerge from the field when averaged over time.  However, an alternative interpretation of classical-field trajectories as representative of single experimental trajectories can be made, but, as noted in \cite{Blakie08}, this interpretation must be made with caution.  In particular it is \emph{not correct} to interpret the instantaneous state of the many-body wave function as the Hartree product of the classical field $\psi(\mathbf{x})$ (as is the interpretation of the GP wave function in the zero-temperature GP theory; see section~\ref{subsec:back_GPE})\footnote{It could indeed be said that this is the definitive distinction of interpretation between the time-dependent GP theory and the time-dependent classical-field theory.}.  Rather, the field $\psi(\mathbf{x})$ must be interpreted as an approximation to the quantum \emph{field operator}, from which certain correlations can be deduced.  Some averaging of this instantaneous field $\psi(\mathbf{x})$, over an ensemble of trajectories \cite{Norrie06a} or over time \cite{Blakie05}, must be performed to construct the correlations necessary to describe the thermal core-filling of a vortex. 

\begin{figure}
	\begin{center}
	\includegraphics[width=0.65\textwidth]{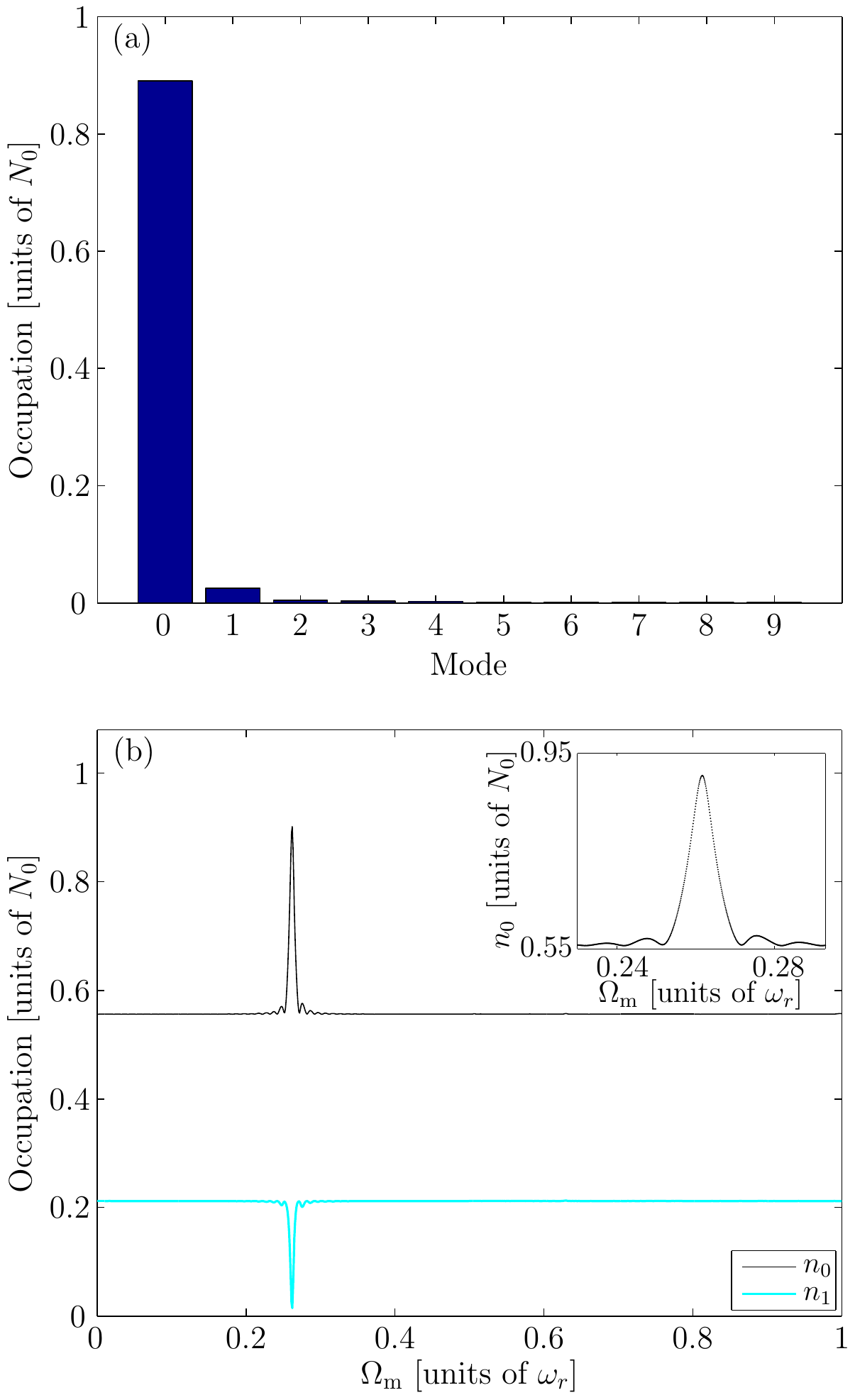}
	\caption{\label{fig:fracs_and_occs_corot}  (a) Occupations of the 10 most highly occupied eigenmodes of the covariance matrix determined in the frame with $\Omega_\mathrm{m}=0.2610\omega_r$. (b) Dependence of mode populations on the rotation frequency of the frame in which the covariance matrix is constructed. The inset shows the behaviour of $n_0$ close to its maximum.}
	\end{center}
\end{figure}

We calculate the time-dependent expansion coefficients of the modes $\chi_i(\mathbf{x})$ in the same manner as described in section~\ref{subsubsec:prec_lab-frame_PO}.  The mean occupations of the modes are plotted in figure~\ref{fig:fracs_and_occs_corot}(a), and reveal that a single mode contains 89.1\% of the field population.  We present in figure~\ref{fig:POsurfs_E110_corot}(d) the coherence functions obtained from the mode coefficients.  The most highly occupied mode is now the only mode with $g^{(2)}$ significantly less than the thermal value $g^{(2)}=2$.  We find in fact $g^{(2)}_0 = 1.0002$, indicating very high coherence of the condensate mode over the averaging period.  The next most highly occupied modes now have values $g^{(2)}\gtrsim2.2$, \emph{greater} than that expected for thermally or incoherently occupied single-particle modes.  This is a signature of the strongly \emph{collective} nature of the low-lying excitations of the condensate.  The form of these excitations is strongly affected by the presence of the interacting, macroscopically occupied condensate, which induces anomalous (pairing) correlations in the field (see section~\ref{sec:pairing_correlations}), which in turn manifest as anomalous values for the coherence functions \cite{Gardiner01}.  We note that the Goldstone mode has $g^{(2)}=2.90$, close to the upper bound $g^{(2)}=3$ \cite{Gardiner01}, as we might expect given that a Goldstone mode, with $u_\mathrm{G}=-v_\mathrm{G}^*$, is the most strongly collective excitation possible.  The fact that these pairing correlations are manifest in the one-body density matrix formed in the co-rotating frame is further evidence that this frame gives the correct resolution of the condensate.  It is interesting to note that the anomalous nature of these correlations is not apparent in coordinate-space representations of the field \cite{Blakie05}, or in the lab-frame analysis of section~\ref{subsubsec:prec_lab-frame_PO}.  A more natural characterisation of the low-lying excitations would exploit these anomalous one-body correlations in order to construct an appropriate quasiparticle basis \cite{Blaizot86}.

The choice of measurement-frame frequency ($\Omega_\mathrm{m}=0.2610\omega_r$) optimises the resolution of the condensate as we now discuss.  From the discussion of section~\ref{subsubsec:prec_lab-frame_PO}, we expect (neglecting for now the role of the Goldstone mode) that the classical field can be resolved into a single condensate mode coexisting with a bath of thermal atoms, in some rotating frame.  In general the rotation frequency of the measurement frame differs from the equilibrium rotation frequency of the condensate, and this difference leads to dephasing of the distinct angular-momentum components of the condensate mode, and hence to a spurious fragmentation of the condensate into a set of temporally coherent angular-momentum eigenmodes.  We propose as a model for analysis of classical-field data, that in the frame in which condensation occurs, the field should decompose into a single condensate mode plus thermal material upon performing the PO procedure.  This frame depends on a single parameter (the measurement-frame rotation frequency $\Omega_\mathrm{m}$), and the magnitude of the largest eigenvalue, regarded as a function of the measurement-frame frequency ($n_0=n_0(\Omega_\mathrm{m})$) forms the corresponding \emph{optimality function}.  Our procedure then is to maximise the optimality function to determine the optimal measurement-frame frequency $\Omega_\mathrm{c}$, which is the frequency of the rotating frame in which the condensate mode is (most) stationary.  This provides a sensitive measure of the (lab-frame) rotation frequency of the vortical condensate mode, and provides a unified basis for the calculation of the properties of the condensed and noncondensed material in the classical field.  In particular, we will regard the frame of vortex precession and the frame of condensation as interchangeable, though in practice we will use the latter exclusively in the remainder of this chapter.
%%%%%%%%%%%%%%%%%%%%%%%%%%%%%%%%%%%%%%%
%%%%%%%%%%%%%%%%%%%%%%%%%%%%%%%%%%%%%%%
In figure~\ref{fig:fracs_and_occs_corot}(b) we present results of this optimisation procedure as applied to our simulation with $E=1.10E_\mathrm{g}$ over a $100$ cyc period starting from $t=9000$ cyc.  We observe a prominent peak in $n_0(\Omega_\mathrm{m})$ at $\Omega_\mathrm{m}=0.2610\omega_r$.  We also plot the magnitude of the second largest eigenvalue $n_1(\Omega_\mathrm{m})$, and observe that it exhibits a dip at this same frequency, indicating the sensitivity of the one-body correlations to the frequency $\Omega_\mathrm{m}$.  We take the location of the peak in $n_0(\Omega_\mathrm{m})$ as an estimate for the rotation frequency $\Omega_\mathrm{c}$ of the condensate\footnote{As we vary $\Omega_\mathrm{m}$ in increments of $\delta \Omega = 10^{-4}\omega_r$, smaller than the frequency resolution afforded by the sampling period ($\Delta\Omega = 1/T_\mathrm{samp}=2\pi\times0.01\omega_r$), we resolve the profile $n_0(\Omega_\mathrm{m}) \sim\mathrm{sinc}^2[(\Omega_\mathrm{m}-\Omega_\mathrm{m}^0)/\Delta\Omega]$ of the peak of $n_0(\Omega_\mathrm{m})$ (inset to figure~\ref{fig:fracs_and_occs_corot}(b)) resulting from its convolution with the boxcar function representing the finite sample time $T_\mathrm{samp}$ \cite{Press92}.  The location of the central peak of this form provides the best estimate for $\Omega_\mathrm{c}$ in an effective least-squares sense \cite{Fox09}.}.   
%%%%%%%%%%%%%%%%%%%%%%%%%%%%%%%%%%%%%%%%%%%%%%%%%%%%%%%%%%%%%%%%%%%%%%%%%%%%%%%%%%%%%%%%
\subsection{Temporal decoherence and sample length}\label{subsec:prec_temporal_decoherence}
\begin{figure}
	\begin{center}
	\includegraphics[width=0.65\textwidth]{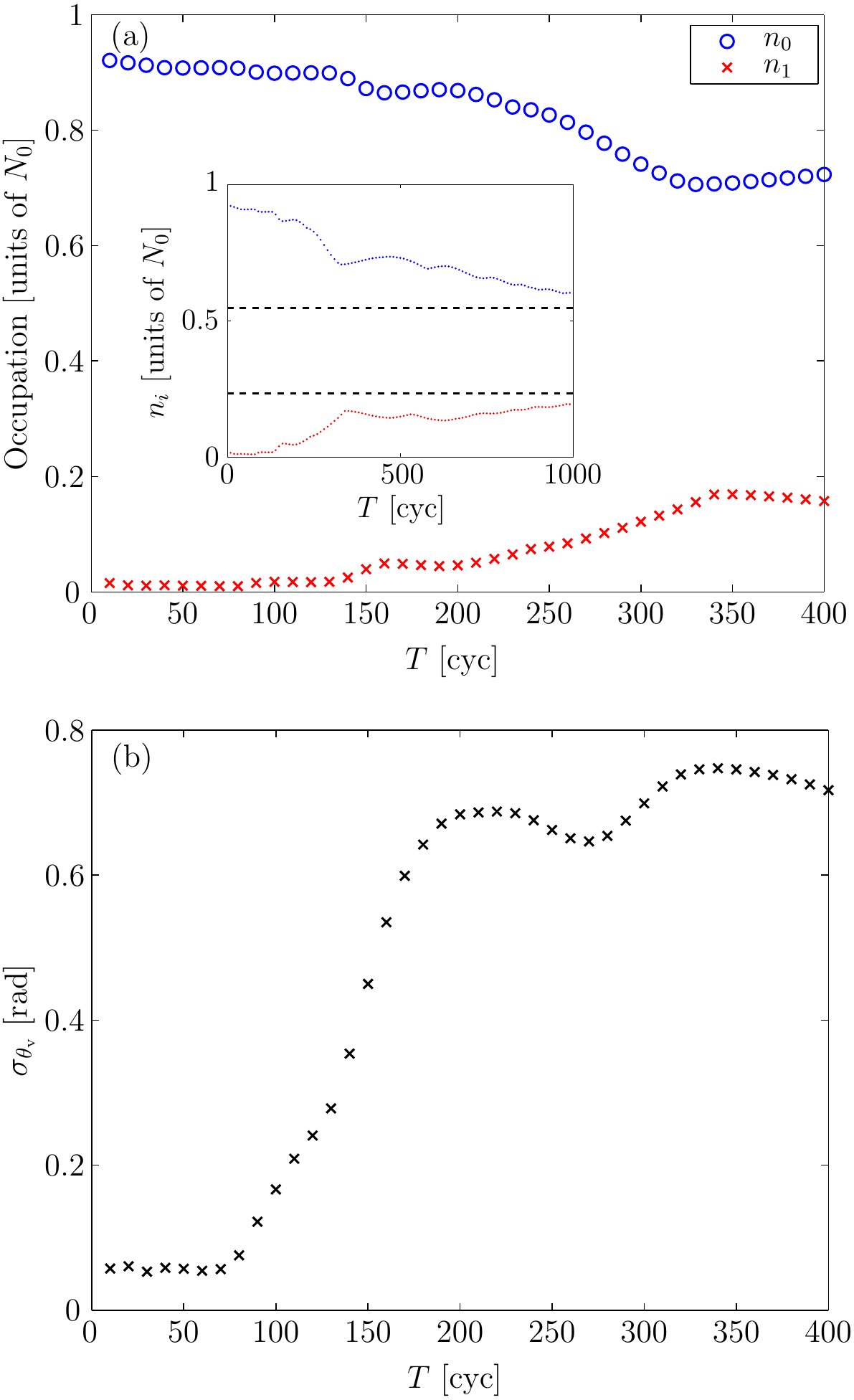}
	\caption{\label{fig:temporal_decay}  (a) Largest eigenvalues ($n_0$ and $n_1$) of the covariance matrix as a function of averaging time. The inset shows the behaviour of $n_0$ and $n_1$ (upper and lower dotted lines, respectively) over longer times, where the eigenvalues approach their values obtained in the laboratory frame (dashed lines). (b) Standard deviation of vortex phases measured in a frame rotating at $\Omega=0.2610\omega_r$.}
	\end{center}
\end{figure}

We now consider how the condensate mode determined by the above procedure decays as a function of the time over which the averaging procedure is performed.  As already noted in section~\ref{subsec:prec_Vortex_precession}, the phase of the vortex (with respect to any uniform rotation) diffuses over time.  This diffusion is intimately related to the breaking of the rotational symmetry of the system by the presence of the off-axis vortex, as we now discuss.  The condensate orbital formed in the classical field is an off-axis vortex mode, which is not an eigenstate of angular momentum and thus not rotationally invariant.  This breaks the rotational ($\mathrm{SO}(2)$) symmetry of the Hamiltonian equation~(\ref{eq:cfield_HCF})\footnote{The rotational symmetry broken here is the invariance of the Hamiltonian under real-space rotations of the field $\psi\rightarrow e^{i\Theta L_z}\psi$.  The phase symmetry of the Hamiltonian (invariance under transformations $\psi\rightarrow\psi e^{i\theta}$) is also broken by \emph{any} condensate mode that forms, as discussed in chapter~\ref{chap:anomalous}.  However, the global phase of the classical field has no effect on the density-matrix calculations we perform here.}.  A well-known consequence of the breaking of a continuous symmetry by a dynamical equilibrium in classical mechanics is the appearance of a zero-energy normal mode \cite{Goldstein50}, and this result persists in the quantum theory, where broken symmetries are similarly accompanied by so-called spurious or Goldstone modes \cite{Blaizot86}. 
Such excitations are associated with a collective motion without restoring force \cite{Blaizot86,Lewenstein96}.  In the context of classical-field theory, we expect that the erratic evolution of the field results in random excitation of this motion, corresponding to fluctuations of the vortex phase.  As there is no `restoring force' associated with this excitation, over time the phase can drift arbitrarily far from any initial value, invalidating the interpretation of the equilibrium as a configuration in which energy is equipartitioned over normal excitations of the symmetry-broken condensate orbital.  As a consequence, the decomposition of the field into condensate and noncondensate by the PO procedure also deteriorates over time.  This is of course consistent with our expectation that as the averaging time becomes large, the statistics we measure converge to those of the ergodic density of the microcanonical system, which must reflect the rotational symmetry of the Hamiltonian \cite{Lebowitz73}, and is entirely analogous to the diffusion of the condensate phase considered in chapter~\ref{chap:anomalous}, except in this case the Goldstone mode is distinct from the condensate mode itself.
In figure~\ref{fig:temporal_decay}(a) we plot the values of the two largest eigenvalues of the covariance matrix formed by averaging for varying lengths of time, each average starting at $t=9000$ cycles.  For each averaging period we have optimised the condensate fraction over the frequency of frame rotation, which causes only small variation in the optimal frequency ($\lesssim0.003\hbar\omega_r$).  We observe that the largest eigenvalue gets smaller as the averaging period increases, while the second largest eigenvalue gets larger.  Over longer periods (inset to figure~\ref{fig:temporal_decay}(a)) the two eigenvalues approach the values obtained from the density matrix constructed in the laboratory frame (dashed lines), indicating that the orientation of the condensate mode has diffused to such an extent that the density matrix describes a fragmented condensate in all uniformly rotating frames.  In figure~\ref{fig:temporal_decay}(b) we plot the standard deviation of the vortex phases as a function of the averaging period, in a frame rotating at $\Omega=0.2610\omega_r$.  We observe that for the first $\approx75$ trap cycles the standard deviation remains reasonably steady at a small value $\sigma_{\theta_\mathrm{v}}\sim0.05$ radians.  After this the standard deviation begins to increase dramatically, and this is ultimately reflected in the measured eigenvalues (figure~\ref{fig:temporal_decay}(a)).  We note that the standard deviation we have presented is simply that of the distribution of vortex phases measured in a single trajectory, as a function of the length of time over which vortex phases are accumulated.  To properly characterise the rate of rotational diffusion would require the calculation of the variance of the vortex-phase change over an ensemble of similarly prepared classical-field trajectories (cf. \cite{Sinatra08}), a numerically heavy task which we do not pursue here.  We note further that the initial period of comparative stability of the vortex phase is not a generic feature of the system; in general the excursions of the vortex phase relative to its mean precessional motion occur unpredictably.  We note however that the condensate fraction obtained by averaging for only 10 cyc already agrees with the values throughout this period of stability to within $2-3\%$, in agreement with the findings of \cite{Blakie05}.  We therefore adopt the following approach to determining the condensate in the remainder of this chapter: we form the covariance matrix $G_{ij}$ from samples over a period of $T=10$ cyc, in a frame rotating at frequency $\Omega_\mathrm{m}$.  We vary this frequency so that the largest eigenvalue of the corresponding covariance matrix is maximised.  We take the frequency of this optimal frame as an estimate of the angular velocity of the condensate, and the largest eigenvalue as an estimate of the condensate fraction, and similarly for subsidiary quantities obtained from the PO decomposition (see section~\ref{subsec:prec_rotational_properties}).  We do this for 100 consecutive periods of 10 cycles, over the period $t=9000$--$10\,000$ cyc and average the estimates for $\Omega_\mathrm{m}$ and $f_\mathrm{c}\equiv n_0/\sum_k n_k$.  In this way we sample the short-time dynamics which determine the correlations of interest, while exploiting the ergodic character of the classical-field evolution over longer times.  For the simulation with $E=1.10E_\mathrm{g}$, we find from this procedure $\Omega_\mathrm{c}=0.2598\omega_r$ and $f_\mathrm{c}=0.9110$, in reasonably close agreement with the values obtained from a single averaging over a longer time period.
%%%%%%%%%%%%%%%%%%%%%%%%%%%%%%%%%%%%%%%
%%%%%%%%%%%%%%%%%%%%%%%%%%%%%%%%%%%%%%%
We conclude this discussion by considering the interpretation of the condensate fraction we measure, in light of our abandonment of formal ergodic averaging.  We interpret the results of the short-time averaging procedure in terms of a symmetry-broken representation \cite{Castin04} of the many-body system.  The one-body density matrix we calculate corresponds to a particular (mean) orientation of the condensate mode (vortex phase $\theta_\mathrm{v}$).  Labelling this density matrix by $G_{\theta_\mathrm{v}}$, the full one-body density matrix is obtained by integrating over all possible vortex phases $G = (1/N)\int\!d\theta_\mathrm{v}\,G_{\theta_\mathrm{v}}$ \cite{Castin04,Seiringer08}, with $N$ a normalisation factor.  This one-body density matrix does not exhibit condensation in the Penrose-Onsager sense, and this is true of the physical $N$-body system also.  Nevertheless, the individual one-body density matrix $G_{\theta_\mathrm{v}}$ exhibits a condensate, and we interpret the correlations it describes as characterising the behaviour of a corresponding physical system \emph{observed} to have a particular vortex orientation.  With this in mind, we will continue to refer to the largest eigenvalue of the density matrix $G_{\theta_\mathrm{v}}$ constructed from the classical field trajectory as the condensate fraction.
%%%%%%%%%%%%%%%%%%%%%%%%%%%%%%%%%%%%%%%%%%%%%%%%%%%%%%%%%%%%%%%%%%%%%%%%%%%%%%%%%%%%%%%%%%%%%%%%%%%%%%%%%%%%%%%%%%%%%%%%%%%%%%%%%%%%
\section{Rotational properties of the field}\label{subsec:prec_rotational_properties}
A characteristic feature of superfluids is that they resist any attempt to impart a rotation to them, and they acquire angular momentum through the nucleation of vortices\footnote{Condensates formed in dilute Bose gases may possess some angular momentum without containing vortices due to their surface oscillations (see, e.g., \cite{Stringari96}).}, which endow the superfluid with \emph{quantised} flow circulation (see section~\ref{subsec:back_vortices}).  Thus while a classical fluid in equilibrium with a container rotating at angular velocity $\Omega$ satisfies the equality
\begin{equation}
	\langle L_z\rangle = \Omega \langle\Theta_\mathrm{c}\rangle,
\end{equation}
where the classical moment of inertia is defined
\begin{equation}\label{eq:classical_inertia}
	\Theta_\mathrm{c} \equiv mr^2,
\end{equation}
a superfluid in equilibrium with such a container will in general fail to do so.  That is to say, the (quantum) moment of inertia 
\begin{equation}\label{eq:quantum_inertia}
	\Theta\equiv \frac{\langle L_z\rangle}{\Omega},
\end{equation}
will not equal the classical value $\langle\Theta_\mathrm{c}\rangle$ in such a state. We note that for pure superfluids, such as a dilute Bose gas at $T=0$ (which is well described as a ground state of the GP equation), the steady-state angular momentum is typically much less than the classical value determined by its mass distribution \cite{Zambelli01}, and this is true also for condensates containing small numbers of vortices~\cite{Feder01}.  At finite temperatures, we expect the condensed superfluid component of the field to exhibit a suppressed moment of inertia, while the complementary thermal component rotates as a classical fluid.
%%%%%%%%%%%%%%%%%%%%%%%%%%%%%%%%%%%%%%%%%%%%%%%%%%%%%%%%%
\subsection{Condensate and noncondensate observables}
The prescription we have developed for decomposing the classical field into condensed and noncondensed parts allows us to extract quantities which characterise the rotation of the two components individually, by calculating appropriate expectation values.  We recall from section~\ref{subsec:back_2nd_Q} that the extension of a single-body operator $J$ to a second-quantised operator $\hat{J}=\int\!d\mathbf{x}\,\hat{\psi}^\dagger J \hat{\psi}$ on the many-particle Fock space \cite{Fetter71a} has expectation value in the many-body state $\langle\hat{J}\rangle = \mathrm{Tr}\{\hat{\rho}_1J\}$.  We thus define an expectation value of a single-body operator in the classical field analogously by 
\begin{equation}
	\langle J \rangle_\mathrm{c} \equiv  \mathrm{Tr}\{G J\},
\end{equation}
where $G$ is the covariance matrix defined in equation~\reff{eq:covar_mtx}. 
The decomposition into a condensed mode and noncondensed modes introduced in section~\ref{subsec:prec_PO_rotating_frames} allows us to write\footnote{In the short-time averages (section~\ref{subsec:prec_temporal_decoherence}) the Goldstone mode is typically not differentiated from the proper noncondensed modes, as its `occupation' is likely to be small.  We therefore simply regard all modes orthogonal to the condensate mode $\chi_0(\mathbf{x})$ as comprising the noncondensate field here.} (using Dirac notation for vectors in the projected single-particle space)
\begin{equation}
	G = n_0|\chi_0\rangle\langle \chi_0| + \sum_{k>0} n_k |\chi_k\rangle\langle \chi_k| \equiv G_0 + G_\mathrm{th}.
\end{equation}
We thus define averages in the condensate and noncondensate by 
\begin{equation}
	\langle J\rangle_0 = \mathrm{Tr}\{G_0J\} = n_0\sum_m\langle m|\chi_0\rangle\langle \chi_0|J|m\rangle,
\end{equation}
and
\begin{equation} 
	\langle J\rangle_\mathrm{th} = \mathrm{Tr}\{G_\mathrm{th}J\} = \sum_{k>0}n_k\sum_m\langle m|\chi_k\rangle\langle \chi_k|J|m\rangle,
\end{equation}
respectively.
It is clear from the form of $G_0$ and $G_\mathrm{th}$ that they are (proportional to) mutually orthogonal projectors, and so we have the additivity property $\langle J\rangle_0 + \langle J\rangle_\mathrm{th} = \langle J \rangle_\mathrm{c}$.
%%%%%%%%%%%%%%%%%%%%%%%%%%%%%%%%%%%%%%%%%%%%%%%%%%%%%%%%%%%%%%%%%%%%%%%%%%%%%%%%%%%%%%%%
\subsection{Angular momentum of the condensate}
As noted in section~\ref{subsec:prec_PO_rotating_frames}, in the precessing-vortex scenario the condensate determined by our method is a non-axisymmetric state with an off-centre vortex.  Analogous condensate orbitals appear in the zero-temperature GP theory as intermediates between the rotationally invariant $\ell\equiv\langle L_z \rangle/N\hbar =0$ and $\ell=1$ modes, with fractional angular-momentum (per atom) expectation values $0 < \ell < 1$ \cite{Butts99,Papanicolaou05}, and are mechanically unstable (see the discussion in \cite{Komineas05}).  

Applying the above averaging procedure to our simulation with $E=1.10E_\mathrm{g}$ we obtain $L_\mathrm{c} \equiv \langle L_z\rangle_0 / n_0 = 0.63\hbar$.  We recall that the condensate fraction obtained for this state was $f_\mathrm{c}=0.911$.  Turning to the noncondensate we find $L_\mathrm{th}/N_\mathrm{th} = 4.80\hbar$.  While at this energy, only $9\%$ of the atoms have been excited out of the condensate, in order to maintain rotational equilibrium of the system, the thermal atoms carry nearly half of the angular momentum of the field.  This is a consequence of the suppressed moment of inertia of the condensate: in order to have the same angular velocity, the thermal cloud must possess much more angular momentum per particle than the condensate \cite{Guery-Odelin00,Haljan01a}.
%%%%%%%%%%%%%%%%%%%%%%%%%%%%%%%%%%%%%%%%%%%%%%%%%%%%%%%%%%%%%%%%%%%%%%%%%%%%%%%%%%%%%%%%
\subsection{Angular velocity of the thermal cloud}\label{subsubsec:prec_cloud_rotation}
Now we consider the angular velocity of the thermal cloud.  The procedure for separating the condensate and thermal components developed in this chapter allows us analyse the rotation and inertia of both of these two components individually.  In particular our approach includes the contribution of thermal fluctuations that traverse the central condensed region, as well as those which fill the vortex core, to the total thermal component of the field.  To estimate the cloud rotation rate, we assume the expectation value of the classical moment of inertia equation~\reff{eq:classical_inertia} evaluated in the thermal cloud as an estimate of the cloud's true moment of inertia, and thus obtain 
\begin{equation}
	\Omega_\mathrm{th} \approx \frac{\langle L_z \rangle_\mathrm{th}}{\langle \Theta_\mathrm{c} \rangle_\mathrm{th}},
\end{equation}
In this way we find $\Omega_\mathrm{th} = 0.262\omega_r$,  in fair agreement with the precession frequency of the vortex.  
%%%%%%%%%%%%%%%%%%%%%%%%%%%%%%%%%%%%%%%%%%%%%%%%%%%%%%%%%%%%%%%%%%%%%%%%%%%%%%%%%%%%%%%%%%%%%%%%%%%%%%%%%%%%%%%%%%%%%%%%%%%%%%%%%%%%
\section{Dependence on internal energy}\label{sec:prec_Energy_dependence}
We now consider the effect of varying the energy of the classical field on its equilibrium properties.  We expect quite generically for Hamiltonian classical-field simulations such as those performed here that an increase in the classical-field energy will result in an increase in the field temperature and a suppression of the condensate fraction \cite{Blakie05}.  In this scenario with finite conserved angular momentum, we also expect the rotational properties of the system to depend strongly on the field energy.
%%%%%%%%%%%%%%%%%%%%%%%%%%%%%%%%%%%%%%%%%%%%%%%%%%%%%%%%%%%%%%%%%%%%%%%%%%%%%%%%%%%%%%%%
\subsection{Density}
\begin{figure}
	\begin{center}
	\includegraphics[width=0.9\textwidth]{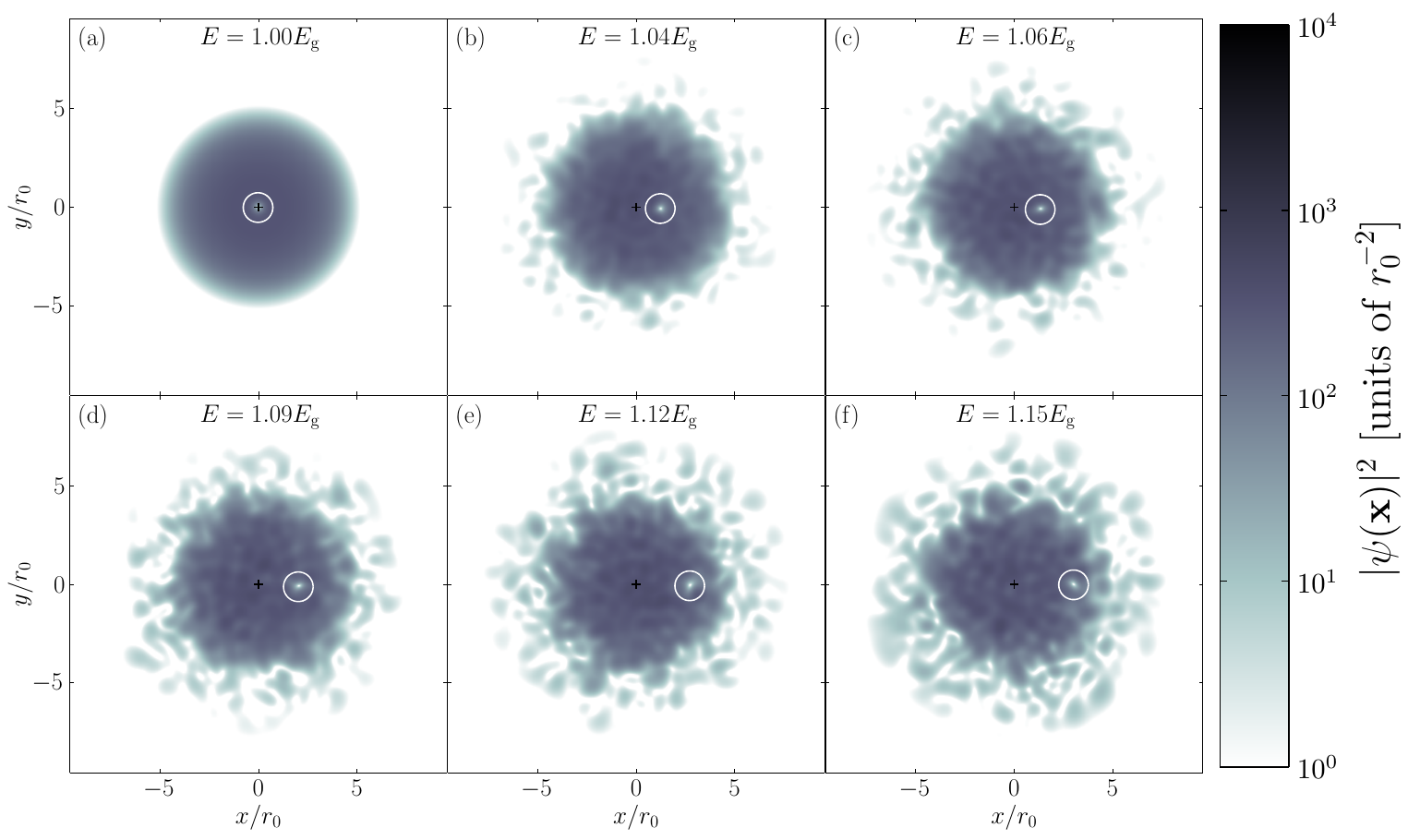}
	\caption{\label{fig:density_vary_E}  (a-f) Representative densities of equilibrium classical fields of different energies. The white circle indicates the vortex position, and $+$ marks the coordinate origin (trap axis).} 
	\end{center}
\end{figure}

The most obvious and generic consequence of increasing the energy of the microcanonical system is an increase in the entropy of the system.  The qualitative effects of this increase can be readily observed in position-space representations of the classical field density.  In figures~\ref{fig:density_vary_E}(a-f) we plot representative densities of the classical field at equilibrium, for various values of the field energy.  
%%%%%%%%%%%%%%%%%%%%%%%%%%%%%%%%%%%%%%%
%%%%%%%%%%%%%%%%%%%%%%%%%%%%%%%%%%%%%%%
These densities are chosen at times in the range $t\in[9000,10\,000]$ cyc such that the vortex is displaced along the $x$ axis, for ease of comparison. We clearly observe from these images an increase in the surface excitations of the condensate and the development of a turbulent outer cloud as the energy is increased.  The images also reveal an increase in the vortex precession radius as the energy is increased.   
%%%%%%%%%%%%%%%%%%%%%%%%%%%%%%%%%%%%%%%%%%%%%%%%%%%%%%%%%%%%%%%%%%%%%%%%%%%%%%%%%%%%%%%%
\subsection{Condensate fraction and angular momentum}\label{subsec:prec_cfrac_and_angmom}
\begin{figure}
	\begin{center}
	\includegraphics[width=1.0\textwidth]{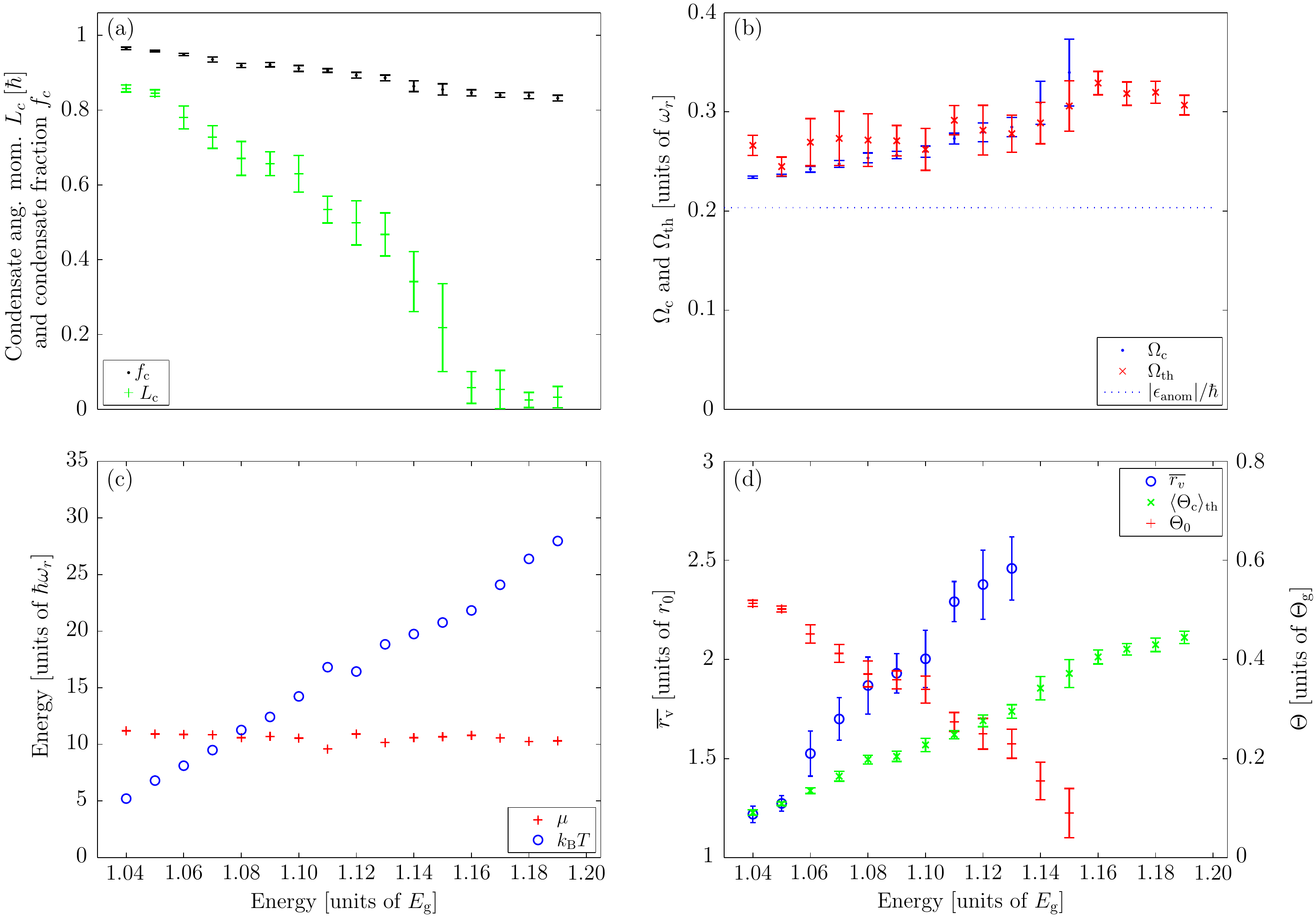}
	\caption{\label{fig:fns_of_energy}  Dependence of equilibrium parameters on field energy. (a) Condensate fraction and condensate angular momentum. (b) Angular velocities of the condensate and thermal cloud. (c) Temperature and chemical potential. (d) Vortex precession radius, classical moment of inertia of the thermal cloud, and nonclassical moment of inertia of the condensate mode.  Error bars indicate the statistical spread in values calculated (see main text).}
	\end{center}
\end{figure}

Using the procedure introduced in section~\ref{subsec:prec_temporal_decoherence}, we find the condensate fraction (figure~\ref{fig:fns_of_energy}(a)) drops gradually from $f_\mathrm{c}=0.96$ to $f_\mathrm{c}=0.83$ over the energy range $E\in[1.04,1.19]E_\mathrm{g}$.  
%%%%%%%%%%%%%%%%%%%%%%%%%%%%%%%%%%%%%%%
%%%%%%%%%%%%%%%%%%%%%%%%%%%%%%%%%%%%%%%
However, the angular momentum (per particle) of the condensate drops dramatically, falling from $L_\mathrm{c}=0.86\hbar$ at $E=1.04E_\mathrm{g}$ to nearly zero by $E=1.16E_\mathrm{g}$.  This comparatively severe drop in angular momentum occurs because we are varying the energy of the field at fixed angular momentum; varying the energy at fixed angular velocity we would expect the drop in angular momentum to be commensurate with the depletion of atoms from the condensate.

At higher energies the condensate is vortex-free and thus irrotational, and exists in thermal and diffusive equilibrium with the thermal cloud which now contains all the angular momentum of the field.  The small residual angular momentum of these high-energy condensates shown in figure~\ref{fig:fns_of_energy}(a) results from surface excitations which are not differentiated from the condensate by the short-time averaging.  We note however that the condensate fraction is essentially continuous across this transition from a symmetry-broken condensate orbital to a non-symmetry-broken (i.e., vortex-free) one, supporting the definition introduced in section~\ref{subsec:prec_temporal_decoherence} as an appropriate measure of condensation in the symmetry-broken regime.
%%%%%%%%%%%%%%%%%%%%%%%%%%%%%%%%%%%%%%%%%%%%%%%%%%%%%%%%%%%%%%%%%%%%%%%%%%%%%%%%%%%%%%%%
\subsection{Condensate and cloud rotation rates}\label{subsec:prec_rotation_rates}
In figure~\ref{fig:fns_of_energy}(b) we plot the condensate rotation frequency, as determined by the frame frequency of optimal condensate resolution (section~\ref{subsec:prec_PO_rotating_frames}).  We find that the rotation frequency increases gradually from $\Omega_\mathrm{c}=0.234\omega_r$ at $E=1.04E_\mathrm{g}$, up to $\Omega_\mathrm{c}=0.340\omega_r$ at $E=1.15E_\mathrm{g}$.  For comparison the frequency of the anomalous mode of the ground vortex state in the Bogoliubov approximation, $|\epsilon_\mathrm{anom}|/\hbar=0.2034\omega_r$, is plotted as a dotted line in figure~\ref{fig:fns_of_energy}(b).  Note the angular velocity of the condensate increases with \emph{decreasing} condensate angular momentum, in agreement with the well-known thermodynamically anomalous behaviour of the zero-temperature (Gross-Pitaevskii) off-axis vortex state \cite{Butts99,Papanicolaou05,Komineas05}.
We note also that the case $E=1.15E_\mathrm{g}$ is very close to the transition from a vortical condensate to a vortex-free one; in $\sim10\%$ of the sampling periods (each of $10$ cyc) the condensate is vortex-free, resulting in maximal values of $n_0(\Omega_\mathrm{m})$ occurring around\footnote{Where the condensate is vortex-free, we expect that the calculated condensate fraction should be independent of the frame in which the covariance matrix is constructed.  However, the condensate fraction often exhibits a small rise around $\Omega_\mathrm{m}=1\omega_r$, which we associate with a persistent dipole excitation of the field.} $\Omega_\mathrm{m}=1\omega_r$.  We discard these values in calculating the mean and standard deviation presented for this simulation in figure~\ref{fig:fns_of_energy}(b).   From figure~\ref{fig:fns_of_energy}(a) we observe that the condensate modes with $E\geq1.16E_\mathrm{g}$ are essentially irrotational, and indeed the condensates in these simulations appear to be largely vortex-free, the majority of samples yielding $\Omega_\mathrm{c}\approx1\omega_r$, with a small minority producing $\Omega_\mathrm{c}\sim0.4\omega_r$, resulting from the transit of short-wavelength surface excitations (so-called \emph{ghost} vortices) which produce a slightly greater calculated condensate fraction in a frame which matches their motion. We thus omit all estimates of the condensate rotation rate for energies $E\geq1.16E_\mathrm{g}$ from figure~\ref{fig:fns_of_energy}(b).    
 
Turning our attention to the cloud rotation frequency (crosses in figure~\ref{fig:fns_of_energy}(b)), determined by the prescription of section~\ref{subsubsec:prec_cloud_rotation},  we find these results in qualitative agreement with the measured condensate rotation frequency over the range $E\in[1.04,1.15]E_\mathrm{g}$.  The bars accompanying the markers for condensate and cloud rotation rates simply represent the statistical dispersion in the measured quantities as discussed in section~\ref{subsec:prec_temporal_decoherence}, and \emph{not} quantitative measures of the errors in these measurements, but it is clear that the measured cloud rotation rate does not match the condensate rotation rate closely.  It is possible that this is associated with pair correlations in the noncondensate, though if anything we expect the presence of pair correlations to suppress the moment of inertia relative to its classical value (cf. \cite{Sorensen73}).  We note finally that the cloud frequency reaches a maximum of $\Omega_\mathrm{th}\approx0.33\omega_r$ at $E=1.16E_\mathrm{g}$ before decreasing as the energy is increased further.  This reduction of $\Omega_\mathrm{th}$ with increasing energy is easily understood:  once the condensate becomes vortex-free (irrotational), as further atoms are depleted from the condensate the thermal cloud population increases while its angular momentum remains fixed, implying a decrease in angular velocity.
%%%%%%%%%%%%%%%%%%%%%%%%%%%%%%%%%%%%%%%%%%%%%%%%%%%%%%%%%%%%%%%%%%%%%%%%%%%%%%%%%%%%%%%%
\subsection{Temperature and chemical potential}\label{subsec:prec_T_and_mu}
We now estimate the thermodynamic parameters (temperature $T$ and chemical potential $\mu$) of the classical field.  Our approach is to construct a fitting function (parameterised by $\mu$ and $T$) and to find the thermal parameters for which it most closely approximates the (time-averaged) density of the classical field.
In general, we expect the density of the field to be well described by a semiclassical Hartree-Fock density profile \cite{Giorgini97}, appropriately modified to the classical-field case.  In such a semiclassical representation, the rotation of the thermal cloud enters simply as a rotational dilation of the trapping potential experienced by the thermal atoms \cite{Stringari99}, which then takes the form $V_\mathrm{rot}(\x) = m(\omega_r^2-\Omega_\mathrm{th}^2)r^2/2$.  However, we would like to avoid performing an iterative calculation to self-consistently solve for the mutually repelling condensed and noncondensed fractions of the field.  A simpler approach to constructing the density of the thermal Bose field is presented by the authors of \cite{Naraschewski98}, in which the mean-field potential due to the noncondensed atoms is neglected.  The thermal cloud is thus constructed as a noninteracting thermal Bose gas which `sits' in the combined potential due to the trap and the mean-field repulsion of the condensate, as in the two-gas model of \cite{Dodd98}.
However, a straightforward application of this model to our 2D system leads to divergences, as can be seen by considering the density distribution of noncondensed bosonic atoms in the \emph{full} self-consistent semiclassical Hartree-Fock model, which in two dimensions becomes (cf. equation~(2) of reference~\cite{Naraschewski98})
\begin{equation}\label{eq:Naraschewski_ntherm}
	n_\mathrm{th}(\mathbf{x}) = \frac{1}{\lambda_\mathrm{dB}^2} g_1\left(e^{-\left[V_\mathrm{rot}(\mathbf{x}) + 2U_\mathrm{2D} (n_\mathrm{c}(\mathbf{x}) + n_\mathrm{th}(\mathbf{x})) - \mu\right]/k_BT}\right)\!,
\end{equation}
where we have introduced the de Broglie wavelength $\lambda_\mathrm{dB} \equiv \sqrt{2\pi\hbar^2/mk_\mathrm{B}T}$ \cite{Pethick02} and the Bose function $g_\nu(z)\equiv\sum_{j=1}^\infty z^j/j^\nu$~\cite{Pathria96}.  Dropping the mean field potential $2U_\mathrm{2D}n_\mathrm{th}(\x)$ due to noncondensed atoms from equation~\reff{eq:Naraschewski_ntherm} leaves us with a local density approximation for the noncondensed atoms in the effective potential $V_\mathrm{eff}(\x) = V_\mathrm{rot}(\x) + 2U_\mathrm{2D}n_\mathrm{c}(\x)$.  As is well known, the local density approximation for the 2D Bose gas yields a divergent atomic density in the limit $\mu \to V_\mathrm{eff}(\x)$, reflecting the absence of thermal-cloud saturation in the homogeneous 2D Bose gas \cite{Hadzibabic10}, due to the logarithmic divergence of the Bose function $g_1(z)$ as $z\rightarrow 1$ \cite{Pathria96}. 

We are thus led to consider a third approach: we develop a fitting function for the noncondensed density $n_\mathrm{th}(\mathbf{x})$ only, which therefore depends only on the \emph{total} field density $n(\mathbf{x})$, which we measure from the classical-field trajectory.  The mean-field self-interaction of the noncondensed atoms then `lifts' the bottom of the effective potential above the chemical potential of these atoms, regularising the divergent behaviour and allowing a solution \cite{Hadzibabic10}.
This has the added benefit that we do not need to distinguish between the condensed and noncondensed components of the field in order to construct the appropriate mean-field potential\footnote{This will be of particular importance in chapter~\ref{chap:stirring} where we consider a disordered vortex-liquid state, in which the central quasi-coherent bulk of the field is not `condensed'.}. 

Our approach to determining $\mu$ and $T$ is thus as follows: 
we assume the noncondensate atoms in the condensate band to be well described by a one-body Wigner distribution, which, for a two-dimensional (Bose) field $\hat{\phi}(\x)$ takes the form
\begin{equation}
	F(\mathbf{x},\mathbf{p}) \equiv \frac{1}{h^2} \int\!d\mathbf{y}\,\left\langle \hat{\phi}^\dagger\left(\mathbf{x}+\frac{\mathbf{y}}{2}\right)\hat{\phi}\left(\mathbf{x}-\frac{\mathbf{y}}{2}\right)\right\rangle e^{i\mathbf{p}\cdot\mathbf{y}/\hbar},
\end{equation}
and we further assume for this Wigner function the approximate semiclassical form \cite{Castin01,Bagnato87}
\begin{equation}\label{eq:semiclassical_Wigner_dist}
	F(\mathbf{x},\mathbf{p}) = \frac{1}{h^2}\frac{1}{\exp[(\epsilon(\mathbf{x},\mathbf{p}) - \mu)/k_\mathrm{B}T] - 1}.
\end{equation} 
The Wigner distribution is most simply constructed in the frame co-rotating with the thermal cloud (with angular velocity $\Omega_\mathrm{th}$), where the semiclassical energy is\footnote{We assume here that the Hartree-Fock potential is invariant under this rotation, as is the case for the azimuthally averaged potential we construct below.}
\begin{equation}
	\epsilon(\mathbf{x},\mathbf{P}) = \frac{\mathbf{P}^2}{2m} + V_\mathrm{eff}^\mathrm{HF}(\mathbf{x}),
\end{equation}
with $\mathbf{P}$ the kinematic momentum in this frame (see appendix~\ref{app:fitting_function}), and the Hartree-Fock effective potential in the frame of the thermal field is thus
\begin{equation}
	V^{\mathrm{HF}}_\mathrm{eff}(\mathbf{x}) = m\left(\omega_r^2 - \Omega_\mathrm{th}^2\right)\frac{r^2}{2} +  2U_\mathrm{2D} n(\mathbf{x}),
\end{equation}
with the total field density the sum of the condensate and noncondensate (thermal) contributions $n(\mathbf{x})=n_\mathrm{c}(\mathbf{x}) + n_\mathrm{th}(\mathbf{x})$.
The factor of two here is the strength of the interaction between pairs of atoms when at least one of the pair is noncondensed \cite{Giorgini97,Naraschewski98,Castin01}, i.e., this is the correct mean-field potential experienced by the \emph{thermal} atoms only.
The classical-field limit of equation~(\ref{eq:semiclassical_Wigner_dist}) is obtained when $F(\mathbf{x},\mathbf{p}) \gg 1/h^2$, which occurs when the parameter $(\epsilon(\mathbf{x},\mathbf{p}) -\mu)/k_\mathrm{B}T \ll 1$, and is given by its first order expansion in this parameter (cf. equation~\reff{eq:cfield_equipartition}),
\begin{equation}\label{eq:classical_limit_dist}
	F_\mathrm{c}(\mathbf{x},\mathbf{p}) = \frac{1}{h^2}\frac{k_\mathrm{B}T}{\epsilon(\mathbf{x},\mathbf{p}) - \mu}.
\end{equation}
It is important to note that although equation~(\ref{eq:classical_limit_dist}) is obtained here as the high occupation approximation to the bosonic distribution equation~(\ref{eq:semiclassical_Wigner_dist}), (describing the classical equipartitioning of energy over phase-space cells of volume $h^2$), we expect it to apply to the equilibrium state of a weakly interacting classical-field system even if the level occupations are $\lesssim 1$ \cite{Sinatra02,Lobo04}.  The spatial density for the fit is then obtained by integrating equation~\reff{eq:classical_limit_dist} over the momentum $\mathbf{p}$, up to the \emph{local} momentum cutoff $P_R(\mathbf{x})$ corresponding to the harmonic-basis cutoff defined by the projector $\mathcal{P}$ (see section~\ref{subsec:num_phase_space_covering}).  This is complicated in the present case by the fact that projector angular velocity $\Omega_\mathrm{p}$ is distinct from the rotation rate of the thermal cloud $\Omega_\mathrm{th}$, and so we relegate the details of the calculation to appendix~\ref{app:fitting_function}, where we derive the form of the fitting function 
\begin{equation}\label{eq:fitting_function}
n(r;\mu,T) = \frac{1}{\lambda_\mathrm{dB}^2} \Bigl[I_1(r;\mu,\Omega_\mathrm{th},\Omega_\mathrm{p}) + I_2(r;\mu,\Omega_\mathrm{th},\Omega_\mathrm{p})\Bigr].
\end{equation}

\begin{figure}
	\begin{center}
	\includegraphics[width=0.65\textwidth]{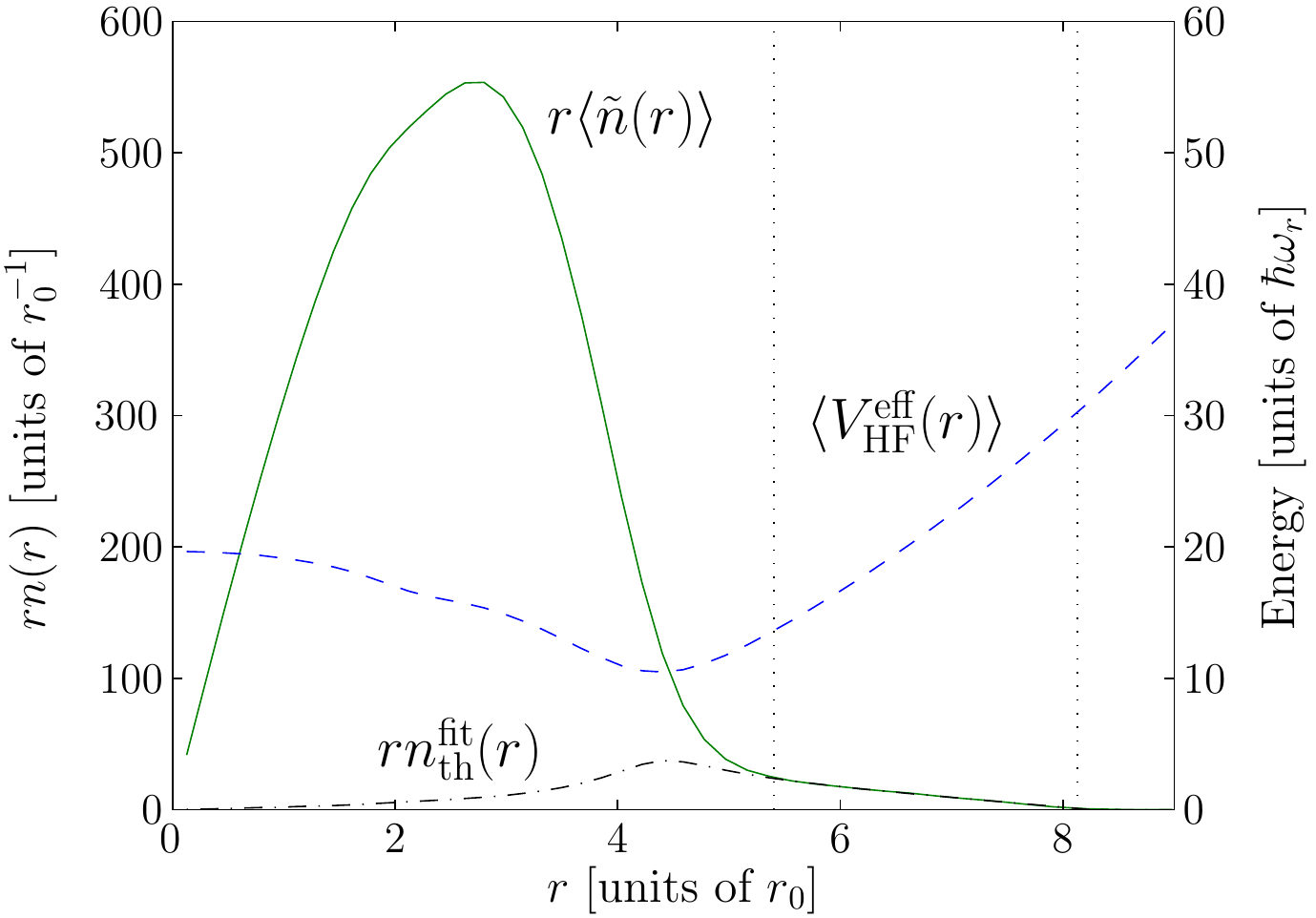}
	\caption{\label{fig:temp_fit} Fitting procedure for the thermodynamic parameters.  The solid line illustrates the averaged density of the field times radius.  The dashed line indicates the effective potential used in the fit, and the dot-dashed line indicates the fit to the wing of the classical-field density distribution (times radius).   The vertical dotted lines indicate the limits of the domain over which the fit is performed.}
	\end{center}
\end{figure}

This expression constitutes an appropriate fitting function for the noncondensed component of the classical field, which requires no explicit knowledge of the distribution of condensed atoms.  To obtain an estimate of $\mu$ and $T$ for our system, we sample the azimuthally averaged field density $\widetilde{n}(r)\equiv\frac{1}{2\pi}\int_0^{2\pi}\!d\theta\,n(r,\theta)$ at $N_t$ equally spaced times over some period $t\in[t_0,t_0+T]$, and average over the sample times to obtain $\langle \widetilde{n}(r)\rangle \equiv \sum_{i=1}^{N_t} \widetilde{n}(r,t_i)/N_t$.  From this we construct the effective potential

\begin{equation}\label{eq:prec_eff_potl}
	\left\langle V_\mathrm{HF}^\mathrm{eff}(r) \right\rangle = m\left(\omega_r^2-\Omega^2\right)\frac{r^2}{2} + 2U_\mathrm{2D}\left\langle\widetilde{n}(r)\right\rangle,
\end{equation}
which enters into the fitting function~\reff{eq:fitting_function} through the factors $I_1$ and $I_2$ (see appendix~\ref{app:fitting_function}).
As our fitting function represents only the noncondensed component of the field, we are obliged to avoid including significant condensate density in our fit to $\langle \widetilde{n}(r)\rangle$, and to therefore restrict our fit to a domain $r\in[r_-,r_\mathrm{tp}]$, where $r_-$ is the location of the minimum of the effective potential, which we expect to approximately mark the condensate boundary, and $r_\mathrm{tp}=\sqrt{2E_R/m(\omega_r^2-\Omega_\mathrm{p}^2)}$ is the semiclassical turning point of the condensate band in the frame of the projector.  In practice the low temperatures of the classical-field configurations we consider make performing such a fit difficult; in particular the fit fails close to $r_-$, in the region of strongly collective excitations of the condensate.  We therefore perform our fits only to the wing of the classical-field density; for definiteness we take $r\in[r_-+1r_0,r_\mathrm{tp}]$.  We take the values $\Omega_\mathrm{th}$ calculated in section~\ref{subsec:prec_rotation_rates} as estimates for the cloud rotation rate in equation~(\ref{eq:fitting_function}).  Although these values may be somewhat inaccurate as discussed in section~\ref{subsec:prec_rotation_rates}, we find that using the vortex precession frequencies (where available) makes little difference to the obtained values of $\mu$ and $T$, and conclude that any discrepancy in $\Omega_\mathrm{th}$ is inconsequential at the level of accuracy of the semiclassical estimates we seek.  An example of such a fit is shown in figure~\ref{fig:temp_fit}.  

The results of this fitting procedure are presented for all field energies $E$ we consider in figure~\ref{fig:fns_of_energy}(c).
%%%%%%%%%%%%%%%%%%%%%%%%%%%%%%%%%%%%%%%
%%%%%%%%%%%%%%%%%%%%%%%%%%%%%%%%%%%%%%%
We observe that the temperature estimate $k_\mathrm{B}T$ increases approximately linearly with the energy $E$, and is of the same order of magnitude as the chemical potential $\mu$.  We note that the chemical-potential estimates are generally higher than the chemical potential of the ground state $\mu_\mathrm{g}=10.35\hbar\omega_r$, whereas we expect the chemical potential of the strictly Hamiltonian PGPE system to decrease with increasing energy (see section~\ref{subsec:phase_freq}).  This suggests that the chemical potential is probably overestimated by the fitting, and some error in the temperature estimate is also expected.  Nevertheless the results suggest that our simulations probe the temperature regime $k_\mathrm{B}T\gtrsim\mu$ in which the thermal friction on the vortex is expected to be appreciable \cite{Fedichev99}, consistent with our observations.
%%%%%%%%%%%%%%%%%%%%%%%%%%%%%%%%%%%%%%%%%%%%%%%%%%%%%%%%%%%%%%%%%%%%%%%%%%%%%%%%%%%%%%%%
\subsection{Moment of inertia and vortex displacement}
In figure~\ref{fig:fns_of_energy}(d) we plot the classical moment of inertia of the noncondensate (crosses), $\langle \Theta_\mathrm{c} \rangle_\mathrm{th}$ (see section~\ref{subsubsec:prec_cloud_rotation}), in units of the classical moment of inertia $\Theta_\mathrm{g}=7.65\times10^4\hbar/\omega_r$ of the principal ground state.  We find that this quantity increases steadily with the field energy as atoms are excited out of the condensate into the thermal cloud.  We also plot the (quantum) moment of inertia of the condensate orbital (plusses), which we define as $\Theta_0\equiv\langle L_z \rangle_0/\Omega_\mathrm{c}$, where the angular momentum $\langle L_z\rangle_0$ and rotation frequency $\Omega_\mathrm{c}$ of the condensate are those discussed in the previous sections.  To give an indication of the variation of estimates in this derived quantity, we include the standard deviation estimated by adding the standard deviations of $\langle L_z\rangle_0$ and $\Omega_\mathrm{c}$ in quadrature\footnote{The variations in these quantities are of course not independent; however, the contribution of the variation in estimates of $\langle L_z\rangle_0$ dominates to such an extent that this is of no concern for the indicative representation we seek here.}.  We note that the moment of inertia of the condensate decays steadily with increasing $E$, following $\langle L_z \rangle_0$, and by $E=1.15E_\mathrm{g}$ is essentially zero.  Above this energy the condensate's angular momentum is basically zero (see the remarks in section~\ref{subsec:prec_cfrac_and_angmom}), i.e. the condensate fails to respond to the rotation imposed by the thermal cloud, and its moment of inertia is thus zero.  

On this plot we also include the displacement of the vortex from the trap centre (circles).  For consistency with the other measurements we calculate this by tracking vortex trajectories over 100 successive intervals of 10 cyc (starting from $t=9000$ cyc) and averaging the measured radii.  Bars on the marker again indicate standard deviations of the estimates over the 100 periods.  The precession radius increases steadily as the energy is increased.  By $E=1.13E_\mathrm{g}$, the precessing vortex is close to the edge of the condensate, and it becomes difficult to track this single vortex in proximity to the proliferation of phase defects undergoing constant creation and annihilation in the condensate periphery.  We therefore do not attempt to include data for these higher energies on the plot.  It is clear however that the vortex displacement increases as the condensate angular momentum decreases, as is the case in the zero-temperature GP theory \cite{Butts99,Ballagh99,Vorov05}.  
%%%%%%%%%%%%%%%%%%%%%%%%%%%%%%%%%%%%%%%%%%%%%%%%%%%%%%%%%%%%%%%%%%%%%%%%%%%%%%%%%%%%%%%%%%%%%%%%%%%%%%%%%%%%%%%%%%%%%%%%%%%%%%%%%%%%
\section{Dependence on frame of projector}\label{sec:prec_projector_frame}
We now consider the effect of the angular velocity $\Omega_\mathrm{p}$ of the frame in which the projector is defined on the equilibrium configurations of the classical-field trajectories.  In irrotational scenarios in which the projector is applied in a laboratory frame, the cutoff energy has a significant effect on the equilibrium attained for given conserved first integrals (energy and normalisation), as discussed in section~\ref{subsec:cfield_thermo}.  In the present case, the cutoff is defined by two parameters: the angular velocity of the frame in which the cutoff is effected, and the (rotating-frame) energy at which the cut is made.  As noted in section~\ref{subsubsec:prec_Choice_of_frame}, the (thermodynamic) angular velocity of the classical field at equilibrium is in general not equal to that of the projector.  Here we consider the effect of the choice of projector angular velocity on the classical-field equilibrium.  We vary the projector frequency over the range $\Omega_p\in[0,0.35]\omega_r$, and keep the cutoff energy (in the rotating frame of the projector) fixed at $E_R = 30\hbar\omega_r$.  There are several consequences of varying the angular velocity $\Omega_\mathrm{p}$.  First, the multiplicity $\mathcal{M}$ (i.e., the number of canonical-coordinate pairs in the Hamiltonian system; see section~\ref{subsec:cfield_proj_cft}) depends strongly on $\Omega_\mathrm{p}$, increasing with increasing $\Omega_\mathrm{p}$ (see section~\ref{subsec:quad_app_lag_nl}).  Moreover, the modal composition of the coherent region $\mathbf{C}$ varies with $\Omega_\mathrm{p}$, with $\mathbf{C}$ becoming increasingly biased towards single-particle modes with $l>0$ as $\Omega_\mathrm{p}$ is increased.  To illustrate the effect varying $\Omega_\mathrm{p}$ has on the classical-field equilibria, we have carried out simulations with $E[\psi]=1.10E_\mathrm{g}$, $E_R=30\hbar\omega_r$, $L[\psi]=N_0\hbar$, and $\Omega_\mathrm{p}\in[0,0.35]\omega_r$.
%%%%%%%%%%%%%%%%%%%%%%%%%%%%%%%%%%%%%%%
\begin{figure}
	\begin{center}
	\includegraphics[width=0.60\textwidth]{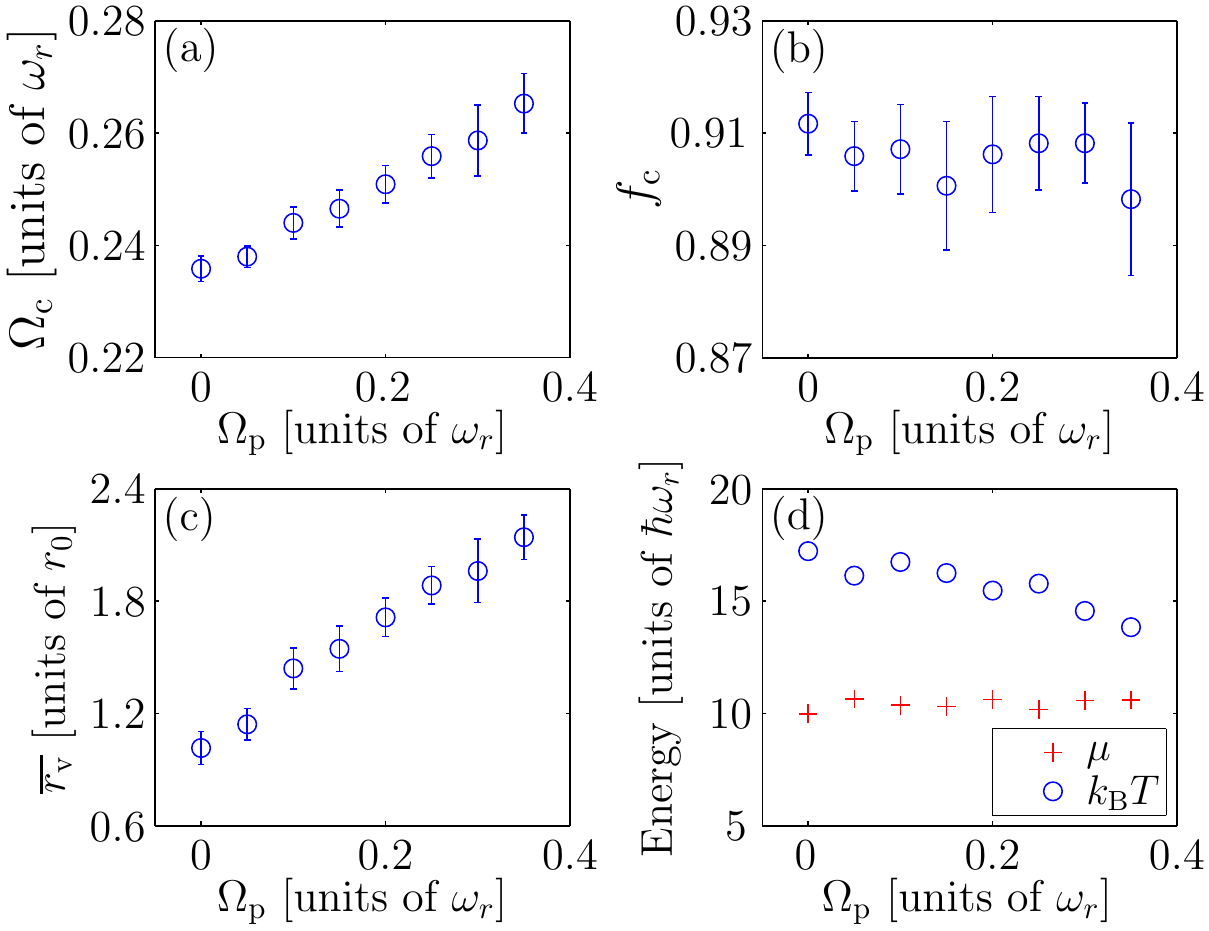}
	\caption{\label{fig:vary_w}  Dependence of equilibrium system parameters on the frame in which the projection is effected. (a) Angular velocity of the condensate, (b) condensate fraction, (c) vortex precession radius, (d) temperature and chemical potential.}
	\end{center}
\end{figure}
%%%%%%%%%%%%%%%%%%%%%%%%%%%%%%%%%%%%%%%
We find that all these simulations exhibit qualitatively similar equilibrium states, each containing a single precessing vortex, as for the simulation with $E=1.10E_\mathrm{g}$ discussed in the main text.  We present the key results in figures~\ref{fig:vary_w}(a-d): in figure~\ref{fig:vary_w}(a) we plot the angular rotation frequency of the condensate mode, as determined by the procedure of section~\ref{subsec:prec_temporal_decoherence}.  The rotation frequency exhibits a clear upward trend as the frame frequency is increased, nevertheless the dependence of the rotation rate on frame frequency is weak, varying by only $\sim0.03\omega_r$ over the range of $\Omega_\mathrm{p}$ considered.  In figure~\ref{fig:vary_w}(b) we plot the corresponding condensate fractions.  Again the dependence on the frame frequency is weak; any trend in the condensate fraction is of the same magnitude as the statistical dispersion of the measured values.  In figure~\ref{fig:vary_w}(c) we plot the vortex precession radii.  We see that the radius increases steadily, approximately doubling between $\Omega_\mathrm{p}=0$ and $\Omega_\mathrm{p}=0.35\omega_r$.  
This is the only quantity which changes significantly over the range of projector angular velocities $\Omega_\mathrm{p}$ we consider, and as in figures~\ref{fig:fns_of_energy}(b,\,d) we find that a small proportional change in precession frequency $\Omega_\mathrm{c}$ corresponds to a much greater proportional change in $\overline{r_\mathrm{v}}$.  
In fact for the simulations in figures~\ref{fig:vary_w}(a,\,c) with the smallest precession frequencies, the precession radius $\overline{r_\mathrm{v}}$ is both smaller and more strongly dependent on $\omega_\mathrm{v}$ than for the condensates in figures~\ref{fig:fns_of_energy}(b,\,d) with comparable vortex precession frequencies, which have greater condensate fractions.  This is in agreement with the results of the Gross-Pitaevskii equation for zero-temperature precessing single-vortex configurations, which show that the frequency of vortex precession in less strongly interacting condensates varies less severely with the precession radius of the vortex \cite{Papanicolaou05}\footnote{Indeed in the `lowest Landau level' limit of very weakly interacting condensates the vortex precession frequency is \emph{independent} of the precession radius \cite{Papanicolaou05,Vorov05}.}.

In figure~\ref{fig:vary_w}(d) we plot the dependence of the chemical potential (plusses) and temperature (circles) on $\Omega_\mathrm{p}$, calculated assuming the condensate angular frequency $\Omega_\mathrm{c}$ as an estimate for the cloud rotation rate.  We observe that the temperature decreases as the frame frequency is increased, consistent with the increase in the number of modes in the condensate band (see discussion of equipartition in section~\ref{subsec:cfield_thermo}).  We conclude then that increasing the frame angular velocity $\Omega_\mathrm{p}$ results in a lowering of the equilibrium temperature, and biases the angular-momentum balance in the system towards the thermal cloud, leading to a higher equilibrium angular velocity.
We emphasise that this dependence is a natural consequence of the projector definition: different values of $\Omega_\mathrm{p}$ define different Hamiltonian systems.  Ultimately, this dependence would be removed by the inclusion of above-cutoff effects \cite{Gardiner03,Davis06,Blakie08}.
%%%%%%%%%%%%%%%%%%%%%%%%%%%%%%%%%%%%%%%%%%%%%%%%%%%%%%%%%%%%%%%%%%%%%%%%%%%%%%%%%%%%%%%%%%%%%%%%%%%%%%%%%%%%%%%%%%%%%%%%%%%%%%%%%%%%
\section{Summary}\label{sec:prec_Conclusions}
\begin{sloppypar}
We have carried out classical-field simulations of a precessing vortex in a finite-temperature Bose-Einstein condensate.  Employing a microcanonical ergodic-evolution approach, we find that randomised states of the classical field constrained to have unit (conserved) angular momentum per atom settle to configurations each containing at most a single vortex precessing under the influence of its own induced velocity field, at thermal and rotational equilibrium with the thermal cloud, so that the mutual friction force between the vortex and the cloud vanishes.
\end{sloppypar}

As the classical field condenses into a rotationally symmetry-broken state, a naive application of the Penrose-Onsager test of one-body coherence in an inertial frame yields a fragmented condensate comprised of a set of coherent angular-momentum eigenmodes.  By transforming the classical field to an appropriate rotating frame, we eliminate the differential phase rotations between the angular momentum components which cause the apparent fragmentation. In this frame we find that the classical field describes a condensate mode with an off-axis vortex coexisting with a core-filling thermal component, and we also observe the Goldstone mode associated with the rotational symmetry breaking.

Due to the ergodic nature of the classical-field evolution, the rotational phase of the vortex diffuses over time.  Thus care must be taken to separate the short-time fluctuation dynamics which define the coherence of the field from the dynamics on much longer time scales, over which ergodic migration of the field configuration occurs.  We found that extracting the condensate by averaging over short times and averaging over successive estimates so obtained allowed us to access the short-time dynamics of interest while exploiting the ergodic nature of the system on longer time scales to better estimate system observables.

We showed that the angular momentum of the condensed mode decreases dramatically as the energy (and thus temperature) are increased at fixed total angular momentum, and correspondingly the vortex precessional radius and frequency increase until the vortex leaves the condensate, and the condensate becomes irrotational.  We proposed a measure of the cloud angular velocity based on the decomposition of the field into condensed and noncondensed components and found qualitative agreement with the vortex precession frequency.
%%%%%%%%%%%%%%%%%%%%%%%%%%%%%%%%%%%%%%%%%%%%%%%%%%%%%%%%%%%%%%%%%%%%%%%%%%%%%%%%%%%%%%%%%%%%%%%%%%%%%%%%%%%%%%%%%%%%%%%%%%%%%%%%%%%%
%%%%%%%%%%%%%%%%%%%%%%%%%%%%%%%%%%%%%%%%%%%%%%%%%%%%%%%%%%%%%%%%%%%%%%%%%%%%%%%%%%%%%%%%%%%%%%%%%%%%%%%%%%%%%%%%%%%%%%%%%%%%%%%%%%%%

\chapter{Nonequilibrium dynamics of vortex arrest}
\label{chap:arrest}
%%%%%%%%%%%%%%%%%%%%%%%%%%%%%%%%%%%%%%%%%%%%%%%%%%%%%%%%%%%%%%%%%%%%%%%%%%%%%%%%%%%%%%%%%%%%%%%%%%%%%%%%%%%%%%%%%%%%%%%%%%%%%%%%%%%%
%%%%%%%%%%%%%%%%%%%%%%%%%%%%%%%%%%%%%%%%%%%%%%%%%%%%%%%%%%%%%%%%%%%%%%%%%%%%%%%%%%%%%%%%%%%%%%%%%%%%%%%%%%%%%%%%%%%%%%%%%%%%%%%%%%%%
In this chapter, we apply the PGPE formalism to a scenario of \emph{nonequilibrium} dynamics of a finite-temperature atomic field.  We begin with a finite-temperature precessing-vortex configuration in a rotationally invariant trap, such as those considered in chapter~\ref{chap:precess}.   We then introduce an elliptical deformation of the trap which breaks this rotational symmetry, leading to the loss of angular momentum from the atomic field and thus the slowing of the rotating cloud, and consequently the decay of the vortex.  Our simulations are the first to describe the arrest of the rotation of both the condensed and noncondensed components of the field in response to the trap anisotropy, and our method has an intrinsic description of the coupled relaxation dynamics of the two components.  Adapting the methodology developed in chapter~\ref{chap:precess}, we identify the condensate and the complementary thermal component of the nonequilibrium field, and compare the evolution of their angular momenta and angular velocities.  We observe novel behaviour in the relaxation dynamics, arising from the thermodynamically anomalous behaviour of a single-vortex state and its coupling to the nonstationary thermal cloud.  By varying the trap anisotropy we alter the relative efficiencies of the vortex-cloud and cloud-trap coupling.  For strong trap anisotropies the angular momentum of the thermal cloud may be entirely depleted before the vortex begins to decay.  For weak trap anisotropies, the thermal cloud exhibits a long-lived steady state in which it rotates at an intermediate angular velocity. 

This chapter is organised as follows.  In section~\ref{sec:decay_procedure} we describe the initial equilibrium states we consider, and our approach to describing their subsequent evolution.  In section~\ref{sec:decay_dynamics} we analyse the dynamics of a representative classical-field simulation.  In section~\ref{sec:decay_Anisotropy_dependence} we discuss the dependence of the dynamics of the vortex and thermal field on the trap anisotropy,  and in section~\ref{sec:decay_Conclusions} we summarise our findings and present our conclusions.
%%%%%%%%%%%%%%%%%%%%%%%%%%%%%%%%%%%%%%%%%%%%%%%%%%%%%%%%%%%%%%%%%%%%%%%%%%%%%%%%%%%%%%%%%%%%%%%%%%%%%%%%%%%%%%%%%%%%%%%%%%%%%%%%%%%%
\section{Simulation procedure}\label{sec:decay_procedure}
%%%%%%%%%%%%%%%%%%%%%%%%%%%%%%%%%%%%%%%%%%%%%%%%%%%%%%%%%%%%%%%%%%%%%%%%%%%%%%%%%%%%%%%%
\subsection{Initial state}\label{subsec:decay_initial_state}
As in chapter~\ref{chap:precess}, we choose physical parameters corresponding to $^{23}\mathrm{Na}$ atoms confined in a strongly oblate trap, with trapping frequencies $(\omega_r, \omega_z)=2\pi\times(10, 2000)$ rad/s.  
We again adopt a 2D representation of the system, formalised in our classical-field method by choosing a cutoff $E_R=30\hbar\omega_r\ll\hbar\omega_z$, such that the low-energy space $\mathbf{C}$ excludes all modes with excitation along the $z$ axis (see section~\ref{subsec:dimless}).  In all our simulations this energy cutoff (projection operation) is effected in an inertial (laboratory) frame, and the low-energy space consists of $465$ single-particle propagation-basis modes (see section~\ref{subsec:cfield_proj_cft}).  

We form a finite-temperature initial state following the procedure of chapter~\ref{chap:precess}:  we construct a (nonequilibrium) randomised classical-field configuration over the modes comprising the space $\mathbf{C}$, with chosen normalisation, energy and angular-momentum first integrals, and we evolve this state for some time ($10^4$ trap cycles), so that the field has time to migrate to an equilibrium configuration.  As in chapter~\ref{chap:precess}, we take as the starting point for forming these configurations the ground Gross-Pitaevskii (GP) eigenstate with $N_0=1.072\times10^4$ atoms in a frame rotating at angular velocity $\Omega_0=0.35\omega_r$, with chemical potential $\mu_\mathrm{g}=10.35\hbar\omega_r$ and energy $E_\mathrm{g} = 7.646\times10^4\hbar\omega_r$ in an inertial frame, and angular momentum $L_0=\hbar N_0$.  To this state we add energy and angular momentum, using the algorithm described in section~\ref{subsec:prec_Microcanonical_evolution}, forming a configuration with $E=1.10E_\mathrm{g}$, and $L=1.20L_0$.  We find that the corresponding equilibrium state is one in which a single vortex precesses very close to the trap axis, i.e., with precession radius $r_\mathrm{v}\lesssim\eta$, where the healing length $\eta=0.20r_0$ is estimated from the density of the ground state (see section~\ref{subsec:prec_Vortex_precession}).  In this configuration the vortex is at equilibrium with the thermal component of the field, and the angular velocity $\Omega_\mathrm{c}$ of the condensate (which is the same as the cloud's at equilibrium) is close to that at the $\ell\equiv\langle L_z\rangle/\hbar N\rightarrow1^-$ limit of the mechanically unstable \cite{Butts99,Komineas05} precessing-vortex branch \cite{Papanicolaou05}.  If additional angular momentum is added to this state, the condensate mode becomes unstable to the nucleation of a second vortex, i.e., the single-vortex state we consider is essentially saturated with angular momentum for the given values of the other conserved integrals (normalisation and energy).  We find for this equilibrium initial state $\Omega_\mathrm{c}\approx0.23\omega_r$, $\mu\approx12\hbar\omega_r$, $k_\mathrm{B}T\approx14\hbar\omega_r$, and condensate fraction $f_\mathrm{c}\approx0.91$. 
%%%%%%%%%%%%%%%%%%%%%%%%%%%%%%%%%%%%%%%%%%%%%%%%%%%%%%%%%%%%%%%%%%%%%%%%%%%%%%%%%%%%%%%%
\subsection{Evolution}
Having formed this initial state, we evolve field trajectories in the laboratory frame in the presence of static trapping-potential anisotropies 
\begin{equation}
	\delta V(\x) = \epsilon m\omega_r^2\left(y^2 - x^2\right),
\end{equation}
with ellipticities in the range $0.005\leq\epsilon\leq0.1$. This procedure corresponds in each case to the sudden introduction of the anisotropy at time $t=0$.  We evolve each of these trajectories with the ARK45-IP algorithm (section~\ref{sec:rk4_app_ark45}), with accuracy chosen such that the relative change in field normalisation is $\leq 10^{-9}$ per time step taken, until the vortex is expelled from the condensate and the total angular momentum of the field has been lost through its interaction with the anisotropy. 
%%%%%%%%%%%%%%%%%%%%%%%%%%%%%%%%%%%%%%%%%%%%%%%%%%%%%%%%%%%%%%%%%%%%%%%%%%%%%%%%%%%%%%%%%%%%%%%%%%%%%%%%%%%%%%%%%%%%%%%%%%%%%%%%%%%%
\section{Arrest dynamics}\label{sec:decay_dynamics}
\begin{figure}
	\begin{center}
	\includegraphics[width=1.0\textwidth]{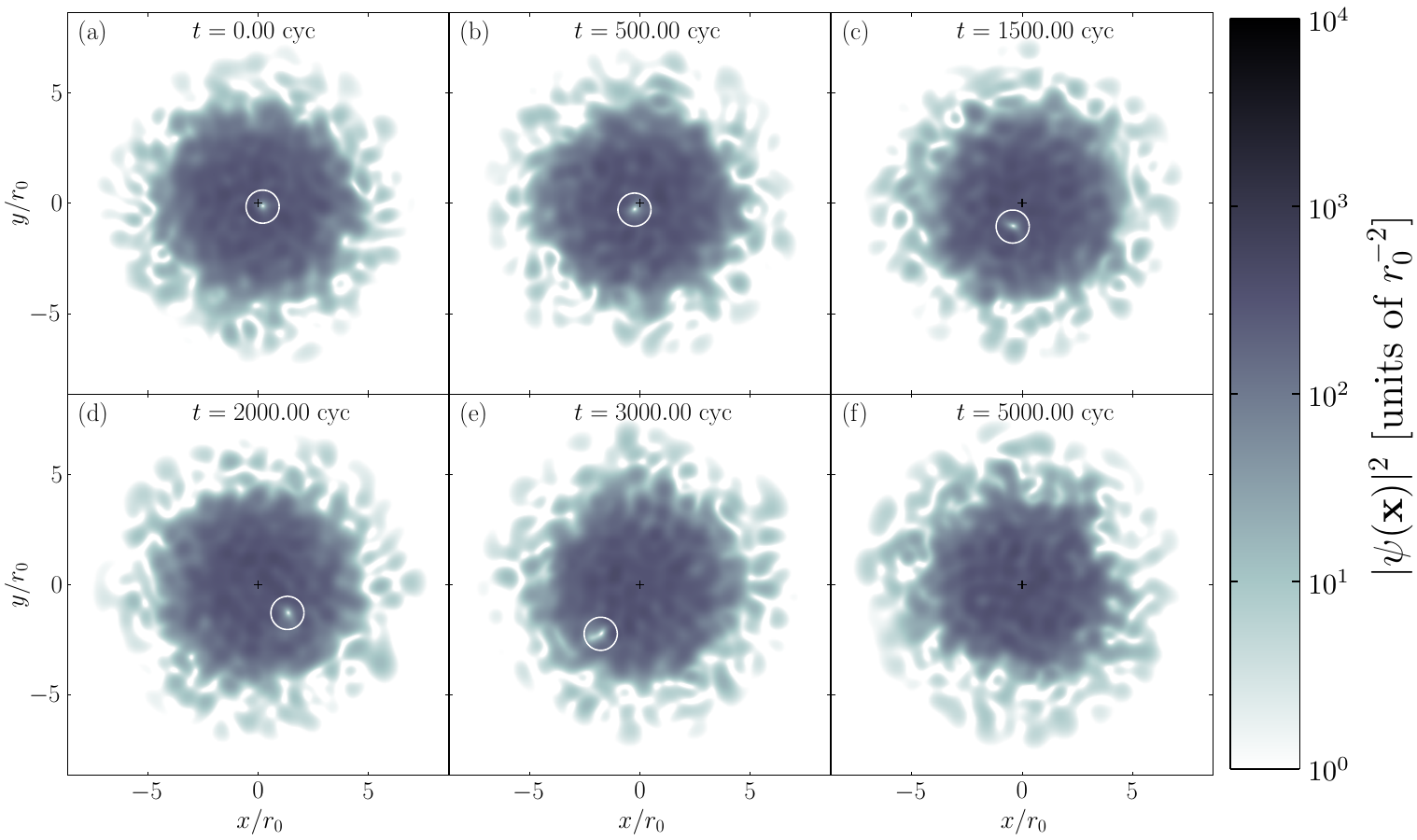}
	\caption{\label{fig:arrest_density_plots1}  (a-f) Classical-field densities at various times during the system evolution.  The white circle indicates the vortex position, and $+$ marks the coordinate origin (trap axis).  Parameters of the simulation are given in the text.}
	\end{center}
\end{figure}

In this section we consider the results of a classical-field simulation with a particular choice of the trap anisotropy, $\epsilon=0.025$.  The relaxation dynamics of the classical field in this case exhibit many features of interest, and serve as a point of comparison for simulations with weaker or stronger anisotropies.  Position-space densities of the classical field at representative times are presented in figures~\ref{fig:arrest_density_plots1}(a-f). 
%%%%%%%%%%%%%%%%%%%%%%%%%%%%%%%%%%%%%%%
%%%%%%%%%%%%%%%%%%%%%%%%%%%%%%%%%%%%%%%
Initially the vortex precesses very close to the trap axis (figure~\ref{fig:arrest_density_plots1}(a)), and it remains close to the trap axis for some time. Figure~\ref{fig:arrest_density_plots1}(b) shows the density at time $t=500$ cyc, in which the vortex core remains within $\sim\eta$ of the trap axis.  By  $t\approx 1000$ cyc the vortex has begun to spiral out of the central density bulk which contains the condensate.  Figure~\ref{fig:arrest_density_plots1}(c) shows that by $t=1500$ cyc the vortex has undergone significant radial displacement.  This increase in radial displacement continues (figure~\ref{fig:arrest_density_plots1}(d)), until the vortex core approaches the violently evolving condensate boundary (figure~\ref{fig:arrest_density_plots1}(e)). At $t\approx3070$ cyc the vortex is lost into the peripheral thermal material, leaving the condensate vortex-free and essentially irrotational (figure~\ref{fig:arrest_density_plots1}(f)). 
%%%%%%%%%%%%%%%%%%%%%%%%%%%%%%%%%%%%%%%%%%%%%%%%%%%%%%%%%%%%%%%%%%%%%%%%%%%%%%%%%%%%%%%%
\subsection{Vortex trajectory}\label{subsec:decay_vortex_trajectory}
\begin{figure}
	\begin{center}
	\includegraphics[width=0.65\textwidth]{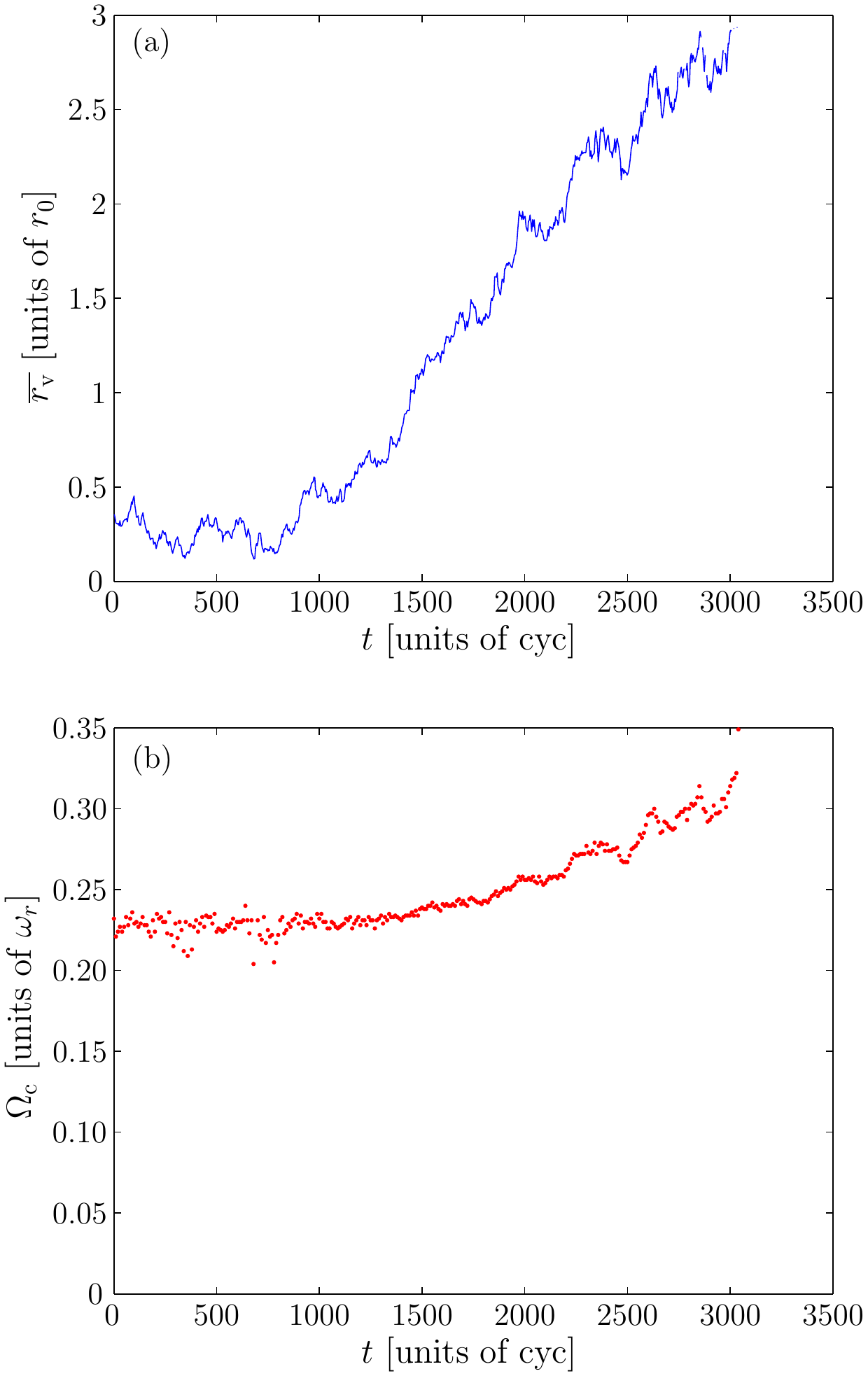}
	\caption{\label{fig:vortex_rad_freq}  Quantities characterising the vortex motion.  (a) Vortex radial displacement and (b) angular velocity of the condensate orbital.  Quantities are for a simulation with trap anisotropy $\epsilon=0.025$, corresponding to the classical-field densities in figure~\ref{fig:arrest_density_plots1}.}
	\end{center}
\end{figure}

We now quantify the behaviour of the vortex during its decay.  We observe that the vortex executes an essentially spiral-like motion as it decays, in agreement with the predictions of \cite{Fedichev99} and the simulations of \cite{Schmidt03,Jackson09}.  However, the vortex trajectory here is strongly stochastic, and the vortex orbits the trap axis many ($\sim 700$) times during its decay, so we do not present the full vortex trajectory.  To characterise the radial drift of the vortex, we track its location at a frequency of 25 samples per trap cycle, and average the resulting series over intervals of 100 samples ($4$ cyc) in order to smooth out rapid fluctuations due to thermal density fluctuations in the background field against which the vortex moves and uncertainties introduced by sampling the vortex location on a Cartesian grid.  In figure~\ref{fig:vortex_rad_freq}(a) we present the smoothed vortex radius ($\overline{r_\mathrm{v}}$) data.
%%%%%%%%%%%%%%%%%%%%%%%%%%%%%%%%%%%%%%%
%%%%%%%%%%%%%%%%%%%%%%%%%%%%%%%%%%%%%%%
It is clear from figure~\ref{fig:vortex_rad_freq}(a) that there is an initial lag of $\sim1000$ cyc between the introduction of the trap anisotropy at $t=0$ cyc and the beginning of the upward trend in vortex displacement.  We note that the displacement radius does not increase monotonically, but exhibits large oscillations during its increase.  The period of these oscillations is often $\gtrsim100$ trap cycles, spanning many periods of the vortex orbit, in contrast to the trajectories presented in \cite{Schmidt03,Jackson09}, in which the vortex generally precesses at a greater radius on each passage about the trap centre.  In figure~\ref{fig:vortex_rad_freq}(b) we plot the vortex precession frequency.  In practice, to evaluate this frequency, we follow the procedure of section~\ref{subsec:prec_PO_rotating_frames}, forming the covariance matrix (classical one-body density matrix)  
\begin{equation}\label{eq:density_matrix}
	G(\mathbf{x},\mathbf{x}') = \left\langle \psi^*(\mathbf{x}) \psi(\mathbf{x}') \right\rangle_\Omega, 
\end{equation}
where $\langle\cdots\rangle_\Omega$ denotes a time average, in a frame rotating at angular velocity $\Omega$ about the trap axis.  We vary $\Omega$ such that the largest eigenvalue of $G(\mathbf{x},\mathbf{x}')$ is maximised.  The value of $\Omega$ at which the maximum occurs is thus that of the rotating frame in which the coherent fraction of the classical field is most stationary, which forms a best estimate for the condensate angular frequency (vortex precession frequency) $\Omega_\mathrm{c}$, as discussed in section~\ref{subsec:prec_PO_rotating_frames}.
As we consider a nonequilibrium scenario, we construct equation~(\ref{eq:density_matrix}) by averaging over short time periods \cite{Blakie05}, in each case calculating the average of 250 consecutive classical-field samples over a ten cycle period.  
Figure~\ref{fig:vortex_rad_freq}(b) shows that the vortex precession frequency determined in this manner increases as the vortex displacement increases, as is well known for a zero-temperature condensate \cite{Fetter01,Komineas05} and observed in the simulations of \cite{Schmidt03,Jackson09}.  Moreover, we observe that the oscillations in vortex radius $r_\mathrm{v}$ at late times $t\gtrsim2000$ cyc are accompanied by oscillations in the condensate angular velocity. The oscillations in the two are positively correlated, i.e., the decreases in vortex radius are associated with periods of slowing of the condensate rotation during its otherwise steady increase with time.
%%%%%%%%%%%%%%%%%%%%%%%%%%%%%%%%%%%%%%%%%%%%%%%%%%%%%%%%%%%%%%%%%%%%%%%%%%%%%%%%%%%%%%%%
\subsection{Rotational dynamics of condensate and cloud}\label{subsec:decay_rotational_dynamics}
\begin{figure}
	\begin{center}
	\includegraphics[width=0.65\textwidth]{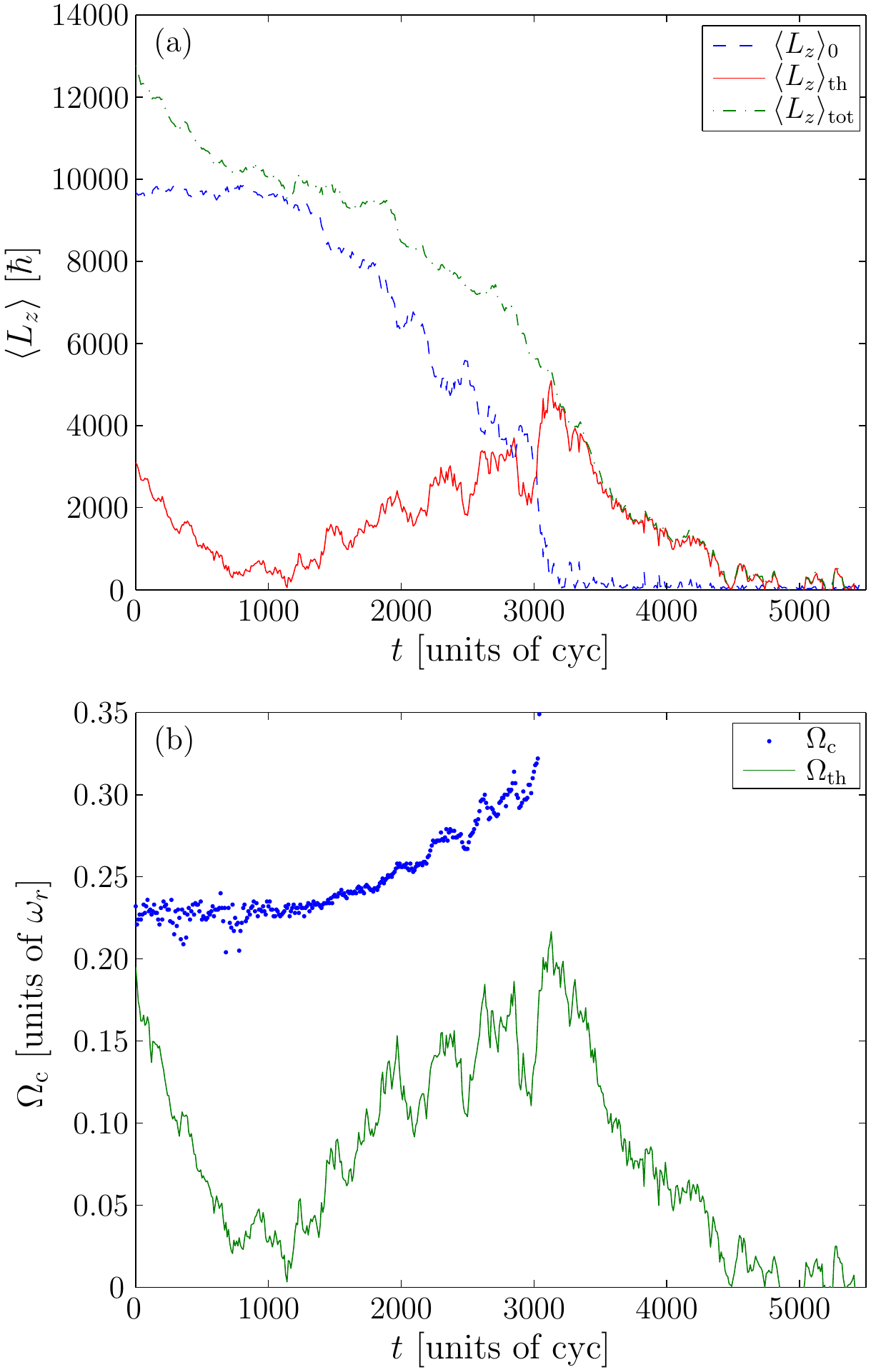}
	\caption{\label{fig:ang_mom_and_vel}  Quantities characterising the temporal evolution of the condensate and thermal cloud.  (a) Angular momenta of the condensate and thermal cloud, and total angular momentum of the field. (b) Angular velocities of the condensate and thermal cloud.}
	\end{center}
\end{figure}

The above definition of the condensate mode in terms of short-time covariance-matrix eigenvectors allows us to resolve the dynamics of the condensed and noncondensed components of the field.  As in section~\ref{subsec:prec_rotational_properties}, we introduce the decomposition of the classical-field one-body density matrix in terms of its eigenvectors $|\chi_i\rangle$ and corresponding eigenvalues $n_i$ 
\begin{equation}
	G = n_0|\chi_0\rangle\langle \chi_0| + \sum_{k>0} n_k |\chi_k\rangle\langle \chi_k| \equiv G_0 + G_\mathrm{th},
\end{equation}
which separates it into condensed and noncondensed parts.
Following section~\ref{subsec:prec_rotational_properties}, we then define the averages of a single-body operator $J$ in the condensed and noncondensed components of the field by $\langle J\rangle_0 = \mathrm{Tr}\{G_0J\}$ and $\langle J\rangle_\mathrm{th} = \mathrm{Tr}\{G_\mathrm{th}J\}$ respectively.
We use this decomposition to calculate the angular momentum of the condensate  ($\langle L_z \rangle_0$) and thermal cloud ($\langle L_z \rangle_\mathrm{th}$).  These quantities are presented in figure~\ref{fig:ang_mom_and_vel}(a), along with the total angular momentum of the field ($\langle L_z\rangle_\mathrm{tot}$).
%%%%%%%%%%%%%%%%%%%%%%%%%%%%%%%%%%%%%%%
%%%%%%%%%%%%%%%%%%%%%%%%%%%%%%%%%%%%%%%
We observe that in the first few hundred trap cycles after the introduction of the trap anisotropy, the angular momentum of the thermal cloud (solid line in figure~\ref{fig:ang_mom_and_vel}(a)) undergoes approximately exponential decay, reaching values close to zero by $t\approx1000$ trap cycles. By contrast, the angular momentum of the condensate (dashed line) is essentially unchanged during this period, which we identify with the initial period of quiescence of the vortex noted in section~\ref{subsec:decay_vortex_trajectory}.  At $t\approx1100$ cyc, the angular momentum of the condensate begins to decay, and after $t\approx3200$ cyc the condensate is essentially irrotational, with some small value of angular momentum remaining in the form of surface excitations.  We note that the angular momentum of the thermal field component \emph{increases} over the period $t\in[1000,3200]$ cyc.  
We understand this as follows: initially the angular momentum of the thermal cloud is lost due to its interaction with the trap anisotropy.  This makes the vortex thermodynamically unstable, and it thus begins to decay, liberating its angular momentum to the cloud.  The cloud loses angular momentum to the trap at a rate proportional to its angular velocity (frictional loss $d\langle L_z\rangle/dt\propto -\Omega_\mathrm{th}$ \cite{Zhuravlev01}), which is small immediately after the initial arrest of the cloud by the trap anisotropy.  Due to its initially slow rotation, the cloud therefore gains angular momentum from the decaying vortex faster than it can dissipate it to the trap.  
As the thermal cloud's rotation rate increases, so does the rate at which it dissipates angular momentum to the trap, which acts to curb the increase in cloud angular velocity.
However, the precessing-vortex condensate is thermodynamically anomalous: as it loses angular momentum, its angular velocity \emph{increases} \cite{Butts99,Papanicolaou05,Komineas05}.  In the present case, the increasing rotation rate of the decaying vortex causes the rate of transfer of angular momentum from the vortex to the cloud (which is proportional to the differential angular velocity between the two \cite{Fedichev99}) to dominate the rate at which the cloud dissipates angular momentum to the trap, throughout the decay of the vortex.  
As shown in figure~\ref{fig:ang_mom_and_vel}(b), this leads to `run-away' spin-up of the vortex precession frequency, with the thermal-cloud rotation rate being driven up in sympathy.  

Closer inspection of the angular momenta $\langle L_z\rangle_0$ and $\langle L_z\rangle_\mathrm{th}$ reveals negatively correlated oscillations in the angular momenta of the two components during the decay, as a result of the detailed nonlinear dynamics of angular momentum exchange and loss during the decay process.  We identify this oscillation in angular-momentum transfer with the oscillations in vortex rotation rate discussed in section~\ref{subsec:decay_vortex_trajectory}.
In figure~\ref{fig:ang_mom_and_vel}(b) we plot the angular velocity of the condensate $\Omega_\mathrm{c}$ (dots) derived in the estimation of the condensate fraction (section~\ref{subsec:decay_vortex_trajectory}), and the angular velocity of the thermal cloud $\Omega_\mathrm{th} = \langle L_z \rangle_\mathrm{th} / \langle \Theta_\mathrm{c} \rangle_\mathrm{th}$ (line), where we assume the expectation value of the classical moment of inertia $\Theta_\mathrm{c}\equiv mr^2$ as an estimate of the cloud's true moment of inertia.  In section~\ref{subsec:prec_rotation_rates} we discussed the level of uncertainty inherent in this procedure, nevertheless it yields a clear qualitative description of the decay dynamics.  The oscillations in rotation rate of the thermal cloud are clearly visible here, and we note that they are \emph{positively} correlated with the oscillations in angular velocity of the condensate, due to its anomalous rotational response.   We note that oscillatory behaviour arises already in a linear analysis of the arrest of a rotating Boltzmann gas \cite{Guery-Odelin00} by a trap anisotropy.  It is therefore not surprising that similar oscillations occur in the transfer of angular momentum from the condensate to the nonequilibrium thermal field, considering the complexity of their coupled dynamics (cf. \cite{Zhuravlev01}), and the anomalous response of the vortical condensate.

Finally, we note that the cloud angular momentum (rotation rate) reaches its peak when the vortex leaves the condensate ($t\approx3200$ cyc), after which it undergoes a second near-exponential decay phase, the angular momentum of the field decaying such that by $t\approx5000$ cyc, only the thermodynamic fluctuations in $\langle L_z \rangle_\mathrm{th}$ exhibited by the finite-temperature field at rest in the anisotropic potential remain.
%%%%%%%%%%%%%%%%%%%%%%%%%%%%%%%%%%%%%%%%%%%%%%%%%%%%%%%%%%%%%%%%%%%%%%%%%%%%%%%%%%%%%%%%
\subsection{Heating of the atomic field}\label{subsec:decay_heating}
We now consider the heating of the atomic field during the arrest of its rotation.  As the system we evolve is Hamiltonian, with a time-independent potential, the total energy of the classical field is a constant of the motion.  Consequently, the trap anisotropy dissipates the angular momentum of the field by converting the rotational kinetic energy of the field into internal energy \cite{Zhuravlev01}, and we therefore expect some heating of the field to occur, due to this redistribution of energy.  We can estimate the heating of the field as follows:  the rotational energy of the gas is initially $(E_\mathrm{rot})_\mathrm{i}=\Omega_\mathrm{i}L_\mathrm{i}\approx2.7\times10^3\hbar\omega_r$.  We compare this to the initial thermal energy, which should be reasonably well estimated by the energy added to the ground vortex state in forming the initial thermal state (section~\ref{sec:decay_procedure}), $E_\mathrm{th}\sim0.1E_\mathrm{g}\approx7.6\times10^3\hbar\omega_r$.  Assuming a linear relationship between thermal energy and temperature (which should be valid for the low temperatures we consider here, see \cite{Davis05}), we might therefore expect an increase in the field temperature of $\sim30\%$ during the arrest of the condensate rotation.  

\begin{figure}
	\begin{center}
	\includegraphics[width=0.65\textwidth]{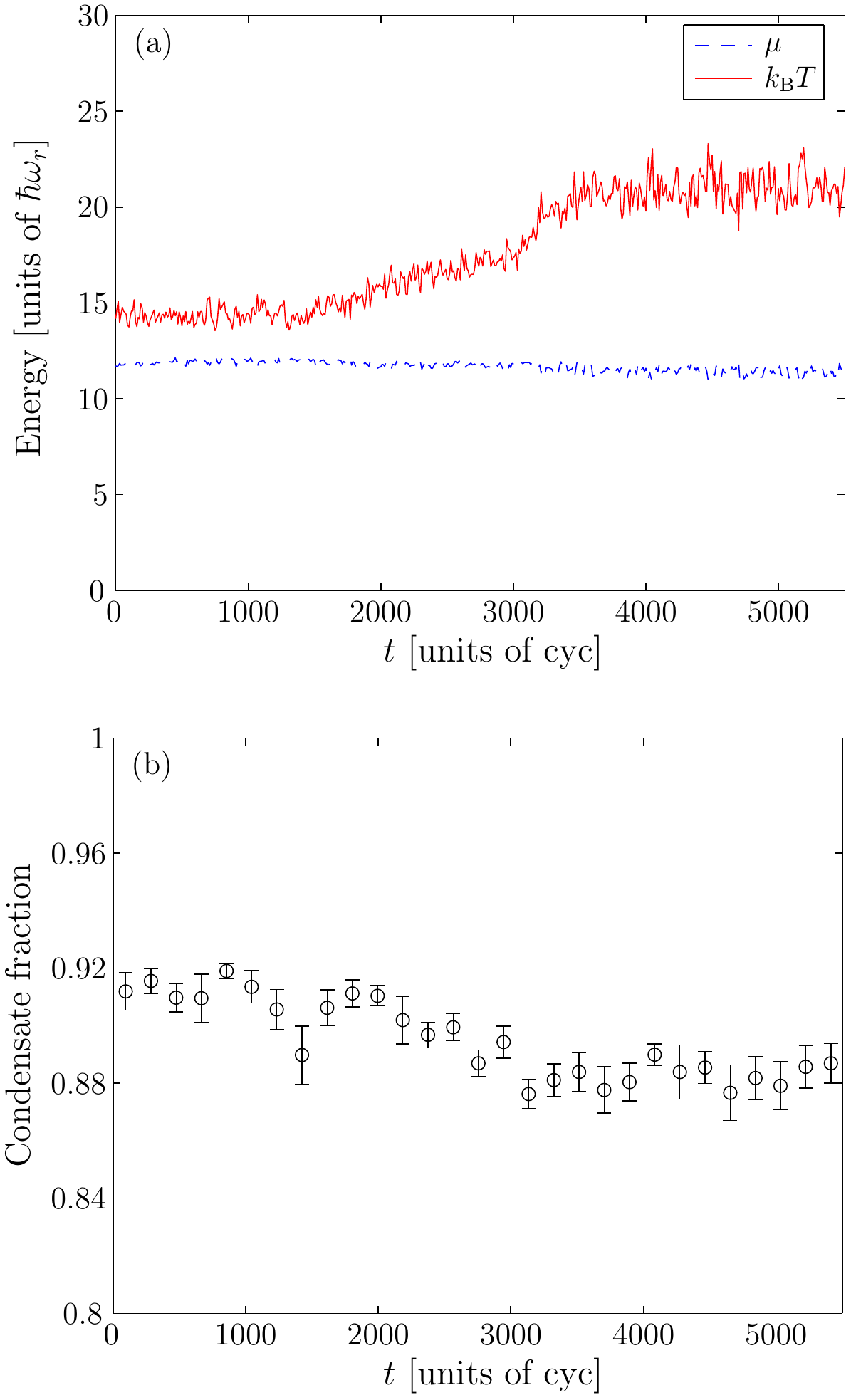}
	\caption{\label{fig:cfrac_and_temp}  Heating of the field.  Evolution of (a) the effective chemical potential and temperature and (b) the condensate fraction.} 
	\end{center}
\end{figure}

In order to quantify the heating and its development during the field evolution, we estimate the (effective) thermodynamic parameters of the field (chemical potential $\mu$ and temperature $T$) using the fitting procedure of appendix~\ref{app:fitting_function}.
As in section~\ref{subsec:prec_T_and_mu}, we allow for the differential rotation between the thermal cloud and the frame in which the classical-field cutoff is effected (the laboratory frame, in the present case). We perform a fit to the wing of the (time-averaged) radial distribution $n(r)$ of the classical field; i.e., we fit over the radii range $r\in[r_-+0.5r_0,r_\mathrm{tp}]$ where $r_-$ marks the minimum of the Hartree-Fock effective potential $V_\mathrm{eff}=(m/2)[\omega_r^2-\Omega_\mathrm{th}^2]r^2 + 2U_\mathrm{2D} n(r)$ \footnote{We refer here to the local minimum associated with the boundary of the condensate, rather than any minimum that may occur due to the presence of the vortex core.}, and $r_\mathrm{tp}$ is the semiclassical turning-point of the low-energy space $\mathbf{C}$.  We fit to radial densities of the classical field averaged over $10$ cyc periods, and assume the values of $\Omega_\mathrm{th}$ calculated in section~\ref{subsec:decay_rotational_dynamics}.  The resulting estimates for $\mu$ and $T$ are presented in figure~\ref{fig:cfrac_and_temp}(a).
%%%%%%%%%%%%%%%%%%%%%%%%%%%%%%%%%%%%%%%
%%%%%%%%%%%%%%%%%%%%%%%%%%%%%%%%%%%%%%%
We find that the temperature (solid line) is approximately constant during the first $\sim1000$ cycles of the field evolution, corresponding to the initial arrest of the thermal field component.  The initial angular momentum of the thermal component is small (see figure~\ref{fig:ang_mom_and_vel}(a)), comprising some $\sim25\%$ of the total angular momentum of the field, and so any heating of the field due to the redistribution of the associated rotational energy is possibly too small to resolve above the uncertainty in the temperature estimates. Beginning at $t\approx1100$ cyc, corresponding to the start of the vortex-decay phase, the temperature exhibits a steady, approximately linear increase.  We associate this increase with the conversion of rotational kinetic energy to thermal energy, due to the action of the trap anisotropy, and presumably also as a result of the scattering of excitations by the vortex which produces the frictional effect \cite{Fedichev99}.  There is a final sharp rise in temperature at $t\approx3000$--$3200$ trap cycles.  This rise is perhaps due to the final decay of the vortex into excitations at the surface of the condensate \cite{Fedichev99}.  We note however that the field, and in particular the condensate surface, undergoes strong fluctuations as the system passes through this transition to the vortex-free state, with, e.g., multiple (ghost) vortices present at the condensate surface at times.  This can be viewed as the system essentially `reversing' through the surface-wave instability arising from the relative motion of condensate and thermal cloud, in which vortices spontaneously grow from surface excitations \cite{Anglin01, Williams2002a, Penckwitt02}.  The thermodynamic parameters may therefore be ill-defined during this period.  After this period of strong surface fluctuations, the temperature levels off at $T\approx 21 \hbar\omega_r/k_\mathrm{B}$, corresponding to heating of $\approx50\%$ during the arrest.  We note that the chemical potential (dashed line in figure~\ref{fig:cfrac_and_temp}(a)) exhibits a slight downward trend beginning around $t\approx1100$ cyc, falling by $\sim0.4\hbar\omega_r$, though we note that this change is of the same order as the variation in estimates for $\mu$ obtained at late times $t\gtrsim3500$ cyc. 

In figure~\ref{fig:cfrac_and_temp}(b) we plot the condensate fraction $f_\mathrm{c} \equiv n_0 / \sum_kn_k$ obtained from the procedure outlined in section~\ref{subsec:decay_vortex_trajectory}.  The fluctuations in estimates of this quantity are large, as is expected given the short time scale ($10$ cyc) over which the appropriate averages are taken, and the nonequilibrium nature of the field.  We therefore group the condensate-fraction estimates into bins each containing 19 consecutive estimates of $f_\mathrm{c}$, and average over the estimates in each bin (cf. section~\ref{subsec:prec_temporal_decoherence}).  We plot the mean and standard deviation of the estimates in each bin (circles and their error bars), and observe a clear downwards trend as time proceeds, with the condensate fraction dropping by $\sim3\%$ during the arrest.
%%%%%%%%%%%%%%%%%%%%%%%%%%%%%%%%%%%%%%%%%%%%%%%%%%%%%%%%%%%%%%%%%%%%%%%%%%%%%%%%%%%%%%%%%%%%%%%%%%%%%%%%%%%%%%%%%%%%%%%%%%%%%%%%%%%%
\section{Dependence on trap anisotropy}\label{sec:decay_Anisotropy_dependence}
We now consider the effect of varying the trap anisotropy $\epsilon$ on the behaviour of the classical-field trajectories.  We intuitively expect the rate at which the trap dissipates the angular momentum of the thermal cloud to depend strongly on the magnitude of the anisotropy, and a quantitative model for this dependence in the case of a classical (Boltzmann) gas was presented in \cite{Guery-Odelin00}.  By contrast, the efficiency with which the thermal cloud extracts angular momentum from the condensate is dictated by the (longitudinal) mutual-friction coefficient, which depends on the temperature and chemical potential of the field, in addition to its microscopic properties \cite{Hall56,Sonin97,Fedichev99}.  We therefore expect, as discussed in the vortex-continuum analysis of \cite{Zhuravlev01}, to explore different regimes of relaxation dynamics as we vary the trap anisotropy and consequently the relative strengths of vortex-cloud and cloud-trap friction.  In this section we consider the evolution of the classical field in the presence of anisotropies of different magnitudes, and observe the resulting differences in the behaviour of the field. 
%%%%%%%%%%%%%%%%%%%%%%%%%%%%%%%%%%%%%%%%%%%%%%%%%%%%%%%%%%%%%%%%%%%%%%%%%%%%%%%%%%%%%%%%
\subsection{Weak damping}\label{subsec:decay_weak_damping}
\begin{figure}
	\begin{center}
	\includegraphics[width=1.0\textwidth]{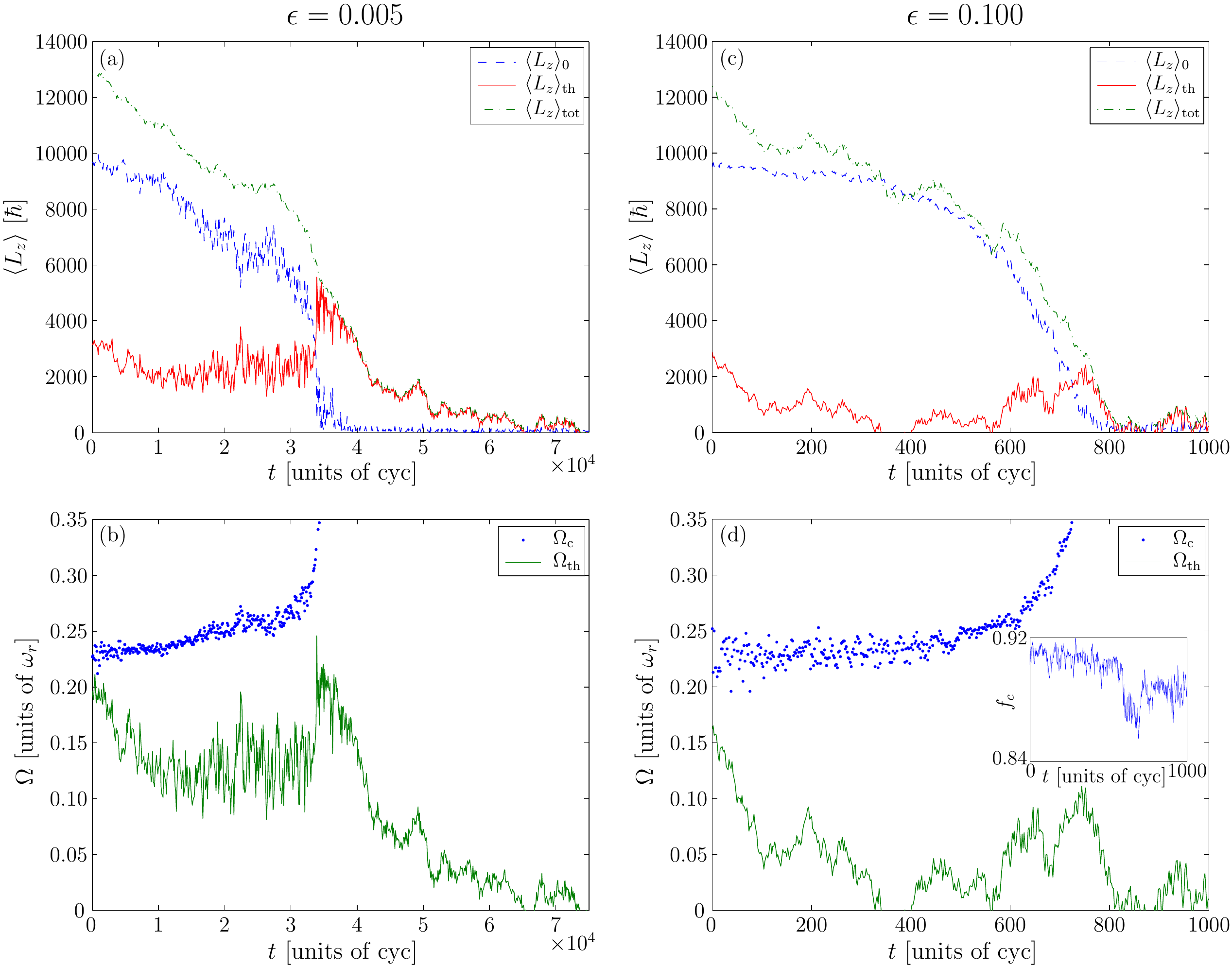}
	\caption{\label{fig:adiabatic_and_not}  (a) Angular momenta and (b) angular velocities of field components in a simulation with trap anisotropy $\epsilon=0.005$. (c) Angular momenta and (d) angular velocities of field components in a simulation with trap anisotropy $\epsilon=0.100$. Inset: Evolution of the condensate fraction in the case $\epsilon=0.100$.}
	\end{center}
\end{figure}

In figure~\ref{fig:adiabatic_and_not}(a) we plot the evolution of the angular momentum of the condensate and thermal cloud in the presence of a weak trap anisotropy ($\epsilon=0.005$).  
%%%%%%%%%%%%%%%%%%%%%%%%%%%%%%%%%%%%%%%
%%%%%%%%%%%%%%%%%%%%%%%%%%%%%%%%%%%%%%%
We observe that after an initial decline over the first $\approx10^4$ trap cycles, the cloud angular momentum, despite exhibiting large fluctuations, maintains a reasonably steady mean value which increases slowly over the period $t\sim[10^4,3\times10^4]$ trap cycles.  During this period the angular momentum of the condensate drops by approximately a factor of $2$, and large fluctuations in the distribution of angular momentum between the two components are visible.  This behaviour is again apparent in the vortex precession frequency and cloud rotation rate presented in figure~\ref{fig:adiabatic_and_not}(b).  This shows that the cloud exhibits a steady-state nonequilibrium behaviour, i.e., its rotation rate remains somewhere between that of the precessing condensate mode, and that of the trap ($\Omega_\mathrm{tr}=0$).  In this regime its angular momentum remains approximately constant, as the rate at which it gains angular momentum from the decaying vortex matches the rate at which it loses angular momentum to the trap.  We expect the angular velocity at which this balance occurs, and indeed whether such a regime occurs at all, to depend strongly on the relative strengths of the vortex-cloud and cloud-trap friction.  We note that the angular velocity of the cloud appears to slowly increase over time, as the condensate loses angular momentum and its angular velocity increases, shifting the cloud rotation rate at which the angular momentum transfer rates are balanced.  At $t\approx3.2\times10^4$ cyc the vortex's precession accelerates as it approaches the condensate boundary, and during the period $t\in[3.2\times10^4,3.4\times10^4]$ cyc the adiabaticity of the vortex dissipation is lost as the field enters the critical regime associated with nucleation of the vortex at the surface (see section~\ref{subsec:decay_heating}).  In this regime the condensate surface is unstable and as noted in section~\ref{subsec:decay_heating}, multiple `ghost' vortices may be present simultaneously at the condensate boundary.  A large amount of angular momentum is transferred nonadiabatically to the thermal cloud during this period of criticality, while small surface-mode oscillations of the condensate persist until $t\approx4\times10^4$ cyc.  The angular momentum is subsequently dissipated from the thermal cloud over a period of $\sim3\times10^4$ cyc, in an oscillatory but approximately exponential decay phase.
%%%%%%%%%%%%%%%%%%%%%%%%%%%%%%%%%%%%%%%%%%%%%%%%%%%%%%%%%%%%%%%%%%%%%%%%%%%%%%%%%%%%%%%%
\subsection{Strong damping}\label{subsec:decay_strong_damping}\enlargethispage{-\baselineskip}
We now turn our attention to a scenario of vortex arrest due to the presence of a strong trap anisotropy ($\epsilon=0.1$).  The dynamics in this case are strongly nonequilibrium and the arrest of the field's rotation occurs on a shorter time scale than the cases already considered, and so here we form the density matrix~\reff{eq:density_matrix} by averaging classical-field samples over a shorter period of two trap cycles.  In this case the angular momentum of the thermal cloud (solid line in figure~\ref{fig:adiabatic_and_not}(c)) is rapidly depleted, and actually fluctuates below zero by $t\approx400$ trap cycles.  The condensate is slower to respond; during this period the condensate angular momentum decreases by $\sim7\%$.  Subsequently, the cloud angular momentum fluctuates strongly, and at $t\approx750$ cyc rises to $\sim2000\hbar$ (close to its initial value), as the vortex is rapidly expelled from the condensate.  The angular momentum is then dissipated from the cloud over a period $\lesssim100$ cyc, and thereafter the angular momentum of the field fluctuates about zero.  The rapid expulsion of the vortex, and strong fluctuations of the cloud rotation, are again visible in the calculated angular velocities (figure~\ref{fig:adiabatic_and_not}(d)).  

The behaviour of the cloud angular momentum in this trajectory suggests that the cloud is overcritically damped by the trap anisotropy \cite{Guery-Odelin00}.  Its response to the anisotropy appears unhindered by its coupling to the condensate via the vortex core, and the angular momentum it acquires from the vortex is quickly yielded to the trap, despite its small angular velocity $\Omega_\mathrm{th}\lesssim0.1$.  The relaxation process in this case is violently nonadiabatic, as evidenced by the strong fluctuations in the cloud angular momentum.  Indeed during the period $t\sim[600,700]$ cyc large, long-wavelength surface oscillations are visible in the field as the vortex precesses rapidly near the condensate boundary.  It appears that the strong trap anisotropy and rapid `stirring' motion of the vortex conspire to strongly perturb the condensate bulk in this regime.  During this period the measured condensate fraction is suppressed (inset to figure~\ref{fig:adiabatic_and_not}(d)) due to the strongly nonequilibrium behaviour of the condensate, in which surface-wave excitations define frames of rotation distinct from that of the vortex.  The decomposition of the field into condensed and noncondensed components must therefore be viewed with some caution in this strongly nonequilibrium scenario.  Nevertheless, it is clear that the effect of the trap on the cloud dominates the vortex-cloud coupling in this scenario, in stark contrast to the near-adiabatic decay scenario of section~\ref{subsec:decay_weak_damping}. 
%%%%%%%%%%%%%%%%%%%%%%%%%%%%%%%%%%%%%%%%%%%%%%%%%%%%%%%%%%%%%%%%%%%%%%%%%%%%%%%%%%%%%%%%
\subsection{Decay times}\label{subsec:decay_decay_times}
\begin{figure}
	\begin{center}
	\includegraphics[width=0.65\textwidth]{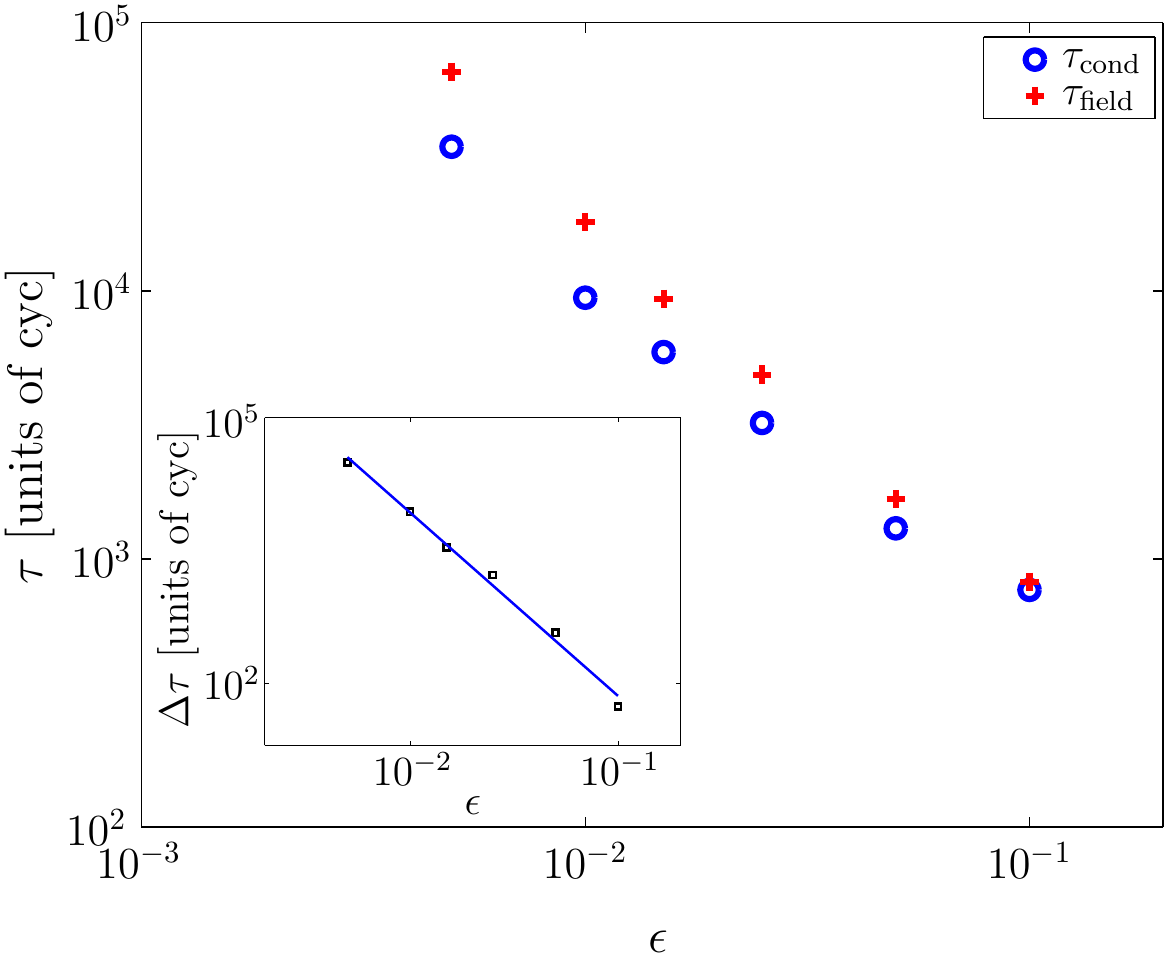}
	\caption{\label{fig:relaxation}  Relaxation of the field.  Circles (plusses) represent the times at which the vortex leaves the condensate (total field angular momentum reaches zero). Inset: Cloud spin-down times and the linear fit performed to extract their scaling with trap ellipticity.}
	\end{center}
\end{figure}

We now consider the dependence of the relaxation times on the strength of the trap anisotropy.  As the relaxation of both the condensate and the thermal cloud is generally nonexponential, we consider the times at which the condensate becomes irrotational, and at which the total angular momentum of the field is lost.  Due to the persistence of surface excitations which prevent the condensate angular momentum from reaching zero, we define the condensate stopping time $\tau_\mathrm{cond}$ as that at which the angular momentum of the condensate first drops below $0.02\hbar$ per particle, and the field stopping time as the first \emph{subsequent} time $\tau_\mathrm{field}$ at which the cloud angular momentum crosses zero.  These times are presented in figure~\ref{fig:relaxation}.  
%%%%%%%%%%%%%%%%%%%%%%%%%%%%%%%%%%%%%%%
%%%%%%%%%%%%%%%%%%%%%%%%%%%%%%%%%%%%%%%
We observe that the condensate arrest times (circles in figure~\ref{fig:relaxation}) vary by two orders of magnitude over the range of trap anisotropies we consider.  Moreover, the total-field relaxation times (plusses) become increasingly longer than the vortex relaxation times as the trap anisotropy is weakened, causing slower dissipation of the cloud angular momentum even in the absence of the vortex.  We therefore consider the time $\Delta \tau = \tau_\mathrm{field}-\tau_\mathrm{cond}$ over which the angular momentum of the cloud dissipates following the expulsion of the vortex.  Although the angular momentum lost in this final damping phase varies over the range $\langle L_z \rangle_\mathrm{th} \sim 3000$--$5000\hbar$, precluding a precise analysis, the times $\Delta \tau$ provide a useful characterisation of the dependence of the cloud relaxation on the trap anisotropy. We perform a linear fit to $\Delta\tau$ as a function of $\epsilon$ in log-log space (inset to figure~\ref{fig:relaxation}), and find the scaling $\Delta \tau \propto \epsilon^{-2.1}$, in good agreement with the scaling $t_\mathrm{down}\propto \epsilon^{-2}$ for thermal-cloud spin-down predicted for weak anisotropies by Gu\'ery-Odelin \cite{Guery-Odelin00}.  The latter calculations considered a simpler scenario in which no condensate was present, however once the condensate becomes vortex-free its primary influence on the thermal cloud is simply to reshape the trapping potential experienced by the thermal atoms.  The quantitative agreement between the spin-down scaling laws predicted by these two very different approaches to describing the thermal-cloud dynamics is a strong indication of the utility of the (superficially Gross-Pitaevskii-like) classical-field model for modelling the dynamics of \emph{thermal} atoms in the degenerate regime.
%%%%%%%%%%%%%%%%%%%%%%%%%%%%%%%%%%%%%%%%%%%%%%%%%%%%%%%%%%%%%%%%%%%%%%%%%%%%%%%%%%%%%%%%%%%%%%%%%%%%%%%%%%%%%%%%%%%%%%%%%%%%%%%%%%%%
\vspace{\baselineskip}
\section{Summary}\label{sec:decay_Conclusions}
We have carried out the first simulations of the arrest of a rotating Bose-Einstein condensate due to the presence of a trap anisotropy which includes the coupled nonequilibrium dynamics of the condensate and thermal cloud.  Our method makes no assumptions of stationarity of a thermal bath, nor are the two components artificially given disparate rotational parameters.  Rather, our approach describes the dynamical migration of an \emph{equilibrium} rotating thermal state to a new, irrotational equilibrium, solely due to the action of the trapping potential.

We observe for all parameters we considered that the rotation rates of the condensate and thermal component are distinct during the decay.  The anomalous rotational response of the precessing-vortex condensate can lead to a counter-intuitive spin-up of the thermal component during the decay, and we observe nonequilibrium oscillations in transfer of angular momentum between the two components as the vortex responds to the dissipative effect of the thermal cloud, which is itself damped by the trapping potential anisotropy.

For trap anisotropies that are weak, the thermal field settles to a nonequilibrium steady state, with rotation rate intermediate between that of the condensate and that of the (static) trap.  In this scenario the angular momentum of the condensate is slowly depleted while that of the thermal cloud remains nearly constant, until the vortex nears the condensate boundary and the linearity of the vortex decay breaks down.  For stronger trap anisotropies, the angular momentum of the thermal cloud may be almost entirely depleted before the condensate responds. 

We quantified the heating of the atomic field during the arrest, and found it to be commensurate with the conversion of rotational kinetic energy into thermal energy by the trap anisotropy.  We also considered the time scales over which the condensate angular momentum and total field angular momentum were dissipated, and found reasonable quantitative agreement between the scaling of the thermal cloud spin-down time with trap anisotropy and the predictions of a Boltzmann-gas model \cite{Guery-Odelin00}. 
%%%%%%%%%%%%%%%%%%%%%%%%%%%%%%%%%%%%%%%%%%%%%%%%%%%%%%%%%%%%%%%%%%%%%%%%%%%%%%%%%%%%%%%%%%%%%%%%%%%%%%%%%%%%%%%%%%%%%%%%%%%%%%%%%%%%
%%%%%%%%%%%%%%%%%%%%%%%%%%%%%%%%%%%%%%%%%%%%%%%%%%%%%%%%%%%%%%%%%%%%%%%%%%%%%%%%%%%%%%%%%%%%%%%%%%%%%%%%%%%%%%%%%%%%%%%%%%%%%%%%%%%%

\chapter{Prior studies of vortex-lattice formation in stirred condensates}
\label{chap:stir_background}
%%%%%%%%%%%%%%%%%%%%%%%%%%%%%%%%%%%%%%%%%%%%%%%%%%%%%%%%%%%%%%%%%%%%%%%%%%%%%%%%%%%%%%%%%%%%%%%%%%%%%%%%%%%%%%%%%%%%%%%%%%%%%%%%%%%%
%%%%%%%%%%%%%%%%%%%%%%%%%%%%%%%%%%%%%%%%%%%%%%%%%%%%%%%%%%%%%%%%%%%%%%%%%%%%%%%%%%%%%%%%%%%%%%%%%%%%%%%%%%%%%%%%%%%%%%%%%%%%%%%%%%%%
In this chapter we review the key prior work on the stirring of Bose-Einstein condensates to produce vortex lattices.  As noted in section~\ref{sec:intro_vortices}, a range of early experiments considered the formation of individual vortices, by techniques including coherent hyperfine-component interconversion \cite{Matthews99,Anderson00}, the `sweeping' of a laser light-shift potential through the atom cloud \cite{Inouye01}, and phase imprinting \cite{Leanhardt02}.  Other experiments considered the formation of vortex lattices by setting a normal thermal cloud into rotation in cylindrically symmetric trap, and cooling it through the condensation transition \cite{Haljan01a}.  However, we here consider the formation of vortices in condensate \emph{stirring} experiments, in which an anisotropic rotating trapping potential defines a rotating frame of equilibrium for the atomic cloud, and induces the formation of vortices in the condensate.
This approach was used to produce the first multiple-vortex configurations and vortex lattices in condensed Bose gases, and as we discuss in this chapter, exhibits a rich phenomenology.  Indeed, the experimental investigation of vortex nucleation and lattice formation by stirring has stimulated substantial theoretical research over recent years, and led to some controversy over the precise mechanisms of the vortex nucleation process, and the role of thermality in the formation of a rotating vortex lattice.  This chapter is structured as follows: in section~\ref{sec:stir_exp} we review the results of experiments on condensate stirring performed by the groups of Dalibard (\'Ecole Normale Sup\'erieure), Ketterle (MIT), and Foot (Oxford).  In section~\ref{sec:stir_theory} we review the various theoretical approaches to explaining the experimental results, and in section~\ref{sec:stir_questions} we formulate the issues on the nature of the vortex formation which remained unanswered by those works. We will address those questions with our own classical-field methodology in chapter~\ref{chap:stirring}.
%%%%%%%%%%%%%%%%%%%%%%%%%%%%%%%%%%%%%%%%%%%%%%%%%%%%%%%%%%%%%%%%%%%%%%%%%%%%%%%%%%%%%%%%%%%%%%%%%%%%%%%%%%%%%%%%%%%%%%%%%%%%%%%%%%%%
\section{Experiments}\label{sec:stir_exp}
%%%%%%%%%%%%%%%%%%%%%%%%%%%%%%%%%%%%%%%%%%%%%%%%%%%%%%%%%%%%%%%%%%%%%%%%%%%%%%%%%%%%%%%%
\subsection{Work of Dalibard's group (ENS)}
The group of Dalibard performed the first experiments aimed at reproducing the `rotating-bucket' paradigm for studies of rotation in superfluid helium.  In the helium experiments (see \cite{Donnelly91} and references therein), the rotating bucket defines the frame of thermal equilibrium through its surface roughness; as the rotation rate of the bucket increases, the superfluid admits vortices which form a regular array or \emph{vortex lattice} at equilibrium.  By analogy, a rotating frame can be imposed on a gaseous condensate by deforming the trapping potential away from cylindrical symmetry, and rotating it.  In the experiments performed on $^{87}\mathrm{Rb}$ atoms by Madison \emph{et al.} \cite{Madison00}, this deformation was provided by the light-shift potential provided by a blue-detuned laser beam (see, e.g., \cite{Caradoc00}), which was \emph{toggled} between two diametrically opposite focus points in the transverse plane of the trap, at a frequency very rapid compared with the dynamics of the atomic mean field.  The resulting effect was that the prolate trapping potential was (to leading order) anisotropically deformed in the $xy$ plane, leading to an elliptical cross-section.  This deformation of the trapping potential was then rotated about the trap axis at a frequency $\Omega$.  In the first experiments \cite{Madison00,Madison00b}, this stirring potential was switched on during the final stages of evaporative cooling, at the end of which, the rf shield was left turned on.  The atomic cloud was left to thermalise in the rotating `bucket', and the resulting structure of the condensate was then studied by time-of-flight analysis.  
The somewhat surprising result of these experiments was the high stirring frequency at which vortices were first observed, which was significantly higher than the \emph{thermodynamic} critical frequency $\Omega_\mathrm{c}$ at which the energy of the vortex state becomes lower than that of the vortex-free state (section~\ref{subsec:back_vortices}).  Dalibard's group then undertook further experiments \cite{Madison02}, in which the stirring was performed on a preformed condensate \emph{after} the completion of the evaporative cooling, and they considered in detail 
the dependence of the vortex nucleation on the ellipticity of the trapping potential, which they varied by adjust the intensity of the stirring laser beam.  They found that for small ellipticities, vortices were formed in a resonant regime centred on the resonant frequency $\omega_\mathrm{Q}=\omega_r/\sqrt{2}$ of the quadrupole surface mode (see section~\ref{subsec:back_GPE}). Furthermore the frequency width of this regime narrowed with decreasing ellipticity $\epsilon$, indicating that the resonant excitation of the quadrupole mode was responsible for the processes leading to vortex nucleation.  Subsequent investigations performed by these authors \cite{Madison01b} into the \emph{vortex-free} rotating ellipsoidal configurations of the condensate predicted by Recati \emph{et al.} \cite{Recati01}, gave further support to this conclusion. 
%%%%%%%%%%%%%%%%%%%%%%%%%%%%%%%%%%%%%%%%%%%%%%%%%%%%%%%%%%%%%%%%%%%%%%%%%%%%%%%%%%%%%%%%
\subsection{Work of Ketterle's group (MIT)}
Ketterle's group at MIT performed similar experiments to those of the ENS group, stirring $^{23}\mathrm{Na}$ condensates and creating large lattices (with $\gtrsim 100$ vortices) \cite{Abo-Shaeer01}.  They used a two-laser pattern to stir the condensate, similarly to the ENS experiments, however the resulting optical potential corresponded to a strongly anharmonic deformation of their trap potential.  In these experiments, the power of the stirring laser was ramped on, held for a variable hold time, and then ramped off.  They observed that for shorter stirring times ($t_\mathrm{stir} \lesssim 500 \mathrm{ms}$) the number of vortices nucleated was resonantly enhanced near the frequency $\omega_\mathrm{Q}$ of the quadrupole mode resonance, as observed in the ENS experiments.  They also observed a broadening of this resonance (with respect to the stirring frequency) with increasing stirring time $t_\mathrm{stir}$, and noted an increase in the number of vortices formed \emph{far} from this resonance as a function of the stirring time.  Vortices were observed to form at drive frequencies as low as $\approx0.27\omega_r$, which was however large compared to the thermodynamic stability frequency $\Omega_\mathrm{c}\approx0.08\omega_r$ for their system.

In the formation of large lattices, these authors noted the emergence of a blurry nonequilibrium structure at short times after the stirring, with density features they attributed to misaligned, disordered vortices.  Taking time-of-flight images after increasing wait times, they observed the increasing alignment and ordering of the vortices as time progressed.  They noted also that off-resonance stirring of the condensate resulted in lattices with fewer vortices, which took longer to equilibrate into a crystalline structure.  These authors argued that the minimum frequencies for vortex nucleation they observed were in fair agreement with the Landau instability frequency (section~\ref{subsec:back_superfluidity}) of their condensate (corresponding to surface modes with angular-momentum projection $m\approx18$, yielding $\Omega_\mathrm{L}\approx0.3\omega_r$).   They noted however that their anharmonic stirring potential leads to strong deformations of the condensate over a wide range of frequencies, and thus allows for the excitation of large-$m$ surface modes, which significantly complicates the analysis of the nucleation process in these experiments.   
%%%%%%%%%%%%%%%%%%%%%%%%%%%%%%%%%%%%%%%%%%%%%%%%%%%%%%%%%%%%%%%%%%%%%%%%%%%%%%%%%%%%%%%%
\subsection{Work of Foot's group (Oxford)}
Foot's group studied the nucleation of vortices by stirring a condensate of $^{87}\mathrm{Rb}$ atoms with a purely magnetic trapping arrangement \cite{Hodby02}.  They achieved the (approximately) elliptic deformation of their oblate time-orbiting-potential trap by modulating the $x$ and $y$ components of the applied magnetic bias fields. 
In practice, they first condensed the Bose cloud in a `static' (cylindrically symmetric) trap not subject to any rotation, achieving temperatures of $\sim0.5T_\mathrm{c}$.  They then ramped on the \emph{ellipticity} of the constantly rotating trap deformation from zero to some final value over a period of time (200 ms).  In contrast to the ENS and MIT experiments, the evaporative rf shield was \emph{not} engaged during the stirring process.  They reported significant heating of the atomic field during the stirring process, with the final temperature rising to approximately $0.8T_\mathrm{c}$ in the absence of the rf shield.  For a given trap ellipticity, they observed that the nucleation of vortices occurred for a narrow range of rotation frequencies centred approximately around the quadrupole resonance frequency $\omega_\mathrm{Q}$.  In agreement with the ENS results, they found the breadth of the resonance increased with increasing ellipticity $\epsilon$.  
In further investigations, they fixed the drive frequency, and ramped the trap anisotropy on adiabatically, and thus measured the critical anisotropy for vortex nucleation at fixed angular velocity.  In contrast to the results of Madison \emph{et al.} \cite{Madison00,Madison02}, they observed the nucleation of vortices for drive frequencies $\Omega < \omega_\mathrm{Q}$, with an increasing ellipticity required for vortex nucleation as the drive frequency was reduced further below the resonance frequency.  They suggested several possible explanations for this phenomenon: that (i) it was due to the thin tails of the instability locus predicted by the theory of Sinha and Castin \cite{Sinha01} (discussed in section~\ref{sec:stir_theory}), or; (ii) higher-order anisotropies or anharmonicities of the trap were present, leading to the excitation of other surface modes (cf. \cite{Madison02}), or; (iii) the presence of the thermal cloud was responsible for facilitating the Landau instability in this regime.  They investigated the effect of temperature by conducting experiments with the rf shield left on, mitigating the heating of the condensate during the stirring process, obtaining a final temperature $T\approx0.5T_\mathrm{c}$, and observed no significant change in the nucleation curve.  They did note, however, that the presence of the shield had a significant impact on the relaxation and decay of the vortex lattices, with stable lattices being much more likely to form and persisting for longer in the presence of the rf shield.
%%%%%%%%%%%%%%%%%%%%%%%%%%%%%%%%%%%%%%%%%%%%%%%%%%%%%%%%%%%%%%%%%%%%%%%%%%%%%%%%%%%%%%%%%%%%%%%%%%%%%%%%%%%%%%%%%%%%%%%%%%%%%%%%%%%%
\section{Theory}\label{sec:stir_theory}
%%%%%%%%%%%%%%%%%%%%%%%%%%%%%%%%%%%%%%%%%%%%%%%%%%%%%%%%%
\subsubsection{Work of Feder \emph{et al.}\,: Anomalous modes, and GP simulations of stirring}
Feder \emph{et al.} \cite{Feder01a} identified that while global thermodynamic considerations show that a vortex-containing state is the ground state for rotation rates exceeding some frequency $\Omega_\mathrm{c}$, the imposition of a rotation at this frequency is not necessarily sufficient to lead to vortex nucleation.  They noted that in general, there is a higher frequency $\Omega_\mathrm{v}$ at which surface excitations become unstable to growth by the Landau instability (section~\ref{subsec:back_superfluidity}).  Furthermore, in the ENS experiments, the frequency at which vortices are first observed is higher again than $\Omega_\mathrm{v}$.  In their paper \cite{Feder01a} they focussed on a third relevant frequency: the frequency $\Omega_\mathrm{m}$ at which a condensate containing a vortex first becomes \emph{metastable} (section~\ref{subsec:back_vortices_in_conds}).  In the Bogoliubov approach to the description of the excitations of the vortex, this is the rotation frequency at which the anomalous mode(s) of the vortex state (which have negative energies in the lab frame \emph{by definition}) are all Doppler shifted to positive values by the rotation.  They noted that for prolate traps (such as those of the ENS experiments), additional anomalous modes (associated with deformations of the vortex line) appear, and can lead to a metastability frequency $\Omega_\mathrm{m}$ which can exceed both $\Omega_\mathrm{v}$ and $\Omega_\mathrm{c}$.  They conjectured that the comparatively high rotation rate required in the ENS experiments was due to the instability of the vortex state below this metastability frequency.  However, subsequent studies of vortex formation in highly prolate condensates by Dalibard's group \cite{Madison01b} found no evidence of the inhibition of vortex nucleation due to anomalous modes.

Feder and his colleagues also performed the first \emph{time-dependent} Gross-Pitaevskii equation simulations of the stirring process, evolving a condensate corresponding to the ENS parameters in the presence of an abruptly introduced, rotating trap anisotropy.  They performed two such simulations (performed in the frame of the rotating trap), with drive frequencies $\Omega=0.7\omega_r$ and $\Omega=0.8\omega_r$ respectively.  In the former case, they found no nucleation of vortices, while in the case $\Omega=0.8\omega_r$, they observed vortices nucleating at the condensate surface and migrating into the centre of the atomic cloud.  However, the motion of the vortices remained turbulent, and the authors concluded that dissipative processes beyond the GPE may be required to facilitate the crystallisation of the vortex array to a rigid lattice.
%%%%%%%%%%%%%%%%%%%%%%%%%%%%%%%%%%%%%%%%%%%%%%%%%%%%%%%%%
\subsubsection{Work of Sinha and Castin\,: Dynamical instabilities of the condensate}
Sinha and Castin made a significant advance \cite{Sinha01} in the understanding of the role of quadrupole oscillations in the stirring experiments, identifying the \emph{dynamical instability} of this surface mode as crucial in the vortex nucleation.  They considered a classical hydrodynamic description \cite{Stringari96} of the condensate flow pattern in the regime of quadrupole oscillations, from which they determined numerically the conditions (rotation frequency and trap anisotropy) for which the quadrupole oscillations became dynamically unstable (see section~\ref{subsec:theory_Bogoliubov}).  They considered two distinct stirring scenarios: (i) an adiabatic scenario, in which the ellipticity of the trap is held constant, while the rotation frequency $\Omega$ of the anisotropy is ramped up adiabatically from zero; (ii) abruptly turning on the stirrer at a constant rotation frequency.  In the first scenario, the condensate adiabatically follows a steady-state rotating-ellipsoid pattern, as previously predicted by Recati \emph{et al.} \cite{Recati01}.  The flow pattern of the equilibrium condensate in this regime is described (in the hydrodynamic limit) by quadratic ans\"atze for the density and phase, and so these authors characterised the dynamic stability of the condensate mode in the rotating trap by considering the (linearised) evolution of small deviations about this flow pattern, employing polynomial ans\"atze for the density and phase deviations of order $n=3,4,\cdots$.  Constructing and diagonalising the evolution operator for these deviations, they identified the parameters for which modes with eigenvalues $\lambda$ with positive real parts, corresponding to dynamically unstable excitations of the condensate, appeared.  They found that the locus of points in the $\Omega$-$\epsilon$~plane for which the condensate is unstable forms a pattern comprised of several crescent structures\footnote{Anglin \cite{Anglin03} has shown that the appearance of such unstable eigenvalues in narrow `finger-like' regions of parameter space is a generic feature of dynamically unstable interacting many-body systems.} with, in particular, a broad base in the limit $\epsilon\to 0^+$,  occurring for $\Omega>\omega_\mathrm{Q}$.
 In the second scenario, which corresponds to the (later) experiments of the ENS group \cite{Madison02}, the ellipticity of the stirrer (with fixed rotation frequency) is abruptly turned on.  Adiabatic following is not applicable in this case, so they calculated the time-dependent evolution operator for the linearised deviations as a function of time and considered the growth of the largest eigenvalue of this operator with increasing time $t$.  They found that in this abrupt scenario, the quadrupole oscillations were dynamically unstable in a narrow range of $\Omega$, given approximately by $\Omega \in [0.71, 0.74]$ for an ellipticity $\epsilon=0.01$.  These results agree well with those of ENS experiments \cite{Madison02}, strongly supporting the hypothesis that the dynamical instabilities of the quadrupole oscillations are instrumental in the formation of vortices.
%%%%%%%%%%%%%%%%%%%%%%%%%%%%%%%%%%%%%%%%%%%%%%%%%%%%%%%%%
\subsubsection{Work of Tsubota \emph{et al.}\,: The role of dissipation}
Tsubota \emph{et al.} \cite{Tsubota02a} noted the necessity of \emph{dissipation} to allow the condensate to relax to the vortical state, which is of a lower energy (in the frame co-rotating with the trap) than the initial vortex-free state.  They therefore considered the solution of a modified GPE, which includes a \emph{phenomenological} damping term, following the earlier work of Choi \emph{et al.} \cite{Choi98}.  In essence, this approach amounts to evolving the time-dependent GP in \emph{complex time}, which hybridises the standard real-time evolution of the equation with the imaginary-time approach to finding ground states of the time-independent GP equation (see, e.g., \cite{Dalfovo96}).  In general, such a damping term leads to a change in normalisation of the wave function, however these authors continuously adjusted the chemical potential appearing in the GP equation throughout the evolution, in order to ensure conservation of the GP-wave-function normalisation.  These authors used this modified GP equation to simulate the stirring of the condensate with an elliptical potential and observed a transition to the vortex-lattice state.  However, they found different driving frequencies for the onset of lattice formation depending on the stirring procedure, namely $\Omega=0.57\omega_r$ for the `abrupt' driving scenario and $\Omega=0.75\omega_r$ in the `adiabatic' scenario.  While it seems clear that the latter case involves the dynamical instability of the quadrupole mode, these authors offered no interpretation for the lower frequency they found in the `abrupt' scenario.
%%%%%%%%%%%%%%%%%%%%%%%%%%%%%%%%%%%%%%%%%%%%%%%%%%%%%%%%%
\subsubsection{Work of Williams \emph{et al.}\,: ZNG theory for surface-mode growth}
Subsequently to the work of Tsubota \emph{et al.} \cite{Tsubota02a}, Williams \emph{et al.} \cite{Williams2002a} considered the dissipative effects of coupling of the condensate mode to the thermal cloud within the `ZNG' Gross-Pitaevskii-Boltzmann theory \cite{Zaremba99}\footnote{We note that this formalism is reviewed (along with many others) in great detail in the tutorial review of Proukakis and Jackson \cite{Proukakis08}.}.  They derived analytical results for the damping of a surface mode of angular-momentum projection $m$ along the trap axis, due to the $C_{12}$ term of the ZNG theory, which describes collisions between condensed and noncondensed atoms which result in the transfer of atoms into (or out of) the condensate.  They found that in the presence of a rotating thermal cloud, the damping of surface modes which `rotate' with the opposite sense to the cloud's circulation increases linearly with the rotation frequency of the cloud, while the damping of modes that rotate with the \emph{same} sense as the cloud \emph{decreases} with increasing $\Omega$.  Moreover, they found that the damping of a given mode with angular momentum projection $m>0$ changes sign (so as to become \emph{growth}) for a cloud driving frequency $\Omega_m = \omega_m/m$, i.e., they derive explicitly the result that the onset of Landau instability (negative excitation frequency) of a surface mode in the GP formalism leads to the spontaneous \emph{growth} of this mode in the presence of a thermal cloud rotating above the corresponding Landau frequency $\Omega_m$ \footnote{We note, however, that their calculation also reveals a shift of the relevant mode frequencies $\omega_m$ from their $T=0$ values due to the coupling to the thermal cloud.}.
%%%%%%%%%%%%%%%%%%%%%%%%%%%%%%%%%%%%%%%%%%%%%%%%%%%%%%%%%
\subsubsection{Work of Penckwitt \emph{et al.}\,: SGPE theory for lattice growth and stabilisation}
Penckwitt \emph{et al.} \cite{Penckwitt02} considered the problem of vortex nucleation within the SGPE framework of Gardiner \emph{et al.} \cite{Gardiner02}.  Their model includes the dissipative effect of collisions between the condensate and noncondensed atoms, similarly to the ZNG model of Williams \emph{et al.} \cite{Williams2002a} , providing a more rigorous basis for a damped-GPE approach such as that of Tsubota \emph{et al.} \cite{Tsubota02a}.  Their thermal atoms are treated as a bath of fixed temperature, chemical potential, and angular velocity, and the latter can be distinct from that of the rotating trapping potential.  In contrast to the work of Tsubota \emph{et al.}, the chemical potential appearing in the evolution equation used in this work has the physical meaning of the chemical potential of the thermal cloud, and the condensate population grows (or shrinks) until the condensate chemical potential matches that of the bath.  Penckwitt \emph{et al.} focussed primarily on the scenario of a vortex lattice forming as a rotating thermal cloud condenses in a cylindrically symmetric trap, which was demonstrated experimentally by Haljan \emph{et al.} \cite{Haljan01a}. In addition to their numerical simulations of this growth scenario, they found analytically from their model that growth occurs for surface modes satisfying the Landau instability criterion, and furthermore derived an analytic result for the gain profile of the surface modes, finding good agreement with their numerical results.  Turning their attention to the condensate stirring, they considered the `abrupt' stirring scenario, with a drive frequency $\Omega=0.65\omega_r$, both in the absence and the presence of a thermal cloud rotating at the same angular velocity.  They found coherent, stable quadrupole oscillations of the condensate in the former case, and the nucleation of vortices in the presence of the cloud.    On the basis of their results, the authors conjectured that the nucleation of vortices and their stabilisation into a rigid lattice is driven by the rotating thermal cloud, and that as such, the principal effect of the rotating trap anisotropy in experiments is to generate such a thermal cloud.  
%%%%%%%%%%%%%%%%%%%%%%%%%%%%%%%%%%%%%%%%%%%%%%%%%%%%%%%%%
\subsubsection{Work of Kasamatsu \emph{et al.}\,: Separating the roles of the instabilities}\enlargethispage{-\baselineskip}
Following on from their earlier work \cite{Tsubota02a}, the group of Ueda revisited the condensate-stirring scenario in a second, longer article \cite{Kasamatsu03}.  In the first part of their paper, they considered again the quadrupole oscillations of the condensate in the presence of the rotating elliptical trap anisotropy.  Adopting the quasiparticle-projection method of \cite{Morgan98}, they considered the evolution of the surface modes of the condensate in the presence of the rotating drive.  The interpretation of the quasiparticle projection technique is somewhat dubious in this regime, as (i) the depletion of the initial circularly symmetric condensate mode is substantial, invalidating a linearised treatment and (ii) the coherent quadrupole oscillations described by the GPE correspond to the \emph{collective} excitation of the condensate, i.e., a deformation of the condensate mode rather than the population of quasiparticle modes about it (see, e.g., \cite{Castin98}).  Nevertheless, their procedure does yield a meaningful projection onto the relevant angular momentum components.  They show that driving of the quadrupole mode at frequencies below its resonance leads to \emph{recurrence} of the initial condensate state, while driving at higher frequencies leads to \emph{stochastisation} of the GP evolution~\cite{Sinatra99}, in which case the trajectories become irregular and chaotic.  They note that this is analogous to the well-known threshold behaviour of chaos and ergodicity in some nonlinear systems, as famously discovered in the pioneering numerical investigations of Fermi, Pasta, and Ulam \cite{Fermi65}.  They therefore conjectured that the role of the dynamical instability identified by Sinha and Castin \cite{Sinha01} is to generate such stochastisation of the condensate evolution.  

Armed with the interpretation of the phenomenological damping term furnished by the work of \cite{Williams2002a} and \cite{Penckwitt02}, they considered again the evolution of the condensate mode in the damped-GPE approach.
They concluded from these investigations that the Landau instability of surface modes in the presence of the rotating thermal cloud is the fundamental mechanism of vortex nucleation (and, in particular, the origin of the critical frequency $\Omega=0.57\omega_r$ observed in their earlier simulations \cite{Tsubota02a}).  They then considered the relation between the Landau instability of surface modes and the dynamical instability of the quadrupole oscillations of the condensate, and conjectured that the role of the dynamical instability is to generate a thermal component of the field, thus explaining the later results of the ENS group \cite{Madison02}.  They noted that this also explains why vortex nucleation was occasionally observed at significantly lower frequencies: it is possible that vortex nucleation may occur simply due to Landau instability driven by the (small) thermal cloud already present in the experiments, but that this process is very slow, and in general the excitation of the dynamical instability is therefore necessary to form a sufficient thermal component to allow the condensate to relax to a vortical state.  In their summation, they identified the need for a framework beyond the Gross-Pitaevskii theory in order to describe the dynamics of vortex nucleation as driven by a dynamically generated thermal component, arising from the dynamical instability of the quadrupole mode\footnote{Indeed, as the dynamical instability of the condensate involves (by definition) the exponential growth of noncondensate modes, the onset of the dynamical instability signifies the breakdown of the Gross-Pitaevskii theory (see section~\ref{subsec:theory_Bogoliubov}).}.
%%%%%%%%%%%%%%%%%%%%%%%%%%%%%%%%%%%%%%%%%%%%%%%%%%%%%%%%%
\subsubsection{Work of Lundh \emph{et al.}\,: GP investigations of sub-resonant nucleation}
Lundh \emph{et al.} \cite{Lundh03} performed systematic investigations of the vortex nucleation with time-dependent GPE calculations.  They considered the scenario of \cite{Hodby02} in which the ellipticity of the constantly rotating trap is ramped on adiabatically, and measured the critical ellipticity for vortex nucleation as a function of the drive frequency.  At frequencies \emph{above} the hydrodynamic resonance frequency $\omega_\mathrm{Q}$ of the quadrupole surface mode, they found good agreement with the hydrodynamic predictions of references~\cite{Recati01,Sinha01}.  They also found reasonable agreement with the results of \cite{Hodby02} for the critical ellipticities at sub-resonant stirring frequencies ($\Omega < \omega_\mathrm{Q}$).  They suggested that vortex nucleation in this regime resulted from a mechanism similar to that discussed by Kr\"amer \emph{et al.} \cite{Kramer02}: under quadrupole deformation of the condensate, the barrier to vortex nucleation (section~\ref{subsec:back_vortices}) admits a saddle point which may become lower in energy than the isotropic vortex-free state, allowing the system to pass to the lower energy vortex state.  However, we note that this effect only addresses the \emph{thermodynamic} stability of the state; i.e., dissipation would still be required in order to allow the condensate to pass to the vortex state.  It is in our opinion more likely that the dynamical instability mechanism \cite{Sinha01} underlies the sub-resonant vortex nucleation observed in their GP simulations and in the experiments of Hodby \emph{et al.} \cite{Hodby02}.

These authors also consider the dynamics of the stirred field over longer time scales, and obtain results similar to those of Feder \emph{et al.} \cite{Feder01a}, with a turbulent field developing after the instability, and the vortices failing to crystallise on the time scales they consider.  
%%%%%%%%%%%%%%%%%%%%%%%%%%%%%%%%%%%%%%%%%%%%%%%%%%%%%%%%%
\subsubsection{Work of Lobo \emph{et al.}\,: Dissipation from turbulence}
The work of Lobo \emph{et al.} \cite{Lobo04} was the first attempt, using a framework beyond the phenomenologically damped Gross-Pitaevskii theory,  to describe the dynamics of vortex nucleation following the stirring of an (initially) zero-temperature condensate.  Following from their earlier work on the truncated-Wigner formalism \cite{Sinatra01,Sinatra02}, these authors considered the evolution of a classical field under the action of the Gross-Pitaevskii \emph{equation}, but interpreted the field as representative of the whole atomic field (see chapter~\ref{chap:cfield}).  In this approach, the turbulent nonlinear dynamics of the classical field provides for the \emph{internal} dissipation of the energy liberated by the condensate during the vortex nucleation.  These authors considered the evolution of a prolate condensate on a 3D Cartesian grid, in the frame of the trapping potential, whose rotation frequency is adiabatically ramped up.  They observed the onset of the dynamical instability of the initial GP mode, and the subsequent evolution of the field into a turbulent state, after which vortex lines were nucleated into the condensate, and ultimately relaxed to a lattice configuration co-rotating with the drive.  The authors estimated the temperature of the field at the end of their simulations, by assuming that the excess energy of the field over that of the ground lattice state was equilibrated over weakly interacting (Bogoliubov) modes. They found that the temperature was very small, of the order of one-tenth of the chemical potential of the field.  This paper thus provided a significant advance in understanding of the problem: by considering the field $\psi$ evolved by the GPE as the whole atomic field, rather than merely the condensate wave function, the dynamical instability of the condensate and the subsequent thermalisation dynamics of the atomic field could be studied within a unified framework.  Moreover, their results suggest that the turbulent states obtained in the simulations of \cite{Feder01a} and \cite{Lundh03} are in fact \emph{finite-temperature} states, beyond the validity of the GP approach.  These authors also considered the stirring of initially finite-temperature classical-field configurations, and observed vortex nucleation for stirring above the thermodynamic critical frequency $\Omega_\mathrm{c}$. 
%%%%%%%%%%%%%%%%%%%%%%%%%%%%%%%%%%%%%%%%%%%%%%%%%%%%%%%%%
\subsubsection{Work of Parker and Adams\,: Turbulence and symmetry breaking}
Subsequent work on the scenario of vortex nucleation by stirring with an elliptical trap anisotropy was performed by Parker and Adams \cite{Parker05a,Parker06b}.  These authors considered the stirring of two-dimensional condensates by a rotating potential, modelled by integrating the time-dependent GPE in the laboratory frame, with a time-dependent rotating potential, using the Crank-Nicolson method (which is a finite-difference scheme; see section~\ref{subsec:numeric_spatial_derivs}).  In their simulations, the field appears to become turbulent following the onset of the instability, and they identified the characteristic \emph{Kolmogorov spectrum} of turbulence \cite{Zakharov92} in the nonequilibrium field. However, contrary to the earlier work of Feder \emph{et al.} \cite{Feder01a} and Lundh \emph{et al.} \cite{Lundh03}, they observed the vortex configuration to ultimately relax to a rigid, crystalline lattice.  Parker and Adams stressed the role of the breaking of the initial twofold rotational symmetry of the GP solution as being essential in allowing the field to relax to the equilibrium lattice configuration.  

They also considered the effect of imposing a momentum space `cutoff' on their simulations, which they implemented by periodically removing the portion of the GP solution outside some radius $\hbar k_\mathrm{c}$ in momentum space\footnote{This operation should be contrasted to the deletive implementations of the projection operator in formally projected classical-field methods (see, e.g., \cite{Norrie05a}), in which the deletion is performed \emph{continuously}, so that no loss of population actually occurs (see, e.g., \cite{Davis_DPhil}).}.
The authors interpreted this process as analogous to an evaporative-cooling type procedure, and noted that despite a reduced final atom number in the presence of this elimination procedure, the final vortex configuration and damping rates were insensitive to the cutoff wave vector $k_\mathrm{c}$.  They therefore suggested that the relaxation process is a zero-temperature phenomenon.
%%%%%%%%%%%%%%%%%%%%%%%%%%%%%%%%%%%%%%%%%%%%%%%%%%%%%%%%%%%%%%%%%%%%%%%%%%%%%%%%%%%%%%%%%%%%%%%%%%%%%%%%%%%%%%%%%%%%%%%%%%%%%%%%%%%%
\section{Outstanding questions}\label{sec:stir_questions}
The description of the condensate-stirring process has presented a strong challenge for many-body theories: to describe the full dynamical evolution of the system, from cold condensate, through strongly turbulent nonequilibrium dynamics, to a rotating thermal equilibrium. 
Although the work of Kasamatsu \emph{et al.} \cite{Kasamatsu03} established the role of the dynamical instability in generating the thermality required to facilitate the nucleation of vortices, and the work of Lobo \emph{et al.} \cite{Lobo04} showed that the requisite dissipative dynamics of vortex-lattice formation can emerge from the deterministic dynamics of a nonlinear-Schr\"odinger classical-field evolution, a number of important questions about the lattice-formation process remained open.  
\newline\newline\emph{Accuracy of classical-field approaches}\newline
The calculations of \cite{Feder01a,Lundh03,Lobo04,Parker05a,Parker06b} each exhibit the nucleation of vortices from the GP equation, while the simulations of \cite{Lobo04} and \cite{Parker05a,Parker06b} additionally result in the relaxation to a vortex-lattice configuration.  Artifacts, inaccuracies, and numerical noise in the solutions of \emph{nonlinear} differential equations such as the GP equation can lead to the solution taking a completely spurious path, and thus produce unphysical results.
It is important to investigate the nucleation and damping phenomena exhibited by these earlier works with the more sophisticated projected-GPE approach, to determine whether the outcomes of these previous numerical studies result from the intrinsic nonlinear dynamics of the classical-field solutions, or if they are due to inaccuracies of the numerical methods\footnote{We follow here a guiding principle of scientific research informally summarised by Feynman \cite{Feynman98}: ``... if there is something really there, and you can't see good because the glass is foggy, and you polish the glass and look clearer, then it's more obvious that it's there, not less."}.
\newline\newline\emph{Emergence of thermal behaviour}\newline
In the classical-field viewpoint of Lobo \emph{et al.} \cite{Lobo04}, it is supposed that the nonlinear dynamics of the classical field lead to an \emph{emergent} thermal component of the field, which provides the dissipation required for vortex nucleation and damping.  The dynamical creation, evolution and development of the thermal field is a theoretical question of fundamental interest.  In the previous calculations, only Lobo \emph{et al.} \cite{Lobo04} attempted any characterisation of the thermal field, and they only considered the final equilibrium thermal field, and only quantified its temperature.  It is therefore important to perform a quantitative and detailed examination of the \emph{development} of the thermal component of the field following the onset of the dynamical instability of the quadrupole oscillations of the condensate. 
\newline\newline\emph{Role of symmetry breaking}\newline
The authors of \cite{Parker06b} stress the importance of the breaking of the twofold rotational symmetry of the system in the processes leading to lattice formation.  In all previous theoretical studies, the rotational symmetry of the condensate was broken either by numerical error \cite{Feder01a,Lundh03,Lobo04}, by \emph{ad hoc} seeding of field modes with noise \cite{Penckwitt02}, or by including perturbations of the trapping potential \cite{Kasamatsu03,Parker05a,Parker06b} intended to mimic the noise sources present in realistic experiments.
None of these approaches to the symmetry breaking is satisfactory, as they rely either on the explicit failure of the integrator in propagating the classical-field solutions, or relate to physical effects in a somewhat arbitrary and uncontrolled way.  The truncated-Wigner theory (section~\ref{sec:cfield_twa}), on the other hand, informs us that any classical-field trajectory must contain a certain irreducible level of fluctuations, corresponding to the quantum uncertainty present in the corresponding quantum field theory, even at zero temperature. Thus by including this representation of \emph{quantum noise} in the initial conditions of field trajectories, we can investigate the dynamical instability and the resulting thermalisation of the field in a \emph{minimal} stirring scenario, in which the system is considered to be otherwise `ideal', and the only source of symmetry breaking is the inescapable quantum-uncertain nature of the true physical system.
\newline\newline\emph{Nature of the equilibrium state in 2D}\newline
Finally, a major question raised by the simulations of \cite{Feder01a,Lundh03,Parker05a,Parker06b} is the nature of the equilibrium configuration of vortices in 2D systems.  Both \cite{Feder01a} and \cite{Lundh03} reported highly turbulent motion of the vortices at the end of their simulations.  The authors of the former paper conjectured that the lattice crystallisation process was dependent on some dissipative process beyond the GP theory, such as coupling to the thermal cloud, while the authors of \cite{Lundh03} suggested (in light of the then preliminary results of \cite{Lobo04}) that the time scale for relaxation in their 2D system may simply be very long.  In contrast, Parker \emph{et al.} \cite{Parker05a,Parker06b} have shown in their 2D calculations the relaxation of the turbulent field subsequent to the instability to a rigid vortex lattice, and suggest that the formation is a zero-temperature phenomenon.  It is therefore important to determine conclusively the nature of the equilibrium state of the stirred condensate in 2D, and whether the lattice-formation scenario observed in \cite{Parker05a,Parker06b} is physical or simply the result of numerical error.  This is of particular importance as all experiments on condensate stirring to date have been performed in fully 3D geometries, while two-dimensional models are often assumed in theoretical investigations of vortex dynamics in Bose condensates.
%%%%%%%%%%%%%%%%%%%%%%%%%%%%%%%%%%%%%%%%%%%%%%%%%%%%%%%%%%%%%%%%%%%%%%%%%%%%%%%%%%%%%%%%%%%%%%%%%%%%%%%%%%%%%%%%%%%%%%%%%%%%%%%%%%%%
%%%%%%%%%%%%%%%%%%%%%%%%%%%%%%%%%%%%%%%%%%%%%%%%%%%%%%%%%%%%%%%%%%%%%%%%%%%%%%%%%%%%%%%%%%%%%%%%%%%%%%%%%%%%%%%%%%%%%%%%%%%%%%%%%%%%

\chapter{Condensate stirring}
\label{chap:stirring}
%%%%%%%%%%%%%%%%%%%%%%%%%%%%%%%%%%%%%%%%%%%%%%%%%%%%%%%%%%%%%%%%%%%%%%%%%%%%%%%%%%%%%%%%%%%%%%%%%%%%%%%%%%%%%%%%%%%%%%%%%%%%%%%%%%%%
%%%%%%%%%%%%%%%%%%%%%%%%%%%%%%%%%%%%%%%%%%%%%%%%%%%%%%%%%%%%%%%%%%%%%%%%%%%%%%%%%%%%%%%%%%%%%%%%%%%%%%%%%%%%%%%%%%%%%%%%%%%%%%%%%%%%
In this chapter, we apply the PGPE formalism to the nonequilibrium dynamics of the condensate stirring scenario.  We consider a highly oblate (quasi-2D) condensate, initially at zero temperature in a cylindrically symmetric trap.  In accordance with the truncated-Wigner interpretation of classical-field trajectories, we include small Gaussian fluctuations of the classical field, representing the initial vacuum occupation of Bogoliubov modes, and explicitly breaking all symmetries of the Hamiltonian.  We then abruptly introduce a weak elliptical trap anisotropy, and simulate the ensuing dynamics of the atomic field in the rotating frame.  We observe the rapid thermalisation of material ejected from the condensate bulk during the dynamical instability of the quadrupole mode.  This material forms a rotating thermal cloud which drives the Landau instability of surface modes and leads to the nucleation of vortices.  We observe the damping of the turbulent vortex motion, and the slow rotational equilibration of the vortices with the rotating trap and thermal cloud.  We find, however, that the equilibrium state of the stirred atomic field in this quasi-2D geometry is not a rigid vortex lattice but a disordered \emph{vortex-liquid} state.  We characterise the thermodynamic parameters of the thermal field, and monitor the change in these parameters during the equilibration of the system.  We find that the final thermodynamic parameters of the field are well predicted by simple arguments based on energy conservation.  We discuss the absence of true Bose condensation in the equilibrium vortex-liquid state, and discuss how the temporal correlations of this phase distinguish it from the thermal component of the field.

This chapter is organised as follows.  In section~\ref{sec:stir_System} we define the parameters of the physical system we model, and our classical-field approach to its evolution.  In section~\ref{sec:stir_Results} we present the results of a typical, representative simulation.  In section~\ref{sec:stir_analysis} we present a detailed analysis of this representative classical-field trajectory, quantifying the rotational response and the heating of the field during its evolution, and characterising the field coherence.  In section~\ref{sec:stir_parameters}, we consider how the behaviour of the field depends on the chemical potential (and thus size and interaction strength) of the initial condensate mode, and the effect of the classical-field cutoff height on the field dynamics.  In section~\ref{sec:stir_summary} we summarise our findings and present our conclusions.
%%%%%%%%%%%%%%%%%%%%%%%%%%%%%%%%%%%%%%%%%%%%%%%%%%%%%%%%%%%%%%%%%%%%%%%%%%%%%%%%%%%%%%%%%%%%%%%%%%%%%%%%%%%%%%%%%%%%%%%%%%%%%%%%%%%%
\section{System and simulation procedure}\label{sec:stir_System}
%%%%%%%%%%%%%%%%%%%%%%%%%%%%%%%%%%%%%%%%%%%%%%%%%%%%%%%%%%%%%%%%%%%%%%%%%%%%%%%%%%%%%%%%
\subsection{The system}
Following chapters~\ref{chap:precess}~and~\ref{chap:arrest}, we choose physical parameters corresponding to $^{23}\mathrm{Na}$ atoms confined in a strongly oblate trap, with trapping frequencies $(\omega_r,\omega_z)=2\pi\times(10,2000)$ rad/s.  We consider an initially zero-temperature, vortex-free condensate characterised by its chemical potential $\mu_\mathrm{i}$ in this trapping potential.  In this chapter we consider condensate chemical potentials $4\hbar\omega_r\leq \mu_\mathrm{i} \leq 20\hbar\omega_r$, corresponding to condensate populations $1600\lesssim N_0\lesssim 42\,000$.
At $t=0$, the trap is deformed into an ellipse in the $xy$ plane which rotates around the $z$ axis with angular frequency $\Omega$.  In the frame co-rotating at frequency $\Omega$, the single-particle Hamiltonian is of the form of equation~\reff{eq:cfield_split}, with trap anisotropy
\begin{equation}
	\delta V(\x) = \epsilon m\omega_r^2\left(y^2 - x^2\right).
\end{equation}
In this chapter we set the trap eccentricity to $\epsilon=0.025$, which is sufficient to excite the quadrupole instability at $\Omega\sim\omega_r/\sqrt{2}$ \cite{Recati01,Sinha01}.  We will always drive the system within the hydrodynamic resonance, choosing $\Omega\equiv0.75\omega_r$.  The sudden turn-on of the rotating potential at fixed $\epsilon$ and $\Omega$ means that the classical-field energy (equation~\reff{eq:cfield_HCF}) of the initial state evaluated in the rotating frame is a conserved quantity of the motion. 
We again adopt a 2D representation of the system, formalised in our classical-field method by choosing a cutoff $E_R=30\hbar\omega_r\ll\hbar\omega_z$, such that the low-energy space $\mathbf{C}$ excludes all modes with excitation along the $z$ axis (see section~\ref{subsec:dimless}).  
%%%%%%%%%%%%%%%%%%%%%%%%%%%%%%%%%%%%%%%%%%%%%%%%%%%%%%%%%%%%%%%%%%%%%%%%%%%%%%%%%%%%%%%%
\subsection{Initial state: inclusion of vacuum noise}
%%%%%%%%%%%%%%%%%%%%%%%%%%%%%%%%%%%%%%%%%%%%%%%%%%%%%%%%%
\subsubsection{Symmetry of the zero-temperature GP eigenstate}
A key issue in studying the condensate-stirring scenario within the framework of classical-field theory is in identifying the \emph{path} that the classical field takes in transforming from a vortex-free, zero-temperature condensate to a rotating (and, as we will show) finite-temperature field configuration.  In general, we expect the field to acquire thermality during its real-time evolution due to the ergodicity of the PGPE evolution, as discussed in section~\ref{sec:cfield_ergodicity}.  However, we face a significant issue in the present case where we seek to describe the thermalisation of an initially \emph{zero-temperature} condensate.  The $T=0$ ground GP state of the isotropic trap is indeed highly excited with respect to the ground GP state (of the same normalisation) in the rotating frame in which we expect equilibrium to result, and at equilibrium we expect the excess energy of the initial state to be distributed throughout the configuration space of the field in the form of thermal fluctuations.  However, as the initial GP state possesses circular rotational symmetry, the exact evolution of the state under the GPE with the elliptic potential we consider would retain unbroken twofold rotational symmetry for all time, and true thermalisation of the field would not occur\footnote{In fact, we can identify the highly symmetric GP eigenstate as a member of the `set of measure zero' for which ergodicity does not apply (section~\ref{sec:cfield_ergodicity}).}.  Moreover, the high degree of symmetry of the GP state impedes the dynamical instability, which results from the exponential \emph{amplification} of deviations about an otherwise stable GP trajectory (section~\ref{subsec:theory_Bogoliubov}).  In the limit of perfect numerical evolution of an initial GP eigenstate under the (projected) GP equation, no such deviations arise, and thus no dynamical instability will occur.  
In \emph{practical} numerical simulations the build-up of numerical error with time will eventually facilitate the dynamical instability and the ergodic thermalisation of the field.  However, this yields somewhat arbitrary results for the time scales for symmetry breaking and the relaxation processes which require it.  It is important then to ensure that the symmetry of the classical-field system is broken in a physically defensible way, rather than to rely on uncertain and uncontrolled numerical inaccuracies.
%%%%%%%%%%%%%%%%%%%%%%%%%%%%%%%%%%%%%%%%%%%%%%%%%%%%%%%%%
\subsubsection{Truncated-Wigner prescription for noise}\enlargethispage{-\baselineskip}
The resolution of the issue of symmetry breaking is found in the truncated-Wigner representation of the quantum-field theory for the Bose gas discussed in section~\ref{sec:cfield_twa}.  In the truncated-Wigner formalism, the PGPE appears as a classical equation of motion for field trajectories which sample the approximate evolution of the density matrix of the quantum-field system.  We recall that in this approach the classical-field trajectories are regarded as stochastic processes, even though in the truncated-Wigner formalism appropriate to the interacting atomic Bose gas, no explicit dynamical noise processes appear.  However, a finite amount of noise does appear in the initial conditions, even at zero temperature, representing quantum uncertainty in the physical state.  This provides a fundamental and quantitatively correct quantum-mechanical mechanism that breaks the symmetry of the Hamiltonian. 
While for finite-temperature states this noise can be safely neglected due to the presence of \emph{thermal} fluctuations, at zero temperature many important processes can only be initiated by quantum fluctuations.  

Thus, although we consider (as in chapter~\ref{chap:arrest}) the relaxation and equilibration dynamics of the classical-field trajectory on long time scales, precluding a formal reconstruction of quantum-field expectation values from the classical-field statistics, we follow the truncated-Wigner prescription in seeding our initial states with the \emph{irreducible} fluctuations of the quantum field theory which are essential for the initiation of the thermalisation dynamics of the field.
We therefore add a representation of the vacuum fluctuations to our initial states in the form of classical (complex) Gaussian noise of mean population equal to $\frac{1}{2}$-quantum per mode.   It is necessary, therefore, to choose appropriate basis modes to which the noise is added, so as to faithfully represent the ground state of the many-body system \cite{Steel98}.  For the interacting system, the appropriate basis modes in which to construct the vacuum will be distinct from the basis modes in which the system is propagated. 
In this work our primary aim is to include the possibility of spontaneous processes in the initial condition, so as to initiate the dynamical instability, and facilitate the ergodicity of the classical-field evolution.  It therefore suffices to populate the Bogoliubov modes orthogonal to the condensate mode with noise~\cite{Gardiner97a,Castin98,Steel98}\footnote{Self-consistent prescriptions for the addition of noise to the truncated-Wigner initial state exist \cite{Sinatra00,Sinatra01,Sinatra02}, but the resulting higher-order corrections to the vacuum state are not important for our application here.}.  The initial state is constructed from the GP ground state of the symmetric trap in the lab frame, $\phi_0(\mathbf{x})$, according to 
\begin{equation}
\psi(\mathbf{x})=\phi_0(\mathbf{x})+\sum_j \alpha_ju_j(\mathbf{x})+\alpha_j^*v_j^*(\mathbf{x}),
\end{equation}
where $(u_j,v_j)$ are the Bogoliubov modes orthogonal to $\phi_0(\mathbf{x})$, and $\alpha_j=(\xi_j+i\eta_j)/2$ for our system where the thermal population is initially zero. The independent (real) Gaussian distributed variables satisfy $\overline{\xi_j}=\overline{\eta_j}=\overline{\eta_i\xi_j}=0$ and $\overline{\xi_i\xi_j}=\overline{\eta_i\eta_j}=\delta_{ij}$.

Population of the quasiparticle basis constructed in this manner ensures that all surface modes of the condensate (that are resolvable within the condensate band) are seeded by noise, including those that are dynamically unstable in the presence of the rotating trap anisotropy.  The noise introduced here plays a role entirely analogous to that played by vacuum electromagnetic field fluctuations in triggering the spontaneous decay of a two-level atom \cite{Scully97}.  
All symmetries of the mean-field state are thus broken \cite{Sinatra00} \emph{before} the exponential growth of noncondensed field density \cite{Castin97} occurs during the dynamical instability.  
%%%%%%%%%%%%%%%%%%%%%%%%%%%%%%%%%%%%%%%%%%%%%%%%%%%%%%%%%
\subsubsection{Validity of adding Wigner noise}
As we have noted in section~\ref{sec:cfield_ergodicity}, due to the ergodicity of the PGPE, classical-field simulations tend to equilibria in which the energy of the field is equipartitioned over its normal modes.  Consequently, the energy (and population) added to the initial states of classical-field simulations according to the truncated-Wigner prescription eventually thermalises, and contributes spuriously to the equilibrium thermal fraction of the field (section~\ref{subsec:cfield_relevance_to_PGPE}). 
Sinatra \emph{et al.} \cite{Sinatra02} thus suggest a validity criterion $T_\mathrm{class} - T \lesssim T$ for the (initial) temperature $T$ of the simulation relative to the equilibrium temperature of the classical field $T_\mathrm{class}$.  We note however that the underlying issue is the relative quantity of energy added in the Wigner noise, as a fraction of the total thermal energy in the field.  In our stirring scenario, as viewed in the rotating frame,  the initial GP state is a highly \emph{collectively} excited state, the energy of which (above that of the ground vortex lattice state in this frame) is converted into thermal energy as the field relaxes (see section~\ref{subsec:stir_Thermo_params}).  The energy added to the initial state according to the truncated-Wigner prescription is therefore dwarfed by the energy liberated from the GP state during its relaxation.  Similarly, the normalisation and angular momentum of the initial vacuum noise constitute very small fractions of the total quantities contained in the equilibrium thermal field, and so we can be confident that the spurious thermalisation of the vacuum noise does not affect the relaxation dynamics of our trajectories significantly.  
%%%%%%%%%%%%%%%%%%%%%%%%%%%%%%%%%%%%%%%%%%%%%%%%%%%%%%%%%
\subsection{Evolution}
Using the initial state described above, we evolve field trajectories in the frame of the rotating trap ellipticity, i.e., we apply the projector in the frame $\Omega_\mathrm{p}=\Omega=0.75$.  Our procedure corresponds to the sudden introduction of the rotating anisotropy at time \mbox{$t=0$}.  We evolve each trajectory with the ARK45-IP algorithm (section~\ref{sec:rk4_app_ark45}), with accuracy chosen such that the relative change in field normalisation is $\leq 4\times 10^{-9}$ per time step taken, and the simulations presented here all have total fractional normalisation changes $\lesssim2\times 10^{-4}$ over their duration ($\sim10^4$ trap cycles).  Importantly, the change in rotating-frame \emph{energy} is of magnitude commensurate with that of the change in normalisation, so that the truncation error represents a loss of population from the system as a whole, rather than a preferential removal of high-energy components as in \cite{Parker06b}, and the relaxation of the condensate-band field is thus due to internal damping processes only. 
%%%%%%%%%%%%%%%%%%%%%%%%%%%%%%%%%%%%%%%%%%%%%%%%%%%%%%%%%%%%%%%%%%%%%%%%%%%%%%%%%%%%%%%%%%%%%%%%%%%%%%%%%%%%%%%%%%%%%%%%%%%%%%%%%%%%
\section{Simulation results}\label{sec:stir_Results}
We present here the results of a simulation with initial chemical potential $\mu_\mathrm{i} = 14\hbar\omega_r$ (corresponding to condensate with $N_0\approx 21~500$ atoms),  whose response to the stirring   is representative of systems throughout the range  $4\hbar\omega_r \leq \mu_\mathrm{i}  \leq 20 \hbar\omega_r$.   Using the criteria discussed in section~\ref{subsec:cfield_proj_cft}, we set the condensate-band cutoff $E_R = 3\mu_\mathrm{i} =42\hbar\omega_r$. The upper limit of the range investigated ($ \mu_\mathrm{i}  \sim 20 \hbar\omega_r $) is set by computational limitations: at the fixed rotation frequency $\Omega=0.75\omega_r$ the size of the Gauss-Laguerre (GL) basis scales approximately as $\mathcal{M} \propto E_R^2$, and the corresponding computational load scales as $\mathcal{O}(E_R^3)$ (section~\ref{sec:numerics_performance}), so that simulations rapidly become numerically expensive.  For the cutoff $E_R=42\hbar\omega_r$ the GL basis consists of $\mathcal{M} = 2028$ modes.  At the lower end of the range ($\mu_\mathrm{i}\lesssim 10\hbar\omega_r$) the simulation results differ somewhat both in the thermalisation process leading to vortex nucleation, and the behaviour of the vortex array, due to the reduced mean-field effects and small numbers of vortices, respectively.  These differences will be discussed in section~\ref{subsec:stir_vary_mu}. 

\begin{figure}
	\begin{center}
	\includegraphics[width=1.0\textwidth]{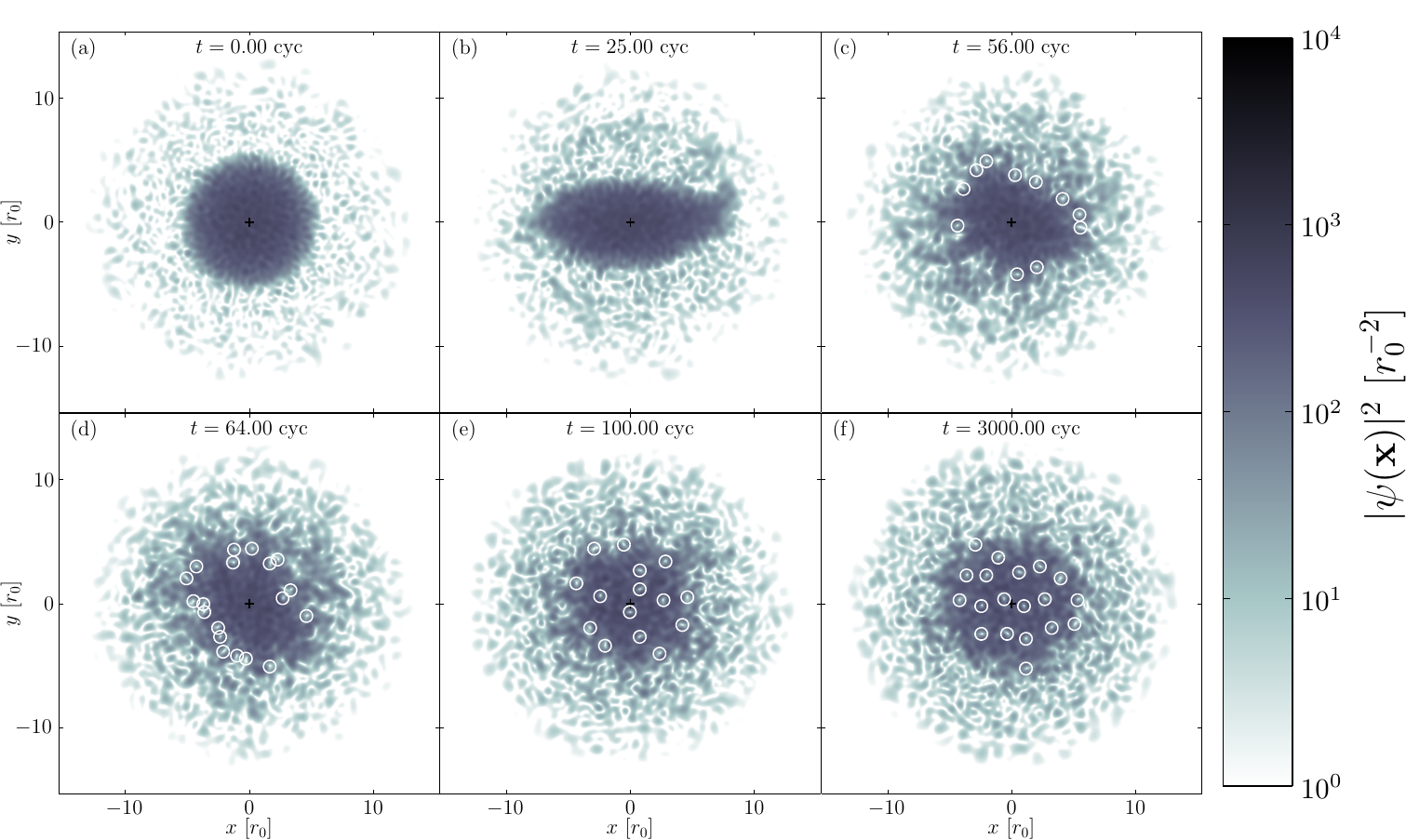}
	\caption{\label{fig:density_plots1}  Classical-field density in the rotating frame at representative times during the evolution.  Shown are (a) initial condition: lab-frame ground state plus vacuum noise, (b) ejection of material during the dynamical instability, (c) nucleation of vortices at the interface between the condensate and the noncondensate material, (d) penetration of vortices into the condensate bulk, (e) nonequilibrium state with rapid vortex motion, and (f) equilibrium state.  Vortices are indicated by white circles, and are shown only where the surrounding density of the fluid exceeds some threshold value, and $+$ marks the coordinate origin (trap axis).  Parameters of the simulation are given in the text.}
	\end{center}
\end{figure}

We will focus primarily on the case of $\mu_\mathrm{i} = 14\hbar\omega_r$, and note for comparison that a pure condensate (i.e., ground GP eigenstate) with the same number of particles rotating at $\Omega = 0.75 \omega_r$ would contain a lattice of $\approx18$ vortices. The response to the stirring is illustrated in the sequence of density distribution plots shown in figure~\ref{fig:density_plots1}, and exhibits the following key features. 
%%%%%%%%%%%%%%%%%%%%%%%%%%%%%%%%%%%%%%%
%%%%%%%%%%%%%%%%%%%%%%%%%%%%%%%%%%%%%%%
\newline\newline\emph{Dynamical instability and formation of thermal cloud}\newline 
The initial state (figure~\ref{fig:density_plots1}(a))  quickly becomes elongated,  with its long axis oscillating  irregularly relative to the  major axis of the trap.  The quadrupole oscillations   are dynamically unstable to the stirring perturbation \cite{Recati01,Sinha01,Parker06}, and since all the Bogoliubov modes of the condensate are effectively populated due to the representation of quantum noise in our simulation, some of the fluctuations  grow exponentially.  The effect is clearly seen in figure~\ref{fig:density_plots1}(b),  where matter streams off the tips of the elliptically deformed condensate in the direction of the trap rotation.  Material  is then ejected more or less continuously,  forming  a ring about the condensate,  with the latter oscillating in its motion and shape.  The ring  soon becomes turbulent and diffuse (figure~\ref{fig:density_plots1}(c)), losing coherence and forming a thermal cloud whose characterisation (section~\ref{subsec:stir_Thermo_params}) is a central theme of this chapter.  In section~\ref{subsubsec:stir_local_rot_prop} we show from an analysis of the angular momentum and density distribution that by  $t=50$ cyc the outer part of the cloud (beyond $r \approx 7r_0$) rotates as a normal fluid in rotational equilibrium with the drive.
\newline\newline\newline\emph{Vortex nucleation}\newline
After the thermal cloud has formed, surface oscillations of the condensate grow in magnitude (figure~\ref{fig:density_plots1}(d)), in accordance with the prediction of Williams \emph{et al.}~\cite{Williams2002a}, that such oscillations are unstable in the presence of a thermal cloud.  These fluctuations form a transition region between the central condensate bulk, characterised by comparatively smooth density and phase distributions (see also figure~\ref{fig:phase_and_correl_time}(a)), and an outer region  of thermally occupied highly energetic excitations. Long-lived vortices are nucleated in this transition region, and then begin to penetrate into the edges of the condensate region. Initially the vortices penetrate only small distances $r\approx r_0=\sqrt{\hbar/m\omega_r}$ with rapid and irregular motion, and visibly lag the rotation of the drive.  This process is one in which the \emph{emergent} thermal component of the field acts to provide the dissipation which was included phenomenologically in the simulations of Tsubota \emph{et al.} \cite{Tsubota02a} and Penckwitt \emph{et al.} \cite{Penckwitt02}.
\newline\newline\emph{Vortex-array formation}\newline
The nucleated vortices remain near the edge of the condensate for a considerable time  before they begin to penetrate substantially into the condensate.  Eventually vortices pass through the central region (by $t \approx 75$ cyc in this simulation), with trajectories that are largely independent, except  when  two vortices closely approach each other\footnote{In this and other simulations we have observed a tendency of vortices to cluster together in (short-lived) co-rotating pairs which orbit their geometric centre, a behaviour characteristic of two-dimensional superfluid turbulence (see reference \cite{Wang07} and references therein).}. The density of vortices within the central region gradually increases (figure~\ref{fig:density_plots1}(e)), and the vortices begin to form a rather disorderly assembly, which still lags behind the drive in its overall rotation about the trap axis.  The erratic motion of individual vortices and their differential rotation with the respect to the drive then gradually slow, and they become more localised, until by $t \approx 150$ cyc they have formed into an array rotating at the same speed as the drive.  This formation process can be interpreted as the damping of vortex motion by the mutual friction between them and the rotating thermal cloud \cite{Sonin87, Donnelly91}.  Alternatively, we can interpret the motion of the vortices as the result of low-energy excitations of the quantum fluid,  which are gradually damped by their interaction with high energy excitations \cite{Fedichev98a, Giorgini98,  Pitaevskii97}. The coupling of the high-energy modes to the low-energy excitations of the fluid drives their distribution towards equilibrium \cite{Sagdeev88, Sinatra99}.  The high energy modes comprising the thermal cloud equilibrate to a distribution with identifiable \emph{effective} thermodynamic parameters (see section~\ref{subsec:stir_Thermo_params}) on a relatively short time scale ($\sim 100$ cyc).  While the effective temperature remains approximately fixed, the effective chemical potential subsequently increases over a longer time scale ($\sim 3000$ cyc), as the distribution is shaped by its interaction with the low energy excitations.  By $t\approx1000$ cyc the vortex motion (as quantified by the mean vortex speed; section~\ref{subsec:stir_motional_damping}) reaches its equilibrium level, although the \emph{density} of vortices in the central bulk of the field continues to growth until $t\approx5000$ cyc (section~\ref{subsec:stir_Rotational_params}).  We find that the equilibrium populations of thermal excitations in this final state are such that the vortex distribution does not `crystallise' into a rigid lattice, and is more appropriately characterised as an equilibrium \emph{vortex liquid}, as we discuss in section~\ref{subsec:stir_Temporal}.   
%%%%%%%%%%%%%%%%%%%%%%%%%%%%%%%%%%%%%%%%%%%%%%%%%%%%%%%%%%%%%%%%%%%%%%%%%%%%%%%%%%%%%%%%%%%%%%%%%%%%%%%%%%%%%%%%%%%%%%%%%%%%%%%%%%%%
\section{Analysis}\label{sec:stir_analysis}
In this section we present the set of measurements we use to analyse the properties of the thermal cloud, and to characterise the coherence of the central region in which the vortices reside.  We begin by making a simple estimation of the energy and temperature the thermal cloud would acquire during an idealised lattice formation scenario.  Next we obtain the chemical potential of the thermal cloud by using a self consistent fitting procedure for the atomic distribution function, which also provides a measure of the temperature.  We characterise the rotational properties of different spatial regions of the  system, which provides evidence that during the vortex nucleation stage, the central part behaves as a superfluid, while the outer cloud behaves as a classical gas, rotating like a rigid body. 

The issue of condensate identification is a very important one,  and is most commonly done in terms of spatial correlation functions, using ensemble averages of quantum mechanical operators \cite{Leggett01,Dalfovo99}.  However, we will show that the varying vortex configurations within the ensemble have a destructive effect on the spatial order of the system we study here. 
This leads us, in section~\ref{subsec:stir_Temporal}, to investigate temporal correlation functions, which we show can be used to characterise the local coherence of the field, and consequently distinguish the quasi-coherent behaviour of the central region from that of the thermal cloud.

Following the first appearance of an identifiable vortex array, there is a long period over which it slowly relaxes to an equilibrium state.   In section~\ref{subsec:stir_motional_damping} we measure the motional damping of the vortices and interpret it in terms of their interaction with the thermal component of the field  \cite{Fedichev99}.  Note that all results and figures in this section (figures~\ref{fig:example_temp_fit}-\ref{fig:vortex_damping1}) correspond to the simulation shown in figure~\ref{fig:density_plots1}.
%%%%%%%%%%%%%%%%%%%%%%%%%%%%%%%%%%%%%%%%%%%%%%%%%%%%%%%%%%%%%%%%%%%%%%%%%%%%%%%%%%%%%%%%
\subsection{Thermodynamic parameters}\label{subsec:stir_Thermo_params}
%%%%%%%%%%%%%%%%%%%%%%%%%%%%%%%%%%%%%%%%%%%%%%%%%%%%%%%%%
\subsubsection{Analytic predictions}
Since our system conserves both normalisation and energy in the rotating
frame, some simple analytic predictions can be made about the development
of the thermal component of the field, using the Thomas-Fermi (TF) approximation. In a frame rotating
at angular velocity $\Omega$, the ground state is a vortex lattice.  In the Thomas-Fermi approximation, the vortices are replaced by a centrifugal dilation of the trapping potential arising from the coarse-grained (rigid-body) flow pattern, yielding the TF wave function
\begin{equation}
\psi_\mathrm{TF}^{\Omega}=\sqrt{\frac{\mu_\Omega-m\left(\omega_r^2-\Omega^{2}\right)r^{2}/2}{U_\mathrm{2D}}}\,\Theta\biggl(\mu_\Omega-m\left(\omega_r^2-\Omega^{2}\right)\frac{r^{2}}{2}\biggr),\label{eq:psiTF}\end{equation}
 where $ $$\mu_\Omega$ is the chemical potential of the rotating frame solution.
From equation~(\ref{eq:psiTF}) we obtain the number of particles $ $$N_\mathrm{TF}^\Omega$
and energy $E_\mathrm{TF}^\Omega$ \footnote{The appropriate Thomas-Fermi energy functional is similarly given by equation~\reff{eq:cfield_HCF} upon neglecting the kinetic energy, and including the appropriate centrifugal dilation of the trapping potential.} for the lattice state
\begin{eqnarray}
N_\mathrm{TF}^\Omega&=&\int\!d\x\,\left|\psi_\mathrm{TF}^\Omega\right|^2 = \frac{\pi\mu_\Omega^{2}}{U_\mathrm{2D}m\left(\omega_r^2-\Omega^{2}\right)},\label{eq:NTF_vortex}\\
E_\mathrm{TF}^\Omega&=& \int\!d\x\left[ \frac{m}{2}\left(\omega_r^2-\Omega^2\right) + \frac{U_\mathrm{2D}}{2} \left|\psi_\mathrm{TF}^\Omega\right|^2\right]\left|\psi_\mathrm{TF}^\Omega\right|^2 = \frac{2}{3}\mu_\Omega N_\mathrm{TF}^\Omega.\label{eq:ETF_vortex}
\end{eqnarray}

The corresponding quantities for the vortex-free inertial-frame ground state are given by setting 
$\Omega=0$. We note that the chemical potential $\mu_0 \equiv \mu_{\Omega=0}$ of the nonrotating state has the same value in both the inertial and rotating frames.  
If we now compare a `lattice' state (i.e. $\psi_\mathrm{TF}^\Omega$) and a vortex-free
state ($\psi_\mathrm{TF}^0$), both fully condensed and with equal occupation, we see from equation~(\ref{eq:NTF_vortex}) that the chemical potentials
are related by 
\begin{equation}\label{eq:reduced_mu}
	\mu_\Omega=\mu_0\sqrt{1-\Omega^{2}/\omega_r^2}, 
\end{equation}
and hence from equation~(\ref{eq:ETF_vortex}) that their rotating-frame energies 
are related by 
\begin{equation}
E_\mathrm{TF}^\Omega=E_\mathrm{TF}^0\sqrt{1-\Omega^{2}/\omega_r^2}.
\label{eq:Ecompare}
\end{equation}
The excess rotating-frame energy of the vortex-free ground state over the lattice
state of the same number of atoms, \begin{equation}
\Delta E\equiv E_\mathrm{TF}^0-E_\mathrm{TF}^\Omega\label{eq:delta_E}\end{equation}
can be significant, and for example is approximately $E_\mathrm{TF}^0/3$
for an angular velocity of $\Omega=0.75\omega_r$. Thus in our stirring scenario,
the rotating equilibrium condensate state reached at the end of the process must
have less energy and less atoms than the initial vortex-free state,
and the excess energy and atoms constitute a thermal cloud\footnote{This energy change $\Delta E$ cannot be 
attributed to the initial trap deformation associated with the stirrer,
as the energy of an axially symmetric state is unchanged by the abrupt introduction of 
a perturbation of quadrupolar symmetry (i.e., within the sudden approximation \cite{Bransden00}
implicit in our procedure).}. The exact
result for the excess energy of the classical field, obtained from
the simulations, will depend upon the depletion of the condensate
mode required to form the thermal cloud and the mutual interaction
of the cloud with the condensate.
For small thermal fractions this energy could be found by means of a calculation
similar to that described in reference~\cite{Sinatra00}. In the simplest approximation
we assume the limit of a small thermal fraction, in which the energies
of the condensate and thermal field are additive, with equation~(\ref{eq:delta_E})
approximating the excess energy the thermal field contains.  We further
assume that in equilibrium this energy will be classically equipartitioned (section~\ref{subsec:cfield_thermo}) over weakly
interacting harmonic oscillators, in the spirit of the Bogoliubov approximation
\cite{Lobo04}. The equilibrium temperature can thus be predicted by  
\begin{equation}\label{eq:temperature_prediction}
k_\mathrm{B}T=\frac{\Delta E}{\mathcal{M}-1},
\end{equation}
 with $\mathcal{M}$ the condensate-band multiplicity. For a given
initial chemical potential we see therefore that the equilibrium temperature will 
be strongly dependent on the basis size. However, we shall see
in section~\ref{sec:stir_parameters}, equation~(\ref{eq:reduced_mu}) provides
a very good estimate of the final chemical potential reached
by the field, which is essentially independent of cutoff. 
%%%%%%%%%%%%%%%%%%%%%%%%%%%%%%%%%%%%%%%%%%%%%%%%%%%%%%%%%
\subsubsection{Self-consistent fitting}\label{subsec:stir_fitting}
We now estimate the thermodynamic parameters of the nonequilibrium classical field using the semiclassical fitting-function approach of section~\ref{subsec:prec_T_and_mu}.   We remind the reader that in this approach we calculate the profile of the thermal cloud in the Hartree-Fock potential due to the \emph{entire} classical field, and thus do not need to explicitly distinguish the condensed component of the cloud from the thermal component in forming the mean-field potential.  This is particularly advantageous in the present scenario as no true condensate exists in the turbulent field, as we discuss in section~\ref{subsec:stir_Temporal}.

\begin{figure}
	\begin{center}
	\includegraphics[width=0.65\textwidth]{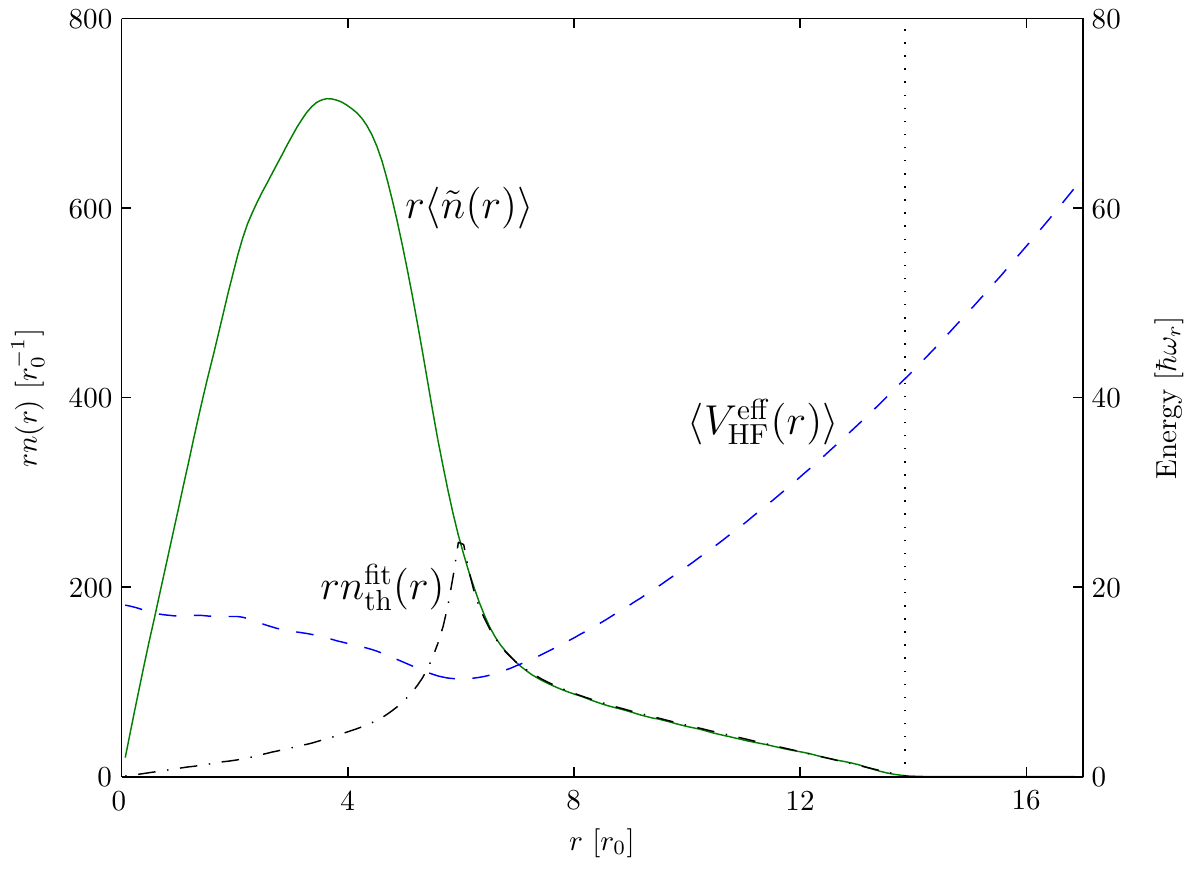}	
	\caption{\label{fig:example_temp_fit}  Fitting procedure for the thermodynamic parameters.  Shown are: the (time-averaged) radial density distribution times radius (solid line), the effective potential experienced by noncondensate atoms (dashed line), and the fitted distribution of noncondensed atoms times radius (dash-dotted line).  The dotted line indicates the classical turning point corresponding to energy cutoff $E_R$.  The data shown corresponds to the period $t=3000$--$3010$ cyc (see figure~\ref{fig:density_plots1}(f)).}
	\end{center}
\end{figure}

Our task is somewhat simplified in the present case, as the thermal cloud rapidly comes to equilibrium with the rotating drive (as we show in section~\ref{subsubsec:stir_local_rot_prop}), and so we assume that the cloud rotation rate\footnote{As noted in section~\ref{subsec:prec_T_and_mu}, the results of the fitting procedure are rather insensitive to the precise value of angular velocity used.} $\Omega_\mathrm{th}=\Omega$.  The fitting function for the radial density distribution thus reduces to (appendix~\ref{app:fitting_function})
\begin{equation}\label{eq:fit_function2}
	n_\mathrm{th}^\mathrm{fit}(r;\mu,T) = \frac{1}{\lambda_\mathrm{dB}^2} \ln\!\left[\frac{E_R - \mu}{\left\langle V^\mathrm{HF}_\mathrm{eff}(r) \right\rangle - \mu}\right],
\end{equation}
where the effective potential
\begin{equation}\label{eq:stir_eff_potl}
	\left\langle V_\mathrm{HF}^\mathrm{eff}(r) \right\rangle = m\left(\omega_r^2-\Omega^2\right)\frac{r^2}{2} + 2U_\mathrm{2D}\left\langle\widetilde{n}(r)\right\rangle, 
\end{equation}
with $\langle\widetilde{n}(r)\rangle$ the azimuthally averaged density of the field (section~\ref{subsec:prec_T_and_mu}), and we present an example of such a fit in figure~\ref{fig:example_temp_fit}. 
%%%%%%%%%%%%%%%%%%%%%%%%%%%%%%%%%%%%%%%
%%%%%%%%%%%%%%%%%%%%%%%%%%%%%%%%%%%%%%%

\begin{figure}
	\begin{center}
	\includegraphics[width=0.65\textwidth]{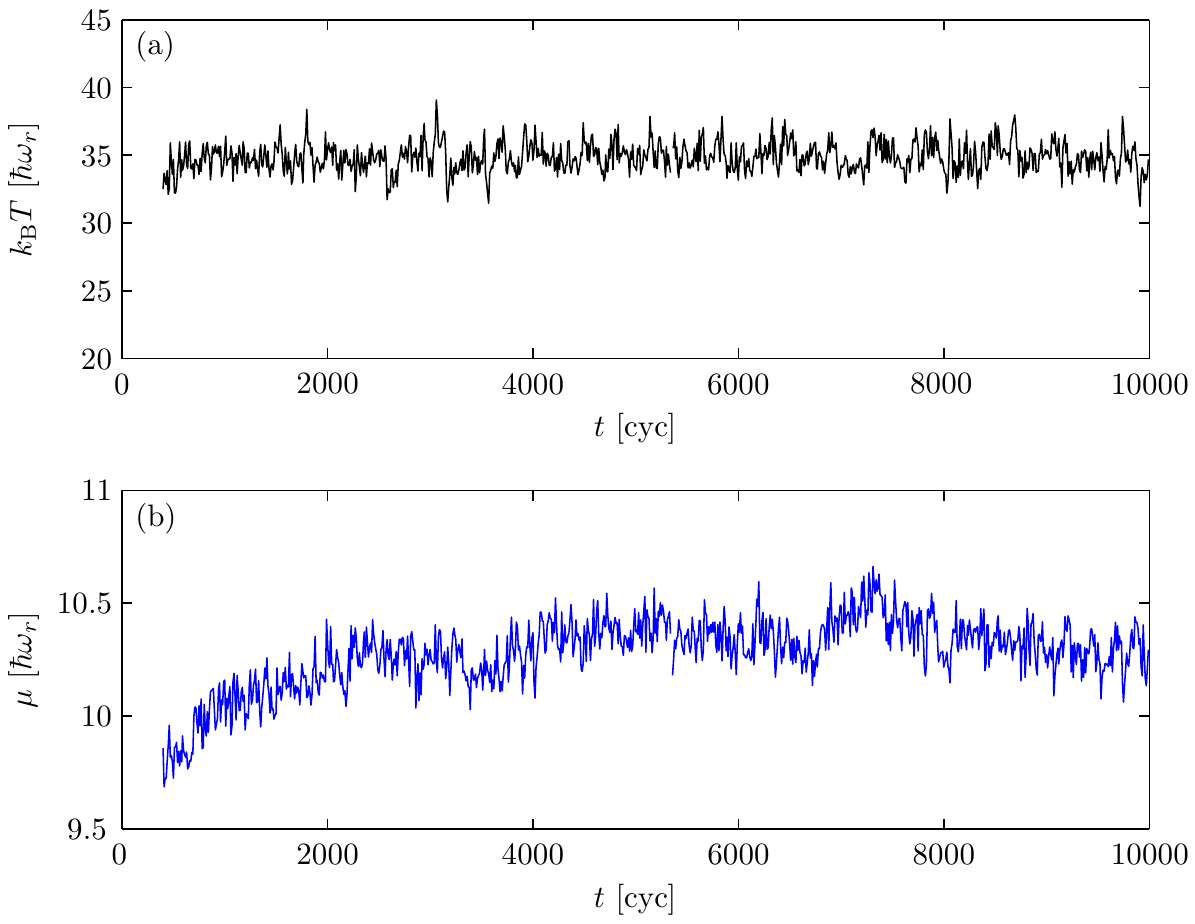}	
	\caption{\label{fig:T_mu_from_fit}  Evolution of the effective (a) temperature and (b) chemical potential of the noncondensed atoms, as measured by the semiclassical fitting procedure.}
	\end{center}
\end{figure}

After the initial strongly nonequilibrium dynamics following the dynamical instability (i.e., by $t\approx400$ cyc), the averaged field densities $\langle \widetilde{n}(r)\rangle$ are well fitted by the function given in equation~(\ref{eq:fit_function2}), and we conclude that the higher energy components of the field have reached a quasistatic equilibrium.  We then follow the evolution of the temperature and chemical potential with time, which we present in figure~\ref{fig:T_mu_from_fit}.  The temperature (figure~\ref{fig:T_mu_from_fit}(a)) appears to be essentially constant within the accuracy of the measurement procedure, while the chemical potential (figure~\ref{fig:T_mu_from_fit}(b)) shows a definite upward trend over the course of the field's evolution.  In section~\ref{subsec:stir_Frequency_dist} we compare the chemical potential of the cloud determined from our fitting procedure with an effective chemical potential of the dense central vortex-containing region, which we estimate by considering the temporal frequency components present in the field.
%%%%%%%%%%%%%%%%%%%%%%%%%%%%%%%%%%%%%%%
%%%%%%%%%%%%%%%%%%%%%%%%%%%%%%%%%%%%%%%
%%%%%%%%%%%%%%%%%%%%%%%%%%%%%%%%%%%%%%%%%%%%%%%%%%%%%%%%%%%%%%%%%%%%%%%%%%%%%%%%%%%%%%%%
\subsection{Rotational parameters}\label{subsec:stir_Rotational_params}
As noted in section~\ref{subsec:prec_rotational_properties}, a rotating superfluid exhibits a moment of inertia which is suppressed below its classical value.  As a superfluid is rotated at an increasing angular velocity, it admits an increasing number of vortices, and its moment of inertia tends towards the classical value~\cite{Hess67}.  In particular, as the rotation frequency $\Omega$ of a zero-temperature condensate (GP state) increases, the (areal) density of vortices increases \cite{Garcia-Ripoll01,Feder01}, and is given to leading order by the Feynman relation (section~\ref{subsec:back_vortices_in_conds}) 
\begin{equation}\label{eq:Feynman_relation}
	\rho_\mathrm{v}^\mathrm{c} = \frac{m\Omega}{\pi\hbar}.
\end{equation}
with corrections arising due to the lattice inhomogeneity and the discrete nature of the vortex array \cite{Sheehy04}.  As the rotation frequency and thus the vortex density increase, both the classical and quantum moments of inertia increase also, and the difference between the two moments vanishes in the limit of rapid rotations $\Omega \rightarrow \omega_r$ \cite{Feder01}. 
%%%%%%%%%%%%%%%%%%%%%%%%%%%%%%%%%%%%%%%%%%%%%%%%%%%%%%%%%
\subsubsection{Vortex density}
\begin{figure}
	\begin{center}
	\includegraphics[width=0.65\textwidth]{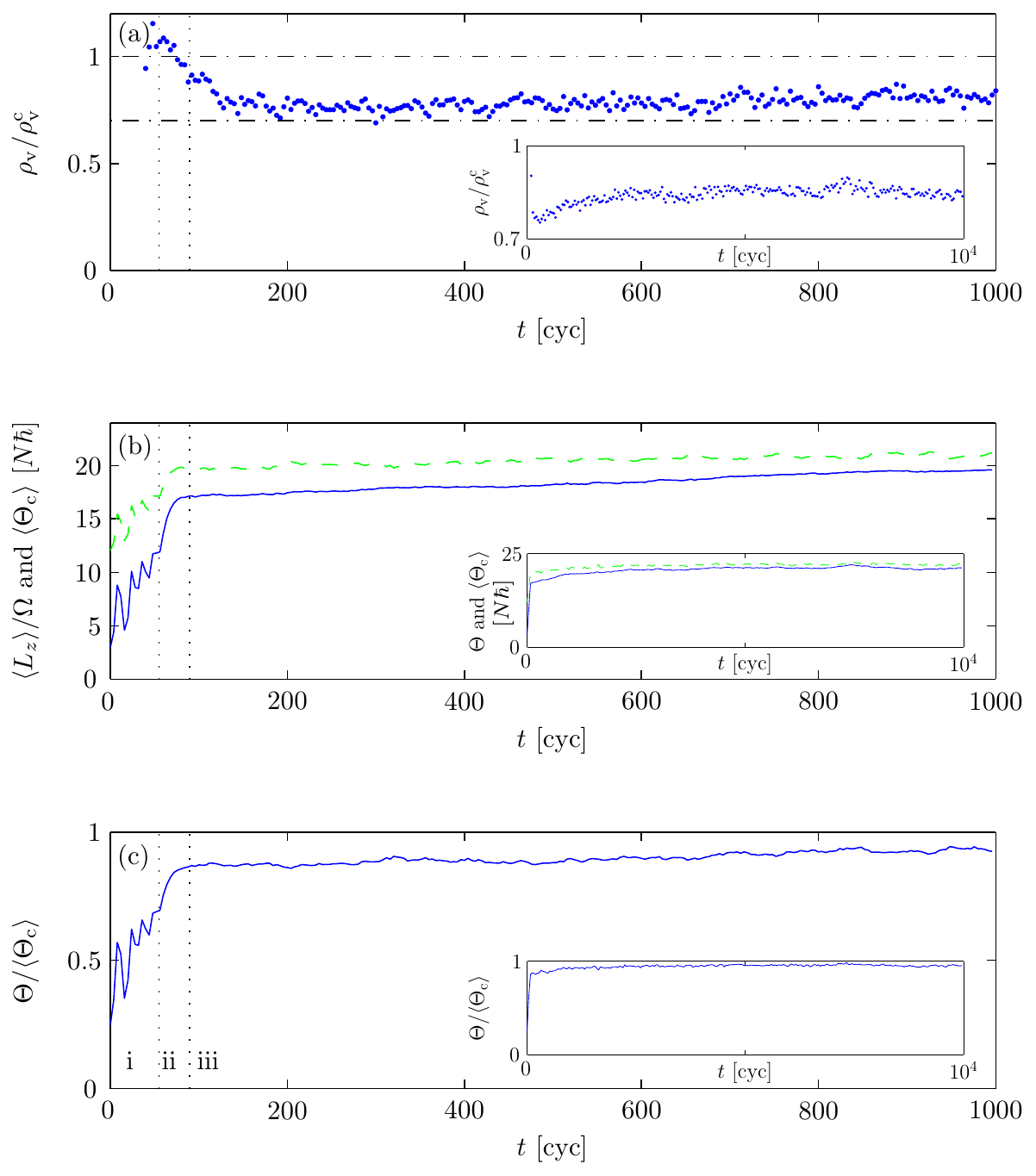}
	\caption{\label{fig:angmom_etc1}  Rotational response of the field. Shown are: (a) Vortex density, with dash-dotted lines indicating $\rho_\mathrm{v}/\rho_\mathrm{v}^\mathrm{c} = 0.7$ and $\rho_\mathrm{v}/\rho_\mathrm{v}^\mathrm{c} =1$ for reference, (b) classical (dashed line) and quantum moments of inertia, and (c) ratio of quantum and classical moments of inertia.  Insets show the behaviour of the same quantities over a longer time scale. Note that the initial $\langle L_z\rangle$ is finite due to the initial vacuum occupation.  For reference dotted lines in (a-c) separate phases of the evolution discussed in the text.}
	\end{center}
\end{figure}

To monitor the increase in vortex density during the system evolution, we count the vortices of positive rotation sense occurring within a circular counting region of radius equal to the TF radius of the initial state, $R_\mathrm{TF} = \sqrt{2\mu_\mathrm{i}/m\omega_r^2}$.  Due to the centrifugal dilation of the rotating condensate this region lies within the central bulk of the condensate, and in this manner we attempt to avoid including in the count the short-lived vortices constantly created and annihilated at the condensate boundary.  At equally spaced times over a period of 4 trap cycles about time $t_i$ we obtain the average number of vortices in the counting radius, $\overline{n_\mathrm{v}(t_i)}$, and from this we determine the average vortex density in the counting region
\begin{equation}\label{eq:vortex_density}
	\rho_\mathrm{v}(t_i) = \frac{\overline{n_\mathrm{v}(t_i)}}{\pi R_\mathrm{TF}^2}.
\end{equation}
In figure~\ref{fig:angmom_etc1}(a) we plot equation~(\ref{eq:vortex_density}) as a fraction of the prediction of equation~(\ref{eq:Feynman_relation}).  During the dynamical instability and initial nucleation of vortices the vortex densities measured in this manner are spuriously high, due to the counting of phase defects in the turbulent nonequilibrium fluid rather than long-lived vortices in the condensate bulk.  After this initial period, i.e. from $t\approx 150$ cyc onwards, we see a gradual increase in $\rho_\mathrm{v}$ as the condensate slowly relaxes and admits more vortices into its interior, and approaches rotational equilibrium with the trap.  By $t\approx 5000$ cyc (inset to figure~\ref{fig:angmom_etc1}(a)), the vortex density appears to have essentially saturated at a level $\rho_\mathrm{v} \approx 0.85\rho_\mathrm{v}^\mathrm{c}$.  Such a value seems reasonable for the rotation rate and condensate size considered here, which are such that inhomogeneity and discreteness effects will be significant \cite{Sheehy04}\footnote{Moreover, as noted by the authors of \cite{Sheehy04}, their idealised $T=0$ model neglects all thermal excitations.  It is therefore likely that thermal effects will further affect the vortex density in a finite-temperature vortex lattice and \emph{a fortiori} the vortex liquid we consider here.}.  
%%%%%%%%%%%%%%%%%%%%%%%%%%%%%%%%%%%%%%%%%%%%%%%%%%%%%%%%%
\subsubsection{Local rotational properties}\label{subsubsec:stir_local_rot_prop}
\begin{figure}
	\begin{center}
	\includegraphics[width=0.65\textwidth]{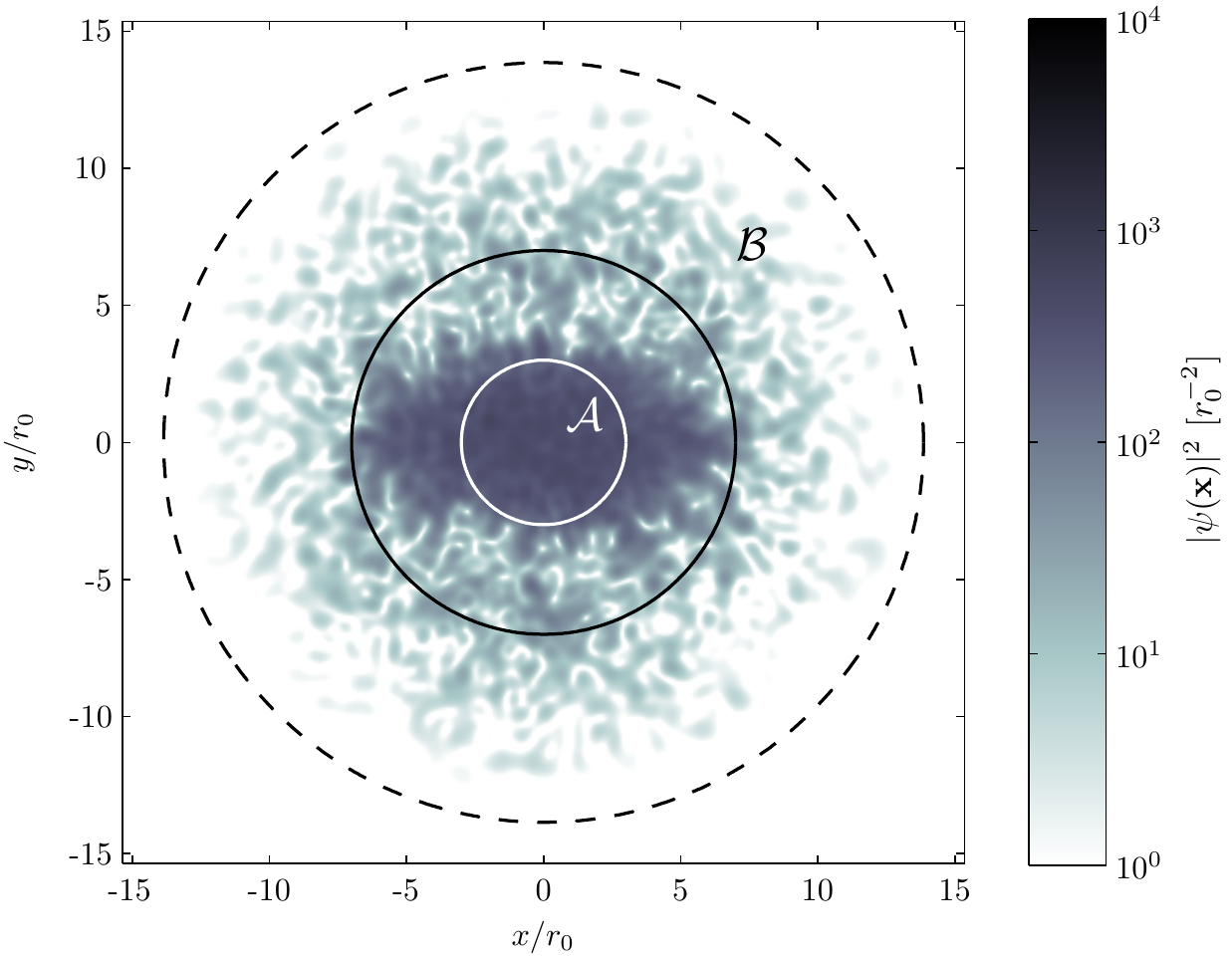}
	\caption{\label{fig:angmom_annulus}  Regions used for the evaluation of angular momenta and moments of inertia.  Disc $\mathcal{A}$ lies entirely within the condensate, where the flow is (initially) irrotational.  Annulus $\mathcal{B}$ (between the solid black line and the dashed line indicating the classical turning point of the condensate band) contains only noncondensed, normal fluid. The field density shown is that at $t=50$ cyc.}
	\end{center}
\end{figure}

In section~\ref{subsec:prec_rotational_properties} we presented an analysis of the rotational properties of the condensate and thermal field in a rotating system, based on a separation of the one-body density matrix into condensed and noncondensed parts.  As we discuss in section~\ref{subsec:stir_Temporal}, the irregular motion of vortices throughout the field we consider in this chapter suppresses the spatial order of the system and thus destroys the condensation in the field.
In order to characterise the rotation of the outer cloud formed from ejected material as well as that of the central bulk of the field we thus measure \emph{localised} expectation values, i.e. expectation values of the classical field evaluated over a restricted spatial region.  In this way we can compare the mass distribution and angular momentum of particular regions, in order to characterise the localised properties of the field.  We define the expectation value of an operator $\mathcal{O}$ on spatial domain $\mathcal{D}$ by
\begin{equation}\label{eq:restricted_expvalue}
	\langle \mathcal{O} \rangle_\mathcal{D} = \int_\mathcal{D}\!d\mathbf{x}\,\psi^*\,\mathcal{O}\,\psi,
\end{equation}
where the $\mathcal{D}$ will be either region $\mathcal{A}$ or $\mathcal{B}$ illustrated in figure~\ref{fig:angmom_annulus}.  Focussing our attention on the outer annulus $\mathcal{B}$, we calculate the classical moment of inertia of material in the annulus
\begin{equation}
	\langle\Theta_\mathrm{c}\rangle_\mathcal{B} = m\left\langle r^2\right\rangle_\mathcal{B},
\end{equation}
and the angular momentum of this material $\langle L_z \rangle_\mathcal{B}$.  From measurements over the period $t=50-51$ cyc we find that 
\begin{equation}
	\frac{\langle L_z\rangle_\mathcal{B}}{\Omega} = \langle \Theta_c \rangle_\mathcal{B},
\end{equation}
to within 4\%, indicating that the cloud in this region is rotating as a normal fluid in rotational equilibrium with the drive.  Averaging samples of these expectation values over a period of a trap cycle (which we denote by an overline $\overline{\cdots}$), we find for the time averages $\overline{\langle L_z \rangle_\mathcal{B}} / \Omega = 1.025 \overline{\langle\Theta_\mathrm{c}\rangle_\mathcal{B}}$.
%%%%%%%%%%%%%%%%%%%%%%%%%%%%%%%%%%%%%%%
%%%%%%%%%%%%%%%%%%%%%%%%%%%%%%%%%%%%%%%
Turning our attention to the central disc $\mathcal{A}$, we find that over the same period, $\overline{\langle L_z\rangle_\mathcal{A}}/\Omega =0.080\overline{\langle\Theta_\mathrm{c}\rangle_\mathcal{A}}$.  The central condensate bulk at this time thus remains approximately stable against vortex nucleation, but possesses some small angular momentum due to its shape oscillations, in contrast to the ejected material which has lost its superfluid character and has come to rotational equilibrium with the drive.  
%%%%%%%%%%%%%%%%%%%%%%%%%%%%%%%%%%%%%%%%%%%%%%%%%%%%%%%%%
\subsubsection{Global rotational properties}
We consider now the rotational properties of the entire field, by evaluating expectation values as above over the full $xy$ plane (i.e., in equation~(\ref{eq:restricted_expvalue}) we set $\mathcal{D}\rightarrow\mathbb{R}^2$).  An obvious measure of the field's rotation is given by the angular momentum $L_z$,  which we here scale by $\Omega$ to form the quantity $\Theta$ (equation~\reff{eq:quantum_inertia}) which, at equilibrium, corresponds to the (quantum) moment of inertia.  We consider the evolution of this quantity with time, and compare it to the classical moment of inertia $\langle\Theta_\mathrm{c}\rangle$ (equation~\reff{eq:classical_inertia}) which characterises the dilation of the field's mass distribution during the stirring process.  
In figure~\ref{fig:angmom_etc1}(b) we plot the evolution of these quantities as time evolves and in figure~\ref{fig:angmom_etc1}(c) we plot the evolution of their ratio.  
Comparing the evolution of these quantities and that of the field's density distribution (cf. figure~\ref{fig:density_plots1}), we identify three roughly distinct phases of the system evolution: (i) Excitation of unstable surface mode oscillations accompanied by permanent increase in angular momentum (figure~\ref{fig:density_plots1}(b)), (ii) further increase in angular momentum and $\Theta/\langle\Theta_\mathrm{c}\rangle$ as vortices are nucleated into the condensate bulk (figures~\ref{fig:density_plots1}(c-e)), and (iii) gradual approach of the field to rotational equilibrium with the drive (figure~\ref{fig:density_plots1}(f)).  Note that although the finite-temperature equilibria here contain a significant \mbox{normal-fluid} component, the quantum moment of inertia is still suppressed below the classical value due to the presence of the superfluid component.
%%%%%%%%%%%%%%%%%%%%%%%%%%%%%%%%%%%%%%%%%%%%%%%%%%%%%%%%%%%%%%%%%%%%%%%%%%%%%%%%%%%%%%%%
\subsection{Temporal analysis}\label{subsec:stir_Temporal}
%%%%%%%%%%%%%%%%%%%%%%%%%%%%%%%%%%%%%%%%%%%%%%%%%%%%%%%%%
\subsubsection{Issues in characterising field coherence}
A major theme of this thesis is the characterisation of correlations in classical-field trajectories, and the identification of condensation in the classical field.  Fundamentally, the correlation functions used to characterise coherence in the formal many-body theory are \emph{ensemble} averages; indeed, these are the only quantities that quantum mechanics (in its most widely accepted interpretations) presumes to describe \cite{Bransden00}.  Thus, by analogy, we expect the statistics of the classical field to be characterised by ensemble averages over classical-field trajectories.  As we have discussed in section~\ref{sec:cfield_ergodicity}, in thermal equilibrium, we can substitute time averages along single classical-field trajectories for averages over multiple trajectories, exploiting the ergodic hypothesis.  Implicit in this approach is the fact that the correlations of an equilibrium field are invariant with respect to displacements in time. 

In nonequilibrium scenarios, the correlations of the field are in general time-dependent, and so the situation is more subtle.  In chapter~\ref{chap:arrest}, we followed \cite{Blakie05} in approximating the time-dependent correlations of a \emph{nonequilibrium} field by taking \emph{short-time} averages over the classical field.  This approach relies on the system being in some sense \emph{quasi-equilibrium}, so that sufficient statistics to characterise the correlations of interest (i.e., those describing condensation), can be accumulated on a time scale short compared to the `macroscopic' relaxation of the field.  However, in taking this approach we must bear in mind that there is no guarantee that the values of correlation functions we measure in this way are representative of the values of the \emph{same} correlation functions evaluated over the appropriate ensemble.  In particular, as noted by Leggett \cite{Leggett01}, one can imagine scenarios in which the trajectories of the system are wildly different, despite each individually exhibiting the coherence properties we might associate with Bose condensation.  In such a case we may obtain a large condensate fraction from short-time averages of the individual trajectories, while averages over the formal ensemble reveal no condensate under the Penrose-Onsager definition (section~\ref{subsec:back_bose_condensation}).  The scenario, considered in chapter~\ref{chap:precess}, of a vortex precessing at equilibrium is similar: while the characterisation of the coherent condensate from short-time averages in the appropriate rotating frame is unambiguous, this condensate mode does not appear in the corresponding one-body density matrix evaluated over the microcanonical ensemble.  The issue is again one of the definition of condensation: the Penrose-Onsager (PO) definition is in some sense the simplest possible measure of the correlations we expect to accompany the presence of condensation in the field, and general scenarios may require the evaluation of more complicated correlation functions in order to appropriately define the `condensate' (see, e.g., references~\cite{Pethick00,Gajda06,Drummond07,Yamada09}).  In general, we expect the appearance of one-body coherence in time averages of single trajectories to correspond to more complicated correlations evaluated over the ensemble.  In this sense, measuring the single-trajectory correlations over short time periods constitutes a kind of short-cut to reveal the coherence effects we are ultimately interested in, without having to develop the appropriate ensemble correlation functions and evaluate them formally.

In this chapter, we consider a scenario in which the classical field contains many, irregularly moving vortices, even at equilibrium.  It is intuitively clear that averaging over distinct, irregular vortex configurations will yield a suppressed measure of one-body coherence, as for example quantified by the PO approach.  Similarly, time averages over a single trajectory of the classical field will also show a suppressed condensate fraction, due to the thermal motion of the vortices over time.   Bradley \emph{et al.} \cite{Bradley08}, considering a similar scenario of thermally excited vortex matter, have shown that a measure of condensation can be obtained by considering short-time averages of the classical field, similarly to those considered in section~\ref{subsec:decay_vortex_trajectory}.  It is clear that this approach gives a useful quantification of the coherence in the system, as evidenced by the behaviour of this condensate fraction as a function of the field temperature, and the results for the growth of coherence in the system with time following a quench of the thermodynamic parameters.  However, it is less clear how the coherence measured by this approach corresponds to formal correlations of the many-body system:  in general, the condensate fraction one measures in such an approach will depend on the time over which the averaging is performed.  In chapter~\ref{chap:precess}, the field coherence was destroyed by the diffusion of the vortex phase resulting from the symmetry breaking; thus the appropriate time scale for averaging was reasonably well defined.  In more general situations, however, the `natural' time scale for averaging may not be clear, and the results will depend strongly on the averaging time, which essentially determines the range of temporal frequency (or \emph{energy}) components which contribute to the correlation function\footnote{Note that in the extreme limit that only one sample of the classical field is taken, motion of all frequencies is excluded, and (e.g.) the PO procedure trivially yields a single coherent condensate mode (namely the instantaneous configuration of the entire classical field).}.

We will show in this section that the irregular motion of vortices in the final state of the field completely destroys condensation as defined by the PO procedure.  Accordingly, there is also no coherently phase-rotating `mean field' as we defined in chapter~\ref{chap:anomalous}.  Nevertheless, we will show that the temporal correlations of the field do give a useful characterisation of the field, and allow us to quantify the time scales over which the field is \emph{locally} coherent.  This allows us to distinguish qualitatively distinct phases of the atomic field in both equilibrium and nonequilibrium conditions.
%%%%%%%%%%%%%%%%%%%%%%%%%%%%%%%%%%%%%%%%%%%%%%%%%%%%%%%%%
\subsubsection{Covariance matrix}\label{subsec:stir_Frequency_dist}
\begin{figure}
	\begin{center}
	\includegraphics[width=0.65\textwidth]{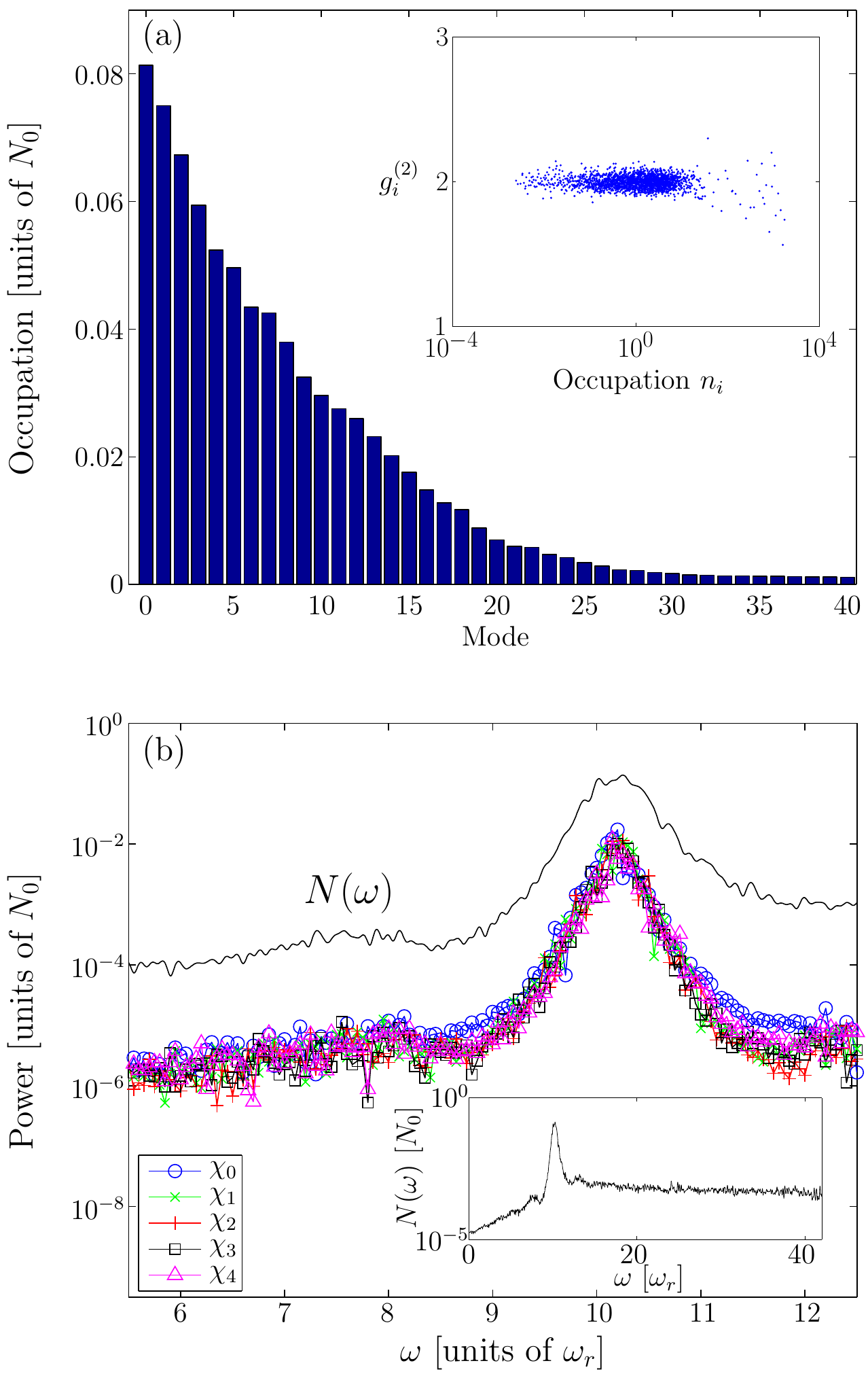}
	\caption{\label{fig:PO_g2_spect} (a) Mean occupations of covariance-matrix eigenmodes. Inset: Second-order coherence functions of the modes versus mode occupation.  (b) Power spectra of the first five covariance-matrix eigenmodes (lines with symbols).  Also indicated is the spatially integrated power spectrum $N(\omega)$ of the whole classical field.  The inset shows the behaviour of $N(\omega)$ over a greater range of frequencies.}
	\end{center}
\end{figure}

We begin by considering the covariance matrix equation~\reff{eq:cfield_density_mtx}, which we form from 2501 samples of the classical field, taken over the period $t\in[9900,10\,000]$ cyc.  From the analysis of section~\ref{subsec:stir_Rotational_params}, we conclude that the trap anisotropy has brought the field to rotational equilibrium in the co-rotating frame, and so we simply construct $G(\mathbf{x},\mathbf{x}')$ in this frame.  Diagonalising this matrix, we find there is no single mode with a large occupation, but rather, the occupations $n_i$ of the $\sim10$ most highly occupied eigenmodes $\chi_i(\mathbf{x})$ of $G(\mathbf{x},\mathbf{x}')$ are comparable, and the occupations of these eigenmodes decay away smoothly with increasing mode number, as shown in figure~\ref{fig:PO_g2_spect}(a).  
%%%%%%%%%%%%%%%%%%%%%%%%%%%%%%%%%%%%%%%
%%%%%%%%%%%%%%%%%%%%%%%%%%%%%%%%%%%%%%%
\enlargethispage{-\baselineskip}
As in section~\ref{subsubsec:prec_lab-frame_PO}, we consider the coherence functions $g_i^{(2)}$ of these modes (inset to figure~\ref{fig:PO_g2_spect}(a)).  In contrast to the case considered in that section, however, we find that these coherence functions indicate incoherent statistics for all modes (even the `most coherent' highly occupied modes have $g_i^{(2)}>1.5$, i.e., closer to thermal statistics than coherent statistics).  We thus conclude that the state of the classical field is not that of a fragmented condensate, but rather that there is no coherent condensate mode at all.  

Following section~\ref{subsubsec:prec_lab-frame_PO}, we proceed to calculate the power spectra of the most highly occupied eigenmodes $\chi_i(\mathbf{x}):0\leq i\leq4$, which we present in figure~\ref{fig:PO_g2_spect}(b).  We find that all these modes exhibit a power-spectrum peak at $\omega\approx10\omega_r$.  However, in contrast to the narrow peaks of the power spectra in figure~\ref{fig:occs_spect_E105}(b), the peaks in this case are very broad (note the logarithmic scale of figure~\ref{fig:PO_g2_spect}(b)).  The breadth of these peaks is perhaps not surprising, given that the modes have essentially incoherent fluctuation statistics.  Nevertheless, the fact that the power in these modes is mostly concentrated in a small frequency range suggests that coherence effects are still important in the field dynamics.  For comparison, we calculate also the spatially integrated power spectrum $N(\omega)$ of the entire classical field (as defined in section~\ref{subsec:id_first_moment}), calculated here from $1401$ samples of the classical field taken over the period $9900-9910$ cyc.  We find that this function also exhibits a broad peak at around the same frequency range as the peaks in the spectra of the individual modes $\chi_i(\mathbf{x})$.  We note that the width of this function is significantly greater than that characteristic of the `windowing' effect of the finite sample time, \mbox{$\Delta\omega = 1/T_\mathrm{samp} \approx 0.016\omega_r$}, in contrast to the power-spectrum peak of the true condensate mode presented in figure~\ref{fig:power_spectrum_in_trap}(b), for which the width was limited by $\Delta\omega$.  However, considering this function over a greater range of frequencies (inset to figure~\ref{fig:PO_g2_spect}(b)), we see that the behaviour of $N(\omega)$ is broadly similar to that of the true condensate studied in section~\ref{subsec:id_first_moment}, with a wide wing of power contained in frequencies greater than that of the peak, with a smaller wing occurring also at frequencies `negative' with respect to the peak.  Although no condensate is present in the field, this structure is not unexpected as, for example, Mora and Castin have shown \cite{Mora03} that a Bogoliubov-like picture can be constructed for the description of \emph{quasicondensates} \cite{Popov83}, in which density fluctuations are suppressed but true condensation is destroyed by phase fluctuations \cite{Petrov00}.
%%%%%%%%%%%%%%%%%%%%%%%%%%%%%%%%%%%%%%%%%%%%%%%%%%%%%%%%%
\subsubsection{Local spectrum of the field}
Here we follow section~\ref{subsec:id_first_moment} in analysing the frequencies present in the field at particular points in position space.  In order to compare the behaviour of the system at different stages of its nonequilibrium evolution, we define the power spectrum of the classical field $\psi$ at position $\mathbf{x}$ about time $t_0$ 
\begin{equation}
	H(\mathbf{x},\omega;t_0) = \left|\mathfrak{F}_{t_0}^T\{\psi(\mathbf{x},t)\}\right|^2,
\end{equation} 
where $\mathfrak{F}_{t_0}^T\{f(t)\}$ denotes the Fourier coefficient
\begin{equation}
	\mathfrak{F}_{t_0}^T\{f(t)\} \equiv \frac{1}{T}\int_{t_0-T/2}^{t_0+T/2} f(t) e^{i\omega t}dt.
\end{equation}
We choose a sampling period $T$ of 10 trap cycles, which is long compared to the time scale $\tau_1=1/\omega_\mathrm{p}\sim 0.1$ cyc corresponding to the location of the broad peaks in figure~\ref{fig:PO_g2_spect}(b) which characterise the phase evolution of the dense central region, while being short compared to that of the relaxation of the field ($\tau_\mathrm{R} \sim 1000$ cyc), and choose the sampling interval so as to resolve all frequencies present due to the combined single-particle and mean-field evolutions ($\hbar\omega_\mathrm{max} = E_R + 2\mu_\mathrm{i}$). As in section~\ref{subsec:id_first_moment}, we average this result over azimuthal angle $\theta$, to form $\widetilde{H}(r,\omega;t_0)$.  In order to gauge the relative strength of the various frequency components at different radii within the classical region, we normalise the resulting data so that the total power at each radius $r_j$ is the same.
%%%%%%%%%%%%%%%%%%%%%%%%%%%%%%%%%%%%%%%
\begin{figure}
	\begin{center}
	\includegraphics[width=1.0\textwidth]{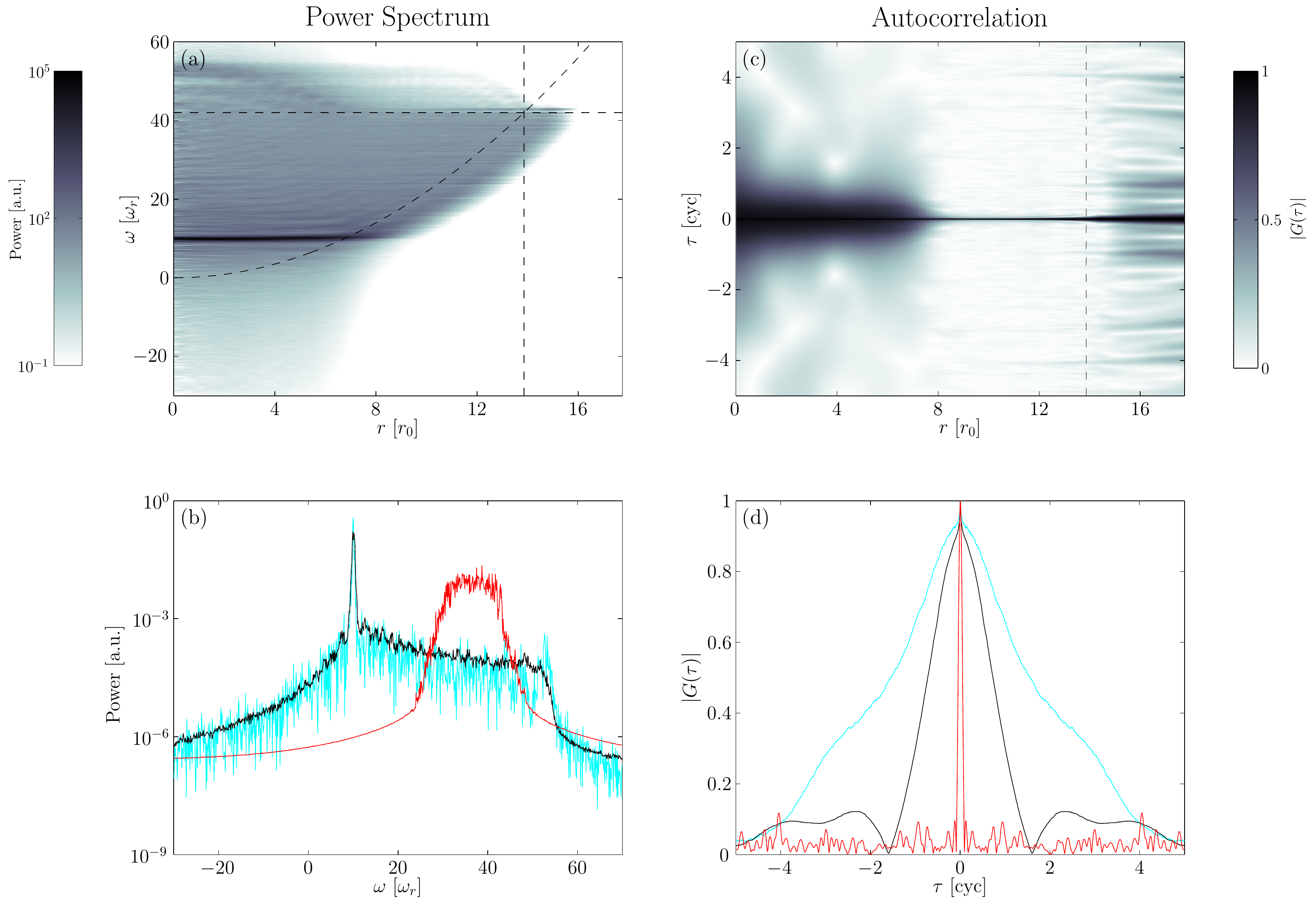}
	\caption{\label{fig:power_and_auto1} (a) Azimuthally averaged power spectral density (PSD).  Dashed lines indicate the cutoff energy, corresponding classical turning point and (centrifugally dilated) trapping potential.  (b) PSD traces at particular radii.  The black and red (dark grey) lines correspond to radii $r=3.1894r_0$ and $r=11.9575r_0$, respectively.  The cyan (light grey) line corresponds to the smallest radius on the quadrature grid (section~\ref{sec:quad_app_laguerre}), $r_\mathrm{min} = 0.0661r_0$. (c)  Autocorrelation function (absolute value) obtained from the power spectrum.  The dashed line indicates the classical turning point of the condensate band.  (d)  Autocorrelation magnitude at radii corresponding to figure~\ref{fig:power_and_auto1}(b). Data corresponds to the period $t=9900-9910$ cyc.}
	\end{center}
\end{figure}
%%%%%%%%%%%%%%%%%%%%%%%%%%%%%%%%%%%%%%%

In figure~\ref{fig:power_and_auto1}(a), we plot the result of such a calculation over the time interval $t=9900-9910$ trap cycles.   For comparison we have overlaid the centrifugally dilated potential $V_\mathrm{rot}(r) = m(\omega_r^2-\Omega^2)r^2/2$ that would be seen by a classical particle in the rotating frame, the cutoff frequency $E_R/\hbar$, and the semiclassical turning point $r_\mathrm{tp}$.  The most immediately noticeable feature of this plot is a prominent peak in the power spectrum at $\omega \approx 10\omega_r$, which occurs for radii $r\lesssim8$, and obviously corresponds to the peaks identified in figure~\ref{fig:PO_g2_spect}(b).  Despite there being no true condensate in the field, the appearance of this peak clearly indicates that the state of the field in the trap centre is not simply thermal.  This result is consistent with analytical results obtained by Graham~\cite{Graham02}, which suggest that (quasi-)long-range spatial order of the Bose field is accompanied by (quasi-)long-range \emph{temporal} correlations which decay in a functionally equivalent way.  
We thus associate this frequency with an \emph{effective} (quasi-)condensate eigenvalue $\lambda_\mathrm{e}\approx 10\omega_r$.  Similarly to figure~\ref{fig:power_spectrum_in_trap}, the spectrum shows distortion in the central region where this quasi-coherent object exists, with power occurring at negative frequencies down to $\omega \approx \lambda_\mathrm{e} - E_R/\hbar$. 
In figure~\ref{fig:power_and_auto1}(b) where the power spectrum is plotted at 3 particular radii, the peak in the spectrum obtained close to the trap centre (cyan (light grey) line) can be seen clearly.  In addition we see that frequencies above and below this are present, and we identify this with the restructuring of the excitation spectrum of the trap due to the presence of the interacting quasicondensate-like structure (cf.~\cite{Mora03}).  %these as due to the thermal occupation of the quasiparticle `particle' and `hole' modes respectively. 
We note that a secondary peak is visible at $\omega\approx52\omega_r=\lambda_\mathrm{e}+E_R/\hbar$, and an examination of the data reveals that a smaller peak is also present at $\omega\approx-32\omega_r=\lambda_\mathrm{e}-E_R/\hbar$.  This is an inevitable artifact of the projected method which occurs as follows:  
the $\mathcal{M}$ Bogoliubov modes which diagonalise the Bogoliubov Hamiltonian in the condensate band in the presence of a stationary condensate mode (GP eigensolution) can be considered as variational approximations to the lowest-lying `true' Bogoliubov modes (obtained as the cutoff $E_R\rightarrow\infty$).  The highest energy modes in the restricted quasiparticle basis are poor representations of the true modes, and, to maintain orthogonality of the set, are spuriously localised to the trap centre, and therefore have spuriously high energies $\epsilon_k + \mu_\mathrm{c} \approx E_R +\mu_\mathrm{c}$ due to the large mean-field interaction there.  Therefore, assuming this behaviour of excitations to hold in a finite-temperature equilibrium of the classical field, a peak is expected to occur at these frequencies at the trap centre, where the density of these modes is comparatively large.  However, the population contained in the peak observed here is $\sim0.5\%$ of the total thermal population and as such we do not expect it to qualitatively affect the relaxation process\footnote{It is important to note that such an artifact must arise in any field theory which is truncated either explicitly or implicitly, and that the defect occurring here is thus close to the minimal defect attainable in such a description.}. 
The black line in figure~\ref{fig:power_and_auto1}(b) indicates the power spectrum at a larger radius $r\approx 3.2r_0$, where the quasicondensate peak is slightly less prominent.  At larger radii still ($r\gtrsim 9r_0$)  the energy distribution returns to approximately that of the noninteracting gas.  As noted in section~\ref{subsec:id_first_moment}, frequencies persist into the classically excluded region as a result of the familiar evanescent decay of energy eigenmodes into the potential wall.  The effect of the mean-field interaction on the energy spectrum at these large radii, as can be seen in the red (dark grey) line in figure~\ref{fig:power_and_auto1}(b), is much less dramatic (e.g., only frequencies $\omega>\lambda_\mathrm{e}$ are present), and the population of frequencies is only significant up to a small increment above the cutoff energy, which we interpret as the mean-field shift of the highest energy single-particle modes.  We measure this shift to be $\approx2.5\%$, indicating that the cutoff we have chosen is sufficiently high to contain all modes significantly modified by the presence of the dense quasi-coherent structure in the field (see section~\ref{subsec:cfield_proj_cft}).  
%%%%%%%%%%%%%%%%%%%%%%%%%%%%%%%%%%%%%%%%%%%%%%%%%%%%%%%%%
\subsubsection{Temporal coherence}\enlargethispage{-\baselineskip}
The temporal analysis of the classical field presented above clearly shows that although there is no true condensate in the field, the field's behaviour in its central region is dramatically different from that at larger radii.  
The presence of the frequency peak is indicative of the temporal coherence of this central region, and we will use the local autocorrelation function
\begin{equation}
	G(\mathbf{x},\tau;t_0) = \left[\mathfrak{F}_{t_0}^T\right]^{-1}\left\{H(\mathbf{x},\omega;t_0)\right\},
\end{equation} 
of the field at point $\mathbf{x}$, at lag $\tau$ relative to the time $t_0$ to quantify the local temporal coherence of the field.  We evaluate the local autocorrelation function by using the Wiener-Khinchin theorem \cite{Press92}.

In practice we take the discrete Fourier transform of $\widetilde{H}(r_k,\omega;t_0)$ to obtain the azimuthally averaged autocorrelation $\widetilde{G}(r_k,\tau;t_0)$, and we normalise it so that $|\widetilde{G}(r_i,\tau=0;t_0)| = 1$ for all radii $r_i$.  We will use $|\widetilde{G}(r_i,\tau;t_0)|$ to characterise the temporal coherence of the classical field, as a function of position, and take the time scale for its decay to represent the time scale over which the field remains temporally coherent at radius $r_k$, near time $t_0$.  Illustrative results are shown in figures~\ref{fig:power_and_auto1}(c) and (d).

At the centre of the trap (cyan (light grey) line in figure~\ref{fig:power_and_auto1}(d)), the peak of the autocorrelation function is broad and decays monotonically with increasing $|\tau|$.  At larger radii ($2r_0 \lesssim r \lesssim 5r_0$), this peak possesses a nontrivial structure (black line), with $|\widetilde{G}|$ vanishing and then increasing again as $\tau$ varies.  Analysing the trajectories of vortices during the sampling period we conclude that this vanishing of the autocorrelation is due to the passage of vortices through these radii\footnote{We expect that averaging the autocorrelation over an ensemble of trajectories we would obtain a smoothly decaying function, due to the random distribution of vortex configurations in the ensemble.}.  
We identify the FWHM of the central peak (figure~\ref{fig:power_and_auto1}(d)) at a radius $r_i$ as (twice) the correlation time of the classical field at this radius.  At the trap centre, the correlation time $t_\mathrm{c} \approx 2.2$ cyc, indicating that despite the presence of highly energetic thermal excitations and mode mixing in the central region, the temporal coherence of this region is significant.  We note also that a small ridge appears in $|\widetilde{G}(\tau)|$ for $\tau$ close to zero, and interpret this as representing the correlations of the thermal component of the field, and note that an analogous behaviour of \emph{spatial} correlations in classical-field simulations was observed in \cite{Bezett08}. 

The width of the broad envelope (figure~\ref{fig:power_and_auto1}(d)) is roughly constant until $r\approx 8r_0$, then decreases with increasing radius, and the strong peak in the power spectrum also rapidly reduces with radius for $r\gtrsim 8r_0$.  Past this point the peak becomes much narrower (red (dark grey) line in figure~\ref{fig:power_and_auto1}(d)); the short correlation time of the outer cloud reflecting the broad range of energies characteristic of the thermally occupied excitations.  Beyond the classical turning point, the evanescent nature of the basis modes yields a spurious amount of temporal coherence and this manifests as an increasing correlation time past this point (a similar phenomenon was observed for spatial correlations in \cite{Bezett08}).  

While the interpretation of the correlation time we calculate in this manner is complicated by the non-monotonic decay of $|\widetilde{G}(\tau)|$ with increasing $|\tau|$, it nevertheless allows us to distinguish two qualitatively different regimes of behaviour in the classical-field equilibrium.  In the outer region of the field, the short time scales over which correlations persist in the field indicate that this material constitutes a truly `thermal' or `normal' fluid.  In contrast, the central region of the field is spatially disordered, yet maintains temporal coherence over a comparatively long time scale.  We identify the phase of the classical field in this central region as a \emph{vortex liquid} \cite{Fisher80,Kragset08,Gifford08,Guillamon09}: although the vortices are themselves disordered in their arrangement and motion, each supports a conserved supercurrent (phase winding of $2\pi$).  These supercurrents are robust in this region, in contrast to the normal phase of the field, where the spontaneous `pair production' of vortices of opposite sign leads to the screening of the vortex interactions and ultimately the destruction of the conserved superflows \cite{Fisher80,Anderson07}.

\begin{figure}
	\begin{center}
	\includegraphics[width=0.75\textwidth]{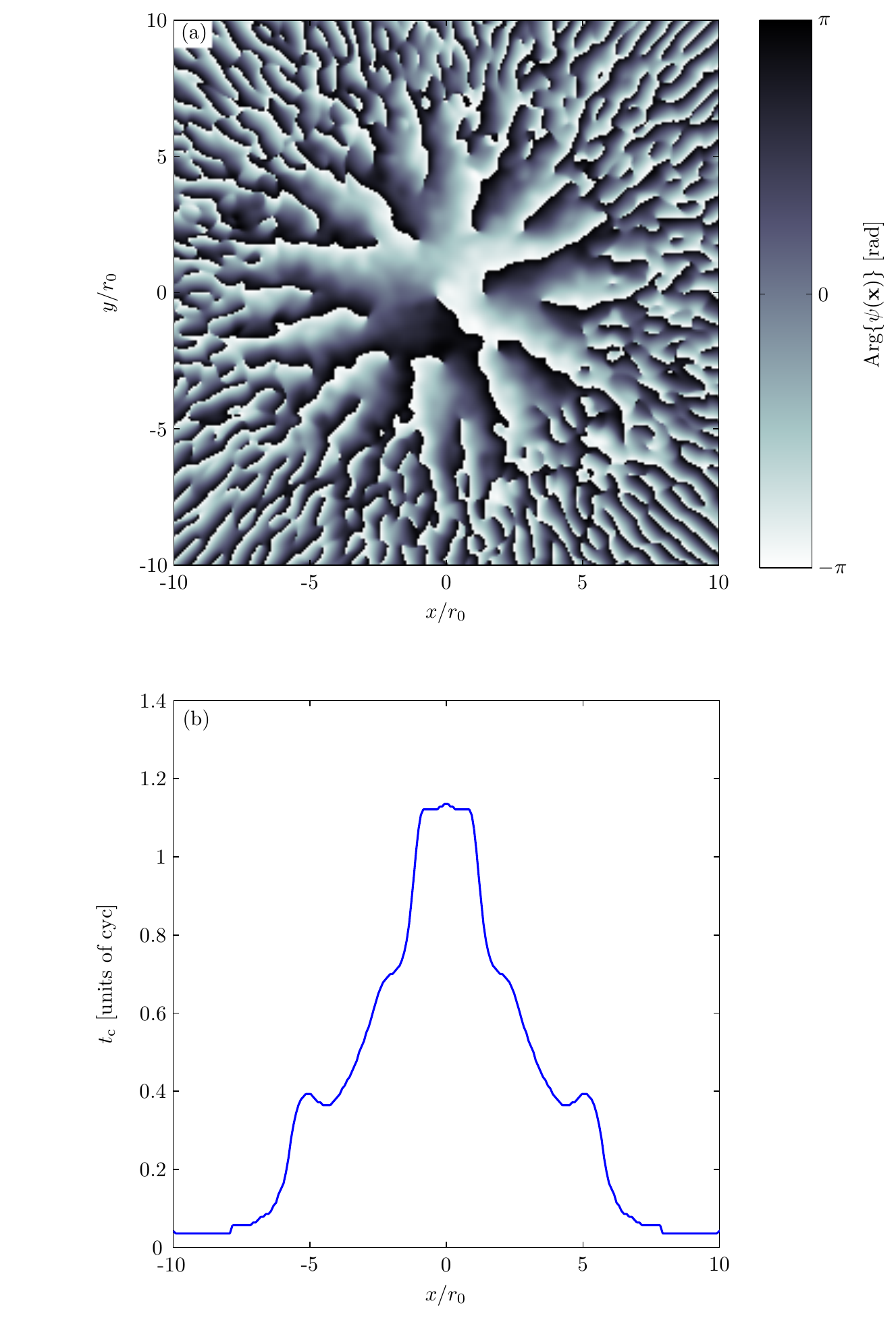}	
	\caption{\label{fig:phase_and_correl_time}  Plots of (a) the condensate-band phase and (b) the corresponding correlation time as a function of radius, near time $t=100$ cyc.}
	\end{center}
\end{figure}

Away from equilibrium, the characterisation of the phases of the classical field is more subtle.  Nevertheless, we can use temporal correlations to distinguish between different regimes of behaviour in the nonequilibrium field.  For example, in figures~\ref{fig:phase_and_correl_time}(a) and (b) we plot the phase profile of the classical field at time $t=100$ cyc and the corresponding correlation time, respectively.  The correlation time maintains a near-constant value of $t_\mathrm{c}\approx 1.1$ cyc as $r$ increases until $r\approx r_0$, where the first vortex (phase singularity in figure~\ref{fig:phase_and_correl_time}(a)) occurs. At this point the correlation time begins to decay steeply with increasing $r$. Throughout the region $r<r_\mathrm{tp}$ the phase structure is complicated, however we see that the correlation time of the central region is at least of order $t_\mathrm{c} \sim 0.5$ trap cycles, indicating that despite the turbulent nature of the field here, interactions serve to maintain its coherence, and we identify the behaviour here as superfluid turbulence \cite{Barenghi01}, in contrast to the outer region ($r\gtrsim6r_0$), which has a very short correlation time ($t_\mathrm{c}\lesssim0.1$ cyc) and thus appears to be truly thermal material.  This is in contrast to the interpretation presented by the authors of \cite{Parker05a,Parker06b}, who considered the turbulence developed during the stirring process to be purely superfluidic (i.e., zero-temperature) in nature.  
%%%%%%%%%%%%%%%%%%%%%%%%%%%%%%%%%%%%%%%
%%%%%%%%%%%%%%%%%%%%%%%%%%%%%%%%%%%%%%%
%%%%%%%%%%%%%%%%%%%%%%%%%%%%%%%%%%%%%%%%%%%%%%%%%%%%%%%%%
\subsubsection{Relaxation of the vortex liquid}
\begin{figure}
	\begin{center}
	\includegraphics[width=0.65\textwidth]{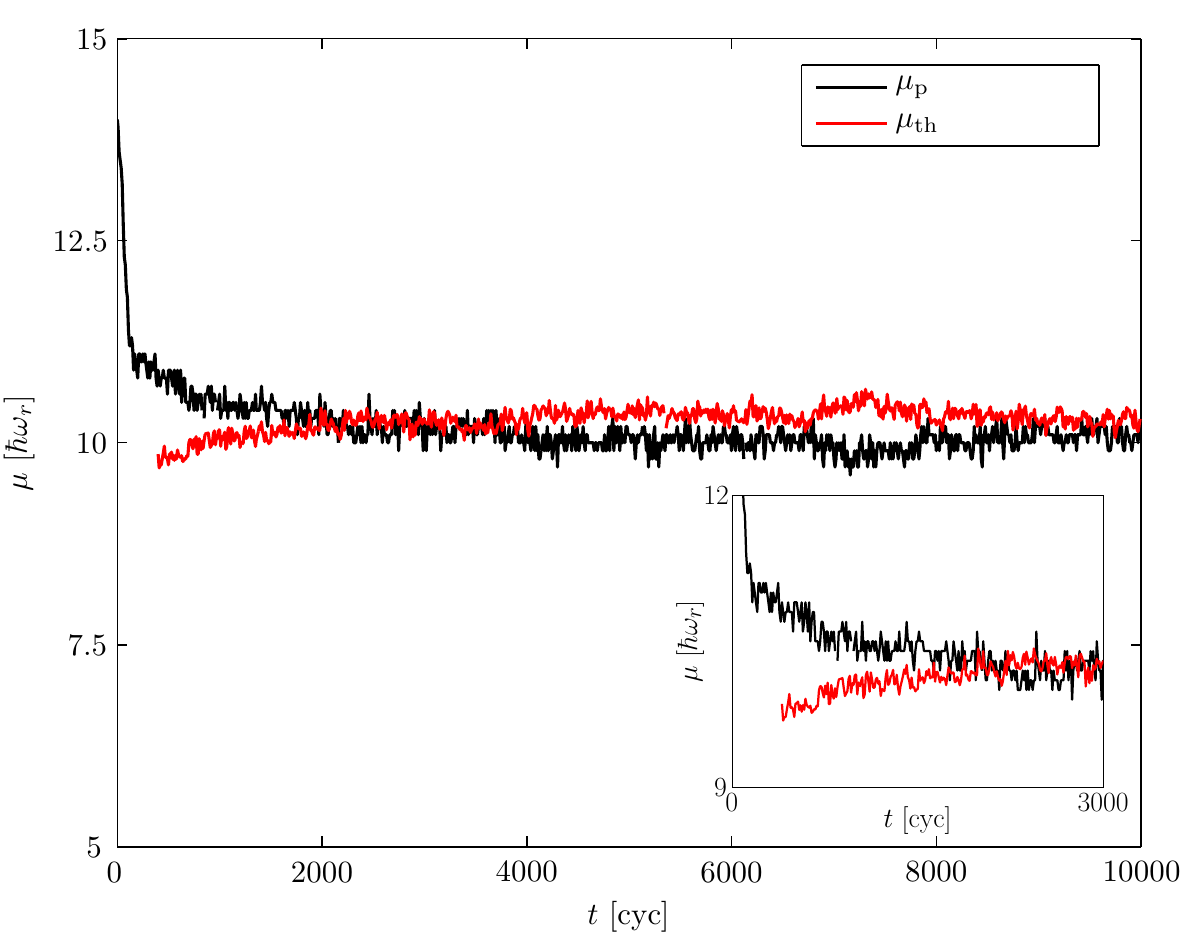}
	\caption{\label{fig:mu_comparison}  Effective chemical potential of the vortex liquid and thermal cloud.  The data displayed is obtained from 10 trap cycle long samples spaced at 100 trap cycle intervals.  Inset: Data with samples spaced at 10 trap cycle intervals over the initial decay (rise) of $\mu_\mathrm{p}$ ($\mu_\mathrm{th}$).}
	\end{center}
\end{figure}

The characterisation of the classical field by its temporal correlations also allows us to quantify the relaxation of the central vortex-liquid material due to its interaction with the thermal cloud.  In order to follow the relaxation of the vortex liquid, we measure the frequency $\omega_\mathrm{p}$ of maximum occupation in the power spectra, for power spectra evaluated over various periods.  Away from equilibrium the interpretation of this frequency is not entirely obvious, 
nevertheless it provides a measure of the energy of the quasi-coherent central structure, and we interpret it as representing an effective (quasi-)condensate chemical potential $\mu_\mathrm{p}=\hbar\omega_\mathrm{p}$ \footnote{At the level of approximation of these arguments, we neglect any correction due to the finite population of the `condensate' (cf. equation~\reff{eq:mu_lambda_relation}).}.  In figure~\ref{fig:mu_comparison} we plot this effective chemical potential of the vortex liquid, and also the chemical potential of the thermal cloud obtained using our fitting procedure (section \ref{subsec:stir_fitting}) for comparison.  We see that the vortex-liquid chemical potential reduces with time, as the liquid is damped by its coupling to the thermal cloud.  By $t\approx 4000$ cyc, the two chemical potentials appear to have reached diffusive equilibrium, to the accuracy of our fitting procedure for $\mu$ and $T$.  
%%%%%%%%%%%%%%%%%%%%%%%%%%%%%%%%%%%%%%%
%%%%%%%%%%%%%%%%%%%%%%%%%%%%%%%%%%%%%%%
%%%%%%%%%%%%%%%%%%%%%%%%%%%%%%%%%%%%%%%%%%%%%%%%%%%%%%%%%%%%%%%%%%%%%%%%%%%%%%%%%%%%%%%%
\subsection{Vortex motion}\label{subsec:stir_motional_damping}
As noted in section~\ref{sec:stir_Results}, the motion of vortices (as viewed in the rotating frame) is initially very rapid, and slows as time goes on.  In order to quantify this motion and its slowing with time, we track vortex trajectories, monitoring the coordinates of all vortices within a circular region of radius equal to the Thomas-Fermi radius of the initial lab-frame condensate.  Due to the rotational dilation of the vortical fluid, this region always lies within the central vortex-liquid region, as defined by the temporal analysis described in section~\ref{subsec:stir_Temporal}.  In general, vortices enter and leave this counting region as time progresses.  We are interested in the motion of vortices which persist in the counting region and can therefore be considered to exist in the condensate, rather than those which occur in the violently evolving condensate periphery. We therefore discard the trajectories of all vortices which do not remain in the region for at least half the counting period of $T_\mathrm{v} = 10$ trap cycles about time $t_i$.  

\begin{figure}
	\begin{center}
	\includegraphics[width=0.65\textwidth]{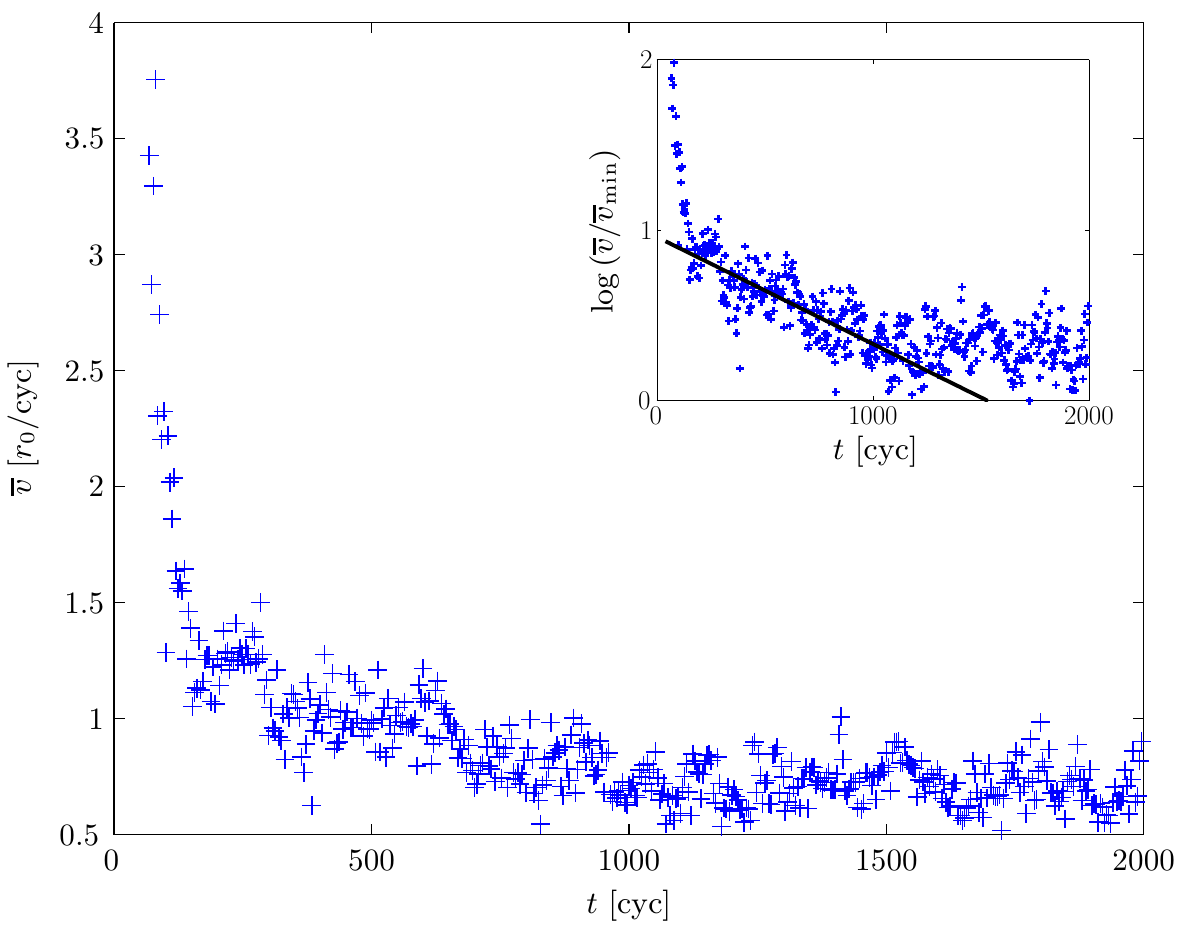}	
	\caption{\label{fig:vortex_damping1}  Mean vortex speed measured from the classical-field trajectory, revealing the rapid slowing of vortices during the initial stage of nucleation ($0\lesssim t\lesssim 120$ cyc), and more gradual damping of the remaining collective excitations ($100\lesssim t\lesssim 1000$ cyc). Inset: Logarithm of the mean vortex velocity (scaled by the minimum velocity measured $\overline{v}_\mathrm{min}$), and linear least-squares fit to the data over the damping period.}
	\end{center}
\end{figure}

The motion of the vortices which remain in the central region long enough to be considered is erratic on a length scale of order of the healing length, which is the origin of the thermal filling of the vortex core discussed in section~\ref{subsec:prec_PO_rotating_frames}.  As this motion occurs on a short time scale, we interpret it as the signature of high-energy excitations of the vortices.  Our interest however is in the low-energy component of vortex motion, which undergoes damping as the field relaxes to equilibrium.  We therefore apply a Gaussian frequency filter centred on $\omega=0$ to the trajectory components $\{x(t),y(t)\}$ to remove their highest \emph{temporal} frequency components.  We choose the width of this filter to be $\sigma_\omega \sim 0.6\omega_r$, and so the frequency components preserved in the trajectories correspond to low-energy excitations ($\omega \sim \omega_r$).  From the $N_t$ coordinate pairs $(x_k,y_k)$ of the $j^\mathrm{th}$ filtered vortex trajectory we extract the mean speed along the trajectory arc,
\begin{equation}\label{eq:arc_speed}
	v_j(t_i) = \frac{1}{T_\mathrm{v}}\int\!ds \approx \frac{1}{T_\mathrm{v}} \sum_{i=1}^{N_t - 1} \sqrt{(x_i-x_{i-1})^2 + (y_i - y_{i-1})^2}.
\end{equation}
We then average this quantity over the vortices included in the count, to find the mean speed per vortex $\overline{v}(t_i) = \sum_{j=1}^{N_\mathrm{v}} v_j(t_i)/N_\mathrm{v}$.  In figure~\ref{fig:vortex_damping1} we plot this quantity as a function of time.  We note first that the mean vortex speed decays dramatically during the initial measuring period $t\approx70-150$ cyc, as vortices enter the condensate with high velocities (with a substantial component opposite to trap rotation, as observed in the rotating frame), and their motion is damped heavily by their interaction with one another and with the thermal component.  Subsequent to this, a secondary, slower damping is observed over the period $t\approx 150-1100$ cyc, after which the vortex motion seems to have reached a lower limit set by thermal excitations.  During this period the condensate contains nearly the number of vortices it contains in equilibrium, and we interpret the damping during this period as the damping of low-energy excitations of the vortex liquid due to their interaction with thermally occupied, high energy modes.  Taking the log of the speed data (inset to figure~\ref{fig:vortex_damping1}), we see that the decay during this period is approximately exponential, and performing a linear fit to the data over this period we extract a damping rate $\gamma_\mathrm{v} \approx 6.3\times10^{-4}$ $(\mathrm{cyc})^{-1}$.

We note that the final distribution of vortices is disordered, with the vortices failing to become localised in a regular lattice.  This result is consistent both with the considerations of section~\ref{subsec:stir_Thermo_params} and with pure-GPE studies of condensate stirring in 2D performed by Feder \emph{et al.} \cite{Feder01a}, and Lundh \emph{et al.} \cite{Lundh03}.  The results presented here (and those of \cite{Lobo04}) suggest that the turbulent states observed in references~\cite{Feder01a,Lundh03} are in fact finite-temperature ones beyond the validity of the GP model employed in those calculations.  By contrast, the calculations of Parker and Adams~\cite{Parker05a,Parker06b} produced fully crystallised vortex lattices in 2D, indicating that additional damping mechanisms were available to the system in their simulations.  It is possible that these are attributable to the spatial differencing technique \cite{Parker_PhD} used in those calculations, however the precise origin of this error is not clear\footnote{While additional numerical procedures (amounting to evaporative cooling) were used in some calculations presented in references~\cite{Parker06b,Parker05a}, those additions were not essential for lattice crystallisation; i.e., relaxation to a lattice state was also observed in simulations performed without such additions \cite{Parker08b}.}.
%%%%%%%%%%%%%%%%%%%%%%%%%%%%%%%%%%%%%%%
%%%%%%%%%%%%%%%%%%%%%%%%%%%%%%%%%%%%%%%
%%%%%%%%%%%%%%%%%%%%%%%%%%%%%%%%%%%%%%%%%%%%%%%%%%%%%%%%%%%%%%%%%%%%%%%%%%%%%%%%%%%%%%%%%%%%%%%%%%%%%%%%%%%%%%%%%%%%%%%%%%%%%%%%%%%%
\section{Dependence on simulation parameters}\label{sec:stir_parameters}
We compare now the behaviour of stirred condensates for different condensate sizes, i.e., for different values of the initial chemical potential $\mu_\mathrm{i}$.  The field density and thus the nonlinear effects of mean-field interaction increase as the chemical potential is increased, and larger `condensates' in rotational equilibrium will also contain larger numbers of vortices.  In section~\ref{subsec:stir_vary_mu} we assess the effect of varying the initial chemical potential on our classical-field solutions.

It is also important in performing classical-field calculations such as those presented here to consider the effect of the cutoff height on the results of simulations.  In section~\ref{subsec:cutoff_dependence} we compare results for simulations with fixed initial chemical potential but varying cutoff heights, and quantify the effect of this purely technical parameter on the physical predictions of our model.
%%%%%%%%%%%%%%%%%%%%%%%%%%%%%%%%%%%%%%%%%%%%%%%%%%%%%%%%%%%%%%%%%%%%%%%%%%%%%%%%%%%%%%%%
\subsection{Chemical potential}\label{subsec:stir_vary_mu}
The initial chemical potential impacts upon the system evolution both qualitatively, affecting the density profile of the condensate mode and thus the nature of the dynamical instability leading to vortex nucleation, and quantitatively, determining the equilibrium parameters of the stationary state of the classical field.  The characteristics of the vortex array and the slowing dynamics of the vortices are also strongly dependent on the chemical potential.  We consider these four aspects in turn. 
%%%%%%%%%%%%%%%%%%%%%%%%%%%%%%%%%%%%%%%%%%%%%%%%%%%%%%%%%
\subsubsection{Effect on dynamical instability}	
\begin{figure}
	\begin{center}
	\includegraphics[width=0.65\textwidth]{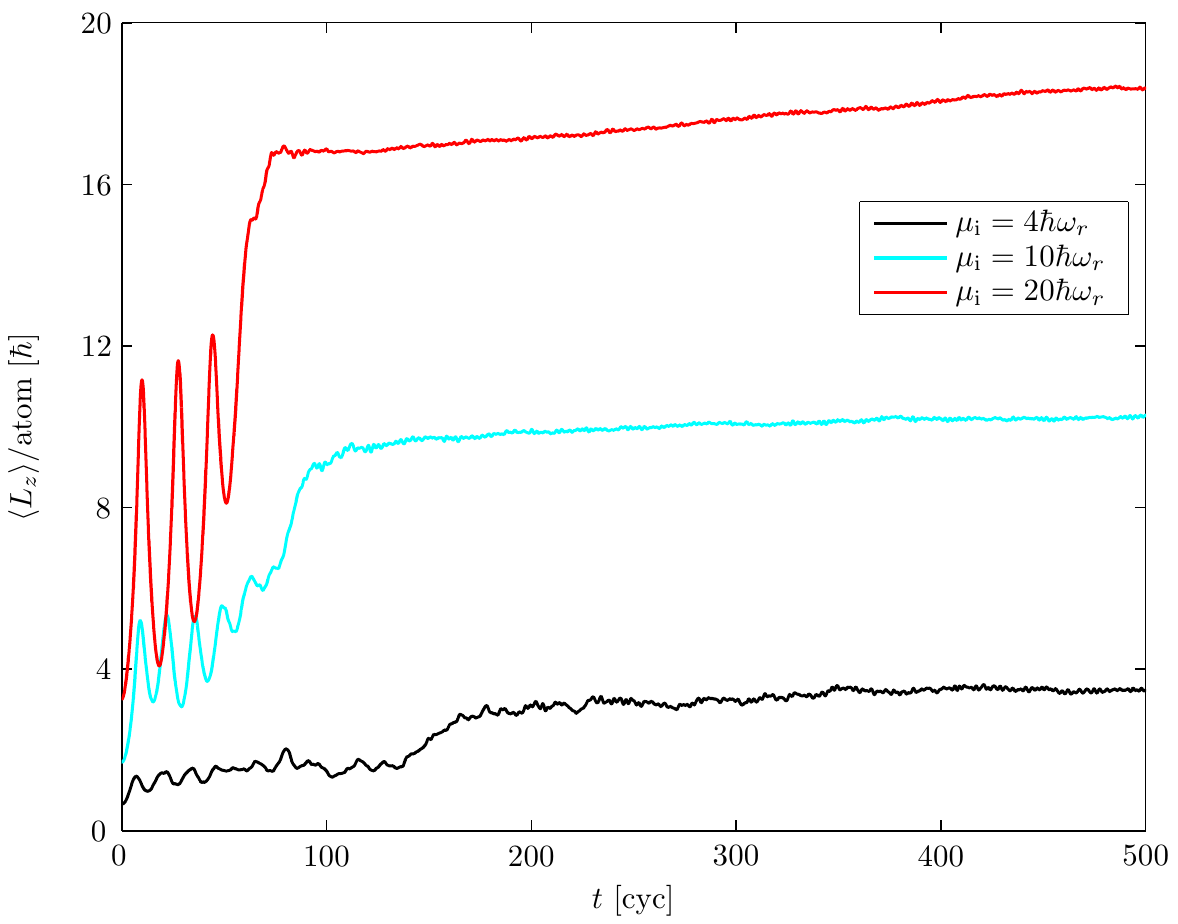}	
	\caption{\label{fig:Lz_different_mu}  Evolution of the field angular momentum for various initial chemical potentials.}
	\end{center}
\end{figure}

The chemical potential determines the strength of the mean-field repulsion influencing the condensate's shape.  As the chemical potential becomes large, the condensate mode approaches that described by the Thomas-Fermi approximation, whereas for smaller chemical potentials the kinetic energy of the condensate becomes important, and the condensate boundary broadens.  Consequently the visual distinction between the condensate mode and its incoherent excitations becomes less clear as the chemical potential is reduced. This affects the early evolution of the system and the onset of the dynamical instability in a pronounced manner.  For the largest chemical potentials considered ($\mu_\mathrm{i} \approx 17-20\hbar\omega_r$), the collective (quadrupole) excitations and their breakdown are clearly visible against a background of incoherent noise resulting from the initial vacuum occupation.  For smaller chemical potentials, although quadrupole oscillations are still visible and the outer cloud becomes visibly more dense as time proceeds, indicating that material is indeed lost from the condensate, any ejection of material from the condensate is obscured by the blurred condensate boundary.  Persistent `ghost' vortices can be seen forming in the incoherent region about the condensate periphery very early in the evolution ($t\approx 20$ cyc for $\mu_\mathrm{i}=4\hbar\omega_r$).  In figure~\ref{fig:Lz_different_mu} we plot the evolution of the field angular momentum per atom for three different initial chemical potentials\footnote{Note that in each case the initial angular momentum of the field is nonzero, due to the contribution of the rotating-frame vacuum.  However, the angular momentum of the vacuum is quickly dominated by that of the nascent thermal material.}.  The most obvious feature is that larger condensates support more angular momentum per atom, due to increased mean-field effects.  The \mbox{evolution} of the field angular momentum for the smallest condensate considered ($\mu_\mathrm{i}=4\hbar\omega_r$) reveals the onset of `irreversible' (see section~\ref{sec:cfield_ergodicity}) coupling of angular momentum into the field at $t\approx 130$ cyc.  Inspection of the field-density evolution shows that vortices begin to be nucleated into the central region of the field at about this time. We therefore conclude that despite the absence of visible break-up of the condensate, the unstable quadrupole oscillations still serve to produce the thermal component required to allow vortices to enter the central (initially) condensed region\footnote{More generally, outside the dynamically unstable regime (e.g., at lower drive frequencies) the thermalisation of the added Wigner noise can eventually lead to the nucleation of vortices due to the \emph{thermodynamic} instability of the condensate.  This pathway to vortex formation is however much slower than that due to the dynamical instability, as the system must fluctuate over the energetic barrier to vortex nucleation (see section~\ref{subsec:back_vortices_in_conds}).  In fact, in the absence of any trap anisotropy the angular momentum of the noise can be sufficient for the condensate to migrate to a state containing a vortex (cf. chapter~\ref{chap:precess}).}.
%%%%%%%%%%%%%%%%%%%%%%%%%%%%%%%%%%%%%%%
%%%%%%%%%%%%%%%%%%%%%%%%%%%%%%%%%%%%%%%
%%%%%%%%%%%%%%%%%%%%%%%%%%%%%%%%%%%%%%%%%%%%%%%%%%%%%%%%%
\subsubsection{Thermodynamic parameters}
\begin{figure}
	\begin{center}
	\includegraphics[width=0.65\textwidth]{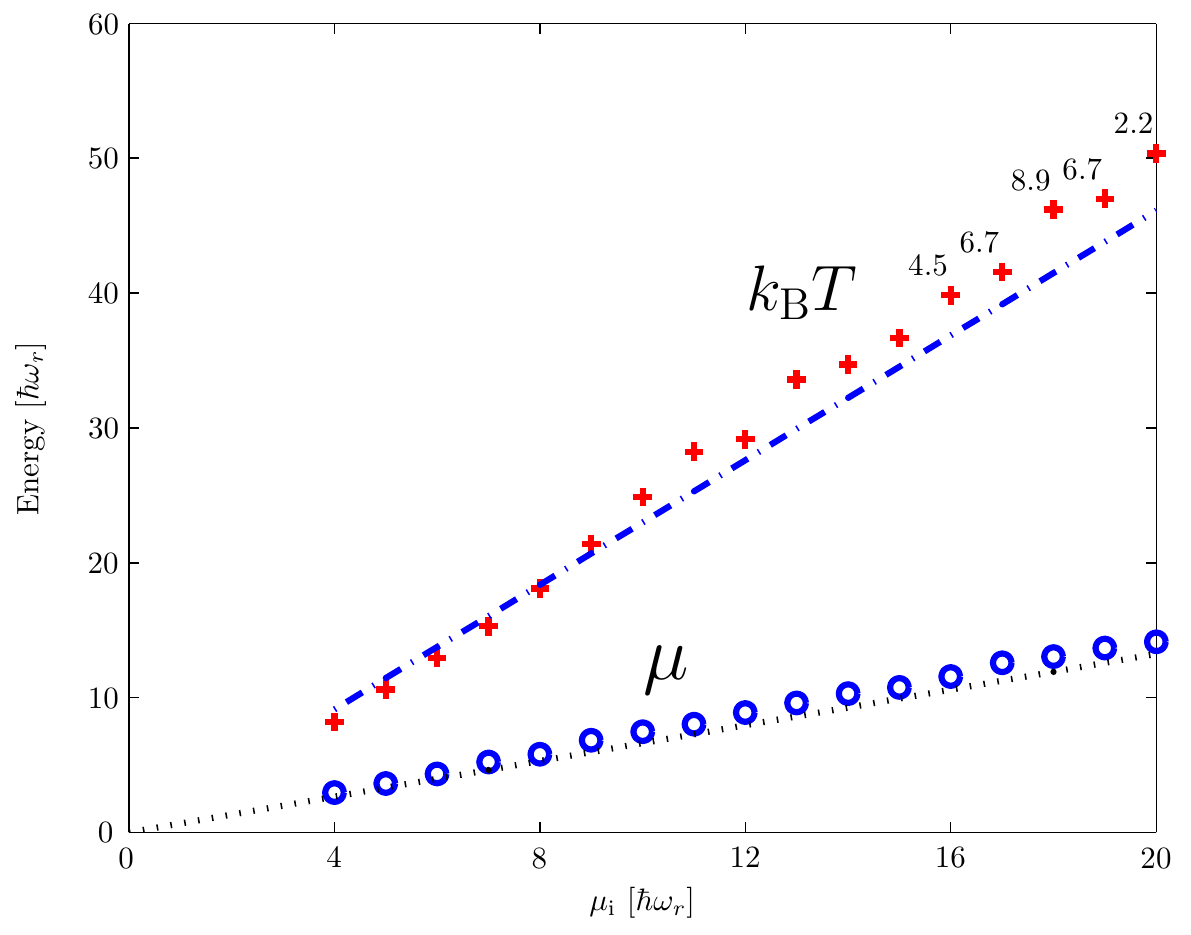}
	\caption{\label{fig:thermo_variation_with_mu}  Equilibrium temperatures and chemical potentials reached as a function of initial chemical potential, as determined by the fitting procedure.  Dotted and dash-dotted lines indicate the predictions of equation~(\ref{eq:reduced_mu}) and equation~(\ref{eq:temperature_prediction}) respectively.  Numbers indicate times at which the measurements were made (thousands of trap cycles) in cases where numerical results were constrained by time.  Points without numbers were measured at $10^4$ cyc.}
	\end{center}
\end{figure}

As described in section~\ref{subsec:stir_Thermo_params}, a simple analysis based on energetic considerations allows us to predict the equilibrium temperatures and chemical potentials of our classical-field solutions, and the dependence of these parameters on the initial chemical potential and the condensate-band multiplicity of our simulations.  In figure~\ref{fig:thermo_variation_with_mu} we plot these analytic predictions alongside the values obtained using the fitting procedure of section~\ref{subsec:stir_Thermo_params}.  The dotted line shows the estimated chemical potential of the (idealised) norm-conserving, rotating-frame condensate mode, $\mu_\Omega = \mu_\mathrm{i}\sqrt{1-\Omega^2/\omega_r^2}$, while the dash-dotted line indicates the estimated equilibrium temperature given by equation~(\ref{eq:temperature_prediction}).
We note that despite its approximate (TF-limit) and asymptotic (zero thermal fraction) nature, the analytical estimate for $\mu$ is a reasonable prediction of the measured values.  The measured value of $\mu$ is slightly higher than the TF prediction, and this discrepancy appears to increase with increasing $\mu_\mathrm{i}$, while the measured temperature appears consistently smaller than the analytical prediction at small $\mu_\mathrm{i}$, and exceeds it as the initial chemical potential is increased. However, the agreement seems very reasonable given the simplicity of the arguments presented in section~\ref{subsec:stir_Thermo_params}, which neglect both the kinetic energy of vortex cores (TF approximation) \cite{Penckwitt03}, the depletion of the condensate population, and all other effects beyond a linear (Bogoliubov) description.  We therefore conclude that our simple analysis captures the essential physics determining the thermodynamic parameters of the equilibrium. That is to say, the temperature is determined by the necessity of the system to redistribute the excess energy of the initial state (as viewed in the rotating frame), and that no additional heating of the system is required for the system to migrate to a state containing vortices.
%%%%%%%%%%%%%%%%%%%%%%%%%%%%%%%%%%%%%%%
%%%%%%%%%%%%%%%%%%%%%%%%%%%%%%%%%%%%%%%
%%%%%%%%%%%%%%%%%%%%%%%%%%%%%%%%%%%%%%%%%%%%%%%%%%%%%%%%%
\subsubsection{Vortex-array structure}
Perhaps the most striking difference between vortex arrays in simulations of differing condensate populations is the number of vortices present.  As noted in section~\ref{subsec:stir_Rotational_params}, the density of vortices is largely determined by the rotation rate of the condensate and is given approximately by equation~(\ref{eq:Feynman_relation}), and (within the TF approximation) the total number of vortices in an equilibrium condensate is approximately proportional to the chemical potential $\mu$. 

The stability of a vortex array in equilibrium is determined by the competition between hydrodynamic and energetic considerations \cite{Donnelly91} (which favour the crystallisation of a rigid lattice), and the perturbing effect of thermal fluctuations in the atomic field.  Close to zero temperature these fluctuations are interpreted as the thermal occupation of Bogoliubov excitations of the underlying vortex-lattice state.  As the temperature of a vortex-lattice state is increased it eventually undergoes a transition to a disordered \emph{vortex-liquid} state \cite{Fisher80,Snoek06,Gifford08,Bradley08} as the long-range order of the lattice is degraded by thermal fluctuations.

After a long time period, $t \sim 10^4$ trap cycles, all our simulated fields appear to describe states on the disordered side of this transition.  All our solutions have vortices migrating within the central bulk, and leaving and entering through its periphery.  In the smallest condensates, with the smallest vortex counts (i.e. $\mu_\mathrm{i} = (4,5,6)\hbar\omega_r$ with counts $N_\mathrm{v} \approx (3,4,5)$), the vortex array is small enough that \emph{quasi-regular} configurations of vortices may occur, in which the vortex positions appear to fluctuate about approximate equilibrium positions on a triangular (or square) lattice.  Such configurations may persist for as long as $\sim 10$ trap cycles, but ultimately break down and give way to new configurations as vortices cycle in and out of the central, high density region of the field.  We interpret this cycling behaviour as the classical field `sampling' different configurations during its ergodic evolution\footnote{An alternate (and contrary) explanation, that the field is exhibiting quasi-recurrent behaviour in the manner of \cite{Fermi65}, seems unlikely given that the trajectories appear to have converged to the temperatures predicted by the simple analysis presented in section~\ref{subsec:stir_Thermo_params}.}.

\begin{figure}
	\begin{center}
	\includegraphics[width=0.65\textwidth]{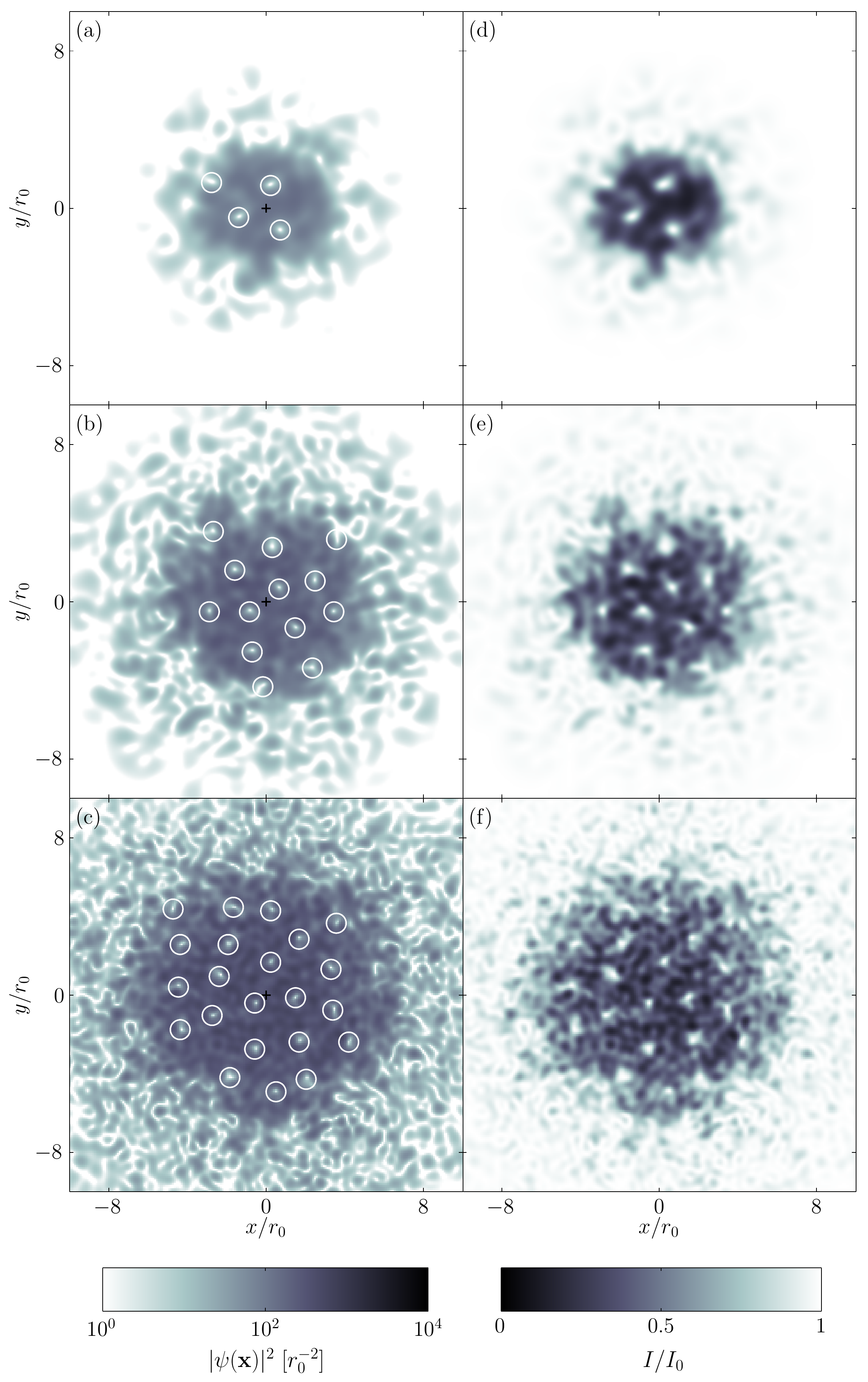}
	\caption{\label{fig:densities_vs_mu}  (a-c) Coordinate-space densities of the classical fields with initial chemical potentials $\mu_\mathrm{i}=5,10,18\hbar\omega_r$ at times $t=9900,9900,8800$ cyc respectively.  (d-f) Simulated absorption images generated from the same densities.}
	\end{center}
\end{figure}

Larger vortex arrays appear to be prohibited from forming regular structures over scales larger than the nearest-neighbour inter-vortex separations (the scale of the apparent order in the smaller condensates). As noted in section~\ref{subsec:stir_Thermo_params}, the equilibrium temperature attained by our simulated atomic fields increases linearly with the initial chemical potential, and so it seems unlikely that increasing tendency to disordered behaviour would abate when the condensate size is increased further.  In figures~\ref{fig:densities_vs_mu}(a-c) we plot the coordinate-space densities of vortex arrays of differing sizes, corresponding to differing initial chemical potentials.  Because of their small size compared to the extent of the condensate, vortex cores are difficult to experimentally image \emph{in situ}, and so their presence is typically detected after free expansion of the atom cloud \cite{Lundh98} by absorptive imaging techniques.  For condensates in the TF regime the expansion is well described by a simple scaling of the position-space density \cite{Kagan96,Castin96} (though the vortex core radius grows in somewhat greater proportion \cite{Lundh98}).  The Beer-Lambert law \cite{Robinson96} for the absorption of an optical probe, and the assumption of a ground-state (Gaussian) profile in $z$ (see section~\ref{subsec:dimless}) yields for the transmitted intensity of the probe
\begin{equation}\label{eq:transmitted_intensity}
	I(x,y) = I_0e^{-\gamma|\psi(x,y)|^2},
\end{equation}
with $I_0$ the input probe intensity and $\gamma$ a constant which depends, in general, on the absorptance of the atomic medium.  In figures~\ref{fig:densities_vs_mu}(d-f) we plot simulated intensity profiles obtained using equation~(\ref{eq:transmitted_intensity}), where we have in each case simply set $\gamma = 2/\{|\psi|^2\}_\mathrm{max}$, with $\{|\psi|^2\}_\mathrm{max}$ the peak value of the density occurring in the distribution, so as to best represent the significant features.  This process accentuates both the disparity in optical depth between the condensate and the outer thermal cloud, and the density fluctuations in the central condensate region, in contrast to the logarithmic plots (figures~\ref{fig:densities_vs_mu}(a-c)) which suppress them.  The images show the thermally distorted vortex-array structure that might be measured in an experiment.  The fact that the formation of rigid vortex lattices here is inhibited by thermal fluctuations, while ordered lattices have been formed experimentally at comparatively high temperatures  \cite{Madison00, Abo-Shaeer02}, suggests that this inhibition of lattice order may be a manifestation of the well-known susceptibility of two-dimensional systems to long-wavelength fluctuations \cite{Posazhennikova06}.  Indeed two theoretical studies \cite{Pogosov06,Snoek06} of vortex-lattice melting suggest that the melting temperature is very low in quasi-2D trapping geometries\footnote{The group of Cornell have performed experiments on the vortex-lattice melting transition in trapped Bose condensates \cite{Lamporesi08}, however the results of these experiments appear to remain unpublished.}.  
%%%%%%%%%%%%%%%%%%%%%%%%%%%%%%%%%%%%%%%
%%%%%%%%%%%%%%%%%%%%%%%%%%%%%%%%%%%%%%%
%%%%%%%%%%%%%%%%%%%%%%%%%%%%%%%%%%%%%%%%%%%%%%%%%%%%%%%%%
\subsubsection{Damping of vortex motion}
\begin{figure}
	\begin{center}
	\includegraphics[width=0.65\textwidth]{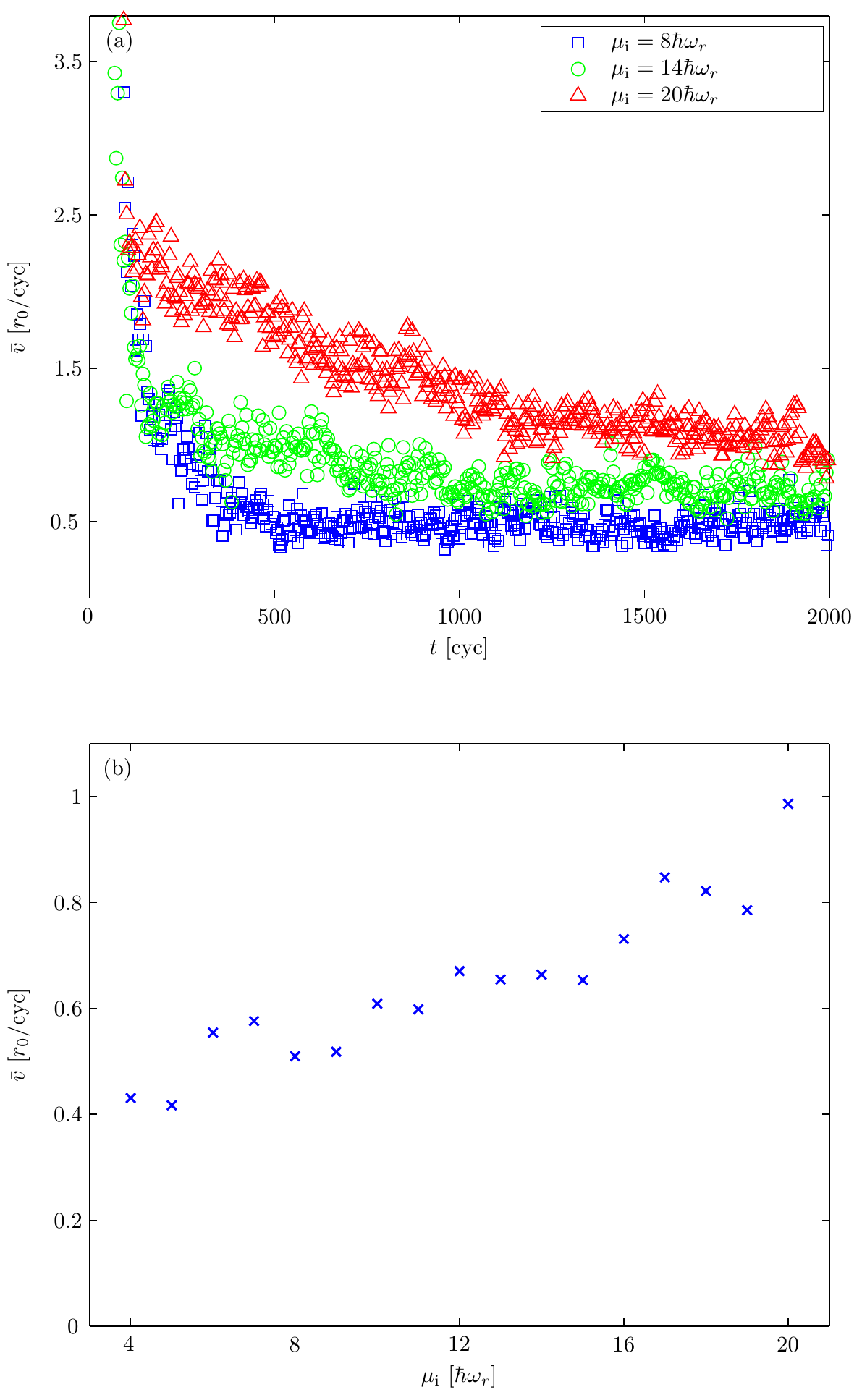}
	\caption{\label{fig:motional_mu}  (a) Vortex motional damping for three representative choices of initial chemical potential $\mu_\mathrm{i}$. (b) Mean vortex speed (averaged over the period $t=1900$--$2000$ cyc) as a function of initial chemical potential $\mu_\mathrm{i}$.)}
	\end{center}
\end{figure}

While the vortex motion as measured by the mean velocity defined in section~\ref{subsec:stir_motional_damping} exhibits the same overall damping behaviour in all our simulations, the damping appears to be generally nonexponential and quite sensitive to the initial noise conditions.  We have observed additional features (e.g. transient and oscillatory behaviours superposed with the damping) peculiar to each trajectory.  It is therefore difficult to unambiguously quantify the rates of motional damping in order to compare the dependence on initial chemical potential, without generating ensemble data for each parameter set, which would be a very heavy task numerically.  We therefore plot in figure~\ref{fig:motional_mu}(a) only the behaviour of three particular trajectories, which indicate at a qualitative level that the vortex motion in the smaller condensates damps to equilibrium more quickly.  We note however that the level of vortex motion that each trajectory damps to shows a clear dependence on the initial chemical potential.  In figure~\ref{fig:motional_mu}(b) we plot the mean velocity of vortices over the period $t=1900$--$2000$ cyc, at which point the vortex motion in each simulation appears to have essentially relaxed to its equilibrium level.  The mean velocity so measured clearly increases with chemical potential. After comparing the results for varying cutoff at fixed initial chemical potential (not shown) we conclude that this is likely due to the increase in equilibrium temperature with increasing initial condensate size.
%%%%%%%%%%%%%%%%%%%%%%%%%%%%%%%%%%%%%%%
%%%%%%%%%%%%%%%%%%%%%%%%%%%%%%%%%%%%%%%
%%%%%%%%%%%%%%%%%%%%%%%%%%%%%%%%%%%%%%%%%%%%%%%%%%%%%%%%%%%%%%%%%%%%%%%%%%%%%%%%%%%%%%%%
\subsection{Cutoff dependence}\label{subsec:cutoff_dependence}
Of crucial importance in classical-field simulations is the effect of the cutoff energy (and thus the basis size) upon the system behaviour.  The dependence of the simulation results on the cutoff can only be eliminated by including above-cutoff physics, and no prescription for doing so in general nonequilibrium scenarios is currently known.  However, the cutoff `height' is chosen on physical grounds (so that all modes significantly modified by interactions are included in the dynamics), and so while we expect (e.g.) the equilibrium temperature to depend on the precise cutoff chosen (see equation~(\ref{eq:temperature_prediction})), we do not expect the dynamics to differ qualitatively between simulations with different cutoff heights.
It might be expected \cite{Lobo04}, however, that the cutoff dependence of the temperature will influence relaxation rates, which depend strongly on the thermal occupation of system excitations (see, e.g., \cite{Fedichev98a}). 
%%%%%%%%%%%%%%%%%%%%%%%%%%%%%%%%%%%%%%%%%%%%%%%%%%%%%%%%%
\subsubsection{Thermodynamic parameters}
%%%%%%%%%%%%%%%%%%%%%%%%%%%%%%%%%%%%%%%
\begin{figure}
	\begin{center}
	\includegraphics[width=0.65\textwidth]{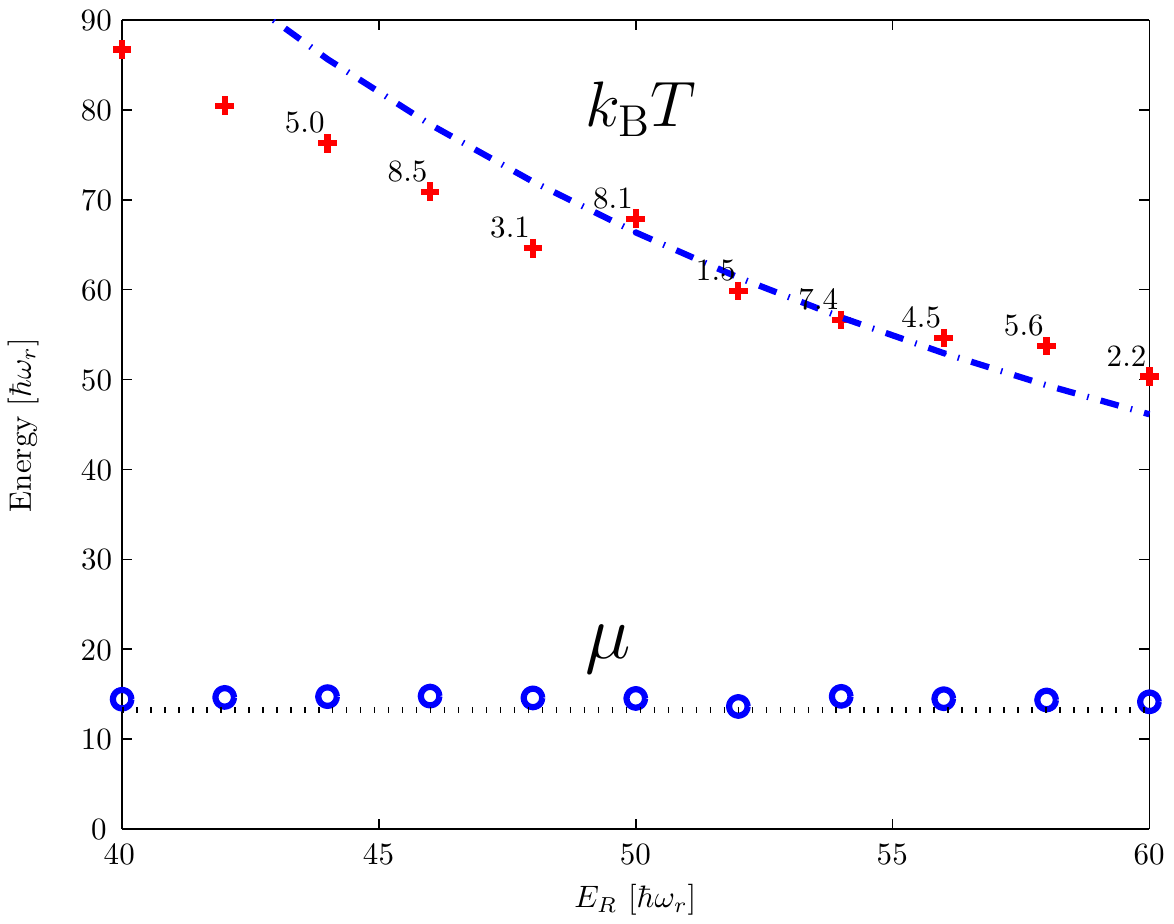}	
	\caption{\label{fig:thermo_variation_with_cutoff}  Field temperatures and chemical potentials reached as a function of cutoff.  Dashed and dash-dotted lines indicate the predictions of equation~(\ref{eq:reduced_mu}) and equation~(\ref{eq:temperature_prediction}), respectively.  Measurement times are indicated as in figure~\ref{fig:thermo_variation_with_mu}.}
	\end{center}
\end{figure}
%%%%%%%%%%%%%%%%%%%%%%%%%%%%%%%%%%%%%%%
In figure~\ref{fig:thermo_variation_with_cutoff} we plot the temperatures and chemical potentials attained in simulations with $\mu_\mathrm{i}=20\hbar\omega_r$ and cutoff heights $E_R$ varying over the range $2\mu_\mathrm{i}\leq E_R\leq3\mu_\mathrm{i}$.  We note first that while the temperature shows a clear dependence on $E_R$, the chemical potential appears relatively insensitive to it.  The system therefore reaches a chemical potential closest to that resulting from the idealised transformation of the condensate to a lattice state without loss of atoms from the condensate mode.   The measured temperature decreases with increasing cutoff as expected, but differs from the $T \propto 1/(\mathcal{M}-1)$ behaviour predicted for a classical field in the Bogoliubov limit.  We attribute this to the constraint imposed by the low cutoff, which artificially constrains the natural modes of the system, leading to spurious thermodynamic behaviour.  Indeed for $E_R=2\mu$, the classical turning point $r_\mathrm{tp}\sim1.7R_\mathrm{TF}$, where $R_\mathrm{TF} \approx 7.8r_0$ is the TF radius of the lattice state, and so the corruption the quasiparticle mode shapes may be quite pronounced.
%%%%%%%%%%%%%%%%%%%%%%%%%%%%%%%%%%%%%%%%%%%%%%%%%%%%%%%%%
\subsubsection{Damping}
\begin{figure}
	\begin{center}
	\includegraphics[width=0.65\textwidth]{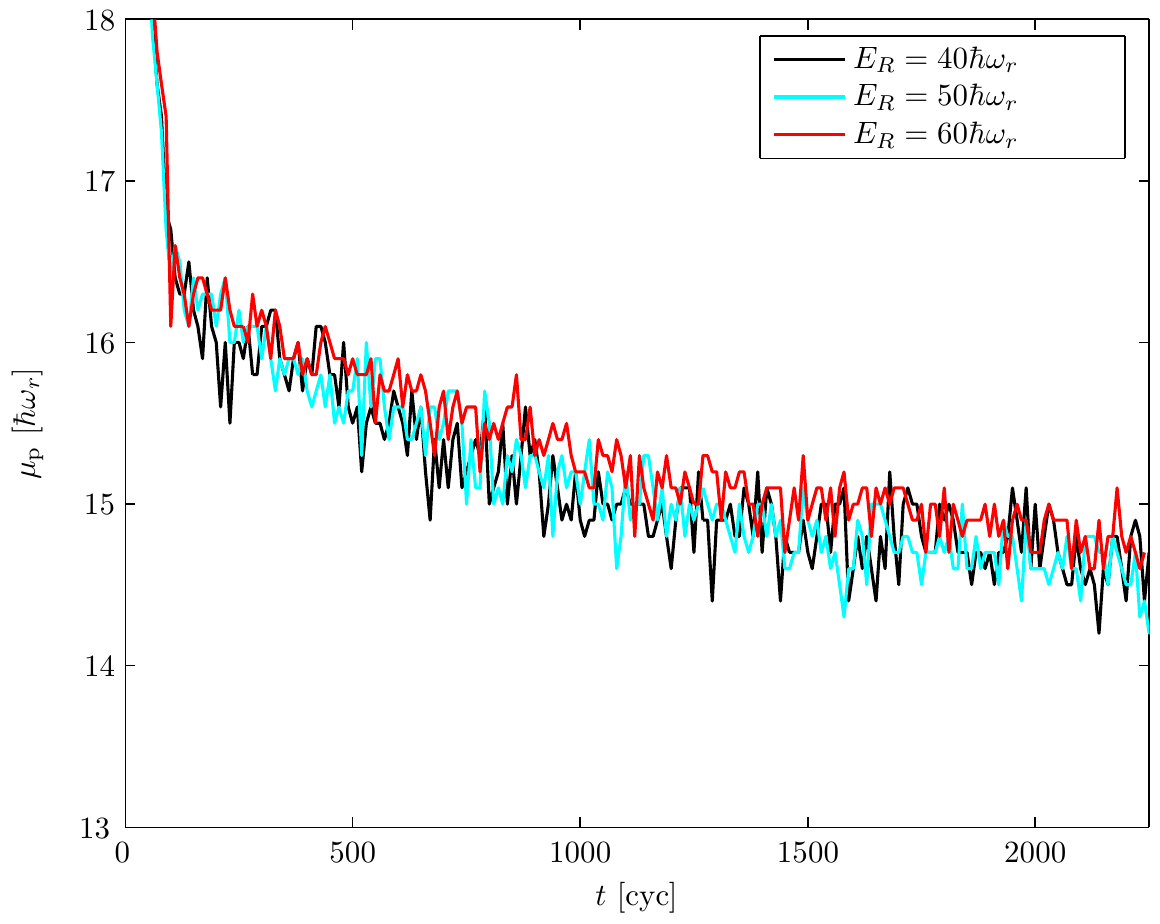}	
	\caption{\label{fig:mu_damp_ER}  Decay of the effective chemical potential of the vortex liquid (as defined in section~\ref{subsec:stir_Temporal}) for trajectories with initial chemical potential $\mu_\mathrm{i}=20\hbar\omega_r$ and three different values of the energy cutoff $E_R$.}
	\end{center}
\end{figure}

As noted in section~\ref{subsec:stir_motional_damping}, it is difficult to extract quantitative comparisons of relaxation rates from the damping of vortex motion.  Here we consider the relaxation of another quantity: the effective chemical potential $\mu_\mathrm{p}$ (power spectrum peak) of the vortex liquid, introduced in section~\ref{subsec:stir_Temporal}.  In figure~\ref{fig:mu_damp_ER}, we show the decay of this quantity with time for three values of the cutoff energy $E_R$.  We observe that the damping is reasonably insensitive to the cutoff.  A possible explanation for this behaviour is that as the effect of lowering the cutoff is to decrease the number of thermally occupied modes in the system and consequently to increase the temperature of these modes, that these two effects serve to approximately cancel one another in determining the rates of thermal damping.  An alternate possibility is that the relaxation dynamics of the low energy modes in the system are relatively insensitive to the precise parameters of the high-energy modes due to the low dimensionality of the system \cite{Polkovnikov09b}.  It is interesting to note, however, that the results of the Ketterle group on the formation of vortices in three dimensions \cite{Abo-Shaeer02} suggest that the rate of relaxation to the lattice state is insensitive to the temperature of the cloud.  Determining the mechanisms underlying this somewhat counterintuitive result would require an analysis of fully three-dimensional classical-field simulations, and we leave this heavy computational task for future work. 
%%%%%%%%%%%%%%%%%%%%%%%%%%%%%%%%%%%%%%%
%%%%%%%%%%%%%%%%%%%%%%%%%%%%%%%%%%%%%%%
%%%%%%%%%%%%%%%%%%%%%%%%%%%%%%%%%%%%%%%%%%%%%%%%%%%%%%%%%%%%%%%%%%%%%%%%%%%%%%%%%%%%%%%%%%%%%%%%%%%%%%%%%%%%%%%%%%%%%%%%%%%%%%%%%%%%
\section{Summary}\label{sec:stir_summary}\enlargethispage{-\baselineskip}
We have carried out the first strictly Hamiltonian simulations of vortex nucleation in stirred Bose-Einstein condensates. Our approach is free from grid-method artifacts such as aliasing and spurious damping at high momenta, and carried out to a very high and well controlled numerical accuracy, enabling a controlled study of field thermalisation and vortex nucleation within an explicitly Hamiltonian classical-field theory. 
\newline\emph{Vacuum symmetry breaking.}---The importance of symmetry breaking has been highlighted in previous works~\cite{Penckwitt02,Parker05a,Parker06b}. Sampling of initial vacuum noise provides an irreducible mechanism for breaking the twofold rotational symmetry of the classical-field solution.   
\newline\emph{Thermalisation.}---Resonant excitation of the quadrupole instability evolves the system from a zero-temperature initial state through a dynamical thermalisation phase in which a rotating thermal cloud forms. Subsequent vortex nucleation in the remaining condensate shows that the cloud provides the requisite dissipation to evolve the condensate subsystem toward a new quasi-equilibrium state in the rotating frame.   
Although the classical-field evolution is Hamiltonian, the stochastically evolving high-energy modes of the classical field provide the thermal-cloud damping that previous authors \cite{Tsubota02a,Penckwitt02} had identified as necessary to drive the nucleation of vortices and the relaxation of the vortex array.
\newline\emph{Identifying the thermal cloud.}---We have quantified the heating caused by the dynamical instability, using a range of measures. Spectral analysis of the nonequilibrium classical-field evolution has been used to identify the condensate chemical potential, and thermal cloud properties were extracted using a semiclassical fit at high energies. While the chemical potentials of the condensate and emerging thermal cloud are slower to equilibrate, the moments of inertia, temperature, and angular momentum of the classical field are rapidly transformed by the instability to those of a rotating, heated Bose gas.  
\newline\emph{Frustrated crystallisation.}---Vortices are seen to nucleate and enter the bulk of the condensate but a regular Abrikosov lattice does not form. Instead the vortices distribute throughout the condensate in spatially disordered vortex-liquid state, consistent with substantial thermal excitation of an underlying regular vortex lattice state. In this respect our results are consistent with previous 2D GPE treatments of similar stirring configurations, which did not observe vortex-lattice crystallisation \cite{Feder01a,Lundh03}.

We conclude that vortex nucleation by stirring is a finite-temperature effect, which arises from dynamical thermalisation when initiated from a zero-temperature Bose condensate.  While we observe vortex \emph{nucleation} in 2D, the asymptotic absence of lattice rigidity suggests that the temperature attained in accommodating the excess (rotating-frame) energy of the initial state is such that vortex lattice \emph{crystallisation} is inhibited in this conserving stirring scenario in 2D. 
%%%%%%%%%%%%%%%%%%%%%%%%%%%%%%%%%%%%%%%%%%%%%%%%%%%%%%%%%%%%%%%%%%%%%%%%%%%%%%%%%%%%%%%%%%%%%%%%%%%%%%%%%%%%%%%%%%%%%%%%%%%%%%%%%%%%
%%%%%%%%%%%%%%%%%%%%%%%%%%%%%%%%%%%%%%%%%%%%%%%%%%%%%%%%%%%%%%%%%%%%%%%%%%%%%%%%%%%%%%%%%%%%%%%%%%%%%%%%%%%%%%%%%%%%%%%%%%%%%%%%%%%%

\chapter{Conclusion}
\label{chap:conclusion}
%%%%%%%%%%%%%%%%%%%%%%%%%%%%%%%%%%%%%%%%%%%%%%%%%%%%%%%%%%%%%%%%%%%%%%%%%%%%%%%%%%%%%%%%%%%%%%%%%%%%%%%%%%%%%%%%%%%%%%%%%%%%%%%%%%%%
%%%%%%%%%%%%%%%%%%%%%%%%%%%%%%%%%%%%%%%%%%%%%%%%%%%%%%%%%%%%%%%%%%%%%%%%%%%%%%%%%%%%%%%%%%%%%%%%%%%%%%%%%%%%%%%%%%%%%%%%%%%%%%%%%%%%
\section{Thesis summary}\label{sec:conc_summary}
In this thesis, we have conducted extensive studies using the \emph{classical-field} method for the description of equilibrium and nonequilibrium finite-temperature degenerate Bose-gas systems.  We have developed new approaches to interpreting the results of classical-field simulations in equilibrium and nonequilibrium regimes, and we have considered the finite-temperature dynamics of quantum vortices, their decay and the dynamics of their formation in the classical-field formalism.

We examined the temporal behaviour of equilibrium classical-field trajectories, and investigated the relationship between temporal correlations and other measures of coherence in the field.  We demonstrated that on short time scales, the condensate in the field can be identified as a nonzero \emph{first moment} of the field in an appropriate phase-rotating frame.
We thus showed how an analogue of the symmetry-breaking assumption typically employed in mean-field theories of Bose condensation emerges naturally from a classical-field trajectory.  We found that the effective eigenvalue of the condensate mode so obtained exhibits the expected temperature-dependent relation to the thermodynamic chemical potential of the field.  We also demonstrated the generality of the symmetry-breaking interpretation, by calculating the \emph{anomalous} thermal-field correlations comprising the pair matrix, and explicitly constructing the anomalous thermal density of the field, which we found to have spatial form and temperature dependence in good agreement with the results of mean-field theories.

Turning our attention to the dynamics of quantum vortices in finite-temperature classical-field theory, we showed that equilibrium precessing-vortex configurations of the classical field could be obtained from the ergodic evolution of classical-field trajectories constrained to a finite conserved angular momentum.  These simulations each exhibited a quantum vortex precessing in rotational equilibrium with the rotating thermal component of the field, naturally realising the balance of coherent and frictional forces on the vortex.  We found that as an off-axis vortex state breaks the rotational symmetry of the Hamiltonian, the vortical condensate mode is fragmented over angular momentum eigenmodes in the formal microcanonical correlations of the field.  We demonstrated by abandoning formal microcanonical (ergodic) averages, and instead considering the field fluctuations on shorter time scales, that the symmetry-broken condensate orbital is revealed by its coherent, number-fluctuation-suppressed statistics.  In addition, this approach revealed the thermal core-filling of the vortex, and produced also the spurious (Goldstone) mode corresponding to the broken rotational symmetry, which we found to have anomalous fluctuation statistics appropriate to its spurious nature.  We showed that the angular momentum of the condensed mode decreases dramatically with increasing field energy (and thus temperature) at fixed overall angular momentum of the field, and that the vortex precessional radius and frequency correspondingly increase until the vortex leaves the condensate.  We demonstrated that the decomposition of the classical field into condensed and noncondensed parts on the basis of our short-time analysis allowed us to calculate the rotational properties of each component, and we found qualitative agreement between the angular velocity of the thermal cloud and the precessional frequency of the vortex.

We then applied this methodology to a nonequilibrium scenario of Bose-field dynamics: the arrest of a finite-temperature precessing-vortex configuration by a fixed trap anisotropy.  Our classical-field approach provided the first simulations of vortex arrest which include the coupled nonequilibrium dynamics of the condensate and thermal cloud, describing the migration of an \emph{equilibrium} rotating finite-temperature field configuration to a new, irrotational equilibrium, due to the action of an externally applied potential.   Decomposing the nonequilibrium field into condensed and noncondensed components, we observed the dynamics of angular momentum exchange between the condensate and thermal cloud, and the loss of angular momentum from the field due to the trap anisotropy.   We found that the anomalous rotational response of the single-vortex state leads to a counter-intuitive spin-up of \emph{both} the condensate and the thermal component during the decay, and we observed nonequilibrium oscillations in the transfer of angular momentum between the two components.  We quantified the heating of the atomic field during its relaxation, and found it to be commensurate with the conversion of rotational kinetic energy into thermal energy during the arrest of the field's rotation.  By varying the severity of the trap anisotropy, we altered the relative strength of the cloud-trap coupling, and observed qualitatively different regimes of decay.  We found that for weak trap anisotropies, the thermal field persists in a long-lived nonequilibrium steady state, with rotation rate intermediate between that of the condensate and that of the static trap.  We observed in this regime the slow loss of angular momentum from the condensate mode while the angular momentum of the thermal field remained approximately constant, until the vortex neared the condensate boundary.  For stronger trap anisotropies, we observed strongly nonequilibrium dynamics, in which the angular momentum of the thermal cloud may be almost entirely depleted before the condensate responds.  Comparing the relaxation of the thermal field following the loss of the vortex from the condensate across simulations with different ellipticities, we found the scaling of the cloud spin-down time with ellipticity to be in good agreement with the predictions of a Boltzmann-gas model, demonstrating that our classical-field treatment correctly describes the dynamics of the thermal material.

Finally, we considered the problem of the stirring of condensates to produce vortex lattices.  While there has been extensive theoretical work to describe the condensate-stirring experiments, there remain many problems with previous treatments.  We therefore considered the problem within our classical-field approach, in order to obtain a more comprehensive and unified treatment.  
We simulated the stirring of quasi-two-dimensional condensates that were initially at zero temperature, and we included a representation of vacuum fluctuations in the initial conditions.  We showed that material spontaneously ejected from the condensate bulk during its dynamic instability quickly comes to a local rotational and thermal equilibrium, forming a rotating thermal cloud which drives the growth of condensate surface modes, leading to the nucleation of vortices.  We showed that the motion of the initially turbulent swarm of vortices admitted into the central bulk of the field in this process is slowly damped by the frictional effect of the thermal field.  We characterised the evolution of the thermodynamic parameters of the field during this relaxation process, and showed that the equilibrium temperature of the field is well predicted by simple energy-conservation arguments.  We found that the heating of the field is such that the vortices in this two-dimensional scenario are prohibited from crystallising into a rigid lattice by thermal fluctuations, and instead form a disordered \emph{vortex liquid}.  We showed that the disordered motion of vortices in this phase destroys true condensation in the equilibrium field, but demonstrated that the quasi-coherent vortex liquid could be distinguished from the remaining thermal component of the field by its \emph{temporal} correlations.
%%%%%%%%%%%%%%%%%%%%%%%%%%%%%%%%%%%%%%%%%%%%%%%%%%%%%%%%%%%%%%%%%%%%%%%%%%%%%%%%%%%%%%%%%%%%%%%%%%%%%%%%%%%%%%%%%%%%%%%%%%%%%%%%%%%%
\section{Directions for future work}\label{sec:conc_future}
The approaches to characterising the coherence of the classical field by its temporal correlations we have developed in this thesis provide a useful extension of the classical-field method.  It would be interesting to further develop this approach to obtain classical-field analogues of temporal correlation functions, such as the one-body Green's functions, that appear in more traditional descriptions of Bose-condensation.  It would also be helpful to derive formal signatures of (e.g.) quasicondensation and other more subtle coherence features in terms of the temporal correlations of classical-field trajectories.  In our approach we have exploited the effective breaking of the phase symmetry by the condensate orbital to define an anomalous averaging operation.  In principle, however, we expect all meaningful spatial (or modal) correlations of the equilibrium field to be able to be expressed in terms of \emph{observable} combinations of classical-field variables; i.e., combinations which are invariant under global gauge rotations of the classical-field phase.  It would therefore be of interest to furnish a characterisation of the anomalous classical-field fluctuations in such terms.  It would also be interesting to see if the normal modes of the classical-field can be obtained from these correlations, and moreover to calculate (for example) the triplet correlations which are related to the scattering of particles in and out of the condensate mode, and the higher cumulants of the field which characterise the beyond-Gaussian corrections to the field fluctuations we might expect to find in the nonperturbative classical-field description.

The simulations of vortex dynamics in this thesis are restricted to quasi-2D systems, and we have considered only a strictly Hamiltonian description of the low-energy coherent region.  The extension of the Gauss-Laguerre quadrature technique for rotating systems to a 3D method would allow us to consider the more complicated dynamics of vortex lines of finite extent.  The extension to three dimensions is conceptually trivial due to the separable nature of the eigenstates of the single-particle Hamiltonian, however, the calculations would be significantly more demanding numerically.  This extension would allow us to study the 3D dynamics of condensate stirring, and to ascertain the role of dimensionality in the vortex-lattice relaxation process.  The inclusion of a mean-field description of above-cutoff atoms would extend the equilibrium rotating-frame PGPE method to allow quantitative comparisons with experiments.  The inclusion of the \emph{dynamics} of the above-cutoff region and its coupling to the coherent region in nonequilibrium regimes such as the vortex arrest and condensate stirring scenarios remains an outstanding goal for the classical-field methodology.  It would also be interesting to further characterise the equilibrium vortex-liquid phase, identify the lattice-liquid and liquid--normal-fluid phase boundaries, and map out the (classical) phase diagram of the rotating Bose gas.  Finally, it would be interesting to see what insights into nonequilibrium and steady-state turbulence in degenerate Bose-gas systems can be obtained from the family of classical-field methods. 
%%%%%%%%%%%%%%%%%%%%%%%%%%%%%%%%%%%%%%%%%%%%%%%%%%%%%%%%%%%%%%%%%%%%%%%%%%%%%%%%%%%%%%%%%%%%%%%%%%%%%%%%%%%%%%%%%%%%%%%%%%%%%%%%%%%%
%%%%%%%%%%%%%%%%%%%%%%%%%%%%%%%%%%%%%%%%%%%%%%%%%%%%%%%%%%%%%%%%%%%%%%%%%%%%%%%%%%%%%%%%%%%%%%%%%%%%%%%%%%%%%%%%%%%%%%%%%%%%%%%%%%%%

%%%%%%%%%%%%%%%%%%%%%%%%%%%%%%%%%%%%%%%%%%%%%%%%%%%%%%%%%%%%%%%%%
\appendix

\renewcommand{\chaptermark}[1]{\markboth{Appendix \thechapter . #1}{}}

\chapter{Condensation in the noninteracting classical field}
\label{app:cfield_transition}
%%%%%%%%%%%%%%%%%%%%%%%%%%%%%%%%%%%%%%%%%%%%%%%%%%%%%%%%%%%%%%%%%%%%%%%%%%%%%%%%%%%%%%%%%%%%%%%%%%%%%%%%%%%%%%%%%%%%%%%%%%%%%%%%%%%%
%%%%%%%%%%%%%%%%%%%%%%%%%%%%%%%%%%%%%%%%%%%%%%%%%%%%%%%%%%%%%%%%%%%%%%%%%%%%%%%%%%%%%%%%%%%%%%%%%%%%%%%%%%%%%%%%%%%%%%%%%%%%%%%%%%%%
In this appendix we illustrate the analogy between the classical-field system and the Bose-field system by considering the thermodynamics of the classical-field distribution.  Specifically, we consider the \emph{noninteracting} limit of the classical field, and derive its transition temperature and the temperature dependence of the condensed fraction of the field.

Our approach here follows the discussion of the noninteracting Bose gas presented in section~\ref{subsec:back_bose_condensation}.  We introduce the classical-field distribution (see section~\ref{subsec:cfield_thermo})
\begin{equation}
	f^\mathrm{CF}(\epsilon_\nu) = \frac{k_\mathrm{B}T}{\epsilon_\nu-\mu} \Theta(E_R - \epsilon_\nu),
\end{equation}
where $\mu$ is the chemical potential, $\epsilon_\nu$ is the energy of a particular single-particle level, $E_R$ is the cutoff energy, and $\Theta(x)$ denotes a Heaviside function.  As in the case of the full Bose distribution, the occupation of the ground level $f^\mathrm{CF}(\epsilon_0\equiv0)$ becomes negative for $\mu>0$, and so we require $\mu \leq 0$.  Consequently the population of any excited state (below the cutoff) is bounded from above: $f^\mathrm{CF}(\epsilon_\nu) \leq k_\mathrm{B}T/ \epsilon_\nu$.  The critical temperature is thus that temperature at which the population of excited states is equal to the total classical-field population $N$, for $\mu=0$.  We follow \cite{Dalfovo99,Pethick02} by making the semiclassical approximation; i.e., we neglect the discrete level structure of the single-particle energy, and treat it as a continuous variable, introducing the general density of states $g(\epsilon)=C_\alpha \epsilon^{\alpha-1}$, and we find\footnote{For $\alpha>1$, the condensate (with $\epsilon=0$) does not contribute to the integral, and so we can safely take the lower bound of the integral to zero (see discussion in reference~\cite{Pathria96}).}
\begin{equation}\label{eq:app1_nbound}
	N = \int_0^{E_R}\!d\epsilon\,g(\epsilon) \frac{k_\mathrm{B}T_\mathrm{c}}{\epsilon} = \frac{C_\alpha E_R^{\alpha-1}k_\mathrm{B}T_\mathrm{c}}{\alpha-1}. 
\end{equation}
The critical temperature of the classical field is therefore
\begin{equation}
	k_\mathrm{B}T_\mathrm{c} = \frac{\alpha-1}{C_\alpha E_R^{\alpha-1}}N,
\end{equation}
which, of course, depends explicitly on the cutoff energy $E_R$.  We note that, as in the case of Bose statistics, the integral diverges for $\alpha\leq1$.  Recalling that $\alpha=d/2-1$ for a $d$-dimensional homogeneous system and $\alpha=d-1$ for a $d$-dimensional harmonic trap \cite{Pethick00}, the classical-field thermodynamics thus preserve the result of the full Bose system that no condensation occurs in a homogeneous system for $d\leq2$, nor in a $1$-dimensional harmonic trap, in the thermodynamic limit.  Moreover, the excited-state population at a temperature $T<T_\mathrm{c}$ is given by expression~\reff{eq:app1_nbound} with $T_\mathrm{c}\to T$, and we thus find
\begin{equation}\label{eq:app1_cfrac}
	\frac{N_0}{N} = 1 - \frac{T}{T_\mathrm{c}},
\end{equation}
as found by Connaughton \emph{et al.} \cite{Connaughton05} for the homogeneous classical-wave system in three dimensions.  We note, however, the somewhat surprising result that the relation~\reff{eq:app1_cfrac} is independent of geometry and dimensionality, provided that the integral in \reff{eq:app1_nbound} converges.   In the limit that the distribution $f^\mathrm{CF}(\epsilon_\nu)$ coincides with the true Bose-Einstein distribution throughout the below-cutoff region, the corresponding \emph{above-cutoff} population of course rapidly increases with temperature \cite{Davis05}, and depends explicitly on the geometry, so that the correct geometry-dependent scaling is regained for the fraction $N_0/(N+N_\mathrm{above})$.
%%%%%%%%%%%%%%%%%%%%%%%%%%%%%%%%%%%%%%%%%%%%%%%%%%%%%%%%%%%%%%%%%%%%%%%%%%%%%%%%%%%%%%%%%%%%%%%%%%%%%%%%%%%%%%%%%%%%%%%%%%%%%%%%%%%%
%%%%%%%%%%%%%%%%%%%%%%%%%%%%%%%%%%%%%%%%%%%%%%%%%%%%%%%%%%%%%%%%%%%%%%%%%%%%%%%%%%%%%%%%%%%%%%%%%%%%%%%%%%%%%%%%%%%%%%%%%%%%%%%%%%%%

\chapter{Interaction-picture integration algorithms}
\label{app:IP_algorithms}
%%%%%%%%%%%%%%%%%%%%%%%%%%%%%%%%%%%%%%%%%%%%%%%%%%%%%%%%%%%%%%%%%%%%%%%%%%%%%%%%%%%%%%%%%%%%%%%%%%%%%%%%%%%%%%%%%%%%%%%%%%%%%%%%%%%%
%%%%%%%%%%%%%%%%%%%%%%%%%%%%%%%%%%%%%%%%%%%%%%%%%%%%%%%%%%%%%%%%%%%%%%%%%%%%%%%%%%%%%%%%%%%%%%%%%%%%%%%%%%%%%%%%%%%%%%%%%%%%%%%%%%%%
In this appendix we review the \emph{interaction-picture} methods for the temporal integration of Gross-Pitaevskii-like equations.  These methods
are all descendants of the original RK4IP algorithm developed by Ballagh \cite{Ballagh00b}, and the adaptive ARK45-IP variant of the algorithm due to Davis \cite{Davis_DPhil}. 
%%%%%%%%%%%%%%%%%%%%%%%%%%%%%%%%%%%%%%%%%%%%%%%%%%%%%%%%%%%%%%%%%%%%%%%%%%%%%%%%%%%%%%%%%%%%%%%%%%%%%%%%%%%%%%%%%%%%%%%%%%%%%%%%%%%%
\section{The RK4IP algorithm}\enlargethispage{-\baselineskip}
The fundamental issue we face in evolving (nonlinear) Schr\"odinger-like equations is the efficient evaluation of operators which are nonlocal in their action on the wave function.  A key observation is that the operators which are nonlocal in \emph{position} space are typically local in \emph{momentum} space; for example, the Laplacian $-\nabla^2$ becomes the multiplication $k^2 \times$ in momentum space.  This observation forms the basis for Fourier-based integration methods, such as the symmetrised split-step method (see \cite{Javanainen06} and references therein), which alternates between the two representations of the wave function, applying operators which are local in each space.   The original RK4IP algorithm \cite{Ballagh00b} is in essence an efficient extension of the split-step method, in which the Fourier transform is used to transform to an \emph{interaction picture} with respect to the Laplacian, and the effect of the remaining operators (which are usually local) is evaluated in this picture, using a standard Runge-Kutta approach.
In general, \emph{projected} Gross-Pitaevskii equations are formulated in a Galerkin approach, i.e., the field is propagated in the representation which diagonalises the Laplacian, or some more general base Hamiltonian $H_0$ (see section~\ref{subsec:galerkin}).  In such cases the basis amplitudes are transformed to the interaction picture with respect to $H_0$, exploiting its diagonal form in this representation, and the effect of the remaining (projected) terms on the basis amplitudes is evaluated in this interaction picture.
%%%%%%%%%%%%%%%%%%%%%%%%%%%%%%%%%%%%%%%%%%%%%%%%%%%%%%%%%%%%%%%%%%%%%%%%%%%%%%%%%%%%%%%%
\subsection{Fourth-order Runge Kutta}
Runge-Kutta (RK) methods form a standard approach \cite{Press92} for the integration of a system of differential equations
\begin{equation}\label{eq:IP_app_ode}
	\frac{d\mathbf{x}}{dt} = \mathbf{f}\bigl(\mathbf{x},t\bigr),
\end{equation}
in steps of $\Delta t$.  To advance the solution of equation~\reff{eq:IP_app_ode} from $\mathbf{x}(t_i)$ to $\mathbf{x}(t_i + \Delta t)$, the fourth-order RK method calculates four estimates of the function change over the interval $[t_i, t_i+\Delta t]$ 
\begin{subequations}
    \begin{eqnarray}
        \mathbf{k}_1 &=& \Delta t \mathbf{f}\bigl(\mathbf{x}(t_i), t_i\bigr) \\
        \mathbf{k}_2 &=& \Delta t \mathbf{f}\bigl(\mathbf{x}(t_i)+\mathbf{k}_1/2, t_i+\Delta t/2\bigr) \\
        \mathbf{k}_3 &=& \Delta t \mathbf{f}\bigl(\mathbf{x}(t_i)+\mathbf{k}_2/2, t_i+\Delta t/2\bigr) \\
        \mathbf{k}_4 &=& \Delta t \mathbf{f}\bigl(\mathbf{x}(t_i)+\mathbf{k}_3, t_i+\Delta t\bigr),
    \end{eqnarray}
\end{subequations}
and averages these to produce the final estimate
\begin{equation}
	\mathbf{x}(t_i+\Delta t) = \mathbf{x}(t_i) + \bigl[\mathbf{k}_1 + 2(\mathbf{k}_2+\mathbf{k}_3) + \mathbf{k}_4\bigr]/6.
\end{equation}
The error term in this estimate is of order $(\Delta t)^5$, and the method is therefore fourth-order accurate.
%%%%%%%%%%%%%%%%%%%%%%%%%%%%%%%%%%%%%%%%%%%%%%%%%%%%%%%%%%%%%%%%%%%%%%%%%%%%%%%%%%%%%%%%
\subsection{RK4IP}\label{subsec:rk4ip}
The projected-GP equations we wish to solve are of the general form
\begin{equation}
	i\frac{d\mathbf{c}}{dt} = H_0 \mathbf{c} + \hat{G}(t)\mathbf{c},
\end{equation}
where the vector $\mathbf{c}$ lists all the mode coefficients $c_j$, and the nonlinear operator $\hat{G}(t)$ includes the projected nonlinear interaction term and possibly some other perturbing term (see section~\ref{subsec:cfield_proj_cft}).  The integration relies on the transformation of the mode coefficients to an interaction picture with origin at some time $t_0$, i.e.
\begin{equation}
	\mathbf{c}^\mathrm{I}(t) = e^{iH_0(t-t_0)}\mathbf{c}(t).
\end{equation}
The resulting differential equation in the interaction picture is
\begin{equation}
	i\frac{d\mathbf{c}^\mathrm{I}}{dt} = \hat{G}^\mathrm{I} \mathbf{c}^\mathrm{I},
\end{equation}
where the interaction-picture nonlinear operator
\begin{equation}
	\hat{G}^\mathrm{I}(t) = e^{iH_0(t-t_0)}\hat{G}e^{-iH_0(t-t_0)}.
\end{equation}
For each step $\mathbf{c}(t_i)\to\mathbf{c}^\mathrm{I}(t_i+\Delta t)$ of the integration, we choose a new interaction-picture origin $t_0$, transform to the interaction picture, evaluate the four Runge-Kutta slopes $\mathbf{k}_i$, and transform back.  A considerable simplification is obtained by choosing the interaction picture origin $t_0=t_i+\Delta t/2$, so that the nonlinear operator need not be transformed to evaluate the midpoint slopes $\mathbf{k}_2$ and $\mathbf{k}_3$ \footnote{In the original RK4IP implementation the solution was propagated in real space, and so the interaction-picture transformation $e^{iH_0t}$ was the most numerically demanding task.  By contrast, the modern implementations propagate the solution in the spectral basis so that the application of $\hat{G}$ is the numerically demanding operation.}.  The resulting algorithm is
\begin{subequations}
    \begin{eqnarray}
        \mathbf{c}^\mathrm{I}(t_i) &=& e^{-i(\Delta/2)H_0} \mathbf{c}(t_i) \\
        \mathbf{k}_1 &=& e^{-i(\Delta t/2)H_0} \bigl[ -i \Delta \hat{G}(t_i)\bigr]\mathbf{c}(t_i) \\
        \mathbf{k}_2 &=& -i\Delta t \hat{G}(t_i+\Delta t/2) \bigl[\mathbf{c}^\mathrm{I}(t_i) + \mathbf{k}_1/2\bigr] \\
        \mathbf{k}_3 &=& -i\Delta t \hat{G}(t_i+\Delta t/2) \bigl[\mathbf{c}^\mathrm{I}(t_i) + \mathbf{k}_2/2\bigr] \\
        \mathbf{k}_4 &=& -i\Delta t \hat{G}(t_i+\Delta t) e^{-i(\Delta t/2)H_0} \bigl[\mathbf{c}^\mathrm{I}(t_i) + \mathbf{k}_3] \\
        \mathbf{c}(t_i+\Delta t) &=& e^{-i(\Delta t/2)H_0}\bigl\{\mathbf{c}^\mathrm{I}(t_i) + \bigl[\mathbf{k}_1 + 2(\mathbf{k}_2 + \mathbf{k}_3)\bigr]/6\bigr\} + \mathbf{k}_4/6.
    \end{eqnarray}
\end{subequations}
We note also that this fixed-step-size interaction-picture RK approach can be adapted to the integration of stochastic differential equations \cite{Gardiner04}. 
%%%%%%%%%%%%%%%%%%%%%%%%%%%%%%%%%%%%%%%%%%%%%%%%%%%%%%%%%%%%%%%%%%%%%%%%%%%%%%%%%%%%%%%%%%%%%%%%%%%%%%%%%%%%%%%%%%%%%%%%%%%%%%%%%%%%
\section{The ARK45-IP algorithm}\label{sec:rk4_app_ark45}
One issue with the fixed-step-size RK4IP method presented in section~\ref{subsec:rk4ip} is that of choosing the appropriate step size $\Delta t$.  A step size which is too large leads to numerical inaccuracy and ultimately to the failure of the integration algorithm to produce the correct dynamics for the classical field.  However, the necessity of choosing a sufficiently short time step must be contrasted with the dependence of the total simulation time on the step size.  In general, one would like to take the largest step size which is still consistent with accurate integration of the field trajectory.  Choosing such a step size is particularly difficult in nonequilibrium scenarios such as we consider in chapters~\ref{chap:arrest} and \ref{chap:stirring}, in which the composition of the field (e.g., the occupation of high-energy modes) and thus the appropriate step size may change during the evolution of the field.  This issue can be solved by using an \emph{adaptive-step-size} algorithm, which allows one to choose an accuracy tolerance at the beginning of the integration.  Throughout the integration the adaptive algorithm then shortens or lengthens the step size so as to always take the largest step consistent with the prescribed accuracy requirement.
%%%%%%%%%%%%%%%%%%%%%%%%%%%%%%%%%%%%%%%%%%%%%%%%%%%%%%%%%%%%%%%%%%%%%%%%%%%%%%%%%%%%%%%%
\subsection{Adaptive Runge-Kutta}
This algorithm is a generalisation of the RK method, based on the notion of calculating slopes of the function at trial midpoints that can be combined to give \emph{two} estimates of the function value at the step endpoint $\mathbf{x}(t_i+\Delta t)$, of two different orders.  By comparing these two estimates we thus obtain a measure of the truncation error in the higher-order estimate, and this allows us to choose after each step how to proceed with the integration; i.e., whether to accept the step, or reject the step and make a second attempt with a shortened step size. 
The most common adaptive RK algorithm is based on the fifth-order RK formula of form
\begin{subequations}
    \begin{eqnarray}\label{eq:rk4_k1}
        \mathbf{k}_1 &=& \Delta t \mathbf{f}\bigl(\mathbf{x}(t_i), t_i\bigr) \\
        \mathbf{k}_2 &=& \Delta t \mathbf{f}\bigl(\mathbf{x}(t_i)+b_{21}\mathbf{k}_1, t_i+a_2\Delta t\bigr) \\ \nonumber
                    &&\cdots \\
        \mathbf{k}_6 &=&\Delta t \mathbf{f}\bigl(\mathbf{x}(t_i)+ b_{61}\mathbf{k}_1 + \cdots + b_{65}\mathbf{k}_5, t_i+a_6\Delta t\bigr) \\
        \mathbf{x}(t_i+\Delta t) &=& \mathbf{x}(t_i) + d_1\mathbf{k}_1 + d_2\mathbf{k}_2 + d_3\mathbf{k}_3 + d_4\mathbf{k}_4 + d_5\mathbf{k}_5 + d_6\mathbf{k}_6,
    \end{eqnarray}
\end{subequations}
where the $a_i$, $b_{ij}$, and $d_i$ are rational-number coefficients\footnote{The particular values of these coefficients that we use here are the Cash-Karp values, which are tabulated in reference~\cite{Press92}.}.
The key point is that having evaluated the slopes $\mathbf{k}_i$, a fourth-order estimate 
\begin{equation}
	\widetilde{\mathbf{x}}(t_i+\Delta t) = \mathbf{x}(t_i) + \widetilde{d}_1\mathbf{k}_1 + \widetilde{d}_2\mathbf{k}_2 + \widetilde{d}_3\mathbf{k}_3 + \widetilde{d}_4\mathbf{k}_4 + \widetilde{d}_5\mathbf{k}_5 + \widetilde{d}_6\mathbf{k}_6,      
\end{equation}
can also be obtained.  The difference between the two is then taken as an estimate of the truncation error in the fifth-order estimate $\mathbf{x}(t_i+\Delta t)$. Since $\mathbf{x}(t)$ is in general a vector, we choose some summary statistic such as (for example) $\Delta_\mathrm{m} = \max\limits_j\{ |\widetilde{x}_j(t_i+\Delta t)-x_j(t_i+\Delta t)|\}$ as an estimate of the error.  With a given error tolerance $\Delta_\mathrm{tol}$ in mind, we can use the expectation that $\Delta_\mathrm{m}$ scales as $h^5$ to deduce the step size that should be taken.  If the error is too large, the step size is reduced, while if it is smaller than the required tolerance, the step size can be safely expanded.  We are thus lead to choose the step size according to 
\begin{equation}\label{eq:rk4_scale}
	\Delta t_{n+1} = \left\{
	\begin{array}{rl}
		S\Delta t_n \left|\frac{\Delta_\mathrm{tol}}{\Delta_m}\right|^{1/5} &\qquad\Delta_m \leq \Delta_\mathrm{tol} \\
		S\Delta t_n \left|\frac{\Delta_\mathrm{tol}}{\Delta_m}\right|^{1/4} &\qquad\Delta_m > \Delta_\mathrm{tol}.
	\end{array} \right. % \}	
\end{equation}
The safety factor $S$ here is a coefficient a few percent smaller than unity, and the rationale for the choice of exponents in equation~\reff{eq:rk4_scale} is given in reference~\cite{Press92}.
%%%%%%%%%%%%%%%%%%%%%%%%%%%%%%%%%%%%%%%%%%%%%%%%%%%%%%%%%%%%%%%%%%%%%%%%%%%%%%%%%%%%%%%%
\subsection{ARK45-IP}\enlargethispage{-\baselineskip}
The ARK45-IP algorithm is due to Davis \cite{Davis_DPhil}, and provides an adaptive variant of the RK4IP algorithm, by applying the formula of equations~(\ref{eq:rk4_k1}-\ref{eq:rk4_scale}) in the interaction picture with respect to $H_0$.  The resulting algorithm is more complicated than the fixed-step RK4IP algorithm, due to the range of intermediate times $a_i \Delta t$ at which the RK slopes are evaluated.  Thus no simplifications of the interaction-picture transformations can in general be made in this case.  However due to the comparatively light numerical load of applying the operator $e^{i(a_i\Delta t)H_0}$ in the spectral representation (in which it is diagonal), this is of little concern.  The algorithm for the fifth-order estimate is of the general form
\begin{subequations}
    \begin{eqnarray}
        \mathbf{k}_1 &=& -i\Delta t \hat{G}(t_i) \mathbf{c}(t_i) \\
        \mathbf{k}_2 &=&  e^{-i (a_2 \Delta t)H_0} \left[-i\Delta t \hat{G}(t_i+a_2\Delta t)\right] e^{i(a_2\Delta t) H_0} \bigl[\mathbf{c}(t_i) + b_{21}\mathbf{k}_1\bigr] \\ \nonumber
        &&\cdots \\	
        \mathbf{k}_6 &=& e^{-i (a_6 \Delta t)H_0} \left[-i\Delta t \hat{G}(t_i+a_6\Delta t)\right] e^{i(a_6\Delta t)H_0} \nonumber \\
                    &&\quad\times \bigl[\mathbf{c}(t_i) + b_{61}\mathbf{k}_1 + b_{62}\mathbf{k}_2 + b_{63}\mathbf{k}_3 + b_{64}\mathbf{k}_4 + b_{65}\mathbf{k}_5\bigr]  \\
        \mathbf{c}(t_i+\Delta t) &=& \mathbf{c}(t_i) + d_1\mathbf{k}_1 + d_2\mathbf{k}_2 + d_3\mathbf{k}_3 + d_4\mathbf{k}_4 + d_5\mathbf{k}_5 + d_6\mathbf{k}_6,      
    \end{eqnarray}
\end{subequations}
and the fourth-order estimate is
\begin{equation}
	\widetilde{\mathbf{c}}(t_i+\Delta t) = \mathbf{c}(t_i) + \widetilde{d}_1\mathbf{k}_1 + \widetilde{d}_2\mathbf{k}_2 + \widetilde{d}_3\mathbf{k}_3 + \widetilde{d}_4\mathbf{k}_4 + \widetilde{d}_5\mathbf{k}_5 + \widetilde{d}_6\mathbf{k}_6.   
\end{equation}
In practice we form the single error estimate
\begin{equation}
	\Delta_\mathrm{m} = \max\limits_j\left\{\frac{|c_j-\widetilde{c}_j|^2}{|c_j|^2}\right\},
\end{equation}
where we only consider those mode elements $c_j$ such that $|c_j|^2 > 10^{-3} \max\{|c_i|^2\}$, and thus either accept the step and increase $\Delta t$, or reject the step and decrease $\Delta t$, according to~\reff{eq:rk4_scale}, with safety factor $S=0.92$ \cite{Davis_DPhil}.
%%%%%%%%%%%%%%%%%%%%%%%%%%%%%%%%%%%%%%%%%%%%%%%%%%%%%%%%%%%%%%%%%%%%%%%%%%%%%%%%%%%%%%%%%%%%%%%%%%%%%%%%%%%%%%%%%%%%%%%%%%%%%%%%%%%%
%%%%%%%%%%%%%%%%%%%%%%%%%%%%%%%%%%%%%%%%%%%%%%%%%%%%%%%%%%%%%%%%%%%%%%%%%%%%%%%%%%%%%%%%%%%%%%%%%%%%%%%%%%%%%%%%%%%%%%%%%%%%%%%%%%%%

\chapter{Quadrature methods for the PGPE}
\label{app:app_quad}
%%%%%%%%%%%%%%%%%%%%%%%%%%%%%%%%%%%%%%%%%%%%%%%%%%%%%%%%%%%%%%%%%%%%%%%%%%%%%%%%%%%%%%%%%%%%%%%%%%%%%%%%%%%%%%%%%%%%%%%%%%%%%%%%%%%%
%%%%%%%%%%%%%%%%%%%%%%%%%%%%%%%%%%%%%%%%%%%%%%%%%%%%%%%%%%%%%%%%%%%%%%%%%%%%%%%%%%%%%%%%%%%%%%%%%%%%%%%%%%%%%%%%%%%%%%%%%%%%%%%%%%%%
In this appendix we show how Gaussian quadrature rules can be used to efficiently and accurately evaluate the integrals which appear in the spectral-Galerkin representation of the PGPE. We closely follow the development presented by Blakie \cite{Blakie08a} and Bradley \cite{Bradley08,Bradley_PhD}.
%%%%%%%%%%%%%%%%%%%%%%%%%%%%%%%%%%%%%%%%%%%%%%%%%%%%%%%%%%%%%%%%%%%%%%%%%%%%%%%%%%%%%%%%%%%%%%%%%%%%%%%%%%%%%%%%%%%%%%%%%%%%%%%%%%%%
\section{Gaussian quadrature rules}
A Gaussian quadrature rule is a means of approximation an integral by a discrete sum, i.e.
\begin{equation}\label{eq:quad_general}
	\int_a^b\!dx\,W(x)f(x) \approx \sum_{j=1}^N w_j f(x_j),
\end{equation}
where $W(x)$ is some \emph{weight function}, and the $w_j$ are the quadrature \emph{weights} corresponding to the \emph{abscissas} $x_j$ that define the appropriate \emph{quadrature grid}, which is in general not evenly spaced.  A particularly nice feature of such Gaussian quadrature rules is that in the case that $f(x)$ is a polynomial, the quadrature-rule order $N$ can be chosen such that the approximate correspondence of equation~\reff{eq:quad_general} becomes \emph{exact}.  Such quadrature rules have deep connection with the theory of orthogonal polynomials:  given a domain of integration $[a,b]$, weight function $W(x)$, and set of polynomials $\{p_j(\mathbf{x})\}$ which are orthonormal with respect to the weighted integral on this domain, i.e.
\begin{equation}
	\langle p_i | p_j \rangle \equiv \int_a^b\!dx\,W(x)p_i(x)p_j(x) = \delta_{ij},
\end{equation}
the $N$ roots of $p_N(x)$ are precisely the $N$ abscissas $x_j$ for the quadrature rule on $[a,b]$ with weight $W(x)$.  The corresponding weights are given by
\begin{equation}
	w_j = \frac{\langle p_{N-1}|p_{N-1}\rangle }{p_{N-1}(x_j)p_N'(x_j)},
\end{equation}
where $p_N'(x_j)$ is the derivative of $p_N(x)$ evaluated at $x_j$.  For a quadrature rule of order $N$, the substitution~\reff{eq:quad_general} is then exact for all \emph{general} polynomials of degree $2N-1$ or less.

For example, for the particular weight function $W(x)=e^{-x^2}$ and domain of integration $(-\infty,\infty)$, the corresponding orthogonal polynomials are the \emph{Hermite polynomials} defined by
\begin{subequations}\label{eq:quad_herm1} 
    \begin{eqnarray}
        H_0(x) &=& 1 \\
        H_1(x) &=& 2x \\ \label{eq:quad_herm3}
        H_{j+1}(x) &=& 2xH_j(x) - 2jH_{j-1}(x),
    \end{eqnarray}
\end{subequations}
and the corresponding rule~\reff{eq:quad_general} is the Gauss-Hermite quadrature rule.  The quadrature abscissas and corresponding weights for this rule are tabulated in (e.g.) \cite{Abramowitz72}.
%%%%%%%%%%%%%%%%%%%%%%%%%%%%%%%%%%%%%%%%%%%%%%%%%%%%%%%%%%%%%%%%%%%%%%%%%%%%%%%%%%%%%%%%%%%%%%%%%%%%%%%%%%%%%%%%%%%%%%%%%%%%%%%%%%%%
\section{Hermite quadratures for harmonic confinement}
In this section we describe the Gauss-Hermite quadrature method for evaluating the (projected) nonlinear term in the nonrotating PGPE.  This discussion closely follows that of Blakie \cite{Blakie08a}.
%%%%%%%%%%%%%%%%%%%%%%%%%%%%%%%%%%%%%%%%%%%%%%%%%%%%%%%%%%%%%%%%%%%%%%%%%%%%%%%%%%%%%%%%
\subsection{Application to the PGPE}
We begin with the dimensionless PGPE
\begin{equation}
	i\partial_t \psi = H_0\psi + \mathcal{P}\left\{U_0|\psi|^2\psi\right\},
\end{equation}
where the dimensionless base Hamiltonian
\begin{equation}
	H_0 = \frac{-\nabla^2}{2} + \frac{1}{2}\left(\lambda_x^2x^2+\lambda_y^2y^2+\lambda_z^2z^2\right).
\end{equation}
For simplicity, we follow Blakie \cite{Blakie08a} in considering the isotropic case $\lambda_x=\lambda_y=\lambda_z=1$.  The extension to anisotropic harmonic trapping is straightforward.  As the Hamiltonian $H_0$ can be separated into additive parts corresponding to the single-particle degrees of freedom in each of the spatial dimensions, i.e. $H_0=H_{0x}+H_{0y}+H_{0z}$, the eigenstates of $H_0$ are separable, i.e. $\varphi_n(\mathbf{x}) = \phi_\alpha(x)\phi_\beta(y)\phi_\gamma(z)$, where (e.g.) the eigenstates $\phi_\alpha(x)$ of $H_{0x}$ are such that
\begin{equation}
	\left[-\frac{1}{2}\frac{d^2}{dx^2} + \frac{1}{2}x^2\right]\phi_\alpha(x) = \epsilon_\alpha\phi_\alpha(x),
\end{equation}
where the eigenvalues $\epsilon_\alpha= \alpha+\frac{1}{2}$, so that the eigenvalues of $H_0$ are
\begin{equation}
	\epsilon_n = \epsilon_\alpha + \epsilon_\beta + \epsilon_\gamma.
\end{equation}
The eigenfunctions themselves are of form
\begin{equation}
	\phi_\alpha(x) = h_\alpha H_\alpha(x)e^{-x^2/2},
\end{equation}
where the normalisation coefficient
\begin{equation}
	h_\alpha = \frac{1}{\sqrt{2^\alpha \alpha!\sqrt{\pi}}},
\end{equation}
and the $H_\alpha(x)$ are the Hermite polynomials (equations~(\ref{eq:quad_herm1}-\ref{eq:quad_herm3})).
The coherent region can thus be defined
\begin{equation}\label{eq:coh_reg_param}
	\mathbf{C} \equiv \{\alpha,\beta,\gamma:\quad \epsilon_\alpha + \epsilon_\beta + \epsilon_\gamma \leq E_R\},
\end{equation}
and the maximum allowed value of each of $\alpha,\beta,\gamma$ is therefore 
\begin{equation}
	\alpha_{\mathrm{max}} = E_R - \frac{3}{2},
\end{equation}
yielding $M=\alpha_{\mathrm{max}}+1$ distinct eigenstates in each dimension.
The equation~\reff{eq:coh_reg_param} defines the coherent region $\mathbf{C}$ as a pyramidal volume in $\alpha\beta\gamma$ space, which thus encloses a volume (i.e., number of modes) $\mathcal{M}\approx (1/6) M^3$.  Our task in evaluating the nonlinear potential is therefore that of evaluating the integral
\begin{equation}
	G_{\alpha\beta\gamma}[\psi(\mathbf{x})] = \int\!d\mathbf{x}\,\phi^*_\alpha(x)\phi^*_\beta(y)\phi^*_\gamma(z) |\psi(\mathbf{x})|^2\psi(\mathbf{x}).
\end{equation}
The important point is that, given the restriction~\reff{eq:coh_reg_param} on the single-particle modes from which $\psi(\mathbf{x})$ is constructed, we have
\begin{equation}
	\psi(\mathbf{x}) = Q(x,y,z) e^{-(x^2+y^2+z^2)/2},
\end{equation}
where $Q(x,y,z)$ is a polynomial
\begin{equation}
	Q(x,y,z) = \sum_{\alpha\beta\gamma\in\mathbf{C}} c_{\alpha\beta\gamma} h_\alpha h_\beta h_\gamma H_\alpha(x) H_\beta(y) H_\gamma(z),
\end{equation}
of maximum degree $M-1$ in each of $x$, $y$, and $z$.  As a result, the nonlinear integral is of form
\begin{equation}\label{eq:quad_integral}
	G_{\alpha\beta\gamma} = \int\!d\mathbf{x}\,e^{-2(x^2+y^2+z^2)} P_{\alpha\beta\gamma}(x,y,z),	
\end{equation}
where 
\begin{equation}
	P_{\alpha\beta\gamma}(x,y,z) = h_\alpha h_\beta h_\gamma H_\alpha(x) H_\beta(y) H_\gamma(z) |Q(x,y,z)|^2Q(x,y,z),
\end{equation}
is therefore of maximum degree $4(M-1)$ in each of $x$, $y$, and $z$.  Correspondingly, the integral $G_{\alpha\beta\gamma}$ is of the appropriate form for Gauss-Hermite quadrature, and can therefore be evaluated \emph{exactly} using a three-dimensional spatial grid of size $2(M-1)\times2(M-1)\times2(M-1)$ points, i.e.
\begin{equation}
	G_{\alpha\beta\gamma} = \sum_{ijk} w_i w_j w_k P_{\alpha\beta\gamma}(x_i,x_j,x_k),
\end{equation}
where the $x_i$ and $w_i$ are respectively the \emph{roots} and \emph{weights} of the one-dimensional Gauss-Hermite quadrature, with weight function $w(x)=\exp(-2x^2)$ (see e.g., \cite{Abramowitz72}).  

The procedure for evaluating the projection of the nonlinear term is thus as follows:
\begin{enumerate}
	\item{Transform the field from the coefficient representation $\{c_{\alpha\beta\gamma}\}$ to a spatial representation on the \emph{quadrature grid} $\{\mathbf{x}_{ijk}\}$, according to 
\begin{equation}
		\psi(\mathbf{x}_{ijk},t) = \sum_{\alpha\beta\gamma\in\mathbf{C}} U_{i\alpha}U_{j\beta}U_{k\gamma} c_{\alpha\beta\gamma}(t),
\end{equation}
where the (precomputed) transformation matrices are simply the 1D basis states evaluated at the quadrature grid points, i.e.
\begin{equation}
	U_{i\alpha} = \phi_\alpha(x_i).
\end{equation}
}

\item{Construct the integrand on the quadrature grid, by dividing by the weight function and premultiplying by the weights
\begin{equation}
	g(\mathbf{x}_{ijk}) = w_iw_jw_ke^{2|\mathbf{x}_{ijk}|^2}|\psi(\mathbf{x}_{ijk},t)|^2\psi(\mathbf{x}_{ijk},t).
\end{equation}
}

\item{Complete the integration by transforming back 
\begin{equation}
	G_{\alpha\beta\gamma} = \sum_{ijk} U_{i\alpha}^*U_{j\beta}^*U_{k\gamma}^* g(\mathbf{x}_{ijk}).
\end{equation}
}
\end{enumerate}
%%%%%%%%%%%%%%%%%%%%%%%%%%%%%%%%%%%%%%%%%%%%%%%%%%%%%%%%%%%%%%%%%%%%%%%%%%%%%%%%%%%%%%%%%%%%%%%%%%%%%%%%%%%%%%%%%%%%%%%%%%%%%%%%%%%%
\section{Laguerre quadratures for the rotating system}\label{sec:quad_app_laguerre}
In this section we describe the evaluation of the projected terms in the two-dimensional rotating-frame PGPE, including the evaluation of an elliptic perturbing-potential term.  This discussion closely follows that of Bradley \cite{Bradley_PhD,Bradley_elliptic}
%%%%%%%%%%%%%%%%%%%%%%%%%%%%%%%%%%%%%%%%%%%%%%%%%%%%%%%%%%%%%%%%%%%%%%%%%%%%%%%%%%%%%%%%
\subsection{Nonlinear term}\label{subsec:quad_app_lag_nl}
We begin with the dimensionless PGPE
\begin{equation}
	i\partial_t \psi = H_0\psi + \mathcal{P}\left\{U_\mathrm{2D}|\psi|^2 \psi\right\},
\end{equation}
where the dimensionless rotating-frame base Hamiltonian
\begin{equation}
	H_0 = -\frac{1}{2r}\frac{\partial}{\partial r}r\frac{\partial}{\partial r} - \frac{1}{2r^2}\frac{\partial^2}{\partial \theta^2} + \frac{r^2}{2} + i\Omega\frac{\partial}{\partial\theta},
\end{equation}
with rotation frequency $0\leq \Omega < 1$. The eigenstates of $H_0$ are 
\begin{equation}
	Y_{nl}(r,\theta) = \sqrt{\frac{n!}{\pi(n+|l|)!}} e^{il\theta} r^{|l|} e^{-r^2/2} L_n^{|l|}\left(r^2\right)\!,
\end{equation}
where $L_n^m(x)$ denotes the \emph{associated Laguerre polynomials}, defined by 
\begin{subequations}
\begin{eqnarray}
	L_0^{(\alpha)}(x)  &=& 1 \\
	L_1^{(\alpha)}(x)  &=& -x+\alpha+1 \\
	nL_n^{(\alpha)}(x) &=& (2n + \alpha - 1 - x)L_{n-1}^{(\alpha)}(x) - (n + \alpha - 1)L_{n-2}^{(\alpha)}(x).
\end{eqnarray}
\end{subequations}
The eigenvalues of the modes $Y_{nl}(r,\theta)$ are
\begin{equation}
	\omega_{nl} = 2n + |l| - \Omega l + 1.
\end{equation}
Introducing the quantity $\bar{N}$ such that $E_R = \hbar\omega_r(\bar{N}+1)$, the coherent region can be defined by
\begin{equation}
	\mathbf{C} = \left\{n,l : 2n + |l| -\Omega l \leq \bar{N}\right\}.
\end{equation}
The ranges of allowed quantum numbers $(n,l)$ are thus
\begin{eqnarray}
	0 &\leq& n \leq \left[\frac{\bar{N}}{2}\right]\!, \\
	-l_-(n) \equiv \left[\frac{\bar{N}-2n}{1+\Omega}\right] &\leq& l \leq \left[\frac{\bar{N}-2n}{1-\Omega}\right] \equiv l_+(n),
\end{eqnarray}
where $[\cdots]$ represents the \emph{floor function}.  The projected wave function is therefore
\begin{equation}
	\psi(r,\theta) = \sum_{n=0}^{[\bar{N}/2]} \sum_{l=-l_-(n)}^{l_+(n)} c_{nl}Y_{nl}(r,\theta).
\end{equation}
Our task in evaluating the nonlinear potential is thus that of evaluating
\begin{equation}
	F_{nl}[\psi(r,\theta)] = \int_0^{2\pi}\!d\theta\,\int_0^\infty\!dr\,r\,Y_{nl}^*(r,\theta)|\psi(r,\theta)|^2\psi(r,\theta),
\end{equation}
A change of variables $x=r^2$ gives
\begin{equation}
	F_{nl} = \frac{1}{2\pi}\int_0^{2\pi}\!d\theta\,e^{-il\theta} \int_0^\infty\!dx\,\left|\psi\bigl(\sqrt{x},\theta\bigr)\right|^2 \psi\bigl(\sqrt{x},\theta\bigr) \Phi_{nl}(x)\,\pi,
\end{equation}
where
\begin{equation}
	\Phi_{nl}(x) = \sqrt{\frac{n!}{\pi(n+|l|)!}} e^{-x/2}x^{|l|/2}L_n^{|l|}(x),
\end{equation}
and
\begin{equation}
	\psi(\sqrt{x},\theta) = \sum_{nl} c_{nl} e^{il\theta} \Phi_{nl}(x).
\end{equation}
The integral to compute is thus of form
\begin{equation}
	I = \frac{1}{2\pi}\int_0^{2\pi}\!d\theta\,e^{-il\theta}\int_0^\infty\!dx\,e^{-2x}Q(x,\theta),
\end{equation}
where $Q(x,\theta)$ is a polynomial in $x$ and $e^{i\theta}$ of a maximum order constrained by the cutoff.  As $\psi$ is a (Laurent) polynomial in $e^{i\theta}$ of maximum positive order $\overline{l}_+ = l_+(0)$ and maximum \emph{negative} order  $\overline{l}_- = l_-(0)$, the function $Q(\theta)$ will in general contain all powers $(e^{i\theta})^q$ where $-2(\overline{l}_- + \overline{l}_+) \leq q \leq 2(\overline{l}_+ + \overline{l}_-)$.  The angular integral can thus be evaluated \emph{exactly} as a discrete Fourier transform (DFT), discretising $\theta$ as $\theta_j = j\Delta = j2\pi/N_\theta$, for \mbox{$j=0,1,\cdots,N_\theta-1$}, where $N_\theta \equiv 2(\overline{l}_++\overline{l}_+)+1$, to give
\begin{equation}
	\int_0^{2\pi}\!d\theta\,e^{-il\theta} Q(x,\theta) = \frac{2\pi}{N_\theta} \sum_{j=0}^{N_\theta-1} e^{-i2\pi jl/N_\theta}Q(x,\theta_j).
\end{equation}
To perform the radial integral, we note that the radial part of $\psi$ is of the form $e^{-x/2}P(x)$, where $P(x)$ is a polynomial of at most order $\overline{l}_+/2$, and the radial integrand is thus of maximum order $2\overline{l}_+=2[\bar{N}/(1-\Omega)]$.  The radial integral is therefore of form
\begin{equation}
	I_r = \int_0^\infty\!dx\,e^{-2x}Q(x,\theta),
\end{equation}
where $Q(x)$ is of order at most $2[\bar{N}/(1-\Omega)]$.  The integral $I_r$ is thus of the appropriate form for \emph{Gauss-Laguerre} quadrature, and can therefore be evaluated \emph{exactly} on a grid with $N_x\equiv[\bar{N}/(1-\Omega)]+1$ points $x_k$, which are the roots of the (simple) Laguerre polynomial $L_{N_x}(x)$ with corresponding weights $w_k$, which are tabulated in (e.g.) reference~\cite{Cohen73},  and we have
\begin{equation}
	I_r = \frac{1}{2}\sum_{k=1}^{N_x} w_k Q(x_k/2,\theta).
\end{equation}
Combining the two quadrature rules we find
\begin{eqnarray}
	F_{nl} &=& \sum_{k=1}^{N_x} \widetilde{w}_k \Phi_{nl}(x_k/2) \frac{1}{N_\theta}\sum_{j=0}^{N_\theta-1}e^{-i2\pi jl/N_\theta} \nonumber \\
		&&\times \left|\psi\bigl(\sqrt{x_k/2},\theta_j\bigr)\right|^2\psi\bigl(\sqrt{x_k/2},\theta_j\bigr),
\end{eqnarray}
where we have grouped the weights as $\widetilde{w}_k=\pi w_k e^{x_k/2}$.  Once the radial parts of the basis modes are precomputed at the quadrature abscissas
\begin{equation}
	P_{kn}^{|l|} = \Phi_{n|l|}(x_k/2), 
\end{equation}
we have the following procedure:
\begin{enumerate}
	\item{Perform the radial part of the transformation from the coefficient representation $\{c_{nl}\}$ to the spatial representation
\begin{equation}
	\chi_{kl} = \sum_{n=0}^{[\bar{N}/2]} P_{kn}^{|l|} c_{nl}.
\end{equation}
}

\item{Construct the position-space field at the quadrature points $\Psi_{kj}$ using the DFT, after zero padding $\chi_{kl}$ to length $N_\theta$ in the $l$ index, $\widetilde{\chi}_{kl}\equiv\mathrm{pad}_l(\chi_{kl},N_\theta)$,
\begin{equation}
	\Psi_{kj} = \psi\bigl(\sqrt{x_k/2},\theta_j\bigr) = \sum_{l=0}^{N_\theta-1}\widetilde{\chi}_{kl}e^{i2\pi jl/N_\theta}.
\end{equation}
}

\item{Form the integrand
\begin{equation}
	\Xi_{kj} = |\Psi_{kj}|^2\Psi_{kj}.
\end{equation}
}

\item{Perform the azimuthal part of the inverse transform by DFT
\begin{equation}
	\Theta_{kl} = \frac{1}{N_\theta}\sum_{j=0}^{N_\theta-1}\Xi_{kj}e^{-i2\pi jl/N_\theta}.	
\end{equation}
}

\item{Perform the radial part of the inverse transform by Gauss-Laguerre quadrature
\begin{equation}
	F_{nl} = \sum_{k=1}^{N_x} \widetilde{w}_k\Theta_{kl}P_{kn}^{|l|}.
\end{equation}
}
\end{enumerate}
%%%%%%%%%%%%%%%%%%%%%%%%%%%%%%%%%%%%%%%%%%%%%%%%%%%%%%%%%%%%%%%%%%%%%%%%%%%%%%%%%%%%%%%%
\subsection{Elliptic trap anisotropy}
The general form of the elliptic potential anisotropy we need to compute (chapters~\ref{chap:arrest} and ~\ref{chap:stirring}) is
\begin{equation}
	V_\epsilon(r,\theta) = -\frac{\epsilon}{2}r^2\cos(2\theta).
\end{equation}
The corresponding term in the PGPE is 
\begin{equation}
	H_{nl} = -\frac{\epsilon}{2} \int_0^{2\pi}\!d\theta\,\int_0^\infty\!dr\,r^3\,Y_{nl}^*(r,\theta) |\psi(r,\theta)|^2\psi(r,\theta);
\end{equation}
making the change of variables $x=r^2$, we have
\begin{eqnarray}
	H_{nl} &=& -\frac{\epsilon}{4\pi} \int_0^{2\pi}\!d\theta\,e^{-il\theta} \cos(2\theta) \nonumber \\
	&&\quad\times \int_0^\infty\!dx\,x\,\Phi_{nl}(x)\psi\bigl(\sqrt{x},\theta\bigr)\pi.
\end{eqnarray}
The integral to evaluate is therefore of the form
\begin{equation}
	H_{nl} = \frac{1}{4\pi} \int_0^{2\pi}\!d\theta\,e^{-il\theta} \cos(2\theta) \int_0^\infty\!dx\,e^{-x} Q(x,\theta),
\end{equation}
where $Q(x,\theta)$ is a polynomial in $x$ and $e^{i\theta}$ of order determined by the cutoff. Due to the constraint on the powers of $e^{i\theta}$ that appear in $\psi(\sqrt{x},\theta)$, the function $Q(x,\theta)$ will in general contain all powers $(e^{i\theta})^q$ where $-\bar{l}_-\leq q\leq \bar{l}_+$.  As a result the total radial integrand will in general contain all powers $-(\bar{l}_-+\bar{l}_++2)\leq q \leq (\bar{l}_-+\bar{l}_++2)$, and the integral can thus be carried out on an azimuthal grid of $N_\theta^{(2)} = \bar{l}_-+\bar{l}_++3$ points, discretising $\theta$ as $\theta_j^{(2)} = j\Delta\theta^{(2)} = j2\pi/N_\theta^{(2)}$, $0\leq j \leq N_\theta^{(2)}-1$.  The azimuthal integral can thus be evaluated as the DFT
\begin{equation}
	\int_0^{2\pi}\!d\theta\,e^{-il\theta} \cos(2\theta) Q(x,\theta) = \frac{2\pi}{N_\theta^{(2)}} \sum_{j=0}^{N_\theta^{(2)}-1} e^{-i2\pi jl/N_\theta^{(2)}} \cos\bigl(2\theta_j^{(2)}\bigr) Q\bigl(x,\theta_j^{(2)}\bigr).  
\end{equation}
To perform the radial integral, we note that the radial part is of the form $e^{-x} Q(x,\theta)$, where $Q$ is of maximum order $\bar{l}_+$ in $x$.  The radial integral can therefore be computed exactly using a Gauss-Laguerre quadrature rule of order $N_x^{(2)}\equiv [\bar{l}_+/2]+1$.  The radial integral is thus
\begin{equation}
	\int_0^\infty\!dx\,e^{-x}Q(x,\theta) = \sum_{k=1}^{N_x^{(2)}} w_k^{(2)} Q\bigl(x_k^{(2)},\theta\bigr),
\end{equation}
where the $x_k^{(2)}$ are the roots of $L_{N_x^{(2)}}(x)$, and $w_k^{(2)}$ are the corresponding weights \cite{Cohen73}.  Combining the two quadrature rules we thus find
\begin{eqnarray}
	H_{nl} &=& -\frac{\epsilon}{2} \sum_{k=1}^{N_x^{(2)}} \widetilde{w}_k^{(2)} \Phi_{nl}\bigl(x_k^{(2)}\bigr) \frac{1}{N_\theta^{(2)}} \sum_{j=0}^{N_\theta^{(2)}-1} e^{-i2\pi jl/N_\theta^{(2)}} \nonumber \\
&&\quad\times \cos\bigl(4\pi jl/N_\theta^{(2)}\bigr)\psi\bigl(\sqrt{x_k^{(2)}},\theta_j^{(2)}\bigr),
\end{eqnarray}
where we have grouped the weights as $\widetilde{w}_k^{(2)} = \pi w_k^{(2)} e^{x_k^{(2)}/2}$.  Once the modes are precomputed at the quadrature abscissas
\begin{equation}
	P_{kn}^{|l|} = \Phi_{n|l|}\bigl(x_k^{(2)}\bigr),
\end{equation}
we have the following procedure:
\begin{enumerate}

\item{Perform the radial part of the transformation
\begin{equation}
	\chi_{kl} = \sum_{n=0}^{[\bar{N}/2]}P_{kn}^{|l|} c_{nl}.
\end{equation}
}

\item{Construct the position-space field at the quadrature points, after zero padding $\chi_{kl}$ to length $N_\theta^{(2)}$ in the $l$ index, $\widetilde{\chi}_{kl}\equiv\mathrm{pad}_l(\chi_{kl},N_\theta^{(2)})$,
\begin{equation}
	\Psi_{kj} = \sum_{l=0}^{N_\theta^{(2)}-1} e^{i2\pi jl/N_\theta^{(2)}}\widetilde{\chi}_{kl}.
\end{equation}
}

\item{Apply the perturbing potential to form the integrand
\begin{equation}
	\Gamma_{kj} = \Psi_{kj}x_k^{(2)}\cos\bigl(2\theta_j^{(2)}\bigr).	
\end{equation}
}

\item{Perform the azimuthal part of the inverse transform by DFT
\begin{equation}
	\Theta_{kl} = \frac{2}{N_\theta^{(2)}} \sum_{j=0}^{N_\theta^{(2)}-1} \Gamma_{kj} e^{-i2\pi jl/N_\theta^{(2)}}.
\end{equation}
}

\item{Perform the radial part of the inverse transform to evaluate the Gauss-Laguerre quadrature
\begin{equation}
	H_{nl} = -\frac{\epsilon}{2} \sum_{k=1}^{\; N_x^{(2)}} \widetilde{w}_k^{(2)}\Theta_{kl} P_{kn}^{|l|}.
\end{equation}
}

\end{enumerate}
The orders of quadrature rules used here are significantly smaller than those used for the evaluation of the nonlinear term, and consequently the computational load increase due to the inclusion of the trap anisotropy over that of the base method is of order $25-50\%$, depending on the rotation rate and cutoff height (see section~\ref{sec:numerics_performance}).
%%%%%%%%%%%%%%%%%%%%%%%%%%%%%%%%%%%%%%%%%%%%%%%%%%%%%%%%%%%%%%%%%%%%%%%%%%%%%%%%%%%%%%%%%%%%%%%%%%%%%%%%%%%%%%%%%%%%%%%%%%%%%%%%%%%%
%%%%%%%%%%%%%%%%%%%%%%%%%%%%%%%%%%%%%%%%%%%%%%%%%%%%%%%%%%%%%%%%%%%%%%%%%%%%%%%%%%%%%%%%%%%%%%%%%%%%%%%%%%%%%%%%%%%%%%%%%%%%%%%%%%%%

\chapter{Semiclassical fitting function}
\label{app:fitting_function}
%%%%%%%%%%%%%%%%%%%%%%%%%%%%%%%%%%%%%%%%%%%%%%%%%%%%%%%%%%%%%%%%%%%%%%%%%%%%%%%%%%%%%%%%%%%%%%%%%%%%%%%%%%%%%%%%%%%%%%%%%%%%%%%%%%%%
%%%%%%%%%%%%%%%%%%%%%%%%%%%%%%%%%%%%%%%%%%%%%%%%%%%%%%%%%%%%%%%%%%%%%%%%%%%%%%%%%%%%%%%%%%%%%%%%%%%%%%%%%%%%%%%%%%%%%%%%%%%%%%%%%%%%
In order to estimate the thermodynamic parameters of the classical field, we assume an approximate semiclassical description for the thermal atoms.  The description consists of two elements: a classical-field distribution in the semiclassical phase-space coordinates ($\mathbf{x}$ and $\mathbf{p}$), and the specification of an appropriate energy cutoff.  In order to construct this distribution, we must take into account two generally disparate rotations: the (quasi\mbox{-})equilibrium rotation of the thermal atoms (angular velocity $\Omega_\mathrm{th}$) and the rotation of the projector defining the cutoff (angular velocity $\Omega_\mathrm{p}$).
The classical-field distribution takes the general form (cf. equation~\reff{eq:cfield_equipartition})
\begin{equation}\label{eq:semiclassical_dist1}
	F_\mathrm{c}(\mathbf{x},\mathbf{p}) = \frac{1}{h^2}\frac{k_\mathrm{B}T}{\epsilon(\mathbf{x},\mathbf{p})-\mu},
\end{equation}
where the semiclassical Hartree-Fock energy of the thermal atoms (rotating at $\Omega_\mathrm{th}$) is  
\begin{equation}
	\epsilon(\mathbf{x},\mathbf{p}) = \frac{p^2}{2m} + \frac{m\omega_r^2r^2}{2} - \Omega_\mathrm{th}(xp_y-yp_x) + 2U_\mathrm{2D} n(r),
\end{equation}
(cf. \cite{Stringari99}) and where $n(r)$ is the total density of the atomic field, which we will take to be circularly symmetric.  The application of the projector $\mathcal{P}$ defined in a rotating frame enforces a cutoff in the energy, which delimits the accessible region of phase space occupied according to equation~(\ref{eq:semiclassical_dist1}).  This region is defined by 
\begin{equation}\label{eq:semiclassical_cutoff_fundamental}
	\frac{p^2}{2m} + \frac{m\omega_r^2r^2}{2} - \Omega_\mathrm{p}(xp_y-yp_x) + 2U_\mathrm{2D}n(r) \leq E_R.
\end{equation}
The density of thermal atoms at any spatial position $\mathbf{x}$ is obtained by integrating equation~(\ref{eq:semiclassical_dist1})  over the momentum, i.e. $n(\mathbf{x}) = \int_\mathcal{D}\!d\mathbf{p}\,F_\mathrm{c}(\mathbf{x},\mathbf{p})$, where the domain $\mathcal{D}$ in momentum space is determined by equation~(\ref{eq:semiclassical_cutoff_fundamental}), and is itself a function of the position-space coordinate $\mathbf{x}$.  To perform this integration we make the transformation to the \emph{kinematic} momentum measured in the frame of the cloud $\mathbf{P}=\mathbf{p}-m\bm{\Omega}_\mathrm{th}\times\mathbf{x}$ \cite{Stringari99,Landau60}.  In this representation the integrand becomes circularly symmetric in $\mathbf{P}$, i.e., we now have $F_\mathrm{c}'(\mathbf{x},\mathbf{P})=(1/h^2)k_\mathrm{B}T/[\epsilon(\mathbf{x},\mathbf{P})-\mu]$, where 
\begin{equation}
	\epsilon(\mathbf{x},\mathbf{P}) = \frac{P^2}{2m} + \frac{m}{2}\left(\omega_r^2-\Omega_\mathrm{th}^2\right)r^2 + 2U_\mathrm{2D} n(r).
\end{equation}
In these momentum coordinates the domain $\mathcal{D}$ is circular, however its centre is in general displaced from the origin $\mathbf{P}=\mathbf{0}$.  After some simple algebra we find that the domain of integration is
\begin{eqnarray}
	\mathcal{D}(\mathbf{x}): &&\frac{1}{2m}\Bigl\{\bigl[P_x+m(\Omega_\mathrm{p}-\Omega_\mathrm{th})y\bigr]^2 + \bigl[P_y-m(\Omega_\mathrm{p}-\Omega_\mathrm{th})x\bigr]^2\Bigr\}  \nonumber \\
    &&\qquad\;\leq E_R - \frac{m}{2}m\left(\omega_r^2-\Omega_\mathrm{p}^2\right)r^2 -2U_\mathrm{2D} n(r).
\end{eqnarray}
As the integrand is circularly symmetric in $\mathbf{P}$, the result of the integral depends only on the radius $r$ of the chosen point in position space, as azimuthal rotations of the coordinate $\mathbf{x}$ simply induce rotations of $\mathcal{D}(\mathbf{x})$ about the origin of momentum space.  We therefore consider without loss of generality $(x,y)=(0,-r)$.  The domain $\mathcal{D}$ thus becomes a circle centred on $P_x=m(\Omega_\mathrm{p}-\Omega_\mathrm{th})r\equiv a$, of radius $b\equiv \sqrt{2m[E_R - (m/2)(\omega_r^2-\Omega_\mathrm{p}^2)r^2-2U_\mathrm{2D} n(r)]}$.  Depending on the value of $r$, we may have $a\leq b$ or $b>a$. We adopt polar coordinates $(P,\phi)$, and find for the angle $\Delta\phi$ subtended by $\mathcal{D}$ at a given radius $P$ 
\begin{equation}
	\Delta \phi(P) = \left\{
	\begin{array}{ll}
		0 &\qquad P<a-b\;\;(\mathrm{when}\;a>b)\\
        2\pi &\qquad P<b-a\;\;(\mathrm{when}\;a<b) \\
        2\cos^{-1}\Big(\frac{P^2-b^2+a^2}{2aP}\Big)  &\qquad |a-b|<P<a+b \\
		0 &\qquad P>a+b
    \end{array} \right. .
\end{equation}
Integrating in polar coordinates we obtain finally
\begin{eqnarray}
	n(r) &=& \frac{2\pi mk_\mathrm{B}T}{h^2} \int_0^\infty \frac{PdP\Delta\phi(P)}{\frac{P^2}{2m}+\frac{m\omega_r^2r^2}{2} + 2U_\mathrm{2D} n(r) - \mu}  \nonumber \\ 
	&=& \frac{1}{\lambda_\mathrm{dB}^2}\Bigl[I_1(r;\Omega_\mathrm{th},\Omega_\mathrm{p}) + I_2(r;\Omega_\mathrm{th},\Omega_\mathrm{p})\Bigr],
\end{eqnarray}
where
\begin{equation}
	I_1(r) = \Theta\Big(b - a\Big)\ln\left[\frac{\frac{(b-a)^2}{2m} + \frac{m}{2}\left(\omega_r^2-\Omega_\mathrm{th}^2\right)r^2+2U_\mathrm{2D} n(r) - \mu}{\frac{m}{2}\left(\omega_r^2-\Omega_\mathrm{th}^2\right)r^2+2U_\mathrm{2D} n(r)-\mu}\right],  
\end{equation}
and 
\begin{equation}
	I_2(r) = \frac{1}{\pi}\int_{|b-a|}^{a+b}\frac{PdP \cos^{-1}\Bigl(\frac{\frac{p^2}{2m}+\frac{m}{2}\left[\omega_r^2-\Omega_\mathrm{p}^2+\left(\Omega_\mathrm{p}-\Omega_\mathrm{th}\right)^2\right]r^2-E_R}{\left(\Omega_\mathrm{p}-\Omega_\mathrm{th}\right)rP}\Bigr)}{\frac{p^2}{2m}+\frac{m}{2}\left(\omega_r^2-\Omega_\mathrm{th}^2\right)r^2-\mu},
\end{equation}
and the de Broglie wavelength $\lambda_\mathrm{dB} =\sqrt{2\pi\hbar^2/mk_\mathrm{B}T}$.  The integral $I_2$ must be evaluated numerically.  In the limit $\Omega_\mathrm{th}\rightarrow\Omega_\mathrm{p}$ the integral $I_2$ vanishes, and we obtain the form
\begin{equation}
	n(r) = \frac{1}{\lambda_\mathrm{dB}^2} \ln\left[\frac{E_R-\mu}{(\omega_r^2-\Omega_\mathrm{th}^2)r^2+2U_\mathrm{2D} n(r) - \mu}\right],
\end{equation} 
employed in chapter~\ref{chap:stirring}.
%%%%%%%%%%%%%%%%%%%%%%%%%%%%%%%%%%%%%%%%%%%%%%%%%%%%%%%%%%%%%%%%%%%%%%%%%%%%%%%%%%%%%%%%%%%%%%%%%%%%%%%%%%%%%%%%%%%%%%%%%%%%%%%%%%%%
%%%%%%%%%%%%%%%%%%%%%%%%%%%%%%%%%%%%%%%%%%%%%%%%%%%%%%%%%%%%%%%%%%%%%%%%%%%%%%%%%%%%%%%%%%%%%%%%%%%%%%%%%%%%%%%%%%%%%%%%%%%%%%%%%%%%

%%%%%%%%%%%%%%%%%%%%%%%%%%%%%%%%%%%%%%%%%%%%%%%%%%%%%%%%%%%%%%%%%
\cleardoublepage

%%%%%%%%%%%%%%%%%%%%%%%%%%%%%%%%%%%%%%%%%%%%%%%%%%%%%%%%%%%%%%%%%
\fancyhead[RO]{\bfseries\nouppercase{\rightmark}}
\phantomsection
\addcontentsline{toc}{chapter}{References}

\newcommand{\Yu}{Yu}

\bibliographystyle{phdthesis}

\bibliography{thesis}
%%%%%%%%%%%%%%%%%%%%%%%%%%%%%%%%%%%%%%%%%%%%%%%%%%%%%%%%%%%%%%%%%
\end{document}